\documentclass[a4paper,12pt]{article}
\usepackage[bottom=25.00mm,top=25.00mm,inner=25.00mm, outer=25.00mm]{geometry}

\usepackage[utf8]{inputenc}
\usepackage[T1]{fontenc}
\usepackage[english]{babel}
\usepackage{amsfonts,amsbsy,bm,euscript,mathrsfs}
\usepackage{amssymb,stmaryrd,faktor}
\usepackage[tbtags]{amsmath}
\usepackage{bbm}
\usepackage{graphicx}
\usepackage{caption}
\usepackage[title,titletoc]{appendix}
\usepackage[bookmarks=true,colorlinks=true,linkcolor=blue,citecolor=blue,urlcolor=blue,bookmarksnumbered]{hyperref}

\usepackage{soul}
\usepackage[dvipsnames]{xcolor}

\usepackage{dsfont}
\usepackage{collref}
\usepackage{lmodern}
\usepackage{mathrsfs}
\usepackage{mathtools}	
\usepackage{cancel}
\usepackage{cases}

\usepackage{bbm}
\usepackage{braket}
\usepackage{slashed}

\usepackage{booktabs}
\usepackage{subfig}
\usepackage{tikz}
\usetikzlibrary{plotmarks,calc,decorations,decorations.markings,decorations.pathmorphing}
\usetikzlibrary{shapes,arrows.meta,automata,positioning}
\usepackage{tikz-3dplot}
\usepackage{circuitikz}

 \usepackage[usenames,dvipsnames]{pstricks}
 \usepackage{pstricks-add}
 \usepackage{epsfig}
 \usepackage{pst-grad} 
 \usepackage{pst-plot} 
 \usepackage[space]{grffile} 
 \usepackage{etoolbox} 
 \makeatletter 
 \patchcmd\Gread@eps{\@inputcheck#1 }{\@inputcheck"#1"\relax}{}{}
 \makeatother

\usepackage{blkarray}
\usepackage{comment}

\numberwithin{equation}{section}

\makeatletter
\renewcommand\section{\@startsection {section}{1}{\z@}
{-3.5ex \@plus -1ex \@minus -.2ex}
{2.3ex \@plus.2ex}
{\normalfont\Large\bfseries}}
\renewcommand\subsection{\@startsection{subsection}{2}{\z@}
{-3.25ex\@plus -1ex \@minus -.2ex}
{1.5ex \@plus.2ex}
{\normalfont\large\bfseries}}
\makeatother

\newcommand{\al}[1]{\mathfrak {#1}}

\newcommand{\tx}[1]{\text {#1}}
\newcommand{\cl}[1]{\mathcal {#1}}
\newcommand{\bb}[1]{\mathbb {#1}}
\newcommand{\fk}[1]{\mathfrak {#1}}
\newcommand{\mb}[1]{\mathbf {#1}}
\newcommand{\mrm}[1]{\mathrm {#1}}

\def\ads{{$ AdS_{5} \times S^{5} $}}
\def\adscp{{$ AdS_{4} \times \bb{CP}^{3} $}}
\def\adsm{{$ AdS_{3} \times S^{3} \times \cl{M}^{4} $}}
\def\adso{{$ AdS_{3} \times S^{3} \times T^{4} $}}
\def\adst{{$ AdS_{3} \times S^{3} \times S^{3} \times S^{1} $}}
\def\adsT{{$ AdS_{2} \times S^{2} \times T^{6} $}}
\def\Hal{{$ \al{su}(2) \oplus \al{su}(2) $}}
\newcommand{\Rx}{$ R $-matrix}
\newcommand{\Rxs}{$ R $ matrices}
\newcommand{\Sx}{$ S $-matrix}

\newcommand{\tsf}[2]{\frac{#1}{#2}}
\newcommand{\hf}{\tsf{1}{2}}

\newcommand{\cdd}{\mathbf{c}^{\dagger}}
\def\id{{\mathds{1}}}
\DeclareMathOperator{\dd}{d\!}
\newcommand{\Rn}[1]{\MakeUppercase{\romannumeral #1}}
\newcommand{\sn}[1]{\mathrm{sn}(#1)}
\newcommand{\cn}[1]{\mathrm{cn}(#1)}
\newcommand{\dn}[1]{\mathrm{dn}(#1)}

\usepackage{fixmath}

\newcommand{\alg}[1]{\mathfrak{#1}}

\newcommand{\lH}{\mathcal{H}}

\newcommand{\co}{\mathbf{c}}
\newcommand{\no}{\mathbf{n}}

\newcommand{\cdo}{\mathbf{c}^\dagger}

\title{Automorphic Symmetries, String integrable structures and Deformations}
\author{Anton Pribytok\footnote{antons.pribitoks@physik.hu-berlin.de, apribytok@maths.tcd.ie}}
\date{}

\begin{document}

\clearpage\maketitle
\thispagestyle{empty}

\begin{center}
	\begingroup\itshape
	School of Mathematics
	\& Hamilton Mathematics Institute\\
	Trinity College Dublin\\
	Dublin, Ireland
	\par\endgroup
\end{center}

\vspace{1cm}
\begin{abstract}
	We address the novel structures arising in quantum and string integrable theories, as well as construct methods to obtain them and provide further analysis. Specifically, we implement the automorphic symmetries on periodic lattice systems for obtaining integrable hierarchies, whose commutativity along with integrable transformations induces a generating structure of integrable classes. This prescription is first applied to 2-dim and 4-dim setups, where we find the new $ \al{sl}_{2} $ sector, {\Hal} with superconductive modes, Generalised Hubbard type classes and more. The corresponding 2- and 4-dim {\Rxs} are resolved through perturbation theory, that allows to recover an exact result. We then construct a boost recursion that allows to address the systems, whose $ R $-/$ S $-matrices exhibit arbitrary spectral dependence, that also is an apparent property of the scattering operators in $ AdS $ integrability. It is then possible to implement the last for Hamiltonian Ans\"atze in $ D = 2,3,4 $, which leads to new models in all dimensions. We also provide a method based on a coupled differential system that allows to resolve for {\Rxs} exactly. Importantly, one can isolate a special class of models of non-difference form in 2-dim case (6vB/8vB), which provides a new structure consistently arising in $ AdS_{3} $ and $ AdS_{2} $ string backgrounds. We prove that these classes can be represented as deformations of the $ AdS_{\{ 2,3 \}} $ models. We also work out that the latter satisfy free fermion constraint, braiding unitarity, crossing and exhibit deformed algebraic structure that shares certain properties with {\adsm} and {\adsT} models. The embedding and mappings of known $ AdS_{\{ 2,3 \}} $ models to 6vB/8vB deformations are demonstrated, along with a discussion on the associated candidates of sigma models.
\end{abstract}



	

%
%
%
%
%
%
%
%
%

\newpage

\newpage

\section{Summary}

The structure of present dissertation is given as follows, we start by introducing the necessary objects, apparatus and properties that will underlie the essence of the questions we raise, \textit{e.g.} finding the novel integrable structures in various dimensions, the obstacles arising in the classification program, development of a novel method for the latter, identification of the properties, relations to other integrable sectors and many more \ref{Introduction}.

First of all, we develop a method based on automorphic boost symmetries, which allows to investigate the existence of a new integrable sector in two-dimensional lattice systems. To achieve that, we consider a generic Hamiltonian Ansatz for a spin chain with 2-dim local space $ \bb{C}^{2} $. We implement the boost automorphism for closed spin chains and generate the local charges, whose commutativity guarantees integrability. The solution space that follows from resolving a polynomial system of the commutativity constraint requires a separate treatment for inducing a structure on it. This can be achieved by the integrable identification transformations, which isolate associated solution generators, where every such generator defines a class. Then the corresponding {\Rxs} are worked out by a novel perturbative bottom-up approach along with all quantum consistency checks. In this base setting, not only a full known model classification is achieved, but also novel integrable structures are found in the $ \al{sl}_{2} $ sector, whose underlying quantum algebra still remains to be resolved. In addition, the applied technique provided completion for the $ \bb{C}^{1|1} $ sector and 	established graded mapping, Sec. \ref{p1}.

The four-dimensional analysis followed from the boost method discovered in the first part. The Hubbard symmetry algebra has been taken as a base for constructing Hamiltonian Ans\"atze in 4-dim. By appropriate extensions of the approach, 8 new integrable classes have been found, the five of which followed from distinct representations of {\Hal} and the rest three came from the so called Generalised Hubbard Models. The GHM sector was obtained from the separate construction, that contained Hubbard kinetic part, generilised hopping, flipping and potential terms up to quadruple order in number operators. A separate method based on $ \cl{H} $-$ R $ differential system was proposed and proven to find all the underlying {\Rxs} satisfying all consistency conditions. The discussion on resolution of the new models through ABA and QSC is provided, Sec. \ref{D_Hubbard}.

At the next stage, a generalisation to higher dimensions and arbitrary spectral dependence is considered, Sec. \ref{NA_S_SD}. Namely, one can address the charge $ \cl{Q} $ automorphic recursion for the systems, whose $ R $-/$ S $-matrix exhibits non-difference (non-additive) spectral dependence, which is characteristic for an integrable model classes arising in $ AdS_{n} $ backgrounds. In what we derive the boost generator that respects arbitrary spectral dependence and seek for the new structures in $ D = 2,3,4 $. In two dimensions four new models found that are relevant for the $ AdS_{\{ 2,3 \}} $ integrable string backgrounds. The 3-dim sector contains nontrivial pseudo-difference models exceptionally in the sub-15-vertex classes. In dimension four, we find a perturbative support for a new structure to exist in the q-deformed sector of {\ads}. It have been also noted that some of the difference form models do not induce the non-difference form analogues. A comparison analysis of formalism with both spectral dependence types is performed. The $ R $-sector is resolved by constructing an approach of coupled Sutherland type system, that leads to non-linear differential system, solution of which establishes exact {\Rxs}.

Consequently the 2-dim sector of arbitrary spectral form (or pseudo-difference) is addressed for its properties and algebraic structure. Specifically, it is 6- and 8-vertex models, that are identified as B class that are relevant for $ AdS_{3} $ and $ AdS_{2} $ integrable deformations, Sec. \ref{FF_Section}, \ref{p5_Sec_AdS_2_3_Integrable_Deformations}. In the first part we prove that these models satisfy the free-fermion condition, braiding unitarity and crossing symmetry. It is shown that free fermions provide useful constraints for modified ABA of $ AdS $ models with background fluxes. In the last part we prove an algebra of 6vB and 8vB arise in deformed form w.r.t. {\adsm} and {\adsT} models. The 6vB/8vB  were established as $ AdS_{\{ 2,3 \}} $ deformations, in this respect, the corresponding mappings and embeddings of known $ AdS $ models along with their deformations have been resolved.

\newpage

\vspace{-3cm}
\thispagestyle{empty}
\vspace{-1cm}

{\hypersetup{linkcolor=black}
\tableofcontents
}

\setcounter{footnote}{0}

\newpage

\section{Introduction}\label{Introduction}

It is known that many interacting field theoretic setups exhibit number important physical properties, where at the same time computation of their specific observables involves resolution of infinities a.k.a. different regularisation schemes and additional technical tools. One of the ways to regulate classes of field theoretic models is the spatial discretisation. This procedure allows to bring field (continuous) models to finite-volume models with finite degrees of freedom (dof). Another important characteristic of a class of field theory models is \textit{integrability}, which initially stemmed from the classical notion of being (infinite-dimensional) Hamiltonian integrable system \cite{zakharov1971korteweg}, and later quantum which exhibited (in)finite conserved quantum symmetries \cite{Korepin_1997_QISM_CF}, that in turn, define complete \textit{exact solvability} of such models through a variety of techniques. Combining both notions of integrability and space discreteness we obtain a very rich class of integrable models and techniques deriving from them.

\subsection{Automorphism generators and charges}\label{I_Master_Symmetries_Discretisation}
\begin{figure}[!h]
	\centering
	\includegraphics[width=0.40\textwidth]{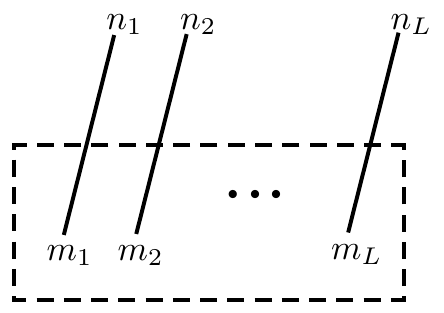}
	\caption{Auxiliary null-space tracing of $ S $ with particle species $ m_{k} $ and $ n_{k} $}
	\label{t-trace}
\end{figure}

It is well known there exists a variety of quantum field and statistical theories that respect factorised scattering in one or the other form. The fundamental characteristic of the last is an infinite tower of conservation laws (quantities), which makes them exactly solvable. It can be also established that if an $ R $- or $ S $-matrix accordingly satisfy YB constraint \eqref{DiffYBE}
\begin{equation}\label{I_YBE_S}
	S_{12}S_{13}S_{23} = S_{23}S_{13}S_{12} \qquad S_{ij} \in \text{End}\left( \bb{V} \otimes \bb{V} \otimes \bb{V} \right)
\end{equation}
then the corresponding transfer matrix $ t(x) $
\begin{equation}\label{I_TransferMatrix}
	t(u) = \tx{Tr}_{0} \prod_{i=1}^{L} S_{0i} \qquad t(u)t(v) = \tx{Tr}_{a}\tx{Tr}_{b} \, \prod_{i=1}^{L} \left[ S_{ai}(u)S_{bi}(v) \right], 
\end{equation}
As indicated in (\ref{I_TransferMatrix}), the trace is taken over the product of $ S $-operators ($ S $-monodromy), which describe scattering of particles of various types Fig. \ref{t-trace}. The commutativity of $ t $ implies its diagonalisation through known integrability methods \cite{Sklyanin_1979quantum,Sklyanin_1982quantum}. 
In general for the quantum transfer matrix $ t $ to exist, there must exist a conserved quantum monodromy based on asymptotic particle states, which satisfies $ RTT $-factorisation and the discrete analogues of $ \cl{P}\cl{T} $ can be realised in the underlying quantum theory. Given prescription along with unitarity relations proposed in \cite{Zamolodchikov:1978xm,Zamolodchikov_1979z,Baxter_1980hardhexagons,Stroganov_1979new,Zamolodchikov:1990bu} is taken to be conventional, since it also permits to construct the corresponding conserved charges by perturbing \eqref{I_TransferMatrix} and obtain induced commutativity \cite{Sklyanin_1979quantum, Faddeev:1996iy}.

Remarkably, from YB relations one can establish a group of transformations $ \cl{G}_{\bar{L}} $ on the hierarchy of conserved charges, which can be proven to constitute the discrete analogue of the Lorentz symmetry. Indeed this group of continuous transformations analogous to the Lorentzian one has direct implications for integrable systems on the lattice. In fact, the Lorentz boosts identify the set of transformations on the pairwise commuting integrals of motion (conserved charges, including Hamiltonian). It is also possible to consider a continuous limit of this lattice Lorentz group, in which case it can decomposed into Lorentz group and a group of transformations of higher integrals of motion \cite{Tetelman}. Specifically, one can also consider parametric series of the given lattice group, where infinite tower of boost generators is mapped to an infinite set of motion integrals.

We can also note this from the continuous setting. If we consider the Lorentz transformation of a certain generating and conserved observable, like a current $ \cl{J}^{\mu}(x) $ of the underlying quantum theory
\begin{equation}\label{key}
	\cl{U} \cl{J}^{\mu}(x) \cl{U}^{-1} = \Lambda_{\nu}^{\mu} \, \cl{J}^{\nu}(\Lambda x)
\end{equation}
where $ \cl{U}\left[ \Lambda(u) \right] $ and $ \Lambda_{\mu}^{\nu} $ is Lorentz operator with fugacity u. In addition, the Lorentzian limit defines the boost $ \cl{B} $ and its commutation properties provide an (integrable) algebraic structure
\begin{equation}\label{I_Boost_Current_Commutator}
	\cl{B} = \dd_{\, u} \Lambda(u 	) \big|_{u = 0} \qquad \left[ \cl{B}, \cl{J}^{\mu}_{a}(x) \right] = \epsilon_{\alpha\beta} \, x^{\alpha} \partial^{\beta} \cl{J}^{\mu}_{a}(x) + \epsilon^{\mu \beta} \cl{J}_{a, \, \beta}^{\mu}
\end{equation}
where $ a $ indicates symmetry indices. By looking at the commutation \eqref{I_Boost_Current_Commutator}, we can get commutativity with the charge \cite{Loebbert:2016cdm}
\begin{equation}\label{key}
	\left[ \cl{B}, Q_{a} \right] = 0 \qquad \tx{with}\quad \cl{J}_{a}(\pm \infty) \equiv 0, \quad Q_{a} = \int_{-\infty}^{+\infty} \dd \, x \cl{J}_{a}^{0}(x) 
\end{equation}
In fact one could achieve an equivalent result with higher currents (accounting for quantum corrections) by looking at commutativity with the stress-energy tensor $ T^{\mu\nu} $ of the system \cite{Curtright_1993}. It can be immediately seen that the above mentioned decomposition and higher symmetries are reflected not only at the level of the Poincar\'e algebra, but would also admit infinite-dimensional extension of internal symmetries, where the boost manifests its automorphic nature, \textit{i.e.}
\begin{equation}\label{key}
	\tx{Poincar\'e:} \qquad \left[ \cl{P}_{-}, \cl{P}_{+} \right] = 0, \quad \left[ \cl{B}, \cl{P}_{\pm} \right] = \pm \cl{P}_{\pm}
\end{equation}
\begin{equation}\label{key}
	\tx{Yangian supplement } \cl{Y} \left[ \fk{g} \right] \tx{: } \quad
	\begin{cases}
		\quad \left[ \cl{P}_{\pm}, Q_{a} \right] = 0 \\
		\quad \left[ \cl{B}, Q_{a} \right] = 0 \\
	\end{cases}
	\quad
	\begin{array}{l}
		\quad \left[ \cl{P}_{\pm}, \, ^{(1)}Q_{a} \right] = 0 \\
		\quad \left[ \cl{B}, \, ^{(1)}Q_{a} \right] = - \fk{C} \dfrac{\hbar}{4 \pi i} Q_{a} \\
	\end{array}
\end{equation}
where $ \cl{P}_{\pm} $ are lightcone translations, $ \fk{C} $ is the Casimir of the enveloping algebra and by $ ^{(n)}Q_{a} $ we shall indicate level-$ n $ Yangian generators (internal generating symmetry). We can notice that the boost $ \cl{B} $ generates nontrivial internal symmetry extension (\ref{B_HDensity})-(\ref{I_BAS}).

As we shall see further, this formalism can be translated into discrete case upon appropriate structure resolutions and conditions. However it is also important to note that boost construction \eqref{I_Boost_Current_Commutator} is characteristic for continuous systems and is distinctive from the boost automorphism for integrable systems in discretised space. Moreover for a number of integrable models it is not clear if there exist mapping limits (\textit{e.g.} irrelevantly deformed 2-dim CFTs \cite{Zamolodchikov:2004ce,Smirnov:2016lqw,Cavaglia:2016oda,Frolov:2019nrr,Jiang:2019epa}). Particularly, it would allow to identify novel integrable structures and provide full control over symmetries. Let us now analyse in greater detail how to implement the automorphic apparatus and what are implications of the above conserved quantities for lattice systems.

\subsection{Quantum system on the lattice and Master Symmetries}

Despite the fact that field theories exhibit very distinct structure, properties and interpretation, it is important to acquire maps from continuous symmetries to their discrete analogues in $ 1+1 $-dim.  Indeed, by construction the symmetry algebra above along with $ \cl{Y}\left[ \fk{g} \right] $ internal extension, possesses compatible discretised structure in the Hopf realisation (algebraic actions, quotients) \cite{Drinfeld:1985rx,Drinfeld1986quantum,Drinfeld_1986DAHAY}. Moreover all operations can be straightforwardly replaced by their discrete counterparts (space stratification, locality, summation etc). Importantly as we shall show below, these discontinuous objects must also close to a symmetry algebra.

Such integrable systems are defined on the discretised spaces or \textit{lattices}, hence the name of the class - \textit{lattice integrable models}. In particular, by lattice models we shall mean Integrable Spin Chains (ISC), which are quantum spin models of diverse group-theoretic and physical properties. The canonical definition of the spin chain includes the collection of lattice sites assembled into spin chain of length $ L $, with local quantum space $ \mathfrak{h}_{i} $ assigned to each site, hence the Hilbert space $ \mathfrak{H}_{L} $ of the system
\begin{equation}\label{key}
	\mathfrak{H}_{L} = \bigotimes_{i=1}^{L} \mathfrak{h}_{i}
\end{equation}
where a variety of spin chain systems can include different dimensionality and representations.
\begin{figure}[!h]
	\centering
	\includegraphics[width=0.5\textwidth]{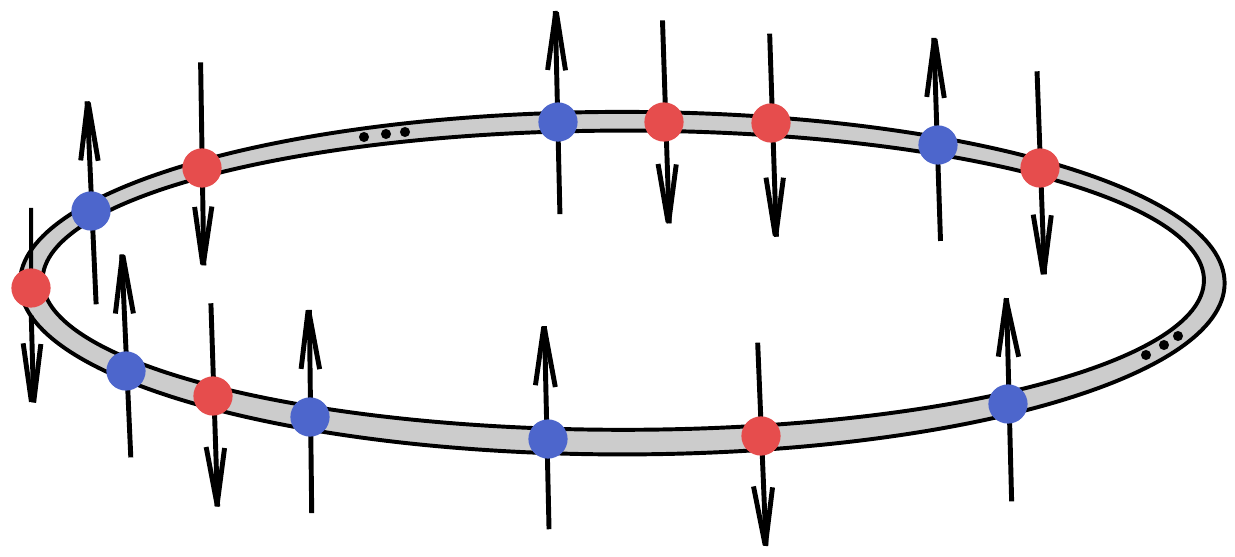}
	\caption{Heisenberg spin chain (magnet class) with Hilbert space $ \mathfrak{H} $ formed by local quantum space $ \mathfrak{h} $ (spins) at site $ i $.}
\end{figure}

Specifically at each local space $ \fk{h}_{i} $ there is some spin orientation $ \fk{s}_{i} $, which forms a state in the system
\begin{equation}\label{key}
	\ket{\psi_{s}} = \ket{\cdots \fk{s}_{i} \, \fk{s}_{i+1} \, \fk{s}_{i+2} \cdots} \quad\qquad \ket{\psi_{s}} \in \cl{H}, \,\, \fk{s}_{k} \in \fk{h}_{k}
\end{equation}
where $ \fk{s}_{k} $ corresponds to a local basis vector assigned to $ \fk{h}_{k} $. It can be clearly illustrated by the example of a spin chain with two-dimensional local quantum space $ \mathfrak{h}_{i} = \mathbb{C}^{2} $. Specifically, one can define generic spin-$ \dfrac{1}{2} $ Heisenberg Hamiltonian of the system in an external magnetic field
\begin{equation}\label{key}
	H_{XYZ}^{\varkappa} = -\dfrac{1}{2} \sum_{i=1}^{L} J_{x}\sigma_{i}^{x}\sigma_{i+1}^{x} + J_{y}\sigma_{i}^{y}\sigma_{i+1}^{y} + J_{z}\sigma_{i}^{z}\sigma_{i+1}^{z} + \varkappa \sigma_{i}^{z}
\end{equation}
where $ J_{i} $ are components of a coupling constant, Pauli matrices $ \sigma_{i} $ and external field $ \varkappa $. In fact, we shall be considering L-site closed spin chains with quantum spin-operators $ S_{i} $ that obey periodicity and algebra on it \ref{Apx_2-dim}
\begin{equation}\label{key}
	S_{L+1}^{\mu} \equiv S_{1}^{\mu} \qquad \left[ S_{j}^{\mu} , S_{k}^{\nu} \right] = i \delta_{jk} \varepsilon_{\mu \nu \rho} S_{j}^{\rho}
\end{equation}
where for spin-$ \dfrac{1}{2} $ case $ S_{j}^{\mu} $ are Pauli operators $ S_{j}^{\mu} = \dfrac{\sigma_{j}^{\mu}}{2} $. One can further proceed with the set of reduced models, which are obtained through identifications of the coupling of the so called XYZ-model from above. Namely, for $ J_{y}=J_{x} $ one obtains XXZ-model or the XXX-model for $ J_{x}=J_{y}=J_{z}=J $. XYZ, XXZ and XXX-model are characterised by Jacobi elliptic, trigonometric and rational function, hence the name of the functional sectors they define. It is possible to reduce even further by removing interaction along specific directions, for instance one can get XY-model for $ J_{z} = 0 $ or quantum Ising chain in the transversal field with $ J_{y} = J_{z} = 0 $.
\begin{figure}[!h]
	\centering
	\includegraphics[width=0.80\textwidth]{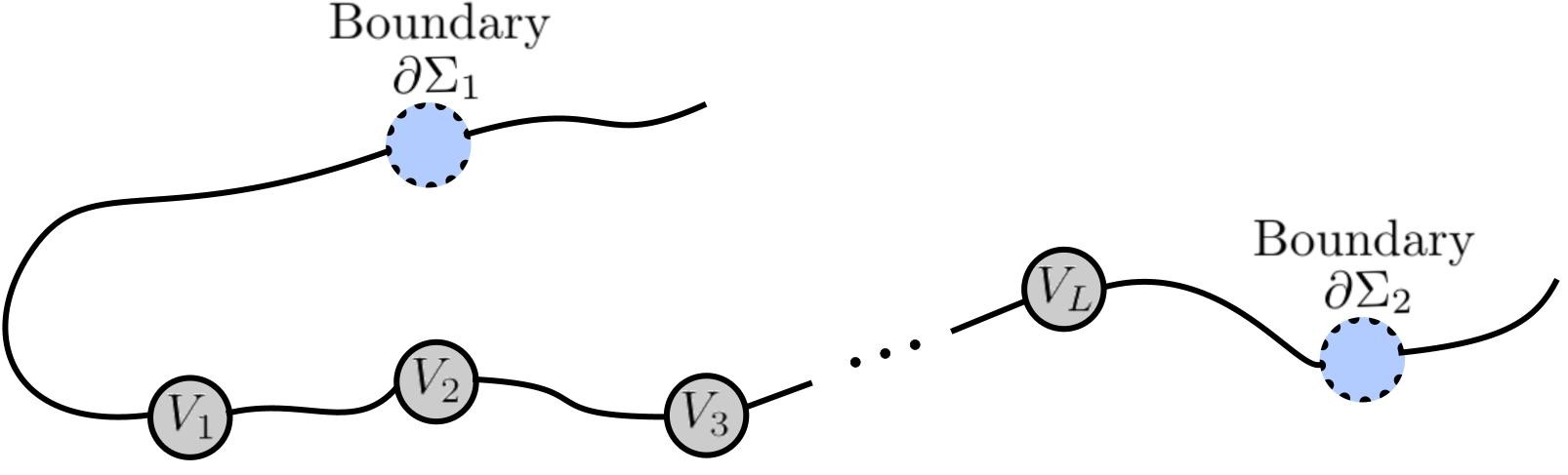}
	\caption{Distinct boundary conditions correspond to different types of integrable spin chains. Here $ \partial \mrm{\Sigma}_{i} $ constitute specific boundary and $ V_{i} \in \bb{C}^{n} $ is $ n $-dimensional local quantum space at site $ i $.}
	\label{I_ISC_BC}
\end{figure}

It is also important to distinguish between types of the spin chains depending on their boundary condition Fig. \ref{I_ISC_BC}. Since different boundary conditions apriori lead to a variety of integrable systems that significantly differ in the spectrum and properties. One can classify 5 main boundary types for the spin chain
\begin{itemize} 
	\item \textit{Infinite} spin chain: $ -\infty \leftarrow \partial \mrm{\Sigma}_{1} $, $ \partial \mrm{\Sigma}_{2} \rightarrow +\infty $
	\item \textit{Semi-infinite} spin chain: $ \partial \mrm{\Sigma}_{1} = V_{0} $, $ \partial \mrm{\Sigma}_{2} \rightarrow +\infty $
	\item \textit{Open} spin chain: $ \partial \mrm{\Sigma}_{1} = V_{0} $, $ \partial \mrm{\Sigma}_{2} = V_{L+1} $
	\item \textit{Closed} spin chain: $ (0) \rightarrow (L+1), \, \, \partial \mrm{\Sigma}_{1} = \partial \mrm{\Sigma}_{2} = V_{L+1}$
	\item \textit{Cyclic} spin chain: $ \partial \mrm{\Sigma}_{1} = \partial \mrm{\Sigma}_{2} = V_{L+1} $ along with shift symmetry\footnote{in general the level-one Yangian does not commute with shift operator $ \left[ \cl{U}, ^{(1)}\cl{Y}[J_{a}] \right] \neq 0 $} $ i \rightarrow i \pm 1 $, and the total momentum of the spin chain excitations vanishes $ P = 0 $.
\end{itemize}

\begin{figure}[!h]
	\centering
	\includegraphics[width=0.58\textwidth]{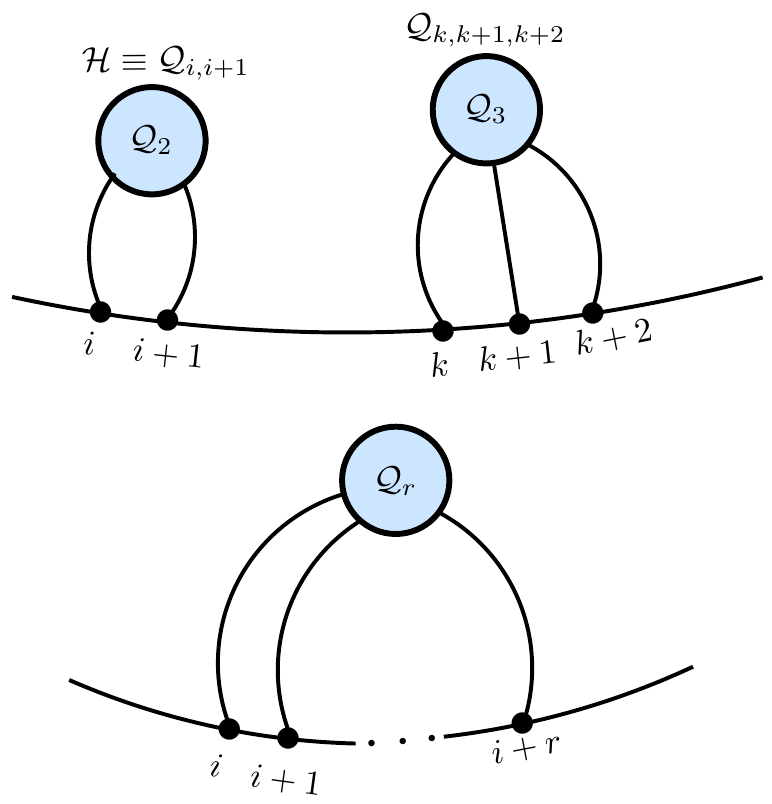}
	\caption{Local charge densities acting on the spin chain}
	\label{Fig_Q_r}
\end{figure}

As it was stated above, the central characteristics of an integrable spin chain are symmetries and associated conserved charges, which can be represented in the hierarchy set $ \{ \bb{Q}_{1}, \bb{Q}_{2} \dots \bb{Q}_{r} \dots \bb{Q}_{n} \} $ and completely define the underlying integrable system. In the context of spin chains the first two charges are usually assigned to translations/momentum $ \bb{Q}_{1} \equiv P $ and Hamiltonian $ \bb{Q}_{2} \equiv \bb{H} $. For the case of a spin chain with interacting adjacent sites, it is identified as a chain with the \textit{nearest-neighbour} (NN) interaction. In the present framework, $ \bb{H} $ is exactly an operator that defines NN-interactions. So the charge $ \bb{Q}_{r} $ that spans $ r $-sites constitutes an interaction range $ r $. It is sometimes referred to as higher charge Fig. \ref{Fig_Q_r} (or \textit{higher rank Hamiltonians} as it will become clear below). It means that on the spin chain we can define local densities of the appropriate operators, the sum of which will correspond to the full operator on the given spin chain
\begin{equation}\label{key}
	\bb{Q}_{2} = \bb{H} = \sum_{i = 1}^{L} \cl{H}_{i,i+1} \qquad \qquad \cl{H}_{L,L+1} = \cl{H}_{L,1}
\end{equation}
with Hamiltonian density $ \cl{H} $ at $ \left( i, i+1 \right) $-sites and the spin chain periodicity (closure condition). So in general, the core integrable structure is encoded in mutual commutativity of the above
\begin{equation}\label{key}
	\left[ \bb{Q}_{r}, \bb{Q}_{s} \right] = 0 \qquad \bb{Q}_{x} \equiv \sum_{k = 1}^{L} \cl{Q}_{k, \, k+1,\, \dots \, ,\, k+x-1}
\end{equation}
where $ \bb{Q}_{x} $ indicate range $ x $ charge. Although we need to remark that such labelling is not always possible. It is a class of spin chains that produces such consistent mapping. It will be explained that automorphism generating property and periodic boundary conditions are important for this. Characteristic properties of these charges are locality and homogeneity. Namely, the $ \cl{Q}_{x} $ densities act locally on a given number of spins $ x $, interaction is isotropic w.r.t. all sites and all charges respect a group $ G $ symmetry with an underlying algebra $ \fk{g} $. However by implementing the boost construction from above, one still needs an additional object to strictly prove integrability and uniqueness of the novel class. 

\begin{figure}[!h]
	\centering
	\includegraphics[width=0.8\textwidth]{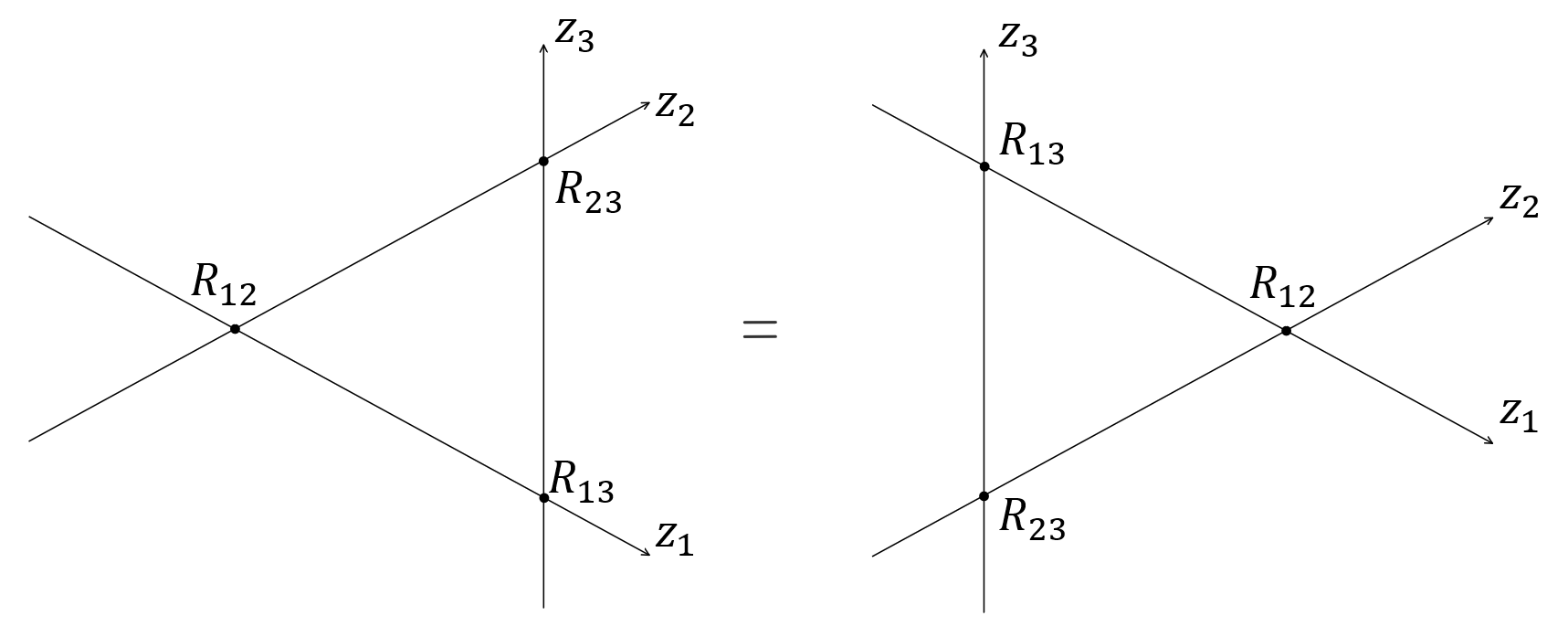}
	\caption{QYBE structure graphical representation, which underlies factorisation of the pairwise consequential scattering.}
	\label{QYBE_Scheme}
\end{figure}

Such a fundamental object turns out to carry scattering data, braiding symmetry and subsequently the spectrum generating information for the quantum integrable model -- quantum $ R $-matrix. For the construction, it is necessary to consider a vector space $ \bb{V} $ over field $ \mathfrak{F} $ and a linear map
\begin{equation}\label{R-lmorphism}
	R: \bb{V} \otimes \bb{V} \rightarrow \bb{V} \otimes \bb{V}
\end{equation}
with endomorphic structure on $ R_{ij} \in \text{End}\left( \bb{V} \otimes \bb{V} \otimes \bb{V} \right) $, where appropriate continuation on identities is assumed
\begin{equation}\label{key}
	R_{12} = R \otimes \mathds{1}_{\bb{V}} \qquad R_{23} = \mathds{1}_{\bb{V}} \otimes R \qquad R_{13} = \left( \mathds{1}_{\bb{V}} \otimes \mathfrak{t}_{\bb{V}\bb{V}} \right).\left( R \otimes \mathds{1}_{\bb{V}} \right).\left( \mathds{1}_{\bb{V}} \otimes \mathfrak{t}_{\bb{V}\bb{V}} \right)
\end{equation}
with twist map\footnote{which in the integrable lattice systems corresponds to permutation operator $ P_{ij} $} $ \mathfrak{t}_{\bb{V}\bb{V}} $ defined as follows
\begin{equation}\label{key}
	\mathfrak{t}_{\bb{V}\bb{V}} : \bb{V} \otimes \bb{V} \rightarrow \bb{V} \otimes \bb{V} \qquad \mathfrak{t}_{\bb{V}\bb{V}} \left[ \bb{W}_{1} \otimes \bb{W}_{2} \right] \equiv \bb{W}_{2} \otimes \bb{W}_{1}
\end{equation}
Structure and properties of the quantum $ R $-matrix are completely identified by the main quantum constraint (or more specifically, characterised by the underlying quantum group structure). The study of braid groups, quantum symmetries and algebra of quantum operators leads us to the main constraint -- Quantum Yang-Baxter equation (QYBE) \ref{QYBE_Scheme}
\begin{equation}\label{YBE_1}
	R_{12}\left( z_{1}, z_{2} \right) R_{13}\left( z_{1}, z_{3} \right) R_{23}\left( z_{2}, z_{3} \right) = R_{23}\left( z_{2}, z_{3} \right) R_{13}\left( z_{1}, z_{3} \right) R_{12}\left( z_{1}, z_{2} \right)
\end{equation}
with $ z_{i} $ to be a spectral parameter associated to space $ i $. In addition, the $ R $-matrix possesses special spectral points, and in particular, reduces to the permutation operator for coinciding parameters
\begin{equation}\label{key}
	R_{12}(z,z) = P_{12}
\end{equation}
which corresponds to the regularity condition. It can be also shown that the Hamiltonian is obtained from $ \log $-derivative of the $ R $-matrix \ref{Apx_2-dim}
\begin{equation}\label{key}
	\bb{H}_{12} = R_{12}\left( z_{1}, z_{2} \right)^{-1} \partial_{z_{1}} R_{12}\left( z_{1}, z_{2} \right) \big|_{z_{2} = z_{1} = z} = P_{12} \partial_{z_{1}} R_{12}\left( z_{1}, z_{2} \right) \big|_{z_{2} = z_{1} = z}
\end{equation}

It is important that for a large class of integrable models it is possible to obtain reduced spectral dependence of the underlying $ R $-matrix. As we shall see, that will result in significant computational reductions, analytic properties and form of the derived operators of the new models. Generically, one can implement the quantum $ R $-matrix in the fundamental $ \mb{GL}\left( N, \bb{C} \right) $ representation $ R_{12}^{*}\left( z \right) \in \text{Mat} \left( N, \bb{C} \right)^{\otimes 2} $
\begin{equation}\label{key}
	R_{12}^{*}\left( z \right) = \sum_{i,j,k,l = 1}^{N} R_{ijkl}^{*}\left( z \right) E_{ij} \otimes E_{kl}
\end{equation}
where $ R_{12}^{*}\left( z \right) $ is $ N^{2} \times N^{2} $ matrix, asterisk indicates presence of a quantum or other characteristic parameter (e.g. $ \hbar $), $ z $ is a spectral parameter and $ E_{ij} $ constitutes a basis in $ \text{Mat}\left( N, \bb{C} \right) $.

Moreover we can impose\footnote{It is derived independently for several of the novel classes below} additional conditions on the {\Rx}, one of them respects \textit{braiding} and \textit{unitarity} symmetric property (or unitarity)
\begin{equation}\label{key}
	R_{12}^{*}\left( z_{1} - z_{2} \right) R_{21}^{*}\left( z_{2} - z_{1} \right) \simeq \id \otimes \id
\end{equation}
with $ R_{21}^{*}\left( u \right) = P_{12} R_{12}^{*}\left( u \right) P_{12} $ and tensor product component permutation $ P_{12} = \sum E_{ij} \otimes E_{ji} $. In the provided setup, one can also generically obtain \textit{antisymmetry}
\begin{equation}\label{key}
	R_{12}^{q_{h}}\left( z_{1} - z_{2} \right) = - R_{21}^{-q_{h}}\left( z_{2} - z_{1} \right)
\end{equation}
with some prescribed quantum parameter $ q_{h} $\footnote{It would also hold for associative YBE in classical integrable systems}. That would lead to the special spectral dependence of \ref{YBE_1}, i.e. YBE of the difference form\footnote{The quantum parameter indication on {\Rx} will be omitted for the rest of the discussion}
\begin{equation}\label{DiffYBE}
	R_{12}R_{13}R_{23} = R_{23}R_{13}R_{12} \qquad R_{mn} = R_{mn}\left( z_{m} - z_{n} \right)
\end{equation}
or reassociating $ R_{13}\left( u \right) $ and $ R_{23}\left( v \right) $
\begin{equation}\label{YBE_2}
	R_{12}(u-v)R_{13}(u)R_{23}(v) = R_{23}(v)R_{13}(u)R_{12}(u-v)
\end{equation}
One of the questions that will be addressed throughout this work is the classification of the quantum integrable models, which from many perspectives is a complex problem on its own. For example, a number of attempts for new integrable models, were based on the resolvability of the Yang-Baxter equation \eqref{YBE_1} or \eqref{YBE_2}, which apriori is not a straightforward task. Moreover, it is known to be directly solvable only for certain subsectors \cite{Kulish1982}. A subsector of integrable lattice vertex models and their relation with Zamolodchikov tetrahedral algebra, as well as constant YBE solutions were shown in \cite{Akutsu:1982,Khachatryan:2012wy,Hietarinta:1992ix,Pourkia:2018}. More recently a classification of $ \{ 4\text{-}8 \} $-vertex via differential approach was given in \cite{Vieira:2017vnw,Vieira:2020lem}.

\begin{figure}[!h]
	\centering
	\includegraphics[width=0.65\textwidth]{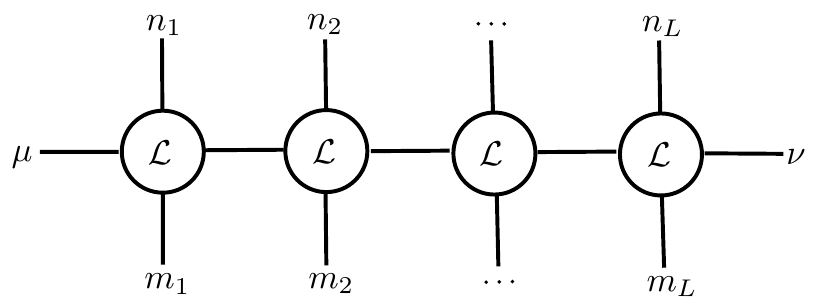}
	\caption{Monodromy construction scheme}
	\label{Fig_T_Construction}
\end{figure}

Other crucial constraint that will appear more efficient for derivation of properties and algebra of models in the non-additive sector \ref{NA_S_SD}, \ref{FF_Section}, \ref{p5_Sec_AdS_2_3_Integrable_Deformations} -- is the braiding algebra realised by intertwining the monodromy $ T(u) $ and the {\Rx},
\begin{equation}\label{I_RTT}
	R_{0\tilde{0}}(u-v) T_{0}(u) T_{\tilde{0}}(v) = T_{\tilde{0}}(v) T_{0}(u) R_{0\tilde{0}}(u-v)
\end{equation}
where the monodromy $ T_{0,\{ 1, \dots, L \}}(u) $ acts on $ L $ physical spaces and one extra space is auxiliary, which we can set through the sequence of the Lax matrices $ \cl{L} $ as can noticed from Fig. \ref{Fig_T_Construction}
\begin{equation}\label{key}
	T_{m_{1}m_{2} \dots m_{L}, \, \nu}^{n_{1}n_{2} \dots n_{L}, \, \mu} = \cl{L}_{s_{2}m_{1}}^{\mu \, n_{1}}(u - \theta_{1}) \cl{L}_{s_{3} m_{2}}^{s_{2} n_{2}}(u - \theta_{2}) \dots \cl{L}_{\nu m_{L}}^{s_{L} n_{L}}(u - \theta_{L})
\end{equation}
with inhomogeneities $ \theta_{i} $ at site $ i $ and $ m_{i}, n_{j}, s_{k} $ correspond to the space position free indices. The notion of $ \cl{L}_{i,a} $ as a linear operator involved in auxiliary spectral problem comes from resolution of integrable systems through Classical Inverse Scattering Method. Specifically, $ \cl{L} $ structure is defined on local quantum $ \fk{h}_{i} $ and auxiliary $ \bb{V}_{a} $ spaces, i.e. its action is defined on $ \fk{h}_{i} \otimes \bb{V}_{a} $. In the case of isotropic \textit{XXX}$ _{\frac{1}{2}} $ spin chain it takes the form
\begin{equation}\label{key}
	\cl{L}_{i,a} (u) = u \id_{i} \otimes \id_{a} + i\sum_{\mu} S_{i}^{\mu} \otimes \sigma^{\mu}
\end{equation}
where $ u $ is a spectral parameter, $ \id_{i} $ and spin operator $ S_{i} $ act in $ \fk{h}_{i} $, whereas $ \id_{a} $ and Pauli operators in $ \bb{V}_{a} $.

The $ RTT $-relation also constitutes one of the main constraints in derivation of the algebraic structures (Bethe Ans\"atze, QISM). After all the monodromic tracing over auxiliary space would result in the transfer matrix $ t(u) $, with 
\begin{equation}\label{key}
	\tx{Tr} \, T_{m_{1}m_{2} \dots m_{L}, \, \nu}^{n_{1}n_{2} \dots n_{L}, \, \mu}(u) = t_{m_{1}m_{2} \dots m_{L}, \, \mu}^{n_{1}n_{2} \dots n_{L}, \, \mu}(u) \qquad t(u) : \bb{V}_{1} \otimes \bb{V}_{2} \otimes \dots \otimes \bb{V}_{L} \rightarrow \bb{V}_{1} \otimes \bb{V}_{2} \otimes \dots \otimes \bb{V}_{L}
\end{equation}
As we shall see below it will become crucial for us to identify the conserved charges following from the conservation laws, whose origin can be read off the aforementioned commutativity and perturbation \cite{Faddeev:1996iy}
\begin{equation}\label{I_t_Commutativity}
	\left[ t(u), t(v) \right] = 0
\end{equation}
along with
\begin{equation}\label{I_TExp}
	t(u+u^{*}) = \cl{U} + i u \, \cl{U} \sum_{i} \cl{Q}_{i,i+1} + \cl{O}(u^{2}) \equiv \cl{U} \exp\left[ i\sum_{r = 2}^{L} \left( u - \dfrac{i}{2} \right)^{r-1} \bb{Q}_{r} \right] 
\end{equation}
\begin{figure}[!h]
	\centering
	\subfloat[\centering $ \cl{U} $]{ \includegraphics[width=9.00cm]{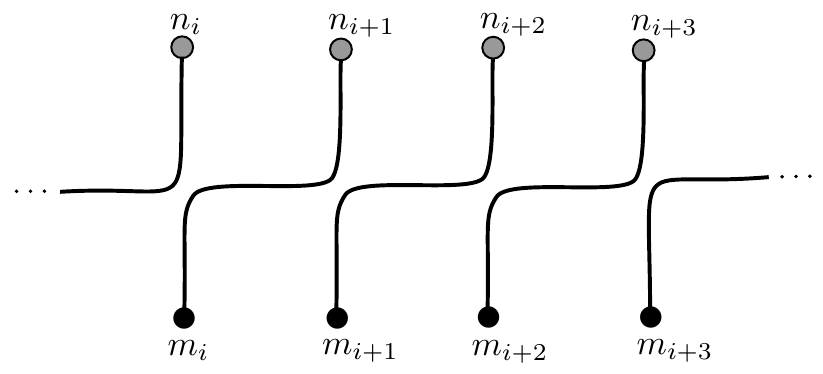} }
	\quad\vspace{1.5cm}
	\subfloat[\centering $ \cl{U} \, \bb{Q}_{2} $]{ \includegraphics[width=9.00cm]{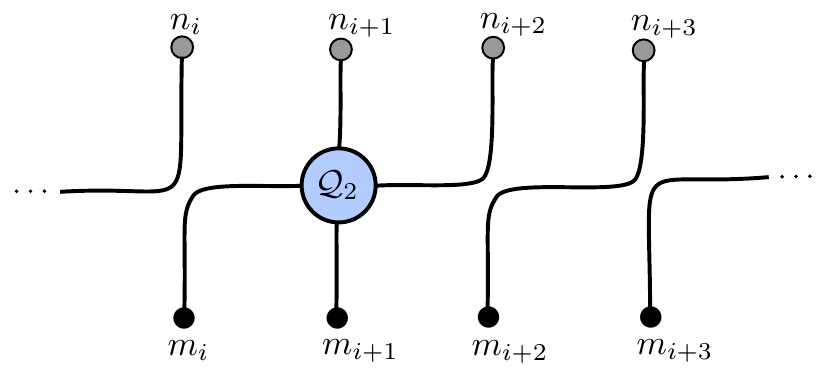} }
	\caption{First and second order of the expansion \eqref{I_TExp}, corresponding to the shift operator $ \cl{U} $ and product of the shift and Hamiltonian of the system}
	\label{Fig_TExp_O1}
\end{figure}
where the $ t $ is expanded around special spectral point $ \frac{i}{2} $ (one can also consider expansion around $ u^{*} = 0 $, depending on existence of the distinguished points and properties of the corresponding {\Rx}), $ \bb{Q}_{r} $ is a range $ r $ conserved charge Fig. \ref{Fig_Q_r} and $ \cl{U} $ is a \textit{shift} operator (recursive product of permutations) Fig. \ref{Fig_TExp_O1}
\begin{equation}\label{key}
	\cl{U} \ket{\psi_{1}, \psi_{2} , \dots , \psi_{L}} = \ket{\psi_{2} , \dots , \psi_{L}, \psi_{1}} \qquad \cl{U} \equiv t \left( \frac{i}{2} \right)
\end{equation}
Hence from \eqref{I_TExp} one can derive a generating expression for range $ r $ local charge ($ r \geq 2 $)
\begin{equation}\label{I_Qr}
	\bb{Q}_{r} = - \dfrac{i}{(r-1)!} \dfrac{\dd^{\, \,r-1}}{\dd u^{r-1}} \log\left[ t(u) \right] \big|_{u = \frac{i}{2}}
\end{equation}
In this respect, if one follows the $ RTT $ or $ R\cl{L}\cl{L} $ analysis along with the local charge properties \eqref{I_RTT}-\eqref{I_TExp} and appropriate boundary limits, one derives an automorphic condition on monodromy $ T(u) $
\begin{equation}\label{I_BT}
	\left[ T, \cl{B} \right] = i \, \dot{T}
\end{equation}
where $ \dot{T}(u) \equiv \dd_{\, \, u} T(u) $ and the boost operator is associated with the weighted generating density sums
\begin{equation}\label{B_HDensity}
	\cl{B} = \sum_{m = -\infty}^{+ \infty} m \, \cl{H}_{m,m+1}
\end{equation}
Based on the definition of the boost, we consider infinite spin chain setting with NN-interaction. From above one can see that there is an emergent symmetry that not only leaves invariant the commutativity of hierarchy, but also implies a generating automorphism property. These two properties characterise the \textit{master symmetry}, which does not affect the commutators after recursive application and generates higher conserved charges by its action on the lower ones \eqref{I_BAS}, hence automorphically producing infinite commutative hierarchy \cite{FOKAS_1981_341,Fuchssteiner_1983}. By taking into account \eqref{I_TExp} we can rewrite \eqref{I_BT} for the transfer matrix
\begin{equation}\label{I_Bt}
	\left[ t, \cl{B} \right] = i \, \dot{t}
\end{equation}
with $ \dot{t}(u) = \dd_{\,\, u} t(u) $ and along with \eqref{I_Qr}, one can straightforwardly identify the generating scheme
\begin{equation}\label{key}
	\begin{array}{c}
		\cl{Q}_{2} = \cl{U}^{-1} \left[ \cl{B}, \cl{U} \right] \\ [2ex]
		\cl{Q}_{3} = \frac{i}{2} \left[ \, \cl{U}^{-1} \left[ \cl{B}, \cl{U} \cl{Q}_{2} \right] - \cl{Q}_{2}\cl{Q}_{2} \, \right] = \frac{i}{2} \big[ \, \underbrace{\left( \cl{U}^{-1} \, \cl{B} \, \cl{U} - \cl{Q}_{2} \right)}_{\cl{B}} \cl{Q}_{2} - \cl{Q}_{2} \cl{B}  \, \big] = \frac{i}{2} \left[ \cl{B}, \cl{Q}_{2} \right]  \\ [2ex]
		\cl{Q}_{4} = \frac{i}{3} \left[ \cl{B}, \cl{Q}_{3} \right] \\ [1ex]
		\dots \\ [1ex]
		\cl{Q}_{r+1} = \frac{i}{r} \left[ \cl{B}, \cl{Q}_{r} \right] \\ [2ex]
		\dots \\ [1ex]
	\end{array}
\end{equation}
so from the first moment $ \cl{B}[\cl{Q}_{2}] $ it possible to generate the complete hierarchy,
\begin{equation}\label{I_BAS}
	\begin{array}{l}
		\qquad\qquad\qquad\quad \mathbf{\cl{B}}\tx{\textbf{oost automorphism scheme}} \\ [2ex]
		\left[ \, \cl{Q}_{i} \, , \, \cl{Q}_{j} \, \right] = 0 : \quad \cl{Q}_{1} \xrightarrow[]{\quad\cl{B}\quad} \cl{Q}_{2} \xrightarrow[]{\quad\cl{B}\quad} \cl{Q}_{3} \xrightarrow[]{\quad\cl{B}\quad} \dots \xrightarrow[]{\quad\cl{B}\quad} \cl{Q}_{r+1} \xrightarrow[]{\quad\cl{B}\quad}
	\end{array}
\end{equation}
which can be noted to be reminiscent of the Lorentz boost for the space and time translations of the two-dimensional Poincar\'e algebra.

\paragraph{Remark} It is important to note, that strictly speaking the above boost lattice formalism have been defined on the infinite spin chains. However for the periodic integrable spin chains, which will be crucial throughout the work the short-range operators are equivalent to the ones on the infinite spin chain, as well as any boundary contributions vanish in both cases. Indeed, that is another natural restriction that charge range must not exceed the length of the spin chain\footnote{Such prescriptions can occur for systems different boundary conditions, where the spectrum admits an asymptotic analysis and in the integrable QFT \cite{Bajnok_2011,Tongeren_2016} or string system can be resolved by TBA \cite{Arutyunov:2009ga,Frolov:2021fmj,Frolov:2021bwp,Frolov:2021zyc}}. Assuming this, one can develop generating recursion scheme for the integrable charges for spin chain with periodic boundary conditions. 

In the context of \ref{I_Master_Symmetries_Discretisation}, it can be noted that this discrete generating automorphism also implies $ \left[ \cl{B}, \, ^{(1)}Q_{a} \right] \eqsim Q_{a} $, which also for closed spin chains is analogous to level-1 Yangian symmetry generators. Later we shall make it clear on how it is implemented for the cases considered in the present work (more detailed on the automorphic structure and implications for generating Yangian symmetries can be found in \ref{Apx_A_NA_Boost_Automorphic_Symmetry}).

\subsection{$ AdS_{n} $ integrability}

Recently significant progress has been achieved in studying integrable systems \cite{Minahan_2003,Beisert:2005tm,Beisert:2006qh,Arutyunov:2006yd,Arutyunov:2009ga} in the context of the Gauge/Gravity duality \cite{Maldacena_1997,Witten_1998,Aharony:1999ti,DHoker:2002nbb,Polchinski:2010hw,Maldacena:2003nj}. Especially important results were in performing various integrability checks of the holographic duality and developing novel integrable structures \cite{Sorokin_2011_SAdS_2,Beisert:2010jr,Arutyunov:2013ega,Borsato:2013qpa} along with their properties. In particular, the main discoveries lie in the planar limit of $ \cl{N} = 4 $ superconformal Yang-Mills theory and free type IIB string on {\ads}. In fact, as in any field theory, one of the most crucial questions is the computation of the spectrum of $ \cl{N} = 4 $ SYM. Since it is a SCFT, there is an analogue to mass spectrum computation, which corresponds to obtaining local operators (fundamental field composites) and their scaling dimension spectrum. Generically, one also needs to account for quantum corrections stemming from interactions, which results contributing to the anomalous dimensions of the theory. The core point is that it becomes possible to determine the spectrum in $ \cl{N} = 4 $ planar limit by a variety of techniques exactly, \textit{e.g.} Asymptotic Bethe Ansatz, Thermodynamic Bethe Ansatz \cite{Arutyunov:2009ax,Arutyunov:2011uz,Arutyunov:2012zt,Arutyunov:2012ai,Frolov:2021bwp}, Y-systems \cite{Bombardelli:2009xz,Alday:2010vh}, Quantum Spectral Curve \cite{Gromov_2014,Gromov:2015wca,Kazakov_2018} and other.

At the level of symmetries, the Gauge/Gravity duality states that global symmetries on both sides must agree, meaning that $ \cl{N} = 4 $ superconformal symmetry and superspace isometries of {\ads} are governed by the covering group $ \widetilde{PSU}(2,2 \vert 4) $. For that reason, the Lie superalgebra $ \al{psu}(2,2 \vert 4) $ as well as its deformed counterparts play a central role in both super-CFT and string integrability. The $ \al{psu} $ superalgebra has dimension $ 30 \vert 32 $, which accordingly constitutes even|odd sectors. It can be represented by the $ \bb{C} $-supermatrices
\begin{equation}\label{key}
	\fk{M} = 
	\left(
	\begin{array}{ c | c}	
		A & B \\ \hline
		C & D
	\end{array}
	\right) 
\end{equation}
where A,D are even and B,C are odd $ 4 \times 4 $ Non-Graßmann $ \bb{C} $-blocks. In turn, the supermatrices must also satisfy
\begin{align}\label{key}
	\left. \left[ \fk{M} , \fk{R} \right. \right\} = \fk{M}\fk{R}-(-1)^{\fk{M}\fk{R}}\fk{R}\fk{M}, \quad S\text{Tr}\; \fk{M} = \text{Tr}A - \text{Tr}D, \quad S\text{Tr} \, \left. \left[ \fk{M} , \fk{R} \right. \right\} = 0 \\
	(-1)^{\fk{M}\fk{T}} \left. \left[ \left. \left[ \fk{M} , \fk{R} \right. \right\} , \fk{T} \right. \right\} + (-1)^{\fk{R}\fk{M}} \left. \left[ \left. \left[ \fk{R} , \fk{T} \right. \right\} , \fk{M} \right. \right\} + (-1)^{\fk{T}\fk{R}} \left. \left[ \left. \left[ \fk{T} , \fk{M} \right. \right\} , \fk{R} \right. \right\} = 0
\end{align}

The complete holographic existence/consistency checks include different background geometries and theories. This manifests in dimensionality, amount of preserved supersymmetry, gauging schemes and many more. The most developed integrable backgrounds include {\ads} \cite{Arutyunov:2006yd,Arutyunov:2009ga}, {\adscp} \cite{Klose:2010ki,Bykov:2010tv}, {\adsm} \cite{Borsato:2013qpa,Borsato:2014exa,Borsato:2014hja,Borsato:2015mma} and {\adsT} \cite{Hoare:2014kma,Hoare:2015kla,Fontanella:2017rvu}. The last two cases are of particular interest as their investigation is still incomplete, especially, on the field theoretic side. These two backgrounds will be analysed in detail in Sec. \ref{p5_Sec_AdS_2_3_Integrable_Deformations}, where a set of novel $ AdS_{\{ 3,2 \}} $ deformations will be discussed. 

For the $ AdS_{3}/CFT_{2} $ integrability one needs to start with the prescription of \textbf{N}on-\textbf{L}inear \textbf{S}igma \textbf{M}odels in order to get an insight into analytic structure and symmetries of the model. In particular, onee can address the bosonic NLSM on {\adso}. 

In such framework with additional constraining (\textit{e.g.} mass sector) and gauge-fixed functional of BNLSM it is more convenient to identify symmetries (for more details \cite{Sfondrini:2014via,Zarembo:2017muf}). For the purpose of underlying integrable structures it is useful to consider the corresponding superstring action as a coset action \cite{Babichenko_2010}
\begin{equation}\label{key}
	\dfrac{PSU(1,1\vert2) \times PSU(1,1\vert2)}{SO(1,2) \times SO(3)} \times U(1)^{4}
\end{equation}
One could also start from the Green-Schwarz action, but in this setting the superisometries are more obscure \cite{Green:1983wt,Grisaru:1985fv}. On the other hand one can start with the coset formulation, but a $ \kappa $-gauge fixing is necessary (identification of the physical region for the fermions). To achieve that, one can consider the closed string bosonic NLSM functional\footnote{Fradkin-Tseytlin term in this prescription is suppressed $ R_{(2)} \rightarrow 0 $}
\begin{equation}\label{key}
	S =- \dfrac{h}{2} \int\limits_{-\frac{L}{2}}\limits^{\frac{L}{2}} \dd \sigma \dd \tau \gamma^{\alpha \beta} \partial_{\alpha} X^{\mu} \partial_{\beta} X^{\nu} G_{\mu \nu}(X)
\end{equation}
along with canonical definition of conjugate momenta
\begin{equation}\label{key}
	p_{\mu} = \dfrac{\delta S}{\delta \dot{X}^{\mu}} =  -h \gamma^{0\beta} \partial_{\beta} X^{\nu} G_{\mu\nu}(X), 
\end{equation}
which for {\adso} result in conserved charges associated with spacetime isometries, \textit{i.e.} time direction in $ AdS_{3} $ and angle $ \phi $ in $ S^{3} $. The last two correspond to translations and rotations along these directions, hence we acquire conserved energy and angular momentum
\begin{equation}\label{key}
	H = -\int\limits_{-\frac{L}{2}}\limits^{\frac{L}{2}} \dd \sigma p_{t}, \qquad J = \int\limits_{-\frac{L}{2}}\limits^{\frac{L}{2}} \dd \sigma p_{\phi}
\end{equation}
where $ p_{i} $ are conjugate momenta, $ H $ is a \textit{target space energy} (spectra contribution after quantisation) and $ J $ is a distinguished angular momentum. It can be found that the corresponding Noether charge $ \cl{P}_{+} $ along with worldsheet energy (Hamiltonian) $ \fk{H} $
\begin{equation}\label{key}
	\cl{P}_{+} = \dfrac{1}{2} \left( H + J \right), \qquad \fk{H} = H - J.
\end{equation}
Since it is not explicit on how to directly quantise $ \fk{H} $ to obtain the spectrum, it is necessary to redefine the fields and coordinates to allow for large $ h $ perturbation for Hamiltonian density
\begin{equation}\label{key}
	\cl{H} = \cl{H}_{2} + \dfrac{1}{h}\cl{H}^{4} + \dfrac{1}{h^{2}}\cl{H}_{6} + \cl{O}(h^{-3})
\end{equation}
and one can consider the BMN limit \cite{Berenstein:2002jq}
\begin{equation}\label{key}
	P_{+} \rightarrow \infty, \quad h \rightarrow \infty, \quad \dfrac{P_{+}}{h} \sim \tx{fixed}
\end{equation}
it is then possible to expand action in $ \frac{1}{P_{+}} $, where $ P_{+} $ is an eigenvalue of $ \cl{P}_{+} $
\begin{equation}\label{key}
	P_{+} = \int\limits_{-\frac{L}{2}}\limits^{\frac{L}{2}} \dd \sigma p_{+} = L, \qquad \tx{in light-cone gauge} \,\, x_{+} = \tau, \tx{ and } p_{+} = 1
\end{equation}
where it is meant to light-cone gauge fix the corresponding coordinates. Or one can proceed with the decompactifying limit
\begin{equation}\label{key}
	P_{+} \rightarrow \infty, \quad h \sim \tx{fixed}
\end{equation}
which admits asymptotic states and construction of the {\Sx}. In fact one can similarly proceed for other related  geometries and find out that {\adsm} contains two non-trivial integrable backgrounds with 16 preserved supercharges (more detailed in Sec. \ref{p5_Sec_AdS_2_3_Integrable_Deformations})
\begin{equation}\label{key}
	\begin{cases}
		\mathcal{M}^{4} = T^{4}, \text{with } \alg{psu}(1,1\vert2)^{2} \\
		\mathcal{M}^{4} = S^{3} \times S^{1} , \text{with } \alg{d}(2,1;\alpha)^{2} \sim \alg{d}(2,1;\alpha)_{L} \oplus \alg{d}(2,1;\alpha)_{R} \oplus \alg{u}(1) \\
	\end{cases}
\end{equation}
where $ \alpha $ is proportional to ratio of sphere radii and bosonic subalgebra of the latter decomposes to left and right sector 
\begin{equation}\label{key}
	\left[ \alg{su}(1,1) \oplus \alg{su}(2) \oplus \alg{su}'(2) \right]_{L} \oplus \left[ \alg{su}(1,1) \oplus \alg{su}(2) \oplus \alg{su}'(2) \right]_{R} \oplus \alg{u}(1)
\end{equation}

The $ AdS_{2}/CFT_{1} $ possesses even less symmetry, a quarter of the total amount of supersymmetry. Still this background has an interesting underlying worldsheet sigma model description. The supercoset formalism is also applicable here, it is the $ AdS_{2} \times S^{2} $ part that can be described by the Metsaev-Tseytlin \cite{Metsaev:1998it} type supercoset
\begin{equation}\label{key}
	AdS_{2} \times S^{2}: \, \dfrac{PSU(1,1\vert2)}{SO(1,1) \times SO(2)}
\end{equation}
\begin{equation}\label{key}
	S _{MT} = \int d^{2}x \; \text{STr}\left[ \left( \tx{Pr}_{+} \mathcal{J}_{+} \right) \mathcal{J}_{-} \right]
\end{equation}
where $ \cl{J}_{\pm} $ are group $ \cl{G} $ valued currents, $ \tx{Pr}_{+} $ are specific projectors. The $ \alg{psu}(1,1\vert2) $ superalgebra \ref{Apx_PSU112_Superalgebra} contains a $ \mathbb{Z}_{4} $ automorphism, which implies classical integrability for the same reason as {\ads} \cite{Bena:2003wd}. As mentioned, also here {\adsT} geometry has suitably truncated form of the Green-Schwarz action \cite{Green:1983wt} for $ AdS_{2} $ supercoset. However in this case one cannot decouple $ AdS_{2} $ dof from the GS fermions, due to the absence of suitable $ \kappa $ gauge choice. So this class is addressed from the light-cone decompactifying gauge, where the {\Sx} would describe the worldsheet excitations above BMN vacuum.

The overall supercoset parts for three backgrounds can be given by generic construction of the factorised supergroup
\begin{equation}\label{key}
	\dfrac{\hat{\fk{F}}}{\fk{f}} = \dfrac{\hat{G} \times \hat{G}}{\fk{f}}
\end{equation}
along with
\begin{equation}\label{key}
	AdS_{n} \times S^{n} = \hat{G}/H
\end{equation}
where $ \fk{f} $ constitutes a bosonic diagonal subgroup of the factorised supergroup $ \fk{F} $. By the given construction, one can get the following supercoset models
\begin{itemize}
		\item $ AdS_{5} \times S^{5} = \dfrac{PSU(2,2\vert4)}{SO(1,4) \times SO(5)} $
		\item $ AdS_{3} \times S^{3} = \dfrac{PSU(1,1\vert2) \times PSU(1,1\vert2)}{SO(1,2) \times SO(3)} $
		\item $ AdS_{2} \times S^{2} = \dfrac{PSU(1,1\vert2)}{SO(1,1) \times SO(2)} $	
\end{itemize}

All of the setups described above will have a deeper investigation in \ref{p5_Sec_AdS_2_3_Integrable_Deformations}, when we shall come to the novel 6- and 8-vertex deformations found by our method, developed in Sec. \ref{Method}, \ref{NA_S_SD}. We shall explicitly elaborate on how 6vB/8vB models given as $ AdS_{\{ 2,3 \}} $ deformations admit the models above and provide the novel structure.

In the next section we shall start our discussion by implementing an approach, which exploits automorphic symmetries of the conserved charges. Firstly, the last will arise in two-dimensional integrable models, however the method will be proven to be universal. In our considerations there will be spin chain Hamiltonians with the nearest-neighbour interaction and YBE with regular solution sector. We shall find novel Hamiltonians that are constant and by bottom-up technique obtain the corresponding {\Rxs}. Each integrable class will be characterised by a certain generating Hamiltonian, which will not only result in the compact presentation, but also will reproduce all the relevant information of the class by virtue of generic integrable symmetry constraints.

Among a number of advantages of the approach, is that one obtains a coupled polynomial system of equations instead of a high degree functional system, which might appear non-solvable in general. One of the aims of the current thesis is to demonstrate a new approach of constructing a novel integrable structure and resolve the classification problem, which usually is not accessible by standard techniques or requires significant restrictions. This method is rather universal and can be applied to any sector of integrable models, since it is based on universal integrability constraints that of course require minor attention when passing to specific subfield (statistical models, integrable spin chains, string integrability, its continuous counterparts for integrable system with irrelevant deformations is currently under investigation). In addition we have developed a coding system, which allows to make number of analytic derivations autonomous, with the requirement that initial system Ansatz is provided. Now it can be stated that for $ D=2,3,4 $ dimensions the Ansatz can be completely generic (technical implementation allows to use it even in high complexity case). Its structure is somewhat reminiscent of analytic bootstrapping, where a set of restricting conditions is applied with an recursive manifestations at the intermediate level.

The current work presentation is structured as follows, first of all, we address the necessary objects involved in the method and show the core steps necessary for its implementation, which will be reflected in finding a new integrable structure and completing classification for 2-dim systems ($ \fk{h}^{2} $), Sec. \ref{p1}. In Sec. \ref{D_Hubbard}, we ask for new integrability and implementation of our technique in the four dimensional sector ($ \fk{h}^{4} $), where a variety of symmetry algebras and representations are studied. The Sec. \ref{NA_S_SD} provides a complete unification of the approach structure ($ \cl{B} $ and $ \cl{H}/R $ resolution schemes) for integrable systems with arbitrary spectral dependence in $ D = 2,3,4 $, whose structure appear relevant for AdS/CFT integrability. Consequently an investigation of string type classes that come from 2-dim setups of Sec. \ref{NA_S_SD} is followed. Special properties, deformations, symmetries and proposed correspondences of these models are discussed in Sec. \ref{FF_Section}, \ref{p5_Sec_AdS_2_3_Integrable_Deformations}. The algebraic apparatus, examples, further proofs and derivations can be found in the corresponding Sections and Appendices.

\newpage    
\section{$ \bb{V}^{2} $-difference sector}\label{p1}

	\subsection{Method}\label{Method}

	\subsubsection{Automorphisms and $ \mathcal{B} $-operator}

	\paragraph{Notation} We begin by looking into integrable spin chains equipped with two-dimensional local quantum spaces $ \mathfrak{h} = \bb{V} \equiv \bb{C}^{2} $ and possess a quantum {\Rx} \ref{R-lmorphism} satisfying the quantum-YBE
	\begin{equation}\label{key}
		R_{12}(u-v)R_{13}(u)R_{23}(v) = R_{23}(v)R_{13}(u)R_{12}(u-v)
	\end{equation}
	in addition $ R(u) $ obeys the regularity condition $ R(0) \equiv P $ and behaves analytically in the $ 0 $-neighborhood. This minimal setting identifies the structure for NN-interacting spin chains of a given length $ L $ and total Hilbert space $ \mathfrak{H} = \left( \bb{C}^{2} \right)^{\otimes L} $. In general, the integrable structure and properties of such system can be obtained from the \textit{transfer} matrix $ t(u) $ through the $ \mathcal{R} $-monodromy
	\begin{equation}\label{key}
		t = \text{Tr}_{0} \mathcal{R}_{0i} \qquad \mathcal{R}_{0i} \equiv \prod_{i=L}^{1} R_{0i}(u)
	\end{equation}
	where $ t = t(u) $ is purely quantum and the trace is over the auxiliary null space \cite{Faddeev:1996iy}, we shall be considering closed spin chains with periodicity $ L+1 \equiv 1 $.

	It is known from an algebraic structure of integrable systems, that one can generate conserved charges \cite{Drinfeld:1985rx,Faddeev:1996iy} via $ t(u) = A(u) + D(u) $\footnote{$ A $, $ D $ are monodromy diag-operators obeying fundamental commutation relations}
	\begin{equation}\label{t-exp}
		\left[ t(u), t(v) \right] = 0 \qquad t(u) = 2 u^{n} + \sum_{i=0}^{n-2} \bb{Q}_{i}u^{i}
	\end{equation}
	that generates the commuting hierarchy of $ n-1 $ charges $ \{ \bb{Q}_{r} \} $, where the range of these operators rises as $ r $ increases. It can be proven that $ \bb{Q}_{1} $, which can be associated with momentum $ P $ and $ \bb{Q}_{2} $ with Hamiltonian $ \bb{H} $ are part of this hierarchy. The permutation shift and log-derivative of the {\Rx} provide accordingly momentum and Hamiltonian
	\begin{align}\label{Q_12}
		& P_{1,2}P_{2,3} \dots P_{L-1,L} = e^{iP} \qquad \qquad \bb{Q}_{1} \eqsim P \\
		& \bb{Q}_{2} \equiv \sum_{n=1}^{L} R_{n,n+1}^{-1} \dfrac{d R_{n,n+1}}{d u} \Bigg|_{u \rightarrow 0} = \sum_{n = 1}^{L} \cl{Q}_{n,n+1} = \bb{H}
	\end{align}
where $ \cl{Q}_{n,n+1} $ are $ \left( n, n+1 \right) $-site local Hamiltonian densities (in general $ \bb{H} $ can be of higher interaction range, but in the present construction $ \bb{Q}_{2} $ will be NN Hamiltonian). Hence along with \ref{t-exp} one gets the aforementioned commuting hierarchy. It can be seen from \ref{Q_12} that Hamiltonian density $ \cl{Q}_{n,n+1} $ is an operator of range two or the nearest-neighbor. Following the algebraic pattern, one can notice for the third charge
	\begin{equation}\label{key}
		\bb{Q}_{3} \equiv \sum_{n} \cl{Q}_{n,n+1,n+2} \qquad \cl{Q}_{n,n+1,n+2} = \left[ \cl{Q}_{n,n+1}, \cl{Q}_{n+1,n+2} \right]
	\end{equation}
	which exhibits exactly range three action. Analogously it is possible to continue the derivation of all higher charges, where one can eventually find the automorphic symmetry structure, as we have indicated in \ref{Introduction}. That can be compactly formulated by means of the \textit{boost} operator $ \cl{B} $ \eqref{I_Bt} \cite{Tetelman,Loebbert:2016cdm}.
	\begin{equation}\label{p1_Boost_Automorphism}
		\cl{B}\left[ \bb{Q}_{2} \right] = \sum_{k = -\infty}^{+ \infty} k \cl{Q}_{k,k+1}
	\end{equation}
	The advantage of such automorphism is reflected in its recursive generating property for all higher charges
	\begin{equation}\label{Q_r+1}
		\bb{Q}_{r+1} = \left[ \cl{B}\left[ \bb{Q}_{2} \right], \bb{Q}_{r} \right]
	\end{equation}
	it follows that the higher part of the hierarchy can be constructed from the Hamiltonian $ \bb{H} $ \cite{Loebbert:2016cdm}. One can also note that by construction, \ref{Q_r+1} contains local densities of range/length $ r + 1 $, which under appropriate finite length closed spin chain prescription defines a consistent $ (r+1) $-site operator. In general, that provides a closed set of commutative constraints on the charges that follow intertwined boost recursion. More detailed argumentation \ref{Apx_A_NA_Boost_Automorphic_Symmetry}, proofs, further restrictions and extensions will be discussed in the following sections.

	\subsubsection{Ansatz and Algebraic system}\label{AA}
	To initiate the implementation of the method, we can start with integrable spin chains with two-dimensional local quantum spaces. To achieve that, one can write a completely generic Ansatz for the Hamiltonian density using some basis. In particular, we can consider the $ 2 \times 2 $ extended Pauli set $ \sigma^{b} $ in the $ b = \{ 0,+,-,z \} $ basis, which is given by
	\begin{equation}\label{key}
		\sigma^{b} \equiv \mathds{1}_{2} \text{, } \quad \sigma^{\pm} \equiv \dfrac{\sigma_{x} \pm i \sigma_{y}}{2} \text{, } \quad \sigma^{z} = \sigma^{z}
	\end{equation}
	$ \mathds{1}_{2} $ is the two dimensional identity and the Ansatz for density is formed by the corresponding range tensor product
	\begin{equation}\label{key}
		\cl{Q}_{ij} = \cl{A}_{ab} \, \sigma^{a} \otimes \sigma^{b}
	\end{equation}
	which in this case contains 16 free parameters $ \cl{A}_{ab} $, however already at this level one could try reducing the number of free parameters by virtue of some generic symmetry transformations. For example, any Hamiltonian normalization does not break integrability, as well as local basis transform
	\begin{equation}\label{LBT}
		\cl{Q}' = \mathfrak{V} \cl{Q} \mathfrak{V}^{-1} \qquad \mathfrak{V} = \cl{V} \otimes \cl{V}
	\end{equation}
	with unimodular restriction on $ \cl{V} $
	\begin{equation}\label{key}
		\cl{V} =
		\begin{pmatrix}
				\fk{a} & \fk{b}  \\
			\fk{g}  & \fk{d}  \\
		\end{pmatrix} \qquad \text{det}\,\cl{V} = 1
	\end{equation}
	Without loss of generality we can demonstrate this by demanding vanishing of 7- and 8-vertex entries, i.e. when $ \cl{A}_{bb} = 0 $ with $ b = \pm $, so that transformed $ \cl{Q}' $ coefficients at $ \sigma^{b} \otimes \sigma^{b} $ are
	\begin{equation}\label{key}
		\begin{array}{c}
			\fk{a} ^4 \cl{A}_{++} +\fk{b} ^4 \cl{A}_{--} -2\fk{a} ^3 \fk{b}  (\cl{A}_{+z}+\cl{A}_{z+})-\fk{a} ^2 \fk{b} ^2 (\cl{A}_{+-}+\cl{A}_{-+}-4 \cl{A}_{zz}) \\
			+2 \fk{a}  \fk{b} ^3 (\cl{A}_{-z}+\cl{A}_{z-}) = \cl{A}'_{++} \\
			\fk{g} ^4 \cl{A}_{++} + \fk{d} ^4 \cl{A}_{--} -2\fk{g} ^3 \fk{d}  (\cl{A}_{+z}+\cl{A}_{z+})-\fk{g} ^2 \fk{d} ^2 (\cl{A}_{+-}+\cl{A}_{-+}-4 \cl{A}_{zz}) \\
			+ 2\fk{g}  \fk{d} ^3 (\cl{A}_{-z}+\cl{A}_{z-}) = \cl{A}'_{-- } \\
		\end{array}
	\end{equation}
	By looking more closely, we can notice that this argument holds in both ways, i.e. if we take $ \cl{A}_{++} = \cl{A}_{--} \equiv 0 $ the Ansatz is trivially of the desired form. Another possibility is if one allows either of $ \cl{A}_{bb} $ to vanish, then there is a nontrivial solution map to $ \cl{A}'_{++} = \cl{A}'_{--} = 0 $ if and only if $ \cl{A}_{\pm z} \neq - \cl{A}_{z \pm} $ and $ \cl{A}_{+-}+\cl{A}_{-+} \neq 4 \cl{A}_{zz} $, and the case $ \cl{A}_{++} = \cl{A}_{--} \neq 0 $ is related to the latter case. This is only an illustration of the possible coefficient reduction scheme and can be implemented in any other form upon demand.
	
	\paragraph{Commutative constraints} Next one would need to impose constraints on the obtained Ansatz. As we have mentioned above these constraints would come from the commutation properties characteristic for the given integrable system, hence $ \left[ \bb{Q}_{p}, \bb{Q}_{r} \right] = 0 $ will restrict the underlying densities of range p and r defined on the spin chain of an appropriate length (which will be always assumed from now on). In other words commutation of densities produces an algebraic system to be solved on the $ \cl{A} $-space.
	
	There is a number of questions related to the type and number of the commutators, their analytic structure and sufficiency, but it will be explained that in all studied cases the first nontrivial commutator $ \left[ \bb{Q}_{2}, \bb{Q}_{3} \right] $ will appear to be necessary and sufficient \cite{Kulish:1982a,Grabowski}. In this respect, let us perform an analysis on the structure and dof properties following from it. As it has been stated, range 3 and 2 charges follow the prescribed Ansatz
	\begin{equation}\label{Com_1}
		\begin{array}{c}
			\begin{split}
				\left[ \bb{Q}_{2}, \bb{Q}_{3} \right] & = \sum_{m,n} \cl{A}_{ab}\cl{A}_{efg} \left[ \dots\sigma_{m}^{a}\sigma_{m+1}^{b}\dots, \dots\sigma_{n}^{e}\sigma_{n+1}^{f}\sigma_{n+2}^{g}\dots \right] \\
				& \equiv \cl{C}_{abef} \sum_{m} \dots\sigma_{m}^{a}\sigma_{m+1}^{b}\sigma_{m+2}^{e}\sigma_{m+3}^{f}\dots
			\end{split}
		\end{array}
	\end{equation}
	with $ \cl{C}_{abef} $ to include combinations of $ \cl{A} $ and algebra structure constants \footnote{dots represent appropriate tensor embedding on the chain, such that the overlaps in the commutator are not trivial, i.e. $ \sum_{m,n} \cl{A}_{ab}\cl{A}_{efg} \left[ \delta_{m+1,n} + \delta_{m,n} + \delta_{m,n+1} + \delta_{m,n+2} \right] \left[ \dots\sigma_{m}^{a}\sigma_{m+1}^{b}\dots, \dots\sigma_{n}^{e}\sigma_{n+1}^{f}\sigma_{n+2}^{g}\dots \right] $}.

	In the general case for the closed spin chain hierarchy, it can be shown that $ \left[ \bb{Q}_{r},\bb{Q}_{s} \right] $ generates a polynomial system of $ \dfrac{1}{2} \left( 3^{r+s-1}-1 \right) $ equations of degree $ r+s-2 $. A significant part of these equations will be trivially satisfied. Again, it is important to note, that $ \left[ \bb{Q}_{2},\bb{Q}_{3} \right] $ is sufficient and for two-dimensional case higher commutators do not impose more constraints on the found solutions - the first commutator appears to contain all non-degenerate unique information on integrable models in two dimensions \cite{Grabowski}. It will be also shown later that a generalisation of this statement can be made.
	
	Another important remark on the boost symmetry includes differences between the initial boost $ \cl{B}\left[ \cdot \right] $ formulation and the closed spin chain setup considered in the current work. Specifically, one needs to pay attention to the fact that the boost formulation on the infinite open spin chains does not apriori admit addition of terms which do vanish on the periodic systems. In particular, densities taking the form of
	\begin{equation}\label{key}
		\delta\cl{Q}_{m} = \cl{Q} \otimes \mathds{1} - \mathds{1} \otimes \cl{Q}
	\end{equation} 
	obviously become vanishing on the closed chains due to periodicity $ \sum_{m} \delta \cl{Q}_{m} = 0 $, hence such symmetry transformations could be performed only for the final result (Hamiltonains), otherwise precaution required, since $ \cl{B}\left[ \bb{Q}_{2} \right] $ will affect the structure of the higher charges in the hierarchy. 
	
	\paragraph{Generating commutators} By observing the commutative structure, it is possible to see that only part of the commutators are actually required in order to understand recursive structure of the whole hierarchy. It means that there must be commutative degeneracy present and not all commutators are independent. For instance, let us consider $ \left[ \bb{Q}_{2},\bb{Q}_{4} \right] $
	\begin{equation}\label{key}
		\begin{aligned}
			& \left[ \bb{Q}_{2},\bb{Q}_{4} \right] = \left[ \bb{Q}_{2}, \left[\cl{B}\left[ \bb{Q}_{2} \right], \bb{Q}_{3} \right] \right] = 
			- \left[
			\left[ \cl{B}\left[ \bb{Q}_{2} \right] , \left[\bb{Q}_{3}, \bb{Q}_{2} \right] \right] +
			\left[ \bb{Q}_{3} , \left[ \bb{Q}_{2}  , \cl{B}\left[ \bb{Q}_{2} \right] \right] 	\right] 
			\right] =\\
			& [ \underbrace{\left[\bb{Q}_{3}, \bb{Q}_{2} \right]}_{\equiv 0} , \cl{B}\left[ \bb{Q}_{2} \right] ] + [ \bb{Q}_{3} , \underbrace{\left[ \cl{B}\left[ \bb{Q}_{2} \right] , \bb{Q}_{2} \right]}_{\bb{Q}_{3}} ] = 0 
		\end{aligned}	
	\end{equation}
	with Jacobi identity for intermediate step. One can analyze recursion and prove by induction that commutators could expressed by means of $ \left[ \bb{Q}_{r}, \bb{Q}_{2r+1} \right] $ and $ \left[ \bb{Q}_{2}, \bb{Q}_{2r+2} \right] $ $ r \geq 1 $. However it is required to have $ \left[ \bb{Q}_{2}, \bb{Q}_{3} \right] = 0 $ and $ \left[ \bb{Q}_{3}, \bb{Q}_{4} \right] = 0 $ as a commutativity initial conditions.

	\subsubsection{Integrable transformations}\label{IT}
	Once an algebraic system that follows from \ref{Com_1} is resolved, one arrives at a significantly large space of solutions. In this respect, it necessary to address if such set of solutions can be reduced to a finite independent subset due to potentially present symmetries. It is indeed true, since there are \textit{integrable} continuous and discrete transformations (part of YB algebra) whose action on a solution of the YBE will generate another set of solutions satisfying it. It is preferable to proceed with such \textit{generators}, as for now we are interested in the novel independent set of solutions, which could be presented in a more compact form. We shall discuss the minimal set of allowed transformations in the current section.
	
	\paragraph{Special cases} The most basic possibility is the \textit{reduction} from the general case to the sub-cases, a conventional illustration of this can be the XYZ $ \rightarrow $ XXZ $ \rightarrow $ XXX, which is achieved by consecutively setting coefficients to zero. In this context, the variety of all special cases of XYZ will be given by the generic 8-vertex model.
	
	\paragraph{Shifts and norms} It is consistent to allow suitable normalisation of the given Hamiltonian $ H $ or perform diagonal shift by adding terms of $ \mathfrak{C} \cdot \mathds{1} $ type.	

	\paragraph{Local basis transformation} As it was noticed above, an integrable spin chain is characterised by the corresponding Hamiltonian $ \bb{H} $ and {\Rx} admits a freedom of local basis change \ref{LBT} in the form $ \cl{O}' = \mathfrak{V} \cl{O} \mathfrak{V}^{-1} $, $ \mathfrak{V} \in \text{End}\left( \bb{V} \right) $, which also produces an equivalence class of integrable spin chains. As it follows from the definition, $ \cl{V} $ assigned to each local space defines a transformation for range $ r $ density $ \cl{Q}_{m_{1},\dots,m_{r}} $ according to
	\begin{equation}\label{key}
		\cl{Q}'_{m_{1},\dots,m_{r}} = \mathfrak{V} \cl{Q}'_{m_{1},\dots,m_{r}} \mathfrak{V}^{-1} \qquad \mathfrak{V} = \bigotimes_{l=1}^{r} \cl{V}_{l} \qquad \det \, \cl{V} = 1
	\end{equation}
	where $ \cl{V}_{l} $ is $ 2 \times 2 $ operator for $ 2 $-dim local spaces. In this form one can map all other models of the specific class, or obtain a necessary form of the Ansatz, which can noticeably lower algebraic system complexity (e.g. removing off-diagonal elements as stated in \ref{AA}).
	
	\paragraph{Discrete maps} Further mapping within the classes can be achieved by the discrete transformations for the {\Rxs}, since it is known that consistent application of transposition-permutation will again lead to the solution of the YBE, analogous arguments can be demonstrated for the underlying Hamiltonians. More explicit reasoning could be seen from perturbation theory of the YBE, which will be presented in the next section, whereas evident transformations are
	\begin{equation}\label{DisTr}
		\begin{matrix}
			R(u) & \qquad \cl{H} \\
			R(u)^{T} & \qquad P \cl{H}^{T} P \\
			P R(u) P & \qquad P \cl{H} P \\	
			P R(u)^{T} P & \qquad \cl{H}^{T} \\
		\end{matrix}
	\end{equation}
	where presentation above implies an agreement at the level of operatorial entries (as two objects are on distinct spaces, satisfy properties etc). It appears that with this minimal amount of identification transformations it becomes possible to find generating representative for each equivalence class.

	\subsubsection{$ H $ $ \Leftrightarrow $ $ R $-matrix and consistency}\label{H-R}
	As we have mentioned, all of the above transformations are implemented at the final stage, however strictly speaking solutions extracted following methodology of \ref{AA} and \ref{IT} could be considered as a necessary step for establishing the space of two-dimensional integrable models. Along the known claims of the sufficiency of the $ \left[ \bb{Q}_{2}, \bb{Q}_{3} \right] = 0 $ condition, it can be noted that relevant checks of the commutators of densities up to range 6 have not further constrained the resulting system. Nevertheless, to be completely exact, one needs to accomplish true quantum integrable checks aka YBE, that require the associated quantum {\Rxs}.
	
	In order to achieve that from our setting, it is necessary to exploit a bottom-up approach, i.e. obtain the {\Rxs} from the Hamiltonians $ \cl{H} $. For that we need to recall quantum YBE solution existence in the form of formal series
	\begin{equation}\label{R_QYBE_Solution}
		R(u,q_{h}) = \id + q_{h} \left( \rho \otimes \rho \right) r(u) + \sum_{k=2}^{\infty} \fk{A}_{k}(u) q_{h}^{k}
	\end{equation}
	where the first two terms constitute the classical YBE part. As known the solution also exists in the form \cite{Drinfeld:1985rx}
	\begin{equation}\label{key}
		R(u) = \id + \rho(\cl{I}_{\mu}) \otimes \rho(\cl{I}_{\mu}) \, u^{-1} + \sum_{k=2}^{\infty} R_{k} u^{-k} \qquad R_{k} \in \text{End} \left( \bb{V} \otimes \bb{V} \right)
	\end{equation}
	with an orthonormal basis $ \{ \cl{I}_{\mu} \} $ in algebra $ \fk{g} $ with a defined representation $ \rho : \fk{g} \rightarrow \text{End} (\bb{V}) $. Since the Hamiltonian has units of energy and from ABA arguments the {\Rx} known to be dimensionless, the expansion can also have inverse reformulation
	\begin{equation}\label{R_InvS}
		R(u) = P + P\cl{H}u + \sum_{k=2}^{\infty} R_{k} u^{k}
	\end{equation}
	Important to note that for the XYZ class all the $ R_{k} $ coefficients take specific perturbative form $ R_{k} = P f_{k}(\cl{H}) $, where $ f_{k} $ is degree $ k $ polynomial in $ \cl{H} $. It can be also noted, that in two-dimensions for $ \cl{H} $ one can get a perturbative reduction, since higher terms of the expansion \ref{R_InvS} can be re-expressed through lower orders $ \{ \id, \cl{H}, \cl{H}^{2}, \cl{H}^{3} \} $ due to Hamilton-Cayley corollary. Taking into account the above statements, one can proceed with the following prescription
	\begin{equation}\label{R_Inv_T}
		R(u) = P \sum_{k=0}^{3} f_{k}(u) \cl{H}^{k} u^{k}
	\end{equation}
	with $ f_{k} $ certain functions on the spectral parameter. It must be emphasized, that such Ansatz might not be applied to any {\Rx}, due to the dimensional limitations of the Hamilton-Cayley argument, analytic properties of the {\Rx} itself and $ f_{k} $ functions. However such construction has proven to work for the current two-dimensional classification problem (including the novel models). Specific constraining of $ f_{k} $ can be drawn immediately from allowed gauging of the {\Rx} by setting either $ f_{0}(u) $ or $ f_{1}(u) $ to be unity. Also from dimensional analysis and expansion structure consistency we can set $ f_{0}(0) = f_{1}(0) = 1 $ and $ f_{0}'(0) = 0 $.
	
	By such minimal perturbative implementation one is able to obtain the {\Rxs} that underlie all models with $ 2 $-dim local spaces. Consistency and existence of the associated {\Rxs} under YBE\footnote{From the modified quantum YBE and of \ref{R_Inv_T} Ansatz one can obtain nontrivial system on intertwined $ \cl{H} $ and $ f_{k} $, instead of order by order perturbation for $ R $, which can be analytically resolved by appropriate convergence only in the very limited number of cases} leads to their quantum integrability proof. In fact there is much more universal procedure for constructing {\Rxs} with reverse approach from the $ \cl{H} $'s, which is distinct from conventional methods, more details are provided in Sec. \ref{NA_S_SD}.

	\subsection{2-dim space}
	\subsubsection{Solution space}
	
	In the setting described above \cite{deLeeuw:2019zsi}, the system constrained by the $ \left[ \bb{Q}_{2},\bb{Q}_{3} \right] $ commutator has proven to be sufficient \cite{Grabowski}. Initially, the current method was checked on the $ 2 $-dim model setup and provided a substrate for further investigation. From the nontrivial results in 2-dim, it becomes clear that it can be also exploited in other frameworks, including string background integrability.
	
	By resolving the first commutator, one obtains a nontrivial coupled cubic system, which after eliminating trivial subsector results in $ \sim 250 $ solutions\footnote{the choice of a suitable algebraic basis leads to the complete solvability}. At the next stage, one needs to address if they could be grouped into certain classes. As it was shown in \ref{IT}, indeed, by the set of \textit{integrable} transformations, one can identify \textit{operatorial} equivalence classes (although it is a matter of presentation). We can illustrate this on the example, which containts XYZ/XXZ reductions
	\begin{align}
		\cl{H} = 
		\begin{pmatrix}
			a & b & -b & 0 \\
			0 & c-a & 2a- c & 0 \\
			0 & 2a+c & -a-c & 0 \\
			0 & 0 & 0 & a
		\end{pmatrix}
	\end{align}
	it is possible to notice that it has a limit to the XXZ model 
	\begin{align}
		\cl{H} =
		\begin{pmatrix}
			a & 0 & 0 & 0 \\
			0 & c-a & 2a-c & 0 \\
			0 & 2a+c & -a-c & 0 \\
			0 & 0 & 0 & a
		\end{pmatrix}
	\end{align}
	which is obvious and no transformation needed. However in order to include generic XYZ content, one needs to apply local basis transformation taking the form
	\begin{align}
		\cl{V} = 
		\begin{pmatrix}
			-\frac{2 \beta  c}{b} & \beta  \\
			0 & -\frac{b}{2 \beta  c}
		\end{pmatrix}
	\end{align}
	Separate analysis is necessary, since when $ b $ or $ c \rightarrow 0 $ due to the apparent singularities given solution can not be brought to the XYZ, and one finds an independent model \ref{n_sl2}. We shall now briefly discuss the structure of a generic elliptic class for the purposes of $ 2 $-dim space completion and demonstration of further grouping schemes. An important note on the grading-related solutions will be provided below.
	
	\subsubsection{The 8-vertex and subspace}\label{XYZ-type}
	The consistency of the method must show completeness of the solution space found. In this regard, one must expect an elliptic and its sub-sectors to be present. The full classification of $ \leq 8 $-vertex models has been provided in \cite{Vieira:2017vnw}. In general 8-vertex Ansatz can be given as follows
	\begin{align}
		\cl{H}^{XYZ} = \begin{pmatrix}
			a_1 & 0 & 0 & d_1 \\
			0 & b_1 & c_1 & 0 \\
			0 & c_2 & b_2 & 0 \\
			d_2 & 0 & 0 & a_2
		\end{pmatrix}.
	\end{align}
	In this category, we can establish 8 independent generating solutions, which include the known $ 2 $-dim solutions of the YBE
	
	\paragraph{Diagonal (4-vertex)} Any main diagonal $ \cl{H} $ is an integrable solution
	\begin{align}
		\mathcal{H}^{XYZ}_1 =
		\begin{pmatrix}
			a_1 & 0 & 0 & 0 \\
			0 & b_1 & 0 & 0 \\
			0 & 0 & b_2 & 0 \\
			0 & 0 & 0 & a_2 
		\end{pmatrix}
	\end{align}
	
	\paragraph{XXZ} There are two families in the 6-vertex sector, which agrees with \cite{Beisert_2013_idXXZ}

	\begin{align}
		&\mathcal{H}^{XYZ}_2 =
		\begin{pmatrix}
			a_1 & 0 & 0 & 0 \\
			0 & b_1 & c_1 & 0 \\
			0 & c_2 & b_2 & 0 \\
			0 & 0 & 0 & a_1 
		\end{pmatrix}
		&&\mathcal{H}^{XYZ}_3 =
		\begin{pmatrix}
			a_1 & 0 & 0 & 0 \\
			0 & b_1 & c_1 & 0 \\
			0 & c_2 & b_2 & 0 \\
			0 & 0 & 0 & -a_1-b_1-b_2 
		\end{pmatrix}
	\end{align}

	\paragraph{7-vertex} There are two generators of the 7-vertex form
	\begin{align}
		&\mathcal{H}^{XYZ}_4 =
		\begin{pmatrix}
			a_1 & 0 & 0 & d_1 \\
			0 & a_1+b_1 & c_1 & 0 \\
			0 & -c_1 & a_1-b_1 & 0 \\
			0 & 0 & 0 & a_1 
		\end{pmatrix}
		&&\mathcal{H}^{XYZ}_5 =
		\begin{pmatrix}
			a_1 & 0 & 0 & d_1 \\
			0 & a_1- c_2 & c_1 & 0 \\
			0 & c_2 & a_1- c_1 & 0 \\
			0 & 0 & 0 & a_1-c_1-c_2 
		\end{pmatrix}
	\end{align}

	\paragraph{8-vertex} The elliptic sector contains three distinct generators in the following form
	\begin{align}
		& \mathcal{H}^{XYZ}_6 =
		\begin{pmatrix}
			a_1 & 0 & 0 & d_1 \\
			0 & b_1 & c_1 & 0 \\
			0 & c_1 & b_1 & 0 \\
			d_2 & 0 & 0 & a_1 
		\end{pmatrix}
		& \mathcal{H}^{XYZ}_7 =
		\begin{pmatrix}
			a_1 & 0 & 0 & d_1 \\
			0 & b_1 & c_1 & 0 \\
			0 & c_1 & b_1 & 0 \\
			d_2 & 0 & 0 & 2b_1-a_1
		\end{pmatrix}
	\end{align}
	\begin{align}
		\mathcal{H}^{XYZ}_8 =
		\begin{pmatrix}
			a_1 & 0 & 0 & d_1 \\
			0 & a_1 & b_1 & 0 \\
			0 & -b_1 & a_1 & 0 \\
			d_2 & 0 & 0 & a_1 
		\end{pmatrix}
	\end{align}

	The corresponding {\Rxs} can be found elsewhere, including classification performed in \cite{Vieira:2017vnw}. The approach \ref{H-R} was cross-checked for XYZ and all subclasses. In addition, we make it explicit on a more nontrivial novel classes presented below.
	
	\newpage
	\subsubsection{The new classes}\label{N-classes}
	\paragraph{Class \Rn{1}} The first novel generator $ \cl{H} $ obtained by virtue of the method \ref{Method}
	\begin{align}
		\cl{H}_{\text{\Rn{1}}} = \begin{pmatrix}
			0 & a_1 & a_2 & 0 \\
			0 & a_5 & 0 & a_3 \\
			0 & 0 & -a_5 & a_4 \\
			0 & 0 & 0 & 0
		\end{pmatrix}
	\end{align} 
	with $a_1 a_3-a_2 a_4 =0$. Along with its associated {\Rx} to be
	\begin{align}\label{aR1}
		R_{\text{\Rn{1}}}(u) =
		\begin{pmatrix}
			1 & \frac{a_1 (e^{a_5 u}-1)}{a_5}&  \frac{a_2(1-e^{-a_5 u}) }{a_5} &  \frac{a_1 a_3+a_2a_4}{a_5^2}(\cosh (a_5 u) -1) \\
			0 & 0 & e^{-a_5 u} & \frac{a_4(1-e^{-a_5 u}) }{a_5} \\
			0 & e^{a_5 u} & 0 & \frac{a_3 (e^{a_5 u}-1)}{a_5} \\
			0 & 0 & 0 & 1
		\end{pmatrix} 
	\end{align}
	It is straightforward that \eqref{aR1} satisfies YBE, regularity and underlying braiding unitarity $ R_{ij}(u)R_{ji}(-u) \sim 1 $. In what follow the novel nontrivial classes below
	\vspace*{5mm}
	
	\paragraph{Class \Rn{2}} The second class constitutes reduced upper-triangular form 
	\begin{equation}
		\cl{H}_{\text{\Rn{2}}}=\left(
		\begin{array}{cccc}
			0 & a_2 & a_3-a_2 & a_5 \\
			0 & a_1 & 0 & a_4 \\
			0 & 0 & -a_1 & a_3-a_4 \\
			0 & 0 & 0 & 0 \\
		\end{array}
		\right)
		\qquad 
		R_{\text{\Rn{2}}}(u)= u P\Big[\,\frac{a_1 }{\sinh (a_1 u)}+ \cl{H}_{\Rn{2}} +\frac{ \tanh(\frac{a_1 u}{2})}{a_1}  \cl{H}^2_{\Rn{2}}  \Big]
	\end{equation}
	which is also regular and satisfies braiding unitarity.
	\vspace*{5mm}

	\paragraph{Class \Rn{3}} The third class is associated to the following generating pair
	\begin{align}
		\cl{H}_{\text{\Rn{3}}} = 
		\begin{pmatrix}
			-a_1 & \left(2 a_1-a_2\right) a_3 & \left(2 a_1+a_2\right) a_3 & 0 \\
			0 & a_1-a_2 & 0 & 0 \\
			0 & 0 & a_1+a_2 & 0 \\
			0 & 0 & 0 & -a_1 \\
		\end{pmatrix}
	\end{align}
	\begin{equation}
		R_{\text{\Rn{3}}}(u)=\left(\begin{array}{cccc}
			e^{-a_1 u} & a_3 \left(e^{(a_1-a_2)u}-e^{-a_1u}\right) & a_3 \left(e^{(a_1+a_2)u}-e^{-a_1u}\right) & 0 \\
			0 & 0 & e^{(a_1+a_2)u} & 0 \\
			0 & e^{(a_1-a_2)u} & 0 & 0 \\
			0 & 0 & 0 & e^{-a_1 u}
		\end{array} \right)
	\end{equation}
	it can be taken as a 4-vertex deformation by $ a_{3} $ parameter, which reproduces it in the $ a_{3} \rightarrow 0 $ limit
	\begin{align}\label{Hr4v}
		\cl{H}'_{4v} = 
		\begin{pmatrix}
			-a_1 & 0 & 0 & 0 \\
			0 & a_1-a_2 & 0 & 0 \\
			0 & 0 & a_1+a_2 & 0 \\
			0 & 0 & 0 & -a_1 
		\end{pmatrix}
	\end{align}
	The corresponding {\Rx} was also provided in the \cite{Vieira:2017vnw} classification problem and appears of the form
	\begin{equation}
		R'_{4v}(u) = P_{12} \left( f_0 (u) + u f_{1}(u) \cl{H} + u^2 f_{2}(u) \cl{H}^2 \right)
	\end{equation}
	where $ f_{i} \equiv f_{i} \left( u, a_{1}, a_{2} \right) $, and the {\Rx} has equivalent $ \cl{H} $-functional representation in both cases of $ a_{3} = 0 $ and $ a_{3} \neq 0 $, since $ a_{3} $ is present only through the $ \cl{H}_{4v} $.
	\vspace*{5mm}
	
	\paragraph{Class \Rn{4}}
	The fourth class again respects braiding unitarity and is identified as follows
	\begin{align}
		\cl{H}_{\text{\Rn{4}}} = 
		\begin{pmatrix}
			a_1 & a_2 & a_2 & a_3 \\
			0 & -a_1 & 0 & a_4 \\
			0 & 0 & -a_1 & a_4 \\
			0 & 0 & 0 & a_1
		\end{pmatrix}
	\end{align}
	\begin{equation}
		R_{\text{\Rn{4}}}(u)=\left(
		\begin{array}{cccc}
			e^{a_1 u } & \frac{a_2 \sinh (a_1 u )}{a_1 } & \frac{a_2 \sinh (a_1 u )}{a_1 } & \frac{e^{a_1 u } (a_2 a_4+ a_1 a_3 \coth (a_1 u )) \sinh ^2(a_1u )}{a_1 ^2} \\
			0 & 0 & e^{-a_1 u } & \frac{a_4 \sinh (a_1 u )}{a_1 } \\
			0 & e^{-a_1 u } & 0 & \frac{a_4 \sinh (a_1 u )}{a_1 } \\
			0 & 0 & 0 & e^{a_1 u } \\
		\end{array}
		\right)
	\end{equation}
	\vspace*{5mm}
	
	\paragraph{Class \Rn{5}}
	The fifth obeys braiding unitarity, regularity and shows distinct non-removable off-diagonal configuration
	\begin{equation}\label{n_sl2}
		\cl{H}_{\text{\Rn{5}}} = 
		\left(\begin{array}{cccc}
			a_1 & a_2 & -a_2 & 0 \\
			0 & -a_1 & 2a_1 & a_3 \\
			0 & 2a_1 & -a_1 & -a_3 \\
			0 & 0 & 0 & a_1
		\end{array}\right)
	\end{equation}
	\begin{align}
		R_{\text{\Rn{5}}} = (1-a_1 u)\left(
		\begin{array}{cccc}
			2 a_1 u+1 & a_2 u & -a_2 u & a_2 a_3 u^2 \\
			0 & 2 a_1 u & 1 & -a_3 u \\
			0 & 1 & 2 a_1 u & a_3 u \\
			0 & 0 & 0 & 2 a_1 u+1 \\
		\end{array}
		\right)
	\end{align}
	\vspace*{5mm}

	\paragraph{Class \Rn{6}}
	The last  class appears as
	\begin{align}
		\cl{H}_{\text{\Rn{6}}} = 
		\begin{pmatrix}
			a_1 & a_2 & a_2 & 0 \\
			0 & -a_1 & 2a_1 & -a_2 \\
			0 & 2a_1 & -a_1 & -a_2 \\
			0 & 0 & 0 & a_1
		\end{pmatrix}
	\end{align}
	\begin{equation}
		R_{\text{\Rn{6}}}(u)=(1-a_1 u)(1+2a_1 u)
		\begin{pmatrix}
			1 & a_2 u & a_2 u & -a_2^2 u^2(2 a_1 u+1 ) \\
			0 & \frac{2 a_1 u}{2 a_1 u+1} & \frac{1}{2 a_1 u+1} & -a_2 u \\
			0 & \frac{1}{2 a_1 u+1} & \frac{2 a_1 u}{2 a_1 u+1} & -a_2 u \\
			0 & 0 & 0 & 1
		\end{pmatrix}
	\end{equation}
	which satisfies braiding unitarity.

	\newpage
	\subsubsection{Structure analysis of the new classes}
	As we have been able to notice above, the proposed approach for models with two-dimensional local space $ \mathbb{C}^{2} $ and generic Ansatz for $ \mathcal{H} $ turned not only to show full agreement with the set of integrable models that are found from the conventional YBE resolution (\textit{i.e.}  Heisenberg class, *-magnets, multivertex etc), but also finds new higher parametric integrable models \ref{N-classes}. Some of these new classes exhibit \textit{non-diagonalisability} and \textit{nilpotency} of the $ \mathcal{H} $, but others develop conserved charges with \textit{non-trivial Jordan blocks}, which leads to important results and corollaries. It also came to our knowledge that some of these classes represent high parametric deformations in the $ \al{sl}_{2} $ sector, which also appears to be a generalisation\footnote{Known one-parameter family of $ \al{sl}_{2} $ models, which also could be related to \cite{Alcaraz_1994}. In the past it was conjectured, that higher-parametric generalisations might exist, however it appeared non-resolvable by $ RTT $-approach, and in our computation we prove their existence and relations.} of \cite{Kulish_1997}. It is important to find all underlying Yangian deformations $ \mathcal{Y}^{*}[\al{sl}_{2}] $ and associated quantum groups for these models, which could also bring novel quantum group structure \cite{Kirillov1990representations,Bytsko:2009a}. It is also known that the Belavin-Drinfeld cohomological classification of quantum symmetries for novel models is an important question on its own. Additionally, it might appear of interest for non-unitary theories, which are studied in the context of conformal fishnet theories \cite{Gromov:2017cja,Caetano:2016ydc} or other systems that develop Jordanian structure \cite{Gainutdinov:2016pxy,Morin-Duchesne:2011a} 
	
	\paragraph{Class \Rn{1} and \Rn{2}} The conserved charges of the first two classes possess nilpotency, similar property have been observed \cite{Caetano:2016ydc}.
	
	\paragraph{Class \Rn{3}-\Rn{6}} It is possible to note that the rest of the classes, in general appear to be non-diagonalisable, however for a set of appropriate parametric restrictions they can be diagonalised. The diagonalisability holds demanding the following
	\begin{itemize}
		\item for $ a_{3} = 0 $ Class \Rn{3} reduces to 4-vertex model \eqref{Hr4v}
		\item $ a_{2} = a_{4} $ and $ a_{3} = \dfrac{a_{4}^{2}}{a_{1}} $ for Class \Rn{4}
		\item $ a_{2} = -a_{3} $ for class \Rn{5}
		\item $ a_{2} = 0 $ for class \Rn{6}
	\end{itemize}
	It is important to mention that since the eigenspectrum seems to depend on $ a_{1} $ in such restricted cases, it means that eigenvalues of classes \Rn{3} and \Rn{4} can be related to the $ z $-component density $ \cl{H} = \sigma^{z} \otimes \sigma^{z} $ and $ \cl{H} = 1 - 2P $ (also part of classes \Rn{5} and \Rn{6}).
	
	\subsubsection{Graded mapping}\label{p1_Graded_Mapping}
	Now we shall discuss the graded structure which initially was implicit in our investigation, in what we provide apparatus relevant for $ \bb{C}^{1|1} $ emerging in our solution space. Let us consider a $ \fk{G} $-graded vector space $ \bb{V} $ over the complex field $ \fk{F} = \bb{C} $, which exhibits the direct decomposition
	\begin{equation}\label{key}
		\bb{V} = \bigoplus_{\fk{a}\in\fk{G}} \bb{V}_{\fk{a}}
	\end{equation}
	If $ \exists \, \fk{G}' $ being nontrivial subgroup of the grading group $ \fk{G} $, then there is a quotient group $ \fk{G}/\fk{G}' $, with a quotient map defined
	\begin{equation}\label{key}
		\pi : \fk{G} \rightarrow \fk{G}/\fk{G}'
	\end{equation}
	then the established grading is odd or even. More specifically, if in the space $ \bb{V} $ we introduce the homogeneous basis $ \{ \fk{e}_{i} : \, i \in \cl{I} \} $, then $ \fk{e}_{i} \in \bb{V}_{\fk{a}} $ has a grade $ p(i) = \fk{a} $, with grading map
	\begin{equation}\label{key}
		p : \cl{I} \rightarrow \fk{G} \qquad \qquad \quad \cl{I} \equiv \{ 1, \dots N \} \text{ and } \dim \bb{V} = N
	\end{equation}
	which can be unified to the basis $ \{ \fk{e}_{i_{\fk{a}},\fk{a}} : \, i \in \cl{I}_{\fk{a}} \} $, with $ \cl{I}_{\fk{a}} = \{ i \in \cl{I} : p(i) = \fk{a} \} $, and one can define \textit{commutation factor} $ \cl{E}_{\fk{a}} $ of $ \fk{a} $
	\begin{equation}\label{key}
		\cl{E}_{\fk{a}} = \left( -1 \right)^{\pi(\fk{a})}
		\qquad \qquad
		\pi(\fk{a}) =
		\begin{cases}
			 + 1 \\
			  0
		\end{cases}
	\end{equation}
	where if $ \cl{E}_{\fk{a}} = + 1 \, \forall \fk{a} \in \fk{G} $ then grading is even, and if $ \exists \, \fk{a} \in \fk{G} $ s.t. $ \cl{E}_{\fk{a}} = -1 $ the grading is odd.
	More specifically, looking at the quotient isomorphic to $ \bb{Z}_{2} $ with $ \bb{V}_{0} \oplus \bb{V}_{1} $ and homogeneous elements $ \fk{v} $ (possessing zero projection onto either of subspaces), we have \textit{parity} $ p $
	\begin{equation}\label{key}
		\begin{cases}
			p(\fk{v}) = 0,\,\, \fk{v} \in \bb{V}_{0},\,\, \dim\bb{V}_{0} = d_{0} \quad \text{\textit{even}} \\
			p(\fk{v}) = 1,\,\, \fk{v} \in \bb{V}_{1},\,\, \dim\bb{V}_{1} = d_{1} \quad \text{\textit{odd}}
		\end{cases} 
	\end{equation}
	with $ \{ \fk{e}_{1} \dots \fk{e}_{d_{0}} \} \in \bb{V}_{0} $, $ \{ \fk{e}_{d_{0}+1} \dots \fk{e}_{d_{1}} \} \in \bb{V}_{1} $ and total space dimension $ \dim \bb{V} = \left( d_{0}, d_{1} \right) $. The linear operators\footnote{Strictly speaking \textit{right} operators having the \textit{right} module in the space $ \bb{V} $} in the chosen basis can be provided by
	\begin{equation}\label{key}
		\cl{O} = \cl{O} (\fk{e}_{i} \fk{v}_{i}) = \fk{e}_{j} \cl{O}_{ij} \fk{v}_{i}
	\end{equation}
	since grading is assigned to the space of $ \cl{O}_{ij} $, one accordingly equips rows and columns with parity and get parity for $ \cl{O} $
	\begin{equation}\label{key}
		p(\cl{O}) = p(i) + p(j) + p(\cl{O}_{ij})
	\end{equation}
	in what follows we shall be considering even operators, which implies
	\begin{equation}\label{key}
		p(\cl{O}_{ij}) = p(i) + p(j)
	\end{equation}
	which is also a central property of graded braid-commutative structures. In this respect, we can see that $ RTT $ relation
	\begin{equation}\label{key}
		T_{a}(u)T_{b}(v) = R_{ab}(u-v) T_{b}(v)T_{a}(u) R_{ab}(u-v)^{-1}
	\end{equation}
	would imply consistent commutativity for the underlying transfer matrices, if and only if we shall consider even {\Rxs} as an operator, i.e.
	\begin{equation}\label{key}
		\left[ t(u), t(v) \right] = 0 \quad \text{along with } t(u) = \text{STr}_{\fk{s}}T_{\fk{s}}(u) \quad \text{STr}\left( \cl{O}_{1} \cl{O}_{2} \right) = (-1)^{p(\cl{O}_{1}) p(\cl{O}_{2})} \text{STr} \left( \cl{O}_{2} \cl{O}_{1} \right)
	\end{equation}
	where $ \fk{s} $ indicates the appropriate space over which the supertrace is taken. On the other hand, it is known that one can derive a one-to-one correspondence (based on the Zamolodchikov graded algebra and its triple monoid consistency)
	\begin{equation}\label{gYBE_1}
		\tilde{R}_{ij}(u) = (-1)^{p(m) p(n)} R_{ij}(u)
	\end{equation}
	where $ \{ m,n \} $ are basis vector indices associated to the action of the $ R(u) $ operator, which in general can act in the tensor product $ \bb{C}^{N} \otimes \bb{C}^{N} $. Namely, by this one can establish 1-1 correspondence between solutions of the qYBE ($ R_{ij} $) and its graded ($ \tilde{R}_{ij} $) analogue \cite{Kulish1982,Kulish:1985bj,Bracken:1994hz,batchelor2008quantum}. More on graded structure \eqref{gYBE_1}, algebra and other relevant properties can be found \ref{Apx1_Graded_Spaces_Algebra}. From here we clearly see that our interest includes only even $ R $ operator and the corresponding YBE solutions. Also from the physical integrable perspective odd structure does not appear to be relevant\footnote{non-mixed structure only, although bosonisation is admissible}, whereas consistent commutativity of the generated charges comes from the even sector.
	
	\paragraph{Mappings} As stated above, we can generalise our tensor algebra to the graded one. By looking at the additional subset of solutions obtained along with \ref{XYZ-type}, one can notice the ones that differ up to a sign in front of the $ E_{21} \otimes E_{12} $ term, which appears in the conventional graded representation
	\begin{equation}\label{gA}
		\cl{A} = \sum (-1)^{p(k) \left( p(i)+p(j) \right)} \cl{A}_{ijkl} E_{ij}\otimes E_{kl}\quad \rightarrow \quad 
		\left(\begin{array}{cccc}
			\cl{A}_{1111} & \cl{A}_{1112} & \cl{A}_{1211} & \cl{A}_{1212} \\
			\cl{A}_{1121} & \cl{A}_{1122} & \cl{A}_{1221} & \cl{A}_{1222} \\
			\cl{A}_{2111} & \cl{A}_{2112} & \cl{A}_{2211} & \cl{A}_{2212} \\
			\cl{A}_{2121} & \cl{A}_{2122} & \cl{A}_{2221} & \cl{A}_{2222}
		\end{array}\right)
	\end{equation}
	where after certain redefinitions the graded prescription results in the equivalent set of generators. This fact could be seen in two independent ways, first by applying method \ref{Method} for \ref{gA} and finding the corresponding $ \cl{H} $, or the second, where one can exploit qYBE and its graded form relation between solution spaces \cite{Kulish1982}. More concrete, having regular XYZ {\Rx} (all free functions assumed to depend on the spectral parameter $ u $)
	\begin{equation}
		R(u) = 
		\left(
		\begin{array}{cccc}
			a_1 & 0 & 0 & d_1  \\
			0 & b_1 & c_1 & 0 \\
			0 & c_2 & b_2 & 0 \\
			d_2 & 0 & 0 & a_2
		\end{array}
		\right)
	\end{equation}
	allows also to have an associated regular $ R $ of the graded-YBE in the modified form
	\begin{equation}
		R(u)=\left(\begin{array}{cccc}
			a_1 & 0 & 0 & \sigma_{1} d_1  \\
			0 & \sigma_{2} b_1 & c_1 & 0 \\
			0 & c_2 & \sigma_{2} b_2 & 0 \\
			- \sigma_{1} d_2 & 0 & 0 & -a_2
		\end{array}\right)
	\end{equation}
	with $ \sigma_{i} = \pm $. From here it is possible to note, that we can establish a bijection mapping between regular non- and graded YBE solutions
	\begin{equation}
		\left(\begin{array}{cccc}
			a_1 & 0 & 0 & d_1  \\
			0 & b_1 & c_1 & 0 \\
			0 & c_2 & b_2 & 0 \\
			d_2 & 0 & 0 & a_2
		\end{array}\right)\quad  \rightarrow\quad  \left(\begin{array}{cccc}
			a_1 & 0 & 0 & d_1  \\
			0 & b_1 & c_1 & 0 \\
			0 & c_2 & b_2 & 0 \\
			- d_2 & 0 & 0 & - a_2
		\end{array}\right),
	\end{equation}
	This also completes the classification problem for models with $ \bb{C}^{2} $ local quantum space in terms generating solutions, which are characterised by their Hamiltonians $ \cl{H} $ and associated {\Rxs}. In the next section we generalise our approach, investigate the models with four-dimensional local space and discover novel models in several sectors, including the Hubbard-Shastry type.

\newpage

\section{$ \bb{V}^{4} $-difference sector: New superconductive integrability and GHM}\label{D_Hubbard}
Another possible extension of integrable models is the study of classes with higher dimensional local spaces. Enlarging the quantum spaces leads to interesting and important physical implications. It can be illustrated by the Hubbard model, which has a 4-dim space and represent a strongly coupled fermionic system. Namely, a site equipped with the 4-dim space admits vacuum, single electron of either spin up or down, or an occupation of two electrons. By such combination it is clear that there is a possibility of superconducting modes or electron pair formation. It can be described by the Hubbard Hamiltonian \cite{Hubbard_1965RSPSA,Essler2005_HB}
\begin{equation}\label{key}
		\bb{H}^{\text{Hubbard}} =  \sum_i  \sum_{ \sigma = \{ \uparrow,\downarrow \} } (\cdd_{\sigma,i}\mb{c}_{\sigma,i+1} + \cdd_{\sigma,i+1}\mb{c}_{\sigma,i} )+ \fk{u} \, \mb{n}_{\uparrow,i}\mb{n}_{\downarrow,i}.
\end{equation}
where from now we shall work in the oscillator basis $ \{ \mb{c}, \mb{c}^{\dagger} \} $, which appears in the kinetic part that provides electron dynamics between neighbouring sites (\textit{hopping term}). The \textit{potential} part counts the number $ \mb{n} $ of electron pairs for each site and coupling $ \fk{u} $ provides the scale in the theory.

It is important that systems with higher local space dimensionality can be also put on a lattice and proven to be integrable. Meaning one needs to deal with 4-dim setup, which results in the $ 16 \times 16 $ {\Rx} that satisfies
\begin{equation}\label{key}
	R_{12}(u,v) R_{13}(u,w) R_{23}(v,w) =  R_{23}(v,w) R_{13}(u,w) R_{12}(u,v)
\end{equation}
The integrability for Hubbard and related models was first proven in \cite{Lieb:1968zza, Shastry:1988_DSTR_IHM,Dye_2002}, which automatically implies the existence of integrable hierarchy of conserved charges. However the construction and investigation of the latter followed very different approaches. As it was indicated in \ref{Method} the commutativity and braiding that is present in such hierarchies is highly constraining. It is important to ask for a broader family of integrable theories in this sector.

As we know, the Hubbard and Shastry class manifest different spectral dependence, symmetry properties and more. This question can be approached by the method above \ref{Method}, which would allow to solve for potential new integrable structures without actually solving qYBE. Moreover the quantum YBE will be only used for quantum consistency checks of the found models (sufficiency condition). At first, we shall consider classes, which are apriori different from the Shastry sector and whose generalisation will be discussed in Sec. \ref{NA_S_SD}.

It has been already shown in \cite{deLeeuw:2019zsi}, one could address this problem from an appropriate Ansatz for the $ \cl{H} $ and eventually obtain the underlying {\Rx}\footnote{There was analogous attempt resolving YBE for nineteen-vertex model \cite{Makoto:1994JPhy1} by means of perturbative {\Rx}} according to the scheme \ref{p2_BoostScheme}. Once this procedure is (if) completed, one ends up with the new integrable structure, and one can further attempt their detailed study by virtue of different techniques, \textit{e.g.} Quantum Inverse Scattering, Bethe Ans\"atze, Quantum Spectral Curve and other. For example a similar classification problem was addressed in \cite{Crampe:2013nha,Fonseca:2014mqa,Crampe:2016she} by applying the Coordinate Bethe Ansatz. It is important to note, that not for all models it is possible to apply the above procedures. We shall discuss this later for the novel models.

\tikzstyle{box1} = [ draw = black, thick, rectangle, inner sep=5pt, inner ysep=5pt]
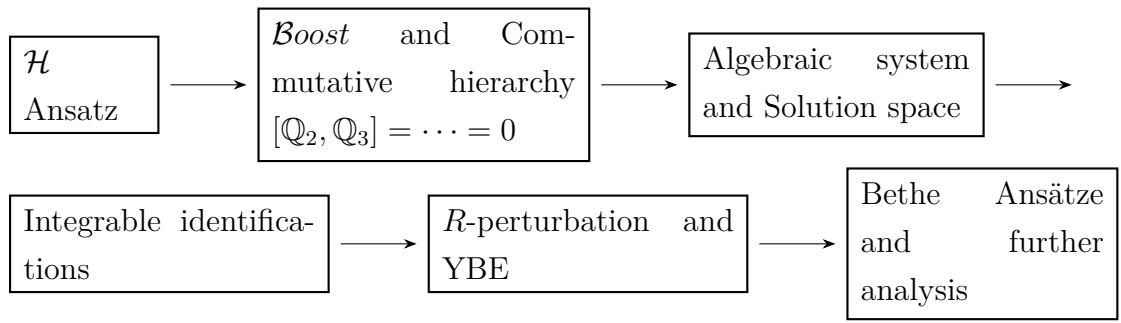
\begin{figure}
	\begin{tikzpicture}[baseline=0cm]
		\node [box1] 
		(box){
			\begin{minipage}{0.10\textwidth}
				$ \cl{H} $ Ansatz
			\end{minipage}
		};
	\end{tikzpicture}
	\begin{tikzpicture}[baseline=0cm]
		\draw [-{Stealth[slant=0]}] (0,0)--(1,0); 
	\end{tikzpicture}
	\begin{tikzpicture}[baseline=0cm]
		\node [box1] 
		(box){
			\begin{minipage}{0.25\textwidth}
				$ \cl{B}oost $ and Commutative hierarchy $ \left[ \bb{Q}_{2},\bb{Q}_{3} \right] = \dots = 0 $
			\end{minipage}
		};
	\end{tikzpicture}
	\begin{tikzpicture}[baseline=0cm]
		\draw [-{Stealth[slant=0]}] (0,0)--(1,0); 
	\end{tikzpicture}
	\begin{tikzpicture}[baseline=0cm]
		\node [box1] 
		(box){
			\begin{minipage}{0.22\textwidth}
				Algebraic system and Solution space
			\end{minipage}
		};
	\end{tikzpicture}
	\begin{tikzpicture}[baseline=0cm]
		\draw [-{Stealth[slant=0]}] (0,0)--(1,0); 
	\end{tikzpicture}
	\newline
	\newline
	\begin{tikzpicture}[baseline=0cm]
		\node [box1] 
		(box){
			\begin{minipage}{0.24\textwidth}
				Integrable identifications
			\end{minipage}
		};
	\end{tikzpicture}
	\begin{tikzpicture}[baseline=0cm]
		\draw [-{Stealth[slant=0]}] (0,0)--(1,0); 
	\end{tikzpicture}
	\begin{tikzpicture}[baseline=0cm]
		\node [box1] 
		(box){
			\begin{minipage}{0.24\textwidth}
				$ R $-perturbation and YBE
			\end{minipage}
		};
	\end{tikzpicture}
	\begin{tikzpicture}[baseline=0cm]
		\draw [-{Stealth[slant=0]}] (0,0)--(1,0); 
	\end{tikzpicture}
	\begin{tikzpicture}[baseline=0cm]
		\node [box1] 
		(box){
			\begin{minipage}{0.20\textwidth}
				Bethe Ans\"atze and further analysis
			\end{minipage}
		};
	\end{tikzpicture}
	\caption{The boost automorphism method scheme for complete resolution of the integrable structure}
	\label{p2_BoostScheme}
\end{figure}
%
%
%
%

\label{HubDiff}
In what follows, we shall solve for novel models with 4-dim local space, where {\Rxs} exhibit difference $ R_{ij}(z_{i},z_{j}) = R_{ij}(z_{i}-z_{j})  $ spectral dependence \eqref{DiffYBE}. Generically, it is a highly nontrivial problem, since an Ansatz for $ \cl{H} $ can contain up to 256 free parameters. In order to make computationally approachable, we shall constrain such generic Ansatz to share some structure with the Hubbard model. Specifically, the first part of the analysis will include the models with {\Hal} symmetry and the second will have kinetic part in common with the Hubbard model. For the latter the algebra is represented by \textit{charge}-\textit{spin} symmetry $ \al{su}_{\cl{C}}(2) \times \al{su}_{\eta}(2) $ \cite{Heilmann:2006,Essler:1991wg}. On the other hand, the Hubbard model itself will not be considered as the part of our solution space, since its spectral structure is different \cite{Shastry:1988_DSTR_IHM,Frolov:2011wg}. We shall return to this fact during the discussion of an arbitrary spectral dependence and integrable structures arising in Gauge/Gravity duality, where one deals with the central extensions leading to superalgebra $ \al{su}(2|2) $ \cite{Beisert:2005tm,Beisert:2006qh,deLeeuw:2015ula}.

We shall perform analysis for all (sub)sectors isomorphic to {\Hal}, among them will include spin chain structure of $ \al{su}(4) $, $ \al{so}(4) $, $ \al{sp}(4) $ and $ \al{su}(2|2) $. Beyond the sectors of conventional magnetics, we find novel structure, where only fermionic pairs have dynamics and their individual dof are not present (nevertheless they do affect the spectrum). 

Another class constitutes free Hubbard deformations, which we call \textbf{G}eneralised \textbf{H}ubbard \textbf{M}odels. For GHM we consider the Hubbard type kinetic part and a generic potential up to quartic order in number operators. This setup is then extended by specific flipping and pair hopping terms. For physical consistency we demand the conservation of the fermion number, which eventually results in spin-up/spin-down decoupled models and a nontrivial model that contains two free parameters. The last one contains spin flipping behaviour (mixing of up-up and down-down states), which resembles elliptic dynamics. Specifically due to the presence of the XYZ-type terms, the inhomogeneous\footnote{also other types of CBA} coordinate Bethe Ansatz is not applicable. It is still in progress to solve this type of models by Generalised ABA or QSC. In what follows we shall now develop the structure \cite{deLeeuw:2019zsi, deLeeuw:2019vdb} of an approach and constraints necessary for the resolution of both groups of models.

\subsection{4-dim setup}\label{dim4_Setup}

In this section we define the structure necessary for the search of novel models and their properties in four dimensions. For this reason we briefly provide core steps of \ref{Method} \cite{deLeeuw:2019zsi} and details appropriate for the extended 4-dim setting. In this respect, along with restrictions required for the ansatz, we shall proceed in two-particle representation to have $ 16 \times 16 $ embedding for the YBE.

\subsubsection{$ \cl{B} $ and $ \cl{H} $} As indicated before we require boost symmetry \cite{Tetelman,Fuchssteiner_1983,Loebbert:2016cdm} as a sum of local densities
\begin{equation}\label{key}
	\cl{B}\left[ \bb{Q}_{2} \right] = \sum_{k = -\infty}^{+ \infty} k \cl{Q}_{k,k+1}
\end{equation}
where the second charge is the Hamiltonian, which now becomes a $ 16 \times 16 $ operator
\begin{equation}\label{key}
	\bb{Q}_{2} \equiv \bb{H} = \sum_{k = -\infty}^{+ \infty} \cl{H}_{k,k+1} \qquad \bb{Q}_{r+1} \sim \left[ \cl{B}\left[ \bb{Q}_{2} \right], \bb{Q}_{r} \right]
\end{equation}
In this case, one can also implement recursion for generating the charge of range $ r $. In general, it is still a requirement to have infinite commuting hierarchy $ \left[ \bb{Q}_{2}, \bb{Q}_{3} \right] = \dots = 0 $. However it can be proven at least for the classes under investigation that it is sufficient to provide the first commutator\footnote{Although it is known that in general that does not need to be true, there is a number of examples in QFT and integrable deformations when higher charges are necessary, \textit{e.g.} Zhiber-Shabat-Mikhailov model \cite{Korepin_1997_QISM_CF}}, so that along with the found {\Rx} the YBE is satisfied.

\subsubsection{Oscillator basis}
The mentioned physical setup can be conveniently formulated in terms of oscillators, which on quantum level are nothing but creation and annihilation operators that satisfy an algebra. As we have already mentioned the Hubbard kinetic and potential parts appear as
\begin{equation}\label{Hub_Model}
	\bb{H}^{\text{Hubbard}} =  \sum_i  \sum_{ \sigma = \{ \uparrow,\downarrow \} } (\cdd_{\sigma,i}\mb{c}_{\sigma,i+1} + \cdd_{\sigma,i+1}\mb{c}_{\sigma,i} )+ \fk{u} \, \mb{n}_{\uparrow,i}\mb{n}_{\downarrow,i}.
\end{equation}
where fermionic operators $ \{ \mb{c}_{\sigma,i}, \mb{c}_{\sigma,i}^{\dagger} \} $ act on $ i $th lattice site with up/down spins $ \sigma $. Accordingly the pair counting is captured by the number operators arising in the potential $ \mb{n}_{\sigma,i} = \mb{c}_{\sigma,i}^{\dagger}\mb{c}_{\sigma,i} $. The local quantum space $ \fk{h}_{i} $ spanned by four states
\begin{equation}\label{Hub_Space}
	\begin{cases}
		\ket{\phi_{1}} = \ket{0} \\
		\ket{\phi_{2}} = \mb{c}_{\uparrow,i}^{\dagger} \mb{c}_{\downarrow,i}^{\dagger} \ket{0} 
	\end{cases}
	\qquad
	\begin{cases}
		\ket{\psi_{1}} = \mb{c}_{\uparrow,i}^{\dagger} \ket{0} \\
		\ket{\psi_{2}} = \mb{c}_{\downarrow,i}^{\dagger} \ket{0} 
	\end{cases}
\end{equation}
with conventional vacuum annihilation $ \mb{c}_{\sigma,i} \ket{0} = 0 $. Because of the fermionic nature of the oscillators, we also consider the anticommutative algebra
\begin{equation}\label{Hub_Algebra}
	\{ \mb{c}_{\rho,i}^{\dagger},\mb{c}_{\sigma,j} \} = \delta_{\rho\sigma}\delta_{ij} \quad \text{and} \quad \{ \mb{c}_{\rho,i}^{\dagger},\mb{c}_{\sigma,j}^{\dagger} \} = \{ \mb{c}_{\rho,i},\mb{c}_{\sigma,j} \} = 0
\end{equation}
It could be also noted from \eqref{Hub_Space} and algebraic structure \eqref{Hub_Algebra}, that one can develop graded structure with such a prescription, i.e. if one allocates $ \{ \phi_{1}, \phi_{2} \} $ as an even sector and has fermions $ \{ \psi_{1}, \psi_{2} \} $ for the odd. With this established, one can identify boson and fermion subspaces, and the total space becomes graded $ \bb{C}^{2|2}$. For that we have graded YBE structure also in the present case \cite{Kulish1982,Essler2005_HB,batchelor2008quantum}, more on the graded algebra and properties is discussed in \ref{Apx1_Graded_Spaces_Algebra}.

\subsubsection{Integrable identifications} \label{IntId_Hub} Since the solution space in four dimensional case can be even larger, we need to implement an identification mechanism through allowed transformations. Such a procedure will again allow to create classes in the solution space and present them in concise form. Here we mention the necessary set of these transformations.
\begin{itemize}
	\item \textbf{Normalisation and Reparametrisation}: A YBE solution multiplied by a scalar function, or some solutions that contain free parameters can be mapped to other solutions by redefining these parameters, which clearly does not produce new integrable structures. \\
	\item \textbf{Local Basis Transform}: 
	\begin{equation}\label{key}
		R \xRightarrow[]{\text{LBT}} R' = \mathfrak{V} \cl{R} \mathfrak{V}^{-1} \qquad \mathfrak{V} = \cl{V} \otimes \cl{V}
	\end{equation}
	where $ \cl{V} $ acts in 4-dim $ \cl{V} : \bb{C}^{4} \rightarrow \bb{C}^{4} $.
	\item  \textbf{Discrete Transform} Transposition and permutation \eqref{DisTr} establish other maps between solutions.
	\item \textbf{Twisting}
	\begin{equation}\label{key}
		R \xRightarrow[]{\cl{W}_{1} \otimes \cl{W}_{2}} R' = \mathfrak{W} \cl{R} \mathfrak{W}^{-1} \qquad \mathfrak{W} = \cl{W}_{1} \otimes \cl{W}_{2} \quad\mathfrak{W}^{-1} = \cl{W}_{2} \otimes \cl{W}_{1}
	\end{equation}
	\begin{equation}\label{key}
		\text{along with } \left[ R, \cl{W}_{i} \otimes \cl{W}_{i} \right] = 0 \qquad \cl{W}_{i} : \bb{C}^{4} \rightarrow \bb{C}^{4}
	\end{equation}
	over both subspaces $ i = 1,2 $. It is important to remark, that in general the \ul{twists do not need to respect $ R $-matrix symmetries, hence they break them}. Moreover by means of twists one can obtain models with different spectrum (change of physical properties). However for now we shall restrict to integrable configurations that can be related by the transformations above
\end{itemize}

\newpage
\subsection{Hubbard type: $ 2 \oplus 1 \oplus 1 $ sector}

Before starting with our prescription, we need to find not only an appropriate parametric restriction due to reasons stated above, but we should also fix a suitable representation to work with. In this respect it is natural to consider that our ansatz respects {\Hal} symmetry algebra similar to Hubbard's spin-charge $ \al{su}_{\cl{C}}(2) \otimes \al{su}_{\eta}(2) $ symmetry \cite{Essler2005_HB}. It is known that there exist two representations  of interest for the current symmetry algebra.

\subsubsection{$ \cl{H}_{\text{Hub}} $ Ansatz}

We shall use the standard notation for nontrivial representations, which exhibit direct decomposition $ \fk{R} = \bigoplus_{\fk{n}} \fk{r}_{\fk{n}} $. In particular, {\Hal} possess a representation in which both copies $ \al{su}(2) $ are two dimensional, \textit{i.e.}
\begin{equation}\label{HH_211_rep}
	\fk{r}_{2\oplus1\oplus1} = 
	\begin{pmatrix}
		\fk{r}_{2} (t_{i}^{L}) &  0 \\
		0 & \fk{r}_{2}(t_{i}^{R}) \\
	\end{pmatrix}
\end{equation}
where $ t_{i} $ are generators and in the index we reflect the form of decomposition, which results in the associated block-diagonal form (more detailed on relevant structure and direct decompositions can be found in \ref{App:4dim_rep_su2}). In this framework, one can consider a two-state representation for the $ \cl{H} $ of the model
\begin{align}\label{HHub_2state_Representation}
	& \cl{H} |\phi_a \phi_b \rangle = A  |\phi_a \phi_b \rangle + B  |\phi_b \phi_a \rangle + C \epsilon_{ab}\epsilon^{\alpha\beta} |\psi_\alpha \psi_\beta \rangle \\
	& \cl{H} |\phi_a \psi_\beta \rangle = G  |\phi_a \psi_\beta \rangle + H  |\psi_\beta \phi_a \rangle,  \\
	& \cl{H} |\psi_\alpha \phi_b \rangle = K  |\psi_\alpha \phi_b \rangle + L  |\phi_b \psi_\alpha \rangle \\
	& \cl{H} |\psi_\alpha \psi_\beta \rangle = D  |\psi_\alpha \psi_\beta \rangle + E  |\psi_\beta \psi_\alpha \rangle + F \epsilon^{ab}\epsilon_{\alpha\beta} |\phi_a \phi_b \rangle.
\end{align}
which results in the conventional form
\setcounter{MaxMatrixCols}{20}
\begin{align}\label{H_su2xsu2}
	\cl{H} = 
	\tiny{
		\begin{pmatrix}
			A+B & 0 & 0 & 0 & 0 & 0 & 0 & 0 & 0 & 0 & 0 & 0 & 0 & 0 & 0 & 0 \\
			0 & A & 0 & 0 & B & 0 & 0 & 0 & 0 & 0 & 0 & F & 0 & 0 & -F & 0 \\
			0 & 0 & G & 0 & 0 & 0 & 0 & 0 & L & 0 & 0 & 0 & 0 & 0 & 0 & 0 \\
			0 & 0 & 0 & G & 0 & 0 & 0 & 0 & 0 & 0 & 0 & 0 & L & 0 & 0 & 0 \\
			0 & B & 0 & 0 & A & 0 & 0 & 0 & 0 & 0 & 0 & -F & 0 & 0 & F & 0 \\
			0 & 0 & 0 & 0 & 0 & A+B & 0 & 0 & 0 & 0 & 0 & 0 & 0 & 0 & 0 & 0 \\
			0 & 0 & 0 & 0 & 0 & 0 & G & 0 & 0 & L & 0 & 0 & 0 & 0 & 0 & 0 \\
			0 & 0 & 0 & 0 & 0 & 0 & 0 & G & 0 & 0 & 0 & 0 & 0 & L & 0 & 0 \\
			0 & 0 & H & 0 & 0 & 0 & 0 & 0 & K & 0 & 0 & 0 & 0 & 0 & 0 & 0 \\
			0 & 0 & 0 & 0 & 0 & 0 & H & 0 & 0 & K & 0 & 0 & 0 & 0 & 0 & 0 \\
			0 & 0 & 0 & 0 & 0 & 0 & 0 & 0 & 0 & 0 & D+E & 0 & 0 & 0 & 0 & 0 \\
			0 & C & 0 & 0 & -C & 0 & 0 & 0 & 0 & 0 & 0 & D & 0 & 0 & E & 0 \\
			0 & 0 & 0 & H & 0 & 0 & 0 & 0 & 0 & 0 & 0 & 0 & K & 0 & 0 & 0 \\
			0 & 0 & 0 & 0 & 0 & 0 & 0 & H & 0 & 0 & 0 & 0 & 0 & K & 0 & 0 \\
			0 & -C & 0 & 0 & C & 0 & 0 & 0 & 0 & 0 & 0 & E & 0 & 0 & D & 0 \\
			0 & 0 & 0 & 0 & 0 & 0 & 0 & 0 & 0 & 0 & 0 & 0 & 0 & 0 & 0 & D+E
	\end{pmatrix}
		}	
\end{align}
where $ \phi_{i} $ and $ \psi_{i} $ are in their fundamental representations of $ \al{su}(2) $. That is also relevant for superstring model in {\ads}, Hubbard model, Heisenberg $ \al{su}(4) $ and other. From the oscillator representation of {\Hal} \ref{App:4dim_rep_su2}, its algebra and commutativity of generic nearest-neighbour operator $ \cl{H}_{ij} $ with each copy of the $ \al{su}(2) $, we can obtain 10-parametric expression for it
\begin{align}\label{H_2xsu2}
	\cl{H}_{ij} =
	&~\sum_{ \rho \neq \sigma }\Big[ ( \cdd_{\rho,1}\mb{c}_{\rho,2} + \mb{c}_{\rho,1} \cdd_{\rho,2}) \left( \cl{A}_1 + \cl{A}_2 (\mb{n}_{\sigma,1} - \mb{n}_{\sigma,2})^2 \right) \nonumber\\
	&\qquad + ( \cdd_{\rho,1}\mb{c}_{\rho,2} - \mb{c}_{\rho,1} \cdd_{\rho,2} )\left(  \cl{A}_3(\mb{n}_{\sigma,1}-\hf) + \cl{A}_4 (\mb{n}_{\sigma,2}-\hf) \right) \Big] \nonumber \\
	&~ +\cl{A}_0 + ( \cdd_{\uparrow,1}\cdd_{\downarrow,1}\mb{c}_{\uparrow,2}\mb{c} _{\downarrow,2} +  \mb{c}_{\uparrow,1}\mb{c}_{\downarrow,1}\cdd_{\uparrow,2}\cdd _{\downarrow,2}) \cl{A}_5 +
	( \cdd_{\uparrow,1}\mb{c}_{\downarrow,1}\cdd_{\downarrow,2}\mb{c} _{\uparrow,2} +  \cdd_{\downarrow,1}\mb{c}_{\uparrow,1}\cdd_{\uparrow,2}\mb{c} _{\downarrow,2}) \cl{A}_6 \nonumber \\
	&~ + \cl{A}_7 (\mb{n}_{\uparrow,1}-\hf)(\mb{n}_{\downarrow,1}-\hf) + \cl{A}_8 (\mb{n}_{\uparrow,2}-\hf)(\mb{n}_{\downarrow,2}-\hf) + \cl{A}_9 (\mb{n}_{\uparrow,1}-\mb{n}_{\downarrow,1})^2 (\mb{n}_{\uparrow,2}-\mb{n}_{\downarrow,2})^2 \nonumber  \\
	&~ + (\cl{A}_5-\cl{A}_6) (\mb{n}_{\uparrow,1}\mb{n}_{\downarrow,1}+\mb{n}_{\uparrow,2}\mb{n}_{\downarrow,2}-1)(\mb{n}_{\uparrow,1}-\mb{n}_{\uparrow,2})(\mb{n}_{\downarrow,1}-\mb{n}_{\downarrow,2})\nonumber \\ 
	&~ +\hf \cl{A}_5 \left[ (\mb{n}_{\uparrow,1}-\mb{n}_{\downarrow,2})^2+(\mb{n}_{\downarrow,1}-\mb{n}_{\uparrow,2})^2 \right]
\end{align}
where the free parameters $ \cl{A}_{n} $ are defined through 2-state coefficients in the following way
\begin{align}
	& \cl{A}_0=\frac{1}{2}(B+G+K),\,\cl{A}_1=\frac{1}{2}(L-H),\,\cl{A}_2=\frac{1}{2}(C-F+H-L), \nonumber\\
	& \cl{A}_3=\frac{1}{2}(H+L-C-F),\,\cl{A}_4=\frac{1}{2}(C+F+H+L),\,\cl{A}_5=-B,\,\cl{A}_6=E,\nonumber\\
	& \cl{A}_7=2A+B-2K,\, \cl{A}_8=2A+B-2G,\,\cl{A}_9=A+B+D+E-G-K .
\end{align}
In what follows we shall present results of the Hubbard type solution space, and emphasize meaning of the new models.

\subsubsection{Hamiltonian solution space}

We start by implementing the steps of the scheme \ref{p2_BoostScheme}, which are modified accordingly for the current system properties. Specifically, we impose Ansatz \eqref{H_su2xsu2} and demand commuting hierarchy of the charges through recursion \cite{deLeeuw:2019zsi}. After solving the system, which leads to 45 solutions, one needs to apply transformations \ref{IntId_Hub} to extract 12 generators at the first stage \cite{deLeeuw:2019vdb}


\setlength{\tabcolsep}{0.5em} 
{\renewcommand{\arraystretch}{1.2}
	\begin{table}[h]
		\begin{center}
			\begin{tabular}{c||c|c|c|c|c|c|c|c|c|c|}
				\textbf{ Model} & \textbf{A} & \textbf{B} & \textbf{C}& \textbf{D} & \textbf{E} &\textbf{F}& \textbf{G}& \textbf{H}& \textbf{K}& \textbf{L}\\
				\hline
				\Rn{1} & $\rho$ &$ -\rho $& 0 & 0 & 0 & 0 & $a $& $\rho e^{-\phi} $& $2 \rho-a$  & $\rho e^\phi$ \\ \hline
				\Rn{2} & 0 & 0 & 0 & $\rho$ & $\rho$ & 0 & $a$ & $\rho e^{-\phi} $& $2 \rho-a$  & $\rho e^\phi$ \\ \hline
				\Rn{3} & $ \rho$ &$ -\rho $ &$ \rho e^{-\phi} $&$ -\rho$&$ \rho $&$ -\rho e^\phi$ & 0 & 0 & 0 & 0 \\ \hline
				\Rn{4} & $ \rho$ &$ -\rho $ &$ \rho e^{-\phi} $&$ \rho$&$ -\rho $&$ \rho e^\phi$ & 0 & 0 & 0 & 0 \\ \hline
				\Rn{5} & $ \frac{7}{4}\rho$ &$ -\rho $ &$ \hf\rho e^{-\phi} $&$ \frac{7}{4}\rho$&$ -\rho $&$ \hf\rho e^\phi$ & 0 & 0 & 0 & 0 \\ \hline
				\Rn{6} & 0 & 0 & 0 & $a$ & 0 & 0 & $b$ & 0 & $c$ & 0 \\ \hline
				\Rn{7} & 0 & 0 & 0 & 0 & 0 & 0 & $a$ & $b$  & $c$ & $d$ \\ \hline
				\Rn{8} & 0 & 0 & 0 & $a+c$ & 0 & 0 & $a$ & $b$ & $c$ & $d$\\ \hline
				\Rn{9} & $\rho$ &$ -\rho $& 0 & $\rho$ & $-\rho$ & 0 & $a $& $\rho e^{-\phi} $& $2 \rho-a$  & $\rho e^\phi$ \\ \hline
				\Rn{10} & $\rho$ &$ -\rho $& 0 & $\rho$ & $\rho$ & 0 & $a $& $\rho e^{-\phi} $& $2 \rho-a$  & $\rho e^\phi$ \\ \hline
				\Rn{11} & $ \rho$ &$ -\rho $ &$ \hf\rho e^{-\phi} $&$ \rho$&$ -\rho $&$ \hf\rho e^\phi$ & $\frac{3}{2}\rho $& $-\frac{3}{2}\rho $ & $\frac{3}{2}\rho$ & $-\frac{3}{2}\rho $ \\ \hline
				\Rn{12} & 0 & 0 & $ - \rho e^{-\phi}$ & 0 & 0 & $\rho  e^{\phi} $ & 0 & $\rho $ & 0 & $ -\rho $\\ \hline
			\end{tabular}
		\end{center}
		\caption{Hubbard type $ \cl{H} $ solution generators in the non-graded sector}
		\label{HHubbard_SolTable}
	\end{table}
Five of the generating solutions above appear to be novel with nontrivial properties.  We shall construct all the underlying {\Rxs}, perform quantum checks, as well as comparison and interpretation for each of the found models.

\subsubsection{{\Rxs}}\label{RHub_Construction}
To avoid misconception, the form of the {\Rx} that follows from the {\Hal} algebra in the fixed basis $ \{ \phi_{a},\psi_{\alpha} \} $ can be presented as follows
\begin{equation}
	{\tiny R_{ij} = 
		\left(
		\begin{array}{cccccccccccccccc}
			r_1+r_2 & 0 & 0 & 0 & 0 & 0 & 0 & 0 & 0 & 0 & 0 & 0 & 0 & 0 & 0 & 0 \\
			0 & r_1 & 0 & 0 & r_2 & 0 & 0 & 0 & 0 & 0 & 0 & -r_8 & 0 & 0 & r_8 & 0 \\
			0 & 0 & r_4 & 0 & 0 & 0 & 0 & 0 & r_{10} & 0 & 0 & 0 & 0 & 0 & 0 & 0 \\
			0 & 0 & 0 & r_4 & 0 & 0 & 0 & 0 & 0 & 0 & 0 & 0 & r_{10} & 0 & 0 & 0 \\
			0 & r_2 & 0 & 0 & r_1 & 0 & 0 & 0 & 0 & 0 & 0 & r_8 & 0 & 0 & -r_8 & 0 \\
			0 & 0 & 0 & 0 & 0 & r_1+r_2 & 0 & 0 & 0 & 0 & 0 & 0 & 0 & 0 & 0 & 0 \\
			0 & 0 & 0 & 0 & 0 & 0 & r_4 & 0 & 0 & r_{10} & 0 & 0 & 0 & 0 & 0 & 0 \\
			0 & 0 & 0 & 0 & 0 & 0 & 0 & r_4 & 0 & 0 & 0 & 0 & 0 & r_{10} & 0 & 0 \\
			0 & 0 & r_7 & 0 & 0 & 0 & 0 & 0 & r_3 & 0 & 0 & 0 & 0 & 0 & 0 & 0 \\
			0 & 0 & 0 & 0 & 0 & 0 & r_7 & 0 & 0 & r_3 & 0 & 0 & 0 & 0 & 0 & 0 \\
			0 & 0 & 0 & 0 & 0 & 0 & 0 & 0 & 0 & 0 & r_5+r_6 & 0 & 0 & 0 & 0 & 0 \\
			0 & -r_9 & 0 & 0 & r_9 & 0 & 0 & 0 & 0 & 0 & 0 & r_5 & 0 & 0 & r_6 & 0 \\
			0 & 0 & 0 & r_7 & 0 & 0 & 0 & 0 & 0 & 0 & 0 & 0 & r_3 & 0 & 0 & 0 \\
			0 & 0 & 0 & 0 & 0 & 0 & 0 & r_7 & 0 & 0 & 0 & 0 & 0 & r_3 & 0 & 0 \\
			0 & r_9 & 0 & 0 & -r_9 & 0 & 0 & 0 & 0 & 0 & 0 & r_6 & 0 & 0 & r_5 & 0 \\
			0 & 0 & 0 & 0 & 0 & 0 & 0 & 0 & 0 & 0 & 0 & 0 & 0 & 0 & 0 & r_5+r_6 \\
		\end{array}
		\right)}
\end{equation}	
We shall also use this form to present all the corresponding {\Rxs} for Table \ref{HHubbard_SolTable}. As was already stated above, the spectral dependence $ r \equiv r(u) $ is assumed.

We should also note that resolution for the {\Rx} in 4-dim case technically differs from the analytic procedure introduced in \ref{H-R}. Concretely, the first two orders can be fixed by regularity and $ \cl{H} $, whereas the rest can follow from the perturbative solution of the YBE. Also in our case, the second order perturbation can provide \textit{initial} restrictions, \textit{i.e.} certain relations on $ r $  and mapping up to a sign. In the next step, with such pre-conditioned ansatz for $ R $ and differentiated YBE with respect to one of the spectral parameters (along with $ u \rightarrow 0 $), it becomes possible to solve for the {\Rx}. The existence of a regular point, braiding unitarity and initial constraints following from perturbation
\begin{equation}\label{key}
	R(0) = P \qquad R_{ij}(u) R_{ji}(-u) \sim \mathds{1} \qquad R(u) = P \left( 1 + \cl{H} u + \cl{H}^{2} \dfrac{u^{2}}{2} + \cl{O}(u^{3}) \right)
\end{equation}
provides the full quantum consistency \cite{Makoto:1994JPhy1}. First of all we present the novel models, new deformations, comparisons, and eventually briefly discuss the produced known models to confirm completeness of the approach.

\subsubsection{The novel models: XXX deformations and Coupled $ \psi\psi $-models}\label{DefXXX_PsiPsi}

\paragraph{Fixing parametrisation for $ \cl{H}/R $} Before presenting the analysis of the models obtained in \ref{HHubbard_SolTable}, we need to note that it is useful to make suitable parametric changes in $ \cl{H} $ and $ R $ by allowed \textit{integrable} transforms. By doing so, we shall benefit in its further treatment and comparison analysis, it is especially relevant for \ref{CCAnalysis}. In particular for models \Rn{1},\Rn{2},\Rn{9} and\Rn{10}, the $ a $=dependence can be decomposed into to the term that vanishes on periodic systems
\begin{equation}\label{key}
	\cl{H} = \cl{H}_{1} + \left( \fk{a} \otimes \mathds{1} - \mathds{1} \otimes \fk{a} \right)
\end{equation}
where $ \cl{H}_{1} $ does not contain $ a $ and $ \fk{a} = \text{diag}\left( 1,1,1,-a,1,-a \right) $. Moreover, we can recover the $ a $-dependence by twist like transformation for the {\Rx}
\begin{equation}\label{RHub_Twist}
	R_{ij} = \left( \cl{T}(u) \otimes \mathds{1} \right) \left( R_{ij, \,a=0}(u) \right) \left( \cl{T}(-u) \otimes \mathds{1} \right) \qquad \cl{T}(u) = \text{diag}\left( 1,1,e^{a u},e^{a u} \right)
\end{equation}
analogously one could treat the $ \phi $-dependence of the Hamiltonians $ \cl{H}_{\text{\Rn{1},\Rn{2},\Rn{9},\Rn{10}}} $, \textit{i.e.}
\begin{equation}\label{HHub_Twist}
	\cl{H} \left( \phi \right) = \left( \cl{W} \otimes \mathds{1} \right) \left( \cl{H}_{\phi=0} \right) \left( \cl{W} 	\otimes \mathds{1} \right) \qquad \cl{W}(\phi) = \text{diag}\left( 1,1,e^{-\phi},e^{-\phi} \right)
\end{equation}
We can see that in this way, we can suitably fix $ \{ a,\phi \} $ and reconstruct their dependence upon demand without loss of generality. It could be also added that for models \Rn{3},\Rn{4},\Rn{5},\Rn{11} and \Rn{12} we can also isolate the $ \phi $ through similarity
\begin{equation}\label{HHub_LBT}
	\cl{H} \left( \phi \right) = \left( \cl{V} \otimes \cl{V} \right) \left( \cl{H}_{\phi=0} \right) \left( \cl{V} 	\otimes \cl{V} \right)^{-1} \qquad \cl{V}(\phi) = \text{diag}\left( 1,1,e^{-\phi},1 \right)
\end{equation}
as we have already indicated, \eqref{RHub_Twist} and \eqref{HHub_Twist} can leave the impact on the spectrum and physical interpretation, whereas basis transformation \eqref{HHub_LBT} does not.

We shall analyse classes \Rn{1}-\Rn{5} and highlight their properties here. In this respect, one can notice that classes \Rn{1} and \Rn{2} represent XXX-type deformations and classes \Rn{3}-\Rn{5} model $ \psi\psi $-pair models. 

\paragraph{Model \Rn{1}} The first class {\Rx} can be given according to \ref{RHub_Construction} as follows
\begin{equation}
	\begin{array}{l}
		r_1=-u\rho\ r_2\\
		r_2=(1-u\rho)^{-1}\\
		r_3=-e^\phi\ r_1\\
		r_6=1\\
	\end{array},\quad
	\begin{array}{l}
		r_4=-e^{-\phi}\ r_1\\
		r_7=e^{u(a-\rho)}\ r_2\\
		r_{10}=e^{u(\rho-a)}\ r_2\\
	\end{array}  
\end{equation}
as it could be noticed it is characterised by $ \{ \rho, a, \phi \} $ and in the limit $ \{ -1, -1, i \pi \} $ it becomes a version of $ \al{su}(4) $. It can be seen as follows
\begin{equation}\label{key}
	\cl{H}_{\text{\Rn{1}}} =  1 - P + \sum_{\{ i,j,k,l \} = 1,2 } \epsilon_{ij} \epsilon^{kl} E_{k}^{i} \otimes E_{l}^{j} = 1 - P + \sum_{\{ i,j \} = 1,2} \left[ E_{i}^{i} \otimes E_{j}^{j} - E_{i}^{j} \otimes E_{j}^{i} \right]
\end{equation}
where by $ E_{a}^{b} $ is a $ 4 \times 4 $ matrix with unit at $ (a,b) $ (the expression above is achieved by applying LBT, that brings basis vectors $ E^{a} $ to $ E^{5-a} $). One can also see that the second term can interpreted as XXX spin chain itself, namely
\begin{equation}\label{key}
	E_{i}^{i} \otimes E_{j}^{j} = \mathds{1}_{2} \qquad E_{i}^{j} \otimes E_{j}^{i} = P_{2}
\end{equation}
where it obviously restricts to the first two basis vectors.

\paragraph{Model \Rn{2}} For the $ R_{\text{\Rn{2}}} $ we get
\begin{equation}
	\begin{array}{l}
		r_2=1\\
		r_6=(1-u \rho)^{-1}\\
		r_5=u \rho\ r_6\\
		r_3=e^{\phi}\ r_5
	\end{array},\quad
	\begin{array}{l}
		r_4=e^{-\phi}\ r_5\\
		r_7= e^{u(a-\rho)}\ r_6\\
		r_{10}=e^{u(\rho-a)}\ r_6
	\end{array} 
\end{equation}
where in the limit $ \{ 1, 1, i \pi \} $, the model \Rn{1} can be related to \Rn{2} by grading transform. It could be written as
\begin{equation}\label{key}
	\cl{H}_{\text{\Rn{2}}} =  1 - P_{g} + \sum_{\{ i,j,k,l \} = 1,2 } \epsilon_{ij} \epsilon^{kl} E_{k}^{i} \otimes E_{l}^{j}
\end{equation}
with graded permutation $ P_{g} $. As a periodic systems, both models \Rn{1} and \Rn{2} possess ferrovacuum $ \ket{\phi_{i} \dots \phi_{i}} = \ket{0} $ and degenerate ground state. We can also note, that $ \al{su}(3) $ chain with a basis $ \{ \phi_{1}, \psi_{1}, \psi_{2} \} $ appears as a subsystem for these models. In general the spectrum depends on the twist parameters, however assuming different $ \{ \rho, a, \phi \} $ limits, one obtains distinct eigenspectra, \textit{e.g.} including similar to $ \al{su}(4) $.

\paragraph{$ \psi\psi $ models} The classes \Rn{3}-\Rn{5} correspond to models, where no isolated fermionic fields are propagating, as it can be seen from \eqref{HHub_2state_Representation} and Table \ref{HHubbard_SolTable} $ G = H = K = L = 0 $. Their {\Rxs} are found to be
\paragraph{Model \Rn{3}}

\begin{equation}
	\begin{array}{l}
		r_1=-r_5=-{\rm tan}(u\ \rho) \\
		r_2=1-r_1\\
		r_6=1+r_1
	\end{array},\quad
	\begin{array}{l}
		r_7=r_{10}=1\\
		r_8 = e^\phi\ r_1 \\
		r_9 = - e^{-\phi}\ r_1\\
	\end{array} 
\end{equation}

\paragraph{Model \Rn{4}}

\begin{equation*}\label{key}
	r_1 = r_{5} = 2+\sqrt{3} \coth \left( \sqrt{3} \rho  u+\log \left(2-\sqrt{3}\right) \right)
\end{equation*}
\begin{equation}
	\begin{array}{l}
		r_2=r_6=1-r_1\\
		r_7=r_{10}=1
	\end{array}
	\quad
	\begin{array}{l}
		r_8 = -e^\phi\ r_1 \\
		r_9 = - e^{-\phi}\ r_1
	\end{array} 
\end{equation}

\paragraph{Model \Rn{5}}

\begin{equation}
	\begin{array}{l}
		r_1=r_5 = \frac{2(e^{\frac{3 \rho  u}{2}}-1)}{e^{\frac{3 \rho  u}{2}}-4} \\
		r_2=r_6=-\frac{e^{\frac{3 \rho  u}{2}}+2}{e^{\frac{3 \rho  u}{2}}-4} \\
	\end{array}
	\quad
	\begin{array}{l}
		r_7=r_{10}=e^{-\frac{1}{4} (3 \rho  u)} \\
		r_8 = e^{-2\phi} r_9 = -\frac{1}{2}e^{\frac{3 \rho  u}{4}+\phi }\ r_1
	\end{array}
\end{equation}

So all boson-fermion scattering is suppressed, nevertheless coupled fermionic dynamics is possible. The problematics that comes up in these models, is that the magnon-form picture is not evident here, hence the issues with Bethe Ans\"atze. For example, appropriately adopted inhomogeneous nested CBA already fails at level 2, but it is important to note that not all possibilities have been tried, including generalised ABA \cite{Slavnov:2020xxj} or QSC. On the other hand, analysis on short spin chains or small number of excitations indicates the highly nontrivial eigenspectrum.

The class \Rn{3} appears as combination of Hermitian and anti-Hermitian operators, whereas \Rn{4} and \Rn{5} are Hermitian. Also \Rn{4} ground state is a ferromagnetic vacuum, on contrary, the \Rn{5} has a nontrivial ground state energy and degeneration pattern. The classes \Rn{1}-\Rn{5} are nontrivially interesting in both their generality and complete analytic resolution.

\subsubsection{Completeness and comparison analysis}\label{CCAnalysis}

The rest of the models \Rn{6}-\Rn{12} and their modified forms already appeared in the literature before. Here we only want to comment on the consistency and completeness of the method for the representation \eqref{HH_211_rep} under discussion. The model list, their structure and properties follow.

\paragraph{Model \Rn{6}} This class is diagonal, hence the integrability is trivially implied. The corresponding {\Rx}
\begin{equation}
	\begin{array}{l}
		r_2=1\\
		r_6=e^{au}\\
	\end{array}
	\quad
	\begin{array}{l}
		r_7=e^{bu}\\
		r_{10}=e^{cu}\\
	\end{array} 
\end{equation}

\paragraph{Model \Rn{7}} This class can be viewed as XXZ $ 16 \times 16 $ embedding (or quadruple embedding - a single vector for each of the doublets $ \{ \phi_{a}, 	\psi_{\beta} \}_{\al{su}(2)} $), where the Hamiltonian arises accordingly to
\begin{equation}\label{key}
	\cl{H}_{\text{\Rn{7}}} = 
	\begin{pmatrix}
		0 & 0 & 0 & 0 \\
		0 & a & b & 0 \\
		0 & d & c & 0 \\
		0 & 0 & 0 & 0 
	\end{pmatrix}
\end{equation}
it can be shown that the spectrum corresponds to XXZ (with the corresponding quadruple eigen-degeneracy). The {\Rx} appears as follows
\begin{equation}\label{key}
	\begin{array}{l}
		r_2=r_6=1\\
		r_3=\frac{a+c}{2b} (\cot (\eta ) \cot (g(u))-1)\\
		r_4=\frac{2 b}{a+c} \sin (\eta ) \csc (g(u)) \cos (g(u)+\eta )
	\end{array},\quad
	\begin{array}{l}
		r_7=e^{\frac{1}{2} u (a-c)} \cos (\eta )  \csc (g(u))\\
		r_{10}=e^{- u (a-c)}\ r_7
	\end{array} 
\end{equation}
where $ g(u) = {\rm arccot}(\tan (\eta ))-\frac{1}{2} u (a+c) \cot (\eta ) $. We can also notice that there are two singularities present, when $ b \rightarrow 0 $, then
\begin{equation}
	\begin{array}{l}
		r_2=r_6=1\\
		r_3=\frac{d \left(e^{u (a+c)}-1\right)}{a+c}\\
	\end{array}
	\quad
	\begin{array}{l}
		r_7=e^{au}\\
		r_{10}=e^{cu}
	\end{array}
\end{equation}
with introduced $ d $
\begin{equation}\label{key}
	d = \frac{(a+c)^2 \csc^2(\eta )}{4 b}
\end{equation}
or when $ (a + c) \rightarrow 0 $, then {\Rx} elements reduce to
\begin{equation}\label{R_Rn7_parsingularity_ac}
	\begin{array}{l}
		r_2=r_6=1\\
		r_3=\frac{d}{b}r_4 = \sqrt{\frac{d}{b}} \tan \left(\sqrt{b\ d}\ u\right)\\
	\end{array},\quad
	\begin{array}{l}
		r_7=e^{a u} \sec \left(\sqrt{b\ d}\ u\right)\\
		r_{10}=e^{-2au}\ r_7
	\end{array} 
\end{equation}
In principle, along $ \eta $ parameter, this analysis allows to control special points and model subcases.

\paragraph{Model \Rn{8}}
In class \Rn{8}, we have staggered XXZ structure, with the $ \cl{H}_{\Rn{8}} $

\begin{align}
	\cl{H}_{\text{\Rn{8}}}^{\text{St}} = 
	\begin{pmatrix}
		0 & 0 & 0 & 0 \\
		0 & a & b & 0 \\
		0 & d & c & 0 \\
		0 & 0 & 0 & a+c 
	\end{pmatrix}
\end{align}
and the associated {\Rx}

\begin{equation}\label{HH_R_St}
	\begin{array}{c}
		\begin{split}
			& r_2=1\\
			& r_3=\frac{2 b}{a+c} g(u)h(u) \text{sech}(\eta )\\
			& r_4 =\frac{a+c}{2 b} g(u)h(u) \cosh(\eta)\\
		\end{split}
	\end{array}
	\quad
	\begin{array}{c}
		\begin{split}
			& r_6=g(u) \sinh \left(\frac{1}{2} u (a+c) \tanh (\eta )+\eta \right)\\
			& r_7=e^{\frac{1}{2} u (a-c)}g(u) \sinh (\eta ) \\
			& r_{10} =e^{\frac{1}{2} u (c-a)}g(u) \sinh (\eta ) \\
		\end{split}
	\end{array}
\end{equation}
where by $ g(u) $ and $ h(u) $ we denote
\begin{equation}\label{key}
	\begin{cases}
		g(u) = \text{csch} \left( \eta - \frac{1}{2} u(a+c) \tanh(\eta) \right) \\
		h(u) = \sinh \left( \frac{1}{2} u(a+c) \tanh(\eta) \right)
	\end{cases}
\end{equation}
Analogously here we observe special points, hence for $ b \rightarrow 0 $ the system \eqref{HH_R_St} becomes

\begin{equation}
	\begin{array}{c}
		\begin{split}
			& r_2=1\\
			& r_3=\frac{d}{a+c} \left(e^{u (a+c)}-1\right)\\
			& r_6=e^{(a+c)u}\\
		\end{split}
	\end{array}
	\quad
	\begin{array}{c}
		\begin{split}
			& r_7=e^{au}\\
			& r_{10}=e^{cu}
		\end{split}
	\end{array}
\end{equation}
with $ d = \frac{(a+c)^2}{4b} \text{sech}^{2}(\eta) $, whereas for $ (a + c) \rightarrow 0 $ the $ d \rightarrow 0 $ and we conclude behaviour equivalent to \eqref{R_Rn7_parsingularity_ac}.

\paragraph{Model \Rn{9}} This class is nothing but the $ \al{su}(4) $ spin chain that contains an additional twist. As we have indicated in \ref{DefXXX_PsiPsi}, when parameters acquire $ \{ -1, -1, i \pi \} $ values, we exactly obtain reduction to $ \al{su}(4) $
\begin{equation}\label{key}
	\cl{H} \rightarrow 1 - P
\end{equation}

The underlying {\Rx} corresponds to
\begin{equation}
	\begin{array}{l}
		r_1=r_5=-u\rho\ r_2\\
		r_2=r_6=(1-u\rho)^{-1}\\
		r_3=-e^\phi\ r_1
	\end{array}
	\quad	
	\begin{array}{l}
		r_4=-e^{-\phi}\ r_1\\
		r_7=e^{u(a-\rho)}\ r_2\\
		r_{10}=e^{u(\rho-a)}\ r_2\\
	\end{array} 
\end{equation}

\paragraph{Model \Rn{10}} Finds the corresponding twisted graded version of the model \Rn{9}, \textit{i.e.} a similar reduction exists under $ \{ 1, 1, i \pi \} $
\begin{equation}\label{key}
	\cl{H} \rightarrow 1 - P_{g}
\end{equation}
a detailed investigation can be found in \cite{Essler2005_HB,Beisert:2005tm} and the {\Rx} can be presented in the following form
\begin{equation}
	\begin{array}{l}
		r_1=-r_5=-u\rho\ r_2\\
		r_2=r_6=(1-u \rho)^{-1}\\
		r_7=e^{u(a-\rho)}r_2
	\end{array}
	\quad
	\begin{array}{l}
		r_3=-e^{\phi}\ r_1 \\
		r_4 = - e^{-\phi}\ r_1\\
		r_{10}=e^{u(\rho-a)}r_2
	\end{array} 
\end{equation}

\paragraph{Model \Rn{11}} In model \Rn{11}, one could again identify well-known integrable model, namely the correspondence to the twisted $ \al{sp}(4) $ spin chain \cite{BERG1978125,KAROWSKI1979244,Kulish:1979cr,Reshetikhin:1986vd}. It can be given as follows

\begin{equation}\label{key}
	\cl{H}_{\text{\Rn{11}}} = \frac{3\rho}{2} 1 - (\cl{T} \otimes \cl{T}) \cl{H}_	{\al{sp}(4)} ( \cl{T} \otimes \cl{T} )^{-1}
	\qquad
	\cl{T}(\phi) = 
	\begin{pmatrix}
		1 & 0 & 0 & 0 \\
		0 & 0 & 0 & e^{\phi /2} \\
		0 & 0 & 1 & 0 \\
		0 & e^{-\phi /2} & 0 & 0 
	\end{pmatrix}
\end{equation}
where we can again obtain distinct family by twisting. The {\Rx} appears as
\begin{equation}
	\begin{array}{l}
		r_1=r_5=\rho  u (3 \rho  u-4)\ f(u) \\
		r_2=r_6= 4(1- \rho  u)\ f(u)\\
		r_3=r_4=-\frac{3}{2}u\rho\ r_7
	\end{array},\quad
	\begin{array}{l}
		r_7=r_{10}=-2( 3 \rho  u-2)^{-1} \\
		e^{2\phi}r_9=r_8= 2 \rho  u e^{\phi }\ f(u)\\
		f(u)^{-1}=(\rho  u-2) (3 \rho  u-2)
	\end{array} 
\end{equation} 
and $ R_{\al{sp}(4)} $ in the graded shifted basis
\begin{align}
	R_{\al{sp}(4)}= u \mathds{1} + P - \frac{u}{u+3}(-1)^{p(i) + p(k)} E^{i}_j\otimes E^{5-i}_{5-j}.
\end{align}
complete analysis and resolution of $ \al{so} $- and $ \al{sp} $-magnetics through BA was provided in \cite{Reshetikhin:1986vd,Li:2018xrb}.

\paragraph{Model \Rn{12}} This class reflects the kinetic part of the Hubbard model (free theory), to establish mapping we need to perform twisting. For Hamiltonian one would find
\begin{equation}\label{key}
	\cl{H}_{\text{\Rn{12}}} = -i \cl{V} \cl{H} \cl{V}^{-1} \qquad \cl{V} = \cl{V}_{0} \otimes \mathds{1} \text{, } \quad \cl{V}_{0} = \text{diag}\{ 1,-1,i,i \}
\end{equation}
and upon appropriate grading and factoring twisting out, one can relate it to the Hubbard model
\begin{equation}\label{key}
	H^{\text{Hub}} \simeq \left( \fk{T} \otimes \fk{T} \right) \cl{V} \cl{H} \cl{V}^{-1} \left( \fk{T} \otimes \fk{T} \right)^{-1} 
	\qquad 
	\fk{T} = 
	\begin{pmatrix}
		0 & e^{-\frac{\phi}{2}} & 0 & 0\\
		e^{-\frac{\phi}{2}} & 0 & 0 & 0\\
		0 & 0 & i & 0\\
		0 & 0 & 0 & i
	\end{pmatrix}
\end{equation}
one can also note here a possible separation into two XX spin chains. The {\Rx} can be presented in the following form
\begin{equation}\label{key}
	\begin{array}{l}
		r_1=r_5=r_4^2 \\
		r_2=r_6=r_7^2 \\
		r_3=-r_4=-{\rm tanh}(u\rho)
	\end{array}
	\quad
	\begin{array}{l}
		r_7=r_{10}={\rm sech}(u\rho) \\
		r_8=e^\phi\ r_4\ r_7 \\
		r_9=-e^{-\phi}\ r_4\ r_7
	\end{array} 
\end{equation}



\subsection{Chains: $ 2 \oplus 2 $ sector}\label{H_2o2sector}

By completing detailed analysis of \ref{HHubbard_SolTable}, one can immediately notice that the $ \al{so}(4) $ spin has not appeared so far on our list, or its form might have been obscure. To make it manifest, one needs to look at all nontrivial representations that arise for the {\Hal} algebra. Indeed, by looking at $ \fk{r}_{2 \oplus 2} : $ {\Hal} and applying our Ansatz accordingly to it, one explicitly finds $ \al{so}(4) $ spin chain. We can characterise the representation $ \fk{r}_{ 2 \oplus 2 } $ by $ L $-$ R $ generators
\begin{equation}\label{key}
	\begin{cases}
		\fk{r}_{ 2 \oplus 2 } (t_{i}^{L}) = \mathds{1} \otimes \fk{r}_{ 2 \oplus 2 } (t_{i}) \\
		\fk{r}_{ 2 \oplus 2 } (t_{i}^{R}) = \fk{r}_{ 2 \oplus 2 } (t_{i}) \otimes \mathds{1}
	\end{cases}
	\qquad t_{i}^{L} \times t_{i}^{R} \in {\text{\Hal}}
\end{equation}
we can further find out that associated $ \fk{r}_{ 2 \oplus 2 } $ gives rise to the following
\begin{equation}\label{key}
	\cl{H} = A + B \underbrace{E_{i}^{j} \otimes E_{j}^{i}}_{P} + C \underbrace{E_{i}^{j} \otimes E_{i}^{j}}_{Q} + D \epsilon_{ijkl} E_{k}^{i} \otimes E_{l}^{j}
\end{equation}
with permutation $ P $, trace operator $ Q $ and $ E_{i}^{j} $ unit-matrices as in \ref{DefXXX_PsiPsi}. Implementing accordingly modified procedure \ref{dim4_Setup}, we find two invariant models
\begin{equation}\label{key}
	\begin{array}{l}
		\cl{H}_{\text{\Rn{13}}} = A Q \\
		R_{\text{\Rn{13}}}= ( 1+ A u ) \left[ \frac{\sqrt{3} \coth \left(\sqrt{3} B u\right)-2}{\sqrt{3} \coth \left(\sqrt{3} B u\right)-1}P +\frac{1}{\sqrt{3} \coth \left(\sqrt{3} B u\right)-1} Q \right]
	\end{array}
\end{equation}
and
\begin{equation}\label{key}
	\begin{array}{l}
		\cl{H}_{\text{\Rn{14}}} = A - B P + B Q + C \epsilon_{ijkl} E_{k}^{i} \otimes E_{l}^{j} \\
		R_{\text{\Rn{14}}} = \frac{(1+ Au)u}{ 1 - B u } \left[ (u \left(B^2-C^2\right)-B)1 + \frac{1-Bu}{u}P +B Q + C \epsilon_{ijkl} E^i_k\otimes E^j_l \right]
	\end{array}
\end{equation}


The model \Rn{13} implies an integrable tracing on its own, however the model \Rn{14} has a $ \al{so} -$-subsector \cite{Reshetikhin:1986vd,Tu_2008}. More specifically, if we suppress one extra integrable antisymmetric contraction by $ C \rightarrow 0 $, we shall obtain regular $ \al{so}(4) $ spin chain \cite{Kulish:1979cr,Reshetikhin:1986vd}. However, generically the dependence of the spectrum on $ C $ appears nontrivial. Due to the aforementioned $ \text{\Hal} \sim \al{so}(4) $ isomorphism containing two independent Casimirs, we can establish decomposition also for \Rn{14}, \textit{i.e.}
\begin{equation}\label{key}
	\cl{H}_{\text{\Rn{14}}} = A + \cl{C}_{+}\fk{r}_{2 \oplus 2} (t_{i}^{L}) \otimes \fk{r}_{2 \oplus 2} (t_{i}^{L}) + \cl{C}_{-}\fk{r}_{2 \oplus 2} (t_{i}^{R}) \otimes \fk{r}_{2 \oplus 2} (t_{i}^{R})
\end{equation}
where summation over repeated indices is assumed and $ \cl{C}_{\pm} = 2 \left( B \pm C \right) $. Indeed, it can be noted that analogously to the regular XXX with $ \cl{H} \sim \sigma^{i} \otimes \sigma^{i} $, the $ \al{so}(4) $ chain does possess XXX-decomposition with the corresponding each copy contribution to the spectrum.

\newpage
\subsection{Generalised Hubbard sector}\label{Hub_GHC}

We shall discuss novel deformation and generlisations of the Hubbard type chains. As we have already discussed, the Hubbard model consists of the kinetic and potential parts \eqref{Hub_Model}. On the other hand, one think of variety of deformations/generalisations when perturbing the free Hubbard (kinetic), as observed for model \Rn{12}.
\begin{equation}\label{FreeHub_Model}
	\bb{H}_{\text{Hub}} = \sum_{ i, \, \sigma = \{ \uparrow,\downarrow \} } (\cdd_{\sigma,i}\mb{c}_{\sigma,i+1} + \cdd_{\sigma,i+1}\mb{c}_{\sigma,i})
\end{equation}
It is obvious that such family does not obey \Hal, but our question would include a class that admits certain Hubbard properties, hence Hubbard \textit{type}. Physically that would be one-dimensional systems of electrons that propagate on the lattice (or conduction band).

In this respect, we can create a generic ansatz with additional allowed terms, and ask if there is a new emerging integrable structure after imposing \ref{dim4_Setup}. The first possibility is the same footing action on pair of electrons, \textit{i.e. hopping terms}. This kinetic sector would include hopping of an fermionic pair
\begin{equation}\label{key}
	\cl{H}_{\text{PH}} = A_1 \cdo_{\uparrow,1}\cdo_{\downarrow,1}\co_{\uparrow,2}\co_{\downarrow,2} + A_2 \cdo_{\uparrow,2}\cdo_{\downarrow,2}\co_{\uparrow,1}\co_{\downarrow,1}
\end{equation}
and the one that performs flipping of spins of adjacent fermions
\begin{equation}\label{key}
	\cl{H}_{\text{Flip}} = A_3 \cdo_{\uparrow,1}\cdo_{\downarrow,2}\co_{\downarrow,1}\co_{\uparrow,2} + A_4 \cdo_{\downarrow,1}\cdo_{\uparrow,2}\co_{\uparrow,1}\co_{\downarrow,2}
	+  A_5 \cdo_{\uparrow,1}\cdo_{\uparrow,2}\co_{\downarrow,1}\co_{\downarrow,2} + A_6 \cdo_{\downarrow,1}\cdo_{\downarrow,2}\co_{\uparrow,1}\co_{\uparrow,2}
\end{equation}
although we are considering fermion number conserving setting, the \textit{flipping} does violate spin conservation since it can mutually map $ \ket{\uparrow\uparrow} \leftrightarrow \ket{\downarrow\downarrow}$. After all we provide prescription for the generic potential of the form
\begin{equation}\label{key}
	\begin{array}{l}
		V = B_1 + B_2\, \no_{\uparrow,1} + B_3\, \no_{\downarrow,1}  + B_4\, \no_{\uparrow,1} \no_{\downarrow,1}  + \\
		B_5\,\no_{\uparrow,2}  + B_6\, \no_{\uparrow,1} \no_{\uparrow,2}  + B_7 \,\no_{\downarrow,1} \no_{\uparrow,2}  + B_8\, \no_{\uparrow,1} \no_{\downarrow,1} \no_{\uparrow,2} + \\
		B_9\,\no_{\downarrow,2}  + B_{10} \,\no_{\uparrow,1} \no_{\downarrow,2}   + B_{11}\, \no_{\downarrow,1} \no_{\downarrow,2}   + B_{12}\, \no_{\uparrow,1} \no_{\downarrow,1} \no_{\downarrow,2} + \\
		B_{13}\,\no_{\uparrow,2} \no_{\downarrow,2}   + B_{14}\, \no_{\uparrow,1} \no_{\uparrow,2} \no_{\downarrow,2}    + B_{15}\, \no_{\downarrow,1} \no_{\uparrow,2} \no_{\downarrow,2}     + B_{16}\, \no_{\uparrow,1} \no_{\downarrow,1} \no_{\uparrow,2} \no_{\downarrow,2} 
	\end{array}
\end{equation}
So that eventually we obtain
\begin{equation}\label{key}
	\cl{H} = \cl{H}_{\text{Hub}} + \cl{H}_{\text{PH}} + \cl{H}_{\text{Flip}} + V 
\end{equation}
where for the first term we assumed nearest-neighbour Hubbard kinetic term. Current ansatz results in total of 22 free parameters.

Following the prescription and resolving integrable hierarchy \cite{deLeeuw:2019vdb}, we getting four solutions of difference (additive) form \ref{HubDiff}. Clearly conventional Shastry type construction is not among these solutions as it does not respect spectral difference form. In what follow Hamiltonians with the nontrivial potential

\begin{align}
	\cl{H}_{\text{\Rn{15}}} & = \cl{H}_{\text{Hub}}  + a_1(\no_{\uparrow,1}+\no_{\uparrow,2}) + a_2(\no_{\uparrow,1}-\no_{\uparrow,2})+ a_3(\no_{\downarrow,1}+\no_{\downarrow,2}) +
	a_4(\no_{\downarrow,1}-\no_{\downarrow,2}) \label{GHM_nzV1}\\
	\cl{H}_{\text{\Rn{16}}} & = \cl{H}_{\text{Hub}}  + a_1(\no_{\uparrow,1}-\no_{\uparrow,2})^2 + a_2(\no_{\uparrow,1}-\no_{\uparrow,2})+ a_3(\no_{\downarrow,1}+\no_{\downarrow,2}) +
	a_4(\no_{\downarrow,1}-\no_{\downarrow,2}) \label{GHM_nzV2} \\
	\cl{H}_{\text{\Rn{17}}} & = \cl{H}_{\text{Hub}}  + a_1(\no_{\uparrow,1}-\no_{\uparrow,2})^2 + a_2(\no_{\uparrow,1}-\no_{\uparrow,2})+ a_3(\no_{\downarrow,1}-\no_{\downarrow,2})^2 +
	a_4(\no_{\downarrow,1}-\no_{\downarrow,2}) \label{GHM_nzV3}
\end{align}
it turned out that there are no classes that would contain a nontrivial pair hopping $ \cl{H}_{\text{PH}} $ term, assuming all the above properties to hold. These classes separate spin wise $ \cl{H} = \cl{H}_{\downarrow} + \cl{H}_{\uparrow} $.

\paragraph{Novel class \Rn{18}} The next model develops nontrivial spin flipping and potential, as it does not preserve spins orientations, it could be associated to an elliptic type deformation of the \eqref{Hub_Model} potential
\begin{equation}\label{key}
	\begin{array}{l}
		\cl{H}_{\text{\Rn{18}}} = \cl{H}_{\text{Hub}} + a \left( \cdo_{\uparrow,1}\cdo_{\downarrow,2}\co_{\downarrow,1}\co_{\uparrow,2} +  \cdo_{\downarrow,1}\cdo_{\uparrow,2}\co_{\uparrow,1}\co_{\downarrow,2}
		+   \cdo_{\uparrow,1}\cdo_{\uparrow,2}\co_{\downarrow,1}\co_{\downarrow,2} + \cdo_{\downarrow,1}\cdo_{\downarrow,2}\co_{\uparrow,1}\co_{\uparrow,2} \right)
		+ \\
		(2a-b) (\no_{\uparrow,1}+\no_{\downarrow,1}) +b(\no_{\uparrow,2}+\no_{\downarrow,2}) - a  (\no_{\uparrow,1}+\no_{\downarrow,1}) (\no_{\uparrow,2}+\no_{\downarrow,2})
	\end{array}
\end{equation}
and the associated {\Rx} appears as
\begin{equation}\label{key}
	R_{\text{\Rn{18}}} = \phi_{0}(u)
	{\tiny{
	\begin{pmatrix}
		r_1 & 0 & 0 & 0 & 0 & 0 & 0 & 0 & 0 & 0 & 0 & 0 & 0 & 0 & 0 & 0 \\
		0 & r_2 & 0 & 0 & r_6 & 0 & 0 & 0 & 0 & 0 & 0 & -r_8 & 0 & 0 & r_8 & 0 \\
		0 & 0 & r_3 & 0 & 0 & 0 & 0 & 0 & r_7 & 0 & 0 & 0 & 0 & 0 & 0 & 0 \\
		0 & 0 & 0 & r_3 & 0 & 0 & 0 & 0 & 0 & 0 & 0 & 0 & r_7 & 0 & 0 & 0 \\
		0 & r_{12} & 0 & 0 & r_2 & 0 & 0 & 0 & 0 & 0 & 0 & -r_9 & 0 & 0 & r_9 & 0 \\
		0 & 0 & 0 & 0 & 0 & r_1 & 0 & 0 & 0 & 0 & 0 & 0 & 0 & 0 & 0 & 0 \\
		0 & 0 & 0 & 0 & 0 & 0 & -r_3 & 0 & 0 & q_{7} & 0 & 0 & 0 & 0 & 0 & 0 \\
		0 & 0 & 0 & 0 & 0 & 0 & 0 & -r_3 & 0 & 0 & 0 & 0 & 0 & q_{7} & 0 & 0 \\
		0 & 0 & q_{7} & 0 & 0 & 0 & 0 & 0 & r_3 & 0 & 0 & 0 & 0 & 0 & 0 & 0 \\
		0 & 0 & 0 & 0 & 0 & 0 & r_7 & 0 & 0 & -r_3 & 0 & 0 & 0 & 0 & 0 & 0 \\
		0 & 0 & 0 & 0 & 0 & 0 & 0 & 0 & 0 & 0 & r_4 & 0 & 0 & 0 & 0 & r_{10} \\
		0 & r_9 & 0 & 0 & r_8 & 0 & 0 & 0 & 0 & 0 & 0 & r_5 & 0 & 0 & r_{11} & 0 \\
		0 & 0 & 0 & q_{7} & 0 & 0 & 0 & 0 & 0 & 0 & 0 & 0 & r_3 & 0 & 0 & 0 \\
		0 & 0 & 0 & 0 & 0 & 0 & 0 & r_7 & 0 & 0 & 0 & 0 & 0 & -r_3 & 0 & 0 \\
		0 & -r_9 & 0 & 0 & -r_8 & 0 & 0 & 0 & 0 & 0 & 0 & r_{11} & 0 & 0 & r_5 & 0 \\
		0 & 0 & 0 & 0 & 0 & 0 & 0 & 0 & 0 & 0 & r_{10} & 0 & 0 & 0 & 0 & r_4 \\
	\end{pmatrix}
	}}
\end{equation}
with the functions defined
\begin{equation}
	\begin{array}{l}
			r_1=\cos (\theta +u \cos (\theta )) \\
			r_2 = \frac{\displaystyle r_3^2}{\displaystyle r_1} \\
			r_3=\sin (u \cos (\theta )) \\
			r_4=\cos (\theta ) \cos (u \cos (\theta )) \\
	\end{array}
	\qquad
	\begin{array}{l}
		r_5=-\cos (\theta ) \tan (\theta +u \cos (\theta ))r_3\\
		r_6 = \frac{\displaystyle r_7^2}{\displaystyle r_1}\\
		r_7=\cos (\theta ) e^{u (a_2+\sin (\theta ))}\\
		r_8=\frac{\displaystyle r_7}{\displaystyle r_1} r_3
	\end{array}
\end{equation}
\begin{equation}\label{key}
	\begin{array}{l}
		r_9=\frac{\displaystyle q_{7}}{\displaystyle r_1} r_3\\
		r_{10}=\sin (\theta ) r_3 \\
		r_{11}=\frac{1}{4} (\cos (2 \theta )-\cos (2 u \cos (\theta ))+\cos (2 (\theta +u \cos (\theta )))+3)\ r_1^{-1}\\
		r_{12}= \frac{\displaystyle q_{7}^2}{\displaystyle r_1} \\
	\end{array}
\end{equation}
where $ q_{7}(u) = r_{7}(-u) $, $ \phi_{0} = \left( 2 a_{2} u + 1 \right) \sec\left( \theta + u \cos \theta \right) $ and braiding unitarity arises in the following form
\begin{equation}\label{key}
	R_{ij}(u)R_{ji}(-u) = 1 - 4 a_{2}^{2}u^{2}
\end{equation}

It could be noticed that current model possesses two free parameters and obviously is not spin-decomposable as \eqref{GHM_nzV1}-\eqref{GHM_nzV3}. For such spin violating model would be to important to investigate its spectral limits or a potential phase diagram. To current knowledge, there is no good prescription for completely resolving this and related classes. However there is a proposal that can be implemented for the systems containing spin-mixing contributions and result in the elliptic type deformations. Due to the absence of the fixed reference state or an established vacuum the standard techniques for the spectra computations are not applicable. Although a similar problem of the XYZ Heisenberg chain has been resolved by virtue of the Quantum Inverse Scattering Method, for the classes above it requires a separate prescription and it is of progress now to resolve this and related models in 2-dim by deriving generalised ABA \cite{Fontanella:2017rvu,Slavnov:2020xxj} or QSC for the spin chains with 4-dim local spaces.


\newpage

\section{Non-difference sector and generalisation for strings}\label{NA_S_SD}

In this section we shall continue with the development of novel integrable structures, generalisation of the technique \ref{Method} and resolution of related questionsthat also appear important for other integrable sectors. As we have already partially noticed, the braided classical/quantum structures along with associated group theoretic methods play a key role in the integrable characteristics of the underlying theories \cite{baxter1972partition,Hubbard_1965RSPSA,jimbo1990yang,Perk2006Yang}. Beyond previously discussed 2-dim and 4-dim models, \textit{e.g.} vertex models, noncompact sector \cite{deLeeuw:2019zsi}, Hubbard model \cite{deLeeuw:2019vdb} and other, there are integrable structures that underlie higher energy sector. A number of problems in this area could be also addressed and solved by means of integrability on the lattice. Significant progress in this context has been achieved in $ N=4 $ SYM \cite{Beisert_2004}, $ AdS_{n} $ integrable backgrounds \cite{Metsaev:1998it,Beisert:2005tm} and integrable properties of Gauge/Gravity duality \cite{Maldacena_1997,Gubser_1998,Witten_1998,Beisert:2010jr} in general. In particular, the spin chain picture and existence of integrable constraints provides a tool for investigation of string integrable models. It will be one of our goals in this section to develop an approach for finding new integrable structures relevant for string worldsheet theories on different backgrounds and perform further studies in this sector.

\subsection{Method generalisation}\label{ND_Technique}
\subsubsection{Setting}

We have discussed in the previous two sections that a special division of models includes the ones that obey regularity, \textit{i.e.} $ R_{i,i+1}(u,u) = P_{i,i+1} $, where $ P_{i,i+1} $ interchanges two local spaces. It is a crucial property of many classes of models, since it implies the existence of special spectral point and underlies shifting dynamics in the lattice models. Generically, the {\Rx} can have undetermined dependence type on the spectral parameter, and then attention is required for the existence of special points and spectral limits.

On the other side, a number of successful methods and unified formalisms that have been developed in the past to address the solution space of the qYBE. An important argument stems from the symmetries of the objects describing the associated integrable classes \cite{Kulish:1981gi,Kulish1982,Jimbo1986}. In this respect, it is natural to demand that a symmetry represented by Lie group element $ \fk{g} \in \cl{G} $ that is commutative with the Hamiltonian, also commutes with the corresponding {\Rx}
\begin{equation}\label{key}
	\left[ \fk{g} \otimes \mathds{1} + \mathds{1} \otimes \fk{g}, R \right] \equiv 0
\end{equation}
One can also determine the structure of $ R(u,v) $ through Hopf \cite{Drinfeld:1985rx,Drinfeld:1987sy} algebraic constraints (\ref{Apx1_Hopf_Algebra}). More specifically, by considering a certain bialgebra $ \cl{G}_{B} $, we can obtain a cocommutativity constraint
\begin{equation}\label{p5_Cocommutativity_1}
	R\Delta(\fk{g}) = \Delta^{\tx{op}}(\fk{g}) R
\end{equation}
where $ R \equiv R(u,v) $, $ \Delta(\cdot) $ is the coproduct and $ \Delta^{\tx{op}}(\cdot) $ is the opposite coproduct, which close under permutation operation in $ \cl{G}_{B} $. In many cases the constraint \ref{p5_Cocommutativity_1} allows to determine the structure of the {\Rx} up to a few free functions and factors, \textit{e.g.} the constraining the $ R $-/$ S $-matrix of $ AdS_{n} $ integrable models \cite{Beisert:2005tm,Borsato:2013qpa,Borsato:2014exa,Borsato:2014hja,Borsato:2015mma,Hoare:2014kma,Sfondrini:2014via}. Without doubt such a prescription requires knowledge of the underlying symmetry, which in a number of cases might not be known or only a part of it might be fixed, or not be present for certain operators. Another important related direction is the construction of the Yang-Baxter solution through the algebra representations, which originates from the works of Drinfeld \cite{Drinfeld:1985rx,Drinfeld_1986DAHAY,Drinfeld:1987sy}. In fact, this representation generating approach, \textit{i.e.} \textit{Baxterisation} \cite{Jimbo:1985vd,cheng1991yang,zhang1991representations,li1993yang,Arnaudon:2002tu,Isaev:1995cs,Kulish:2008a,Kulish:2009cx,Crampe:2016nek,Crampe:2020slf} is strongly intertwined with algebraic structure emerging in knot theory (including Hecke, Temperly-Lieb algebras) \cite{Turaev:1988eb,Jones:1989ed,jones1990baxterization,Wu_1993}.

From the other side, solving the qYBE itself involves cubic functional systems, which in majority of cases is a challenging task. However also here there is a region for which one can implement differentiation and related $ RTT $ arguments and capture the solution space for specific classes. Though even in these cases one might end up with coupled non-linear PDE system, which analytically may not be solvable. Nevertheless there are interesting examples of applying this approach along with reducing constraints, \textit{e.g.} $ 4 \times 4 $ classification of 8-vertex and $ 9 \times 9 $ class below that obeys the ice-rule \cite{Vieira:2017vnw,Vieira:2019vog,Vieira:2020lem}. It is clear that there are objective limitations, since the more free parameters (functions in the non-additive case) the higher the complexity of problem becomes, and one needs to address with distinct techniques.

As we know the integrability constraints could be provided in many ways. However in the present discussion and related ones, it is necessary and sufficient to consider the main quantum constraint - YBE. Undoubtedly, there are other integrable constraints or associated algebras that might appear more convenient, but for now we can keep in mind qYBE and $ RTT $-algebra as the sufficient conditions. As we have established above, the {\Rx}
\begin{equation}\label{key}
	R: \bb{V} \otimes \bb{V} \rightarrow \bb{V} \otimes \bb{V} \qquad R_{ij} \in \text{End}\left( \bb{V} \otimes \bb{V} \otimes \bb{V} \right)
\end{equation}
\begin{equation}\label{YBE_ArbDep}
	R_{12}\left( z_{1}, z_{2} \right) R_{13}\left( z_{1}, z_{3} \right) R_{23}\left( z_{2}, z_{3} \right) = R_{23}\left( z_{2}, z_{3} \right) R_{13}\left( z_{1}, z_{3} \right) R_{12}\left( z_{1}, z_{2} \right) \quad R_{12}(z,z) = P_{12}
\end{equation}
contains some $ z $-dependence, which in general does not need to be of spectral difference/additive form. It is indeed the case, in string integrable sector, \textit{e.g.} Shastry type models, Hubbard chain in {\ads} \cite{Beisert:2008tw,Beisert_2012}, integrable structure on {\adso} and {\adst} backgrounds \cite{Hoare:2014kma,Delduc:2014kha} and other. 

Once the scattering operator {\Rx} or {\Sx} is obtained, one should be able to obtain the spectrum of conserved charges through the monodromy and its expansion
\begin{equation}\label{p3_twy_Rwy}
		t(w, \fk{y}) = \text{Tr}_{a} \left[ R_{aL} \dots R_{a1} \right] \qquad R_{a,i} \equiv R_{a,i}(w,\fk{y})
\end{equation}
\begin{equation}\label{key}
			\log t = \sum_{n = 0}^{+\infty} \bb{Q}_{n+1} \dfrac{(\fk{y} - w)^{n}}{n!}
\end{equation}
where we have extra auxiliary parameter $ \fk{y} $ and a physical $ w $. Commutativity of transfer matrices in general imposes
\begin{equation}\label{key}
	\left[ \bb{Q}_{a}, \bb{Q}_{b} \right] = 0
\end{equation}
which at the same time defines and integrable hierarchy structure. Let us note, similarities and distinctions in the properties of underlying integrable structure and the method that has been developed above \ref{Method}. As we have stated, due to arbitrary spectral dependence setup, in general we shall have extra auxiliary dependence in the conserved charges, that also could be ruled out from \eqref{Q_12}. Namely, now the conserved charges are parametrically dependent, since there is an arbitrary spectral dependence in the {\Rx} \eqref{p3_twy_Rwy}.

Specifically, to describe range two acting operator, we can introduce the nearest-neighbour Hamiltonian densities and define
\begin{equation}\label{ND_Q2}
	\bb{Q}_{2}(w) = \sum_{k} \cl{H}_{k,k+1}(w)
\end{equation}
whereas for the next charge we get through the modified boost automorphism
\begin{equation}\label{key}
	\text{range 3:} \qquad 	\bb{Q}_{3}(w) = \sum_{k} \cl{Q}_{k,k+1,k+2}(w) 
\end{equation}
\begin{equation}\label{Q3_Boost}
	\bb{Q}_{3}(w) = -\sum_{k} \left[ \cl{H}_{k,k+1}(w),\cl{H}_{k+1,k+2}(w) \right] + \partial_{w} \bb{Q}_{2} (w)
\end{equation}
more detailed on emergent differential term can be found in \ref{Apx_A_NA_Boost_Automorphic_Symmetry}. We shall now omit the auxiliary dependence for compactness. Important to remark, that before by $ k $ we have expressed appropriate well-defined sums over the sites of a spin chain. Here we shall consider the 4-site embedding from the beginning. One of the reasons for that is the modified boost action does affect the first commutator $ \left[ \bb{Q}_{2}, \bb{Q}_{3} \right] $ via the third charge generation. Namely, beyond the standard periodicity of the chain, one must apriori consider the corresponding length embedding that is consistent with the associated commutator. Otherwise lower embedding prescription will wind in such a periodic setup and part of the information will be missing due to cancellations. Hence in general the operatorial embedding must correspond to length $ r_{1} + r_{2} - 1 $ for some commutator $ \left[ \bb{Q}_{r_{1}}, \bb{Q}_{r_{2}} \right] $.

So in this prescription instead of addressing the integrable structure through the qYBE or related constraints, we can start with second conserved charge and proceed with generation of the full hierarchy of charges (however the higher commutators would not add extra information in the studied cases). In this bottom-up approach we have functional Ansatz on a Hamiltonian $ \cl{H}_{i,i+1} $, which will generate $ \bb{Q}_{3} $. Given above, we can impose the vanishing of $ \left[ \bb{Q}_{2}, \bb{Q}_{3} \right] $ and address the resolution of the coupled ODE on $ h_{i}(w) $. Once the solutions are found, it is necessary to find the underlying {\Rxs}, which in the existing setting might be a challenging task. Along the mentioned periodic condition $ \cl{H}_{L,L+1} \equiv \cl{H}_{L,1} $, we also know that Hamiltonian could be recovered
\begin{equation}\label{ND_H_R}
	\cl{H}_{12}(u) = P_{12} \left( \partial_{\fk{y}} R(u,\fk{y}) \right) \vert_{\fk{y} \rightarrow u}
\end{equation}	
where we meant the adjacent sites for concreteness. On the other hand, we can work out a coupled system on {\Rxs} via differentiation from qYBE. Taking into account regularity $ R_{12}(z,z) = P_{12} $, existence and consistency of spectral limits and commutativity one endsz up with
\begin{equation}\label{SutherlandEq_1}
	\left[ R_{13} R_{23}, \cl{H}_{12}(u) \right] = R_{13, v}R_{23} - R_{13}R_{23, v}
\end{equation}
\begin{equation}\label{SutherlandEq_2}
	\left[ R_{13} R_{12}, \cl{H}_{23}(v) \right] = R_{13}R_{12, u} - R_{13, u}R_{12}
\end{equation}
where $ R_{ij} \equiv R_{ij}(u,v) $, $ R_{ij, x} \equiv \dfrac{\partial}{\partial x} R_{ij} $ and $ \left( R_{ij, v} \right)\vert_{v \rightarrow u} = P_{ij} \cl{H}_{ij}(u) $. Eventually, one can notice that such a system follows from the Sutherland equation \cite{Sutherland:1970,Sutherland:1975mcqs}. Importantly that initial conditions for this coupled system come from regularity and the expansion of the {\Rx} (with identifiable charges in perturbation).

It is worth noting, that if the {\Rx} of difference form is provided, then Hamiltonians $ \cl{H} $ do not exhibit parametric dependence anymore and the boost, which in non-difference form is given as
\begin{equation}\label{ND_Boost}
	\cl{B}\left[ \bb{Q}_{2} \right] = \sum_{k = -\infty}^{+ \infty} k \cl{Q}_{k,k+1} (w) + \partial_{w}
\end{equation}
\begin{equation}\label{key}
	\bb{Q}_{r+1} \simeq \left[ \cl{B}, \bb{Q}_{r} \right] 
\end{equation}
for the difference spectral dependence reduces to the regular one \eqref{p1_Boost_Automorphism}, \textit{i.e.} all derivatives are suppressed. Namely, \eqref{Q3_Boost} and the first commutator will reduce back to a polynomial system, whereas \eqref{SutherlandEq_1}-\eqref{SutherlandEq_2} will reduce to nonlinear differential system on single spectral parameter. The automorphism expression \eqref{ND_Boost} should be taken as formal infinite series in postulated sense. Since in this prescription as we have shown in Sec. \ref{Method}, it is consistent on the finite spin chains as long as our local charges exhibit finite interaction range and their length does not exceed the length of the spin chain, more can be found in Appendix \ref{Apx_A_NA_Boost_Automorphic_Symmetry}).

Important to note that bottom-up approaches, \textit{i.e.} reconstructing the {\Rxs} form the associated Hamiltonians have developed some time ago, especially for the system, where the $ R(u,v) \equiv R(u-v) $. In a number of recent works \cite{Crampe:2013nha,Fonseca:2014mqa,Crampe:2016she}, the 3-state systems have been resolved exactly by the \textbf{C}oordinate \textbf{B}ethe \textbf{A}nsatz \cite{Bethe_1931_theorie} provided the Hamiltonian and constructed some of the {\Rxs} from them. There is also a recursive reconstruction for difference form {\Rxs} \cite{Idzumi:1994kx,Martins:2013ipa,Martins:2015ufa}. For the case of 2-dim integrable field theories, or theories on the lattice, where scattering operators are manifestly of difference form, the differential system above reduces to an algebraic one. A number of new integrable classes have been found in the latter context for 2- and 4-dimensional settings \cite{deLeeuw:2019zsi,deLeeuw:2019vdb}.

From now on we shall demonstrate the computations for the systems, whose {\Rxs} exhibit generic spectral dependent Ansatz. In what follow we shall be analysing how much of the continues to hold, without actual reduction to the difference form. Also here we shall demonstrate the leading commutativity saturation conjecture \cite{Grabowski}. In particular, for 2-dim case we shall consider completely generic Ansatz, but outline the novel 6- and 8-vertex classes, which will be a cornerstone for our discussion in the sections \ref{FF_Section}, \ref{p5_Sec_AdS_2_3_Integrable_Deformations}. Important that within 2-dim case there are also models beyond 8-vertex, which form upper/lower-triangular forms and conjecturaly are non-difference extension of multi-parameter of the $ \al{sl}_{2} $ sector. The last still requires a separate investigation of structure, however a proposal of modified QISM is made. For the 15-vertex three-state setting there will a restricting requirement that the {\Rx} respects the Cartan subalgebra of the $ \al{su}(3) $ ($ \supset \al{u} \oplus \al{u}(1) \oplus \al{u}(1) $). In the case of four dimensions we shall impose that {\Hal} forms a part of the underlying {\Rx} symmetry, which is an important constituent in a number of models, \textit{e.g.} Heisenberg $ \al{su}(4) $ spin chain, Hubbard model \cite{Hubbard_1965RSPSA,Essler2005_HB}, Shastry model \cite{Shastry_1986,Shastry:1988_DSTR_IHM}, the {\ads} Hubbard chain \cite{Beisert:2005tm,Beisert:2006qh,Beisert:2008tw,Beisert:2011wq,Beisert:2015msa}. We shall also consider relevant Hubbard type models that preserve the fermion number and models with an extended structure. Although these symmetry restrictions are not necessary for our method to be implemented and resolved on generic grounds, it appears useful to have them to constrain for integrable classes with important physical properties. Otherwise one might end up with a very large number of classes, hence unreasonably raising the algebro-differential complexity, but only a part of them will appear of interest \cite{deLeeuw:2019zsi,deLeeuw:2019vdb,deLeeuw:2020ahe}. We shall now describe the integrable identification transforms, necessary to create a structure space of solutions.

\subsubsection{Identification transformations}\label{ND_IT}

We provide an analogous identification scheme, but now relevant for \textit{non-difference} \ref{ND_Boost} solution space, which is also important for structure analysis, properties and mappings for the found models. The transformations are considered for {\Rx} and $ \cl{H} $, which will create a grouping scheme into classes \cite{deLeeuw:2019zsi,deLeeuw:2019vdb,deLeeuw:2020ahe}.

\begin{itemize}	
	\item \textbf{Parametrisation} \quad From the arbitrary spectral dependence, we can see that free parametrisation results in the spectral redefinition (\textit{e.g.} XXX parametrisations for the {\Rx} \cite{Stouten:2017azy,Stouten:2018lkr}). More specifically, one can notice that not only $ R(u,v) $ is a solution of the YBE, but also a functionally reparametrised $ R(f(u),f(v)) $. However it is important to comment that such operation can affect $ \cl{H} $ a.k.a. the {\Rx} reduction \eqref{ND_H_R}
	\begin{equation}\label{key}
		\cl{H} \xrightarrow{f} \cl{H}' = \partial_{u}f \, \cl{H}(f(u))
	\end{equation}
	It could be also noted that in this way we can apply a free reparametrisation of any spectral function that is involved in $ \cl{H}/R $, and one must bring attention according to how the derivative in the boost \eqref{ND_Boost} is involved.
	
	\item \textbf{Normalisation} \quad The normalisation for the {\Rx} can be achieved by the multiplication of an arbitrary function $ f $, whereas for $ \cl{H} $ we can perform a shift $ \cl{H} \xRightarrow[]{\text{Shift}} \cl{H}' = \cl{H} + \partial_{u}f \cdot \mathds{1} $. To preserve regularity, one can fix $ f(u,u) \equiv 1 \text{: } R(u,u) = P $.
	
	\item  \textbf{Discrete Transform} \quad Transposition and permutation
	\begin{equation}\label{key}
		R^{T}, \quad P R(u,v) P, \quad PR^{T}P \quad R \equiv R(u,v)
	\end{equation}
	constitutes solutions of YBE and establishes mapping among them, analogous to \eqref{DisTr}.
	
	\item \textbf{LBT} \quad The Local basis transformation here can represented by the invertible operator $ \cl{V} $ of an appropriate dimension $ \cl{V} : \bb{C}^{n} \rightarrow \bb{C}^{n} $ ($ n $ will appear according to the framework discussed). However we should allow for some spectral dependence for $ \cl{V} $, so the density transformation goes under
	\begin{equation}\label{ND_LBT}
		\cl{H} \xRightarrow[]{\text{LBT}} \cl{H}' = (\cl{V} \otimes \cl{V}) \cl{H} (\cl{V} \otimes \cl{V})^{-1} + \left[ 1 \otimes (\partial_{s}\cl{V})\cl{V}^{-1} - (\partial_{s}\cl{V})\cl{V}^{-1} \otimes 1 \right]
	\end{equation}
	where by $ \partial_{s} $ differential w.r.t. some spectral parameter is assumed, and consequently evaluating the whole expression at some single parameter $ w $, as indicated in \eqref{ND_Q2}.We can also note that cyclic terms $ \cl{F} \otimes 1 - 1 \otimes \cl{F} $ can be cancelled/recast upon demand by $ \partial_{s}\cl{V} = \cl{F} \cl{V} $. It is analogous for the {\Rx},
	\begin{equation}\label{key}
		R \xRightarrow[]{\text{LBT}} R' = \fk{V} R \fk{V}^{-1} \qquad \mathfrak{V} = \cl{V}(u) \otimes \cl{V}(v)
	\end{equation}
	which respects spectral limits and regularity.

	\item \textbf{Twisting} \quad As it was mentioned above, in comparison to transformations above, which do not significantly affect the physical nature of the models, the twists involve change in the spectrum and can even change physical interpretation. Hence twists are called to cause model dependent changes. In current setting, the twisting of {\Rx} can be defined as follows
	\begin{equation}\label{ND_RTwist}
		R \xRightarrow[]{\{ \fk{G}, \fk{W} \}} R' = \fk{G} R \fk{W}^{-1} \qquad \fk{G}(u) = \cl{G}_{1}(u) \otimes \cl{G}_{2}(u), \quad \mathfrak{W}(v) = \cl{W}_{1}(v) \otimes \cl{W}_{2}(v)
	\end{equation}
	where twists also constitute part of the {\Rx} symmetry
	\begin{equation}\label{key}
		\left[ R, \fk{G} \right] = \left[ R, \fk{W} \right] = 0 \qquad \{\cl{G}_{i}, \cl{W}_{i}\} : \bb{C}^{n} \rightarrow \bb{C}^{n}
	\end{equation}
	On the other hand, we can form a more general transform, which essentially is nothing but a composite of known transformations. For example, one can combine LBT and twist transformations, which can be grouped into\footnote{Such combinations could be referred to as of twist transformation class, whenever combined with twists}
	\begin{equation}\label{key}
		R \xRightarrow[]{\{ \fk{W}, \fk{V} \}} R' = \fk{V}^{(u)}\fk{W}^{(u)} R \left( \fk{V}^{(v)} \fk{W}^{(v)} \right)^{-1}
	\end{equation}
	\begin{equation}\label{key}
		\text{along with } \left[ R, \cl{W}(u) \otimes \cl{W}(u) \right] = \left[ R, \cl{V}(u) \otimes \cl{V}(u) \right] = 0
	\end{equation}
	where we indicated the corresponding spectral spaces as $ (u) $ and $ (v) $, we also assume distinct twists $ \fk{W} $ in the general case. At the level of $ \cl{H} $ we can obtain appropriately modified \eqref{ND_RTwist}
	\begin{equation}\label{ND_HTwist}
		\cl{H} \xRightarrow[]{\fk{W}} \cl{H}' = \fk{W} \cl{H} \fk{W}^{-1} + \left( \partial_{u} \fk{W} \right) \fk{W}^{-1}
	\end{equation}
	where we have single parametric evaluation. Moreover, we can get additional twist commutative constraint for the Hamiltonian in the following form
	\begin{equation}\label{key}
		\left[ \cl{H}, \fk{G}\fk{W} \right] = \fk{G}(\partial_{u}\fk{W}) - (\partial_{u}\fk{G})\fk{W}
	\end{equation}
	which could be obtained by combining \eqref{ND_HTwist} or by analysis that follows from Sutherland equation on twisted {\Rx} \eqref{ND_RTwist} along with spectral reduction limit to single parameter. important to note, that one can consider generic Drinfeld twists \cite{drinfeld1983constant,Drinfeld1986quantum}, combinations of their subcases or composites along with general \eqref{ND_H_R}.
\end{itemize}

\subsection{Two dimensions}\label{Sec_ND_2-dim}

We shall start by implementing our technique \cite{deLeeuw:2019zsi} for models with two-dimensional local quantum spaces, where the {\Rx} has an arbitrary spectral dependence. Integrability of provided setup will be explicitly proven, as well as their existence and uniqueness up to free spectral functional form. Consequently, an extension to three- and four-dimensional spaces is provided along with their properties. We shall also keep our discussion along the lines of $ AdS $ integrability \cite{deLeeuw:2020ahe,deLeeuw:2021ufg} in order to show how found classes can arise in certain integrable string sectors.

We shall begin our investigation with a generic Ansatz of 8-vertex class (or lower)
\begin{equation}\label{ND_H8v}
	\begin{split}
		\mathcal{H} & = h_1 \text{ }\mathds{1} + h_2 (\sigma _z\otimes \mathds{1}- \mathds{1}\otimes \sigma _z) + h_3  \sigma _+\otimes \sigma _-  + h_4 \sigma_-\otimes \sigma_+ \\ 
		& + h_5 ( \sigma _z \otimes \mathds{1} +  \mathds{1} \otimes \sigma _z ) + 
		h_6 \sigma _z\otimes \sigma _z  + h_7 \sigma _-\otimes \sigma _- + h_8 \sigma _+\otimes \sigma _+
	\end{split}
\end{equation}
where the corresponding {\Rx} takes conventional form
\begin{equation}\label{p3_8v}
	R_{\tx{8v}}(u,v) = 
	\begin{pmatrix}
		r_{1} & 0 & 0 & r_{8} \\
		0 & r_{2} & r_{6} & 0 \\
		0 & r_{5} & r_{3} & 0 \\
		r_{7} & 0 & 0 & r_{4} \\
	\end{pmatrix}
\end{equation}
where $ r_{k} \equiv r_{k}(u,v) $ are undetermined functions on spectral parameters $ u $ and $ v $. After following procedure \ref{ND_Technique}, \textit{i.e.} taking into account spectral dependence and resolving differentially coupled system, we are finding four distinct generating solutions of 6- and 8-vertex form. In what follows the structure of these classes, which can be described by the characteristic combination of $ h_{i} $ coefficients \eqref{ND_H8v}. 

\paragraph{6-vertex A} The 6-vertex A model (or 6vA) is obtained when we set $ h_{7} = h_{8} =0 $, $ h_{6} \neq 0 $, from where lower coefficients follow as
\begin{equation}\label{key}
	h_3 = c_3 h_6 e^{4 H_5} \qquad h_4 = c_4 h_6 e^{-4 H_5} \qquad c_{i} \in const
\end{equation}
with suitable definitions
\begin{equation}\label{ND_PrimitiveFunctions}
	\dot{h}_{i}(w) = \partial_{w} h_{i}(w), \quad h_{i}(w) \in \cl{H},
	\quad
	\begin{cases}
		H_{i}(w) = \int_{0}^{w} ds \, h_{i}(s) \\
		H_{i}(v, w) = \int_{w}^{v} ds \, h_{i}(s) = H_{i}(v) - H_{i}(w)
	\end{cases}
\end{equation}
where from now on, we shall be suppressing the spectral parameter dependence. We can note, that after appropriate application of LBT/twist/reparametrisation/normalisation transformations \ref{ND_IT}, the 6vA indeed constitutes XXZ brought to
\begin{equation}\label{key}
	\cl{H} = 
	\begin{pmatrix}
		0 & 0 & 0 & 0 \\
		0 & 1 & \fk{c} & 0 \\
		0 & \fk{c} & 1 & 0 \\
		0 & 0 & 0 & 0 \\
	\end{pmatrix}
\end{equation}
which is of the form \eqref{Apx_H6v}, detailed computation algorithm and the underlying {\Rx} can be found in \ref{ND_6vComputation}. It also contains generic diagonal subcase, as off-diagonal elements are fixed. 

\paragraph{6-vertex B}\label{ND_6vB} In 6vB one gets $ h_{\{6,7,8\}} = 0 $, so we result in five free functions, however not all of them remain after \ref{ND_IT} are applied. Specifically, we can reabsorb $ h_{1,3} $ after normalisation and spectral redefinition, as well as $ h_{2} $ by applying LBT.

It suitable to set the {\Rx} normalisation $ r_{5} = 1 $ and redefine $ h_{5} \mapsto \hf h_{4} h_{5} $. Proceeding with the Sutherland equation we obtain
\begin{equation}\label{key}
	\begin{cases}
		r_6=1 \\ 
		r_7=r_8=0 \\
		r_1 r_4 + r_2 r_3 =1
	\end{cases}
	\qquad
	\begin{cases}
		\dot{r_2} = h_4(r_1-h_5 r_2) \\
		\dot{r}_4 =  -h_4(r_3+h_5 r_4) \\
	\end{cases}
\end{equation}
with $ r_{4} $ to satisfy
\begin{equation}\label{key}
	\ddot r_4-\frac{\dot{h}_4}{h_4}\dot{r}_4 = h_4 r_4 \left( h_4h_5^2 - h_3 - \dot{h}_5 \right)
\end{equation}
which corresponds to 2nd order Riccati equation.

So that the system finally solves to
\begin{equation}\label{ND_6vBR}
	\begin{cases}
	r_1(x,y) &= 1 + h_5(x)H_4(x,y) \\
	r_2(x,y) &= H_4(x,y) 
	\end{cases}
	\qquad
	\begin{cases}
		r_3(x,y) & = h_5(x) h_5(y) H_4(x,y) -h_5(x) + h_5(y) \\
		r_4(x,y) & = 1 -h_5(y) H_4(x,y)
	\end{cases}
\end{equation}
along with spectral reparametrisation
\begin{equation}\label{ND_SpectralRedefinition1}
	H_{x}:u_{a} \rightarrow x_{a} = \int^{u_{a}} \dd s \, \frac{\dot h_5}{h_4 h_5^2-h_3}
\end{equation}
as before the $ s $ spectral internal variable (indicated for reference), $ H_{k}(x,y) $ as in \eqref{ND_PrimitiveFunctions}, it was convenient to introduce \eqref{ND_SpectralRedefinition1} due to removal of Riccati inhomogeneity and compact resolution of \eqref{ND_6vBR}. It is useful to look at the {\Rx} in the following form
\begin{equation}\label{key}
	R = H_{4}(x,y)
	\begin{pmatrix}
		h_5(x) & 0 & 0 & 0 \\
		0 & 1 & 0 & 0 \\
		0 & 0 & h_5(x)h_5(y) & 0 \\
		0 & 0 & 0 & -h_5(y)
	\end{pmatrix}
	+
	\begin{pmatrix}
		1 & 0 & 0 & 0 \\
		0 & 0 & 1 & 0 \\
		0 & 1 & h_5(y) - h_5(x) & 0 \\
		0 & 0 & 0 & 1 \\
	\end{pmatrix}
\end{equation}
from we can notice that when $ h_{5} $ becomes constant we get a reduction to standard XXZ. It can be also shown that 6vA can be recovered by twisting the 6vB model\footnote{A full 6-vertex classification of the coloured-YBE was performed in \cite{Sun:1995}, it could be established that by appropriate restrictions 6vA also satisfies colour-YBE, but more detailed checks are required at the level of colour-algebras}. As it was mentioned in \ref{ND_Technique}, along with corresponding boundary conditions it satisfies non-difference YBE.

\paragraph{8-vertex A} The first class of 8-vertex model is obtained for arbitrary $ h_{\{6,7,8\}} \neq 0 $, which results in the following Hamiltonian	
\begin{equation}\label{key}
	\begin{cases}
		h_{3,4} = \fk{c}_3 h_6 \\ 
		h_{5} = 0
	\end{cases}
	\quad
	\begin{cases}
		h_7 = \fk{c}_7 h_6 e^{4H_2} \\
		h_8 = \fk{c}_8 h_6 e^{-4H_2}	
	\end{cases}
\end{equation}
$ \fk{c}_{i} $ are free constants and $ H_{i} $ correspond to primitive functions defined as \eqref{ND_PrimitiveFunctions}. It is obvious that this model covers nothing but an elliptic \cite{Kulish1982,Vieira:2017vnw} non-difference class (or XYZ type models).

\paragraph{8-vertex B}\label{ND_8vBModel} The 8vB allows for one restriction $ h_{6} = 0 $ and $ h_{7,8} \neq 0 $ w.r.t. 8vA and Hamiltonian differential system appears as
\begin{equation}\label{ND_8vB_Sys1}
	\begin{array}{l}
		\frac{\dot{h}_7}{h_7} = 4 h_2 +  \fk{h}_{345} \\ 
		\frac{\dot{h}_8}{h_8} = -4 h_2 +  \fk{h}_{345} \\ 
		\frac{\dot{h}_5}{h_5} =- \frac{h_3^2-h_4^2}{4 h_5}+ \fk{h}_{345} \\ 
		\fk{h}_{345} = \frac{\dot{h}_3+\dot{h}_4}{h_3+h_4}+4\frac{h_3-h_4}{h_3+h_4} h_5
	\end{array}
\end{equation}
The first application of LBT provides us with $ h_{2} = 0 $, after which the system \eqref{ND_8vB_Sys1} solves in 
\begin{equation}\label{key}
	\begin{array}{l}
		h_5 = -\frac{1}{4} (h_3+h_4) \tanh (\fk{h}_{34}) \\
		h_7 = c_7 \frac{h_3+h_4}{ \cosh(\fk{h}_{34})} \\
		h_8 = c_8 \frac{h_3+h_4}{ \cosh(\fk{h}_{34})} \\ 
		\fk{h}_{34} = H_3-H_4+c_5
	\end{array}
\end{equation}
Applying further transformations along with algebraic operations, we can associate constants $ c_{8} = c_{7} $ and resolve $ h_{\{ 3,4 \}} $
\begin{equation}\label{key}
	\begin{cases}
		h_{4} - h_{3} = \dot{\eta} \csc(\eta) \\ 
		\dfrac{h_{4}}{h_{3}} = \dfrac{2 + \dot{\eta}}{2 - \dot{\eta}}
	\end{cases}
\end{equation}
with arbitrary spectral function $ \eta \equiv \eta(u) $. From here one finds $ h_{7} = h_{8} = k $, where $ k = 2 c_{7} $ is constant, eventually part of the R-system gives $ r_{\{ 5,6 \}} = 1 $ and $ r_{7} = r_{8} $. The rest of the R-system that follows from Sutherland system \eqref{SutherlandEq_1}-\eqref{SutherlandEq_2}, can be resolved as
\begin{equation}\label{p3_R_8vB}
	\begin{array}{l}
		r_1 = \frac{1}{\sqrt{\sin \eta(u)}\sqrt{\sin\eta(v)}} \left( \sin\eta_+\frac{\mathrm{cn}}{\mathrm{dn}} - \cos\eta_ + \mathrm{sn} \right) \\
		r_2 = \frac{1}{\sqrt{\sin \eta(u)}\sqrt{\sin\eta(v)}} \left( \cos\eta_-\mathrm{sn} +\sin\eta_-\frac{\mathrm{cn}}{\mathrm{dn}} \right) \\
		r_3 = \frac{1}{\sqrt{\sin \eta(u)}\sqrt{\sin\eta(v)}} \left( \cos\eta_-\mathrm{sn}  - \sin\eta_-\frac{\mathrm{cn}}{\mathrm{dn}} \right) \\ 
		r_4 = \frac{1}{\sqrt{\sin \eta(u)}\sqrt{\sin\eta(v)}} \left( \sin\eta_+\frac{\mathrm{cn}}{\mathrm{dn}} + \cos\eta_+ \mathrm{sn} \right) \\
		r_8 = k \frac{\mathrm{sn} \mathrm{cn}} {\mathrm{dn}}
	\end{array}
\end{equation}
where $ r_{i} \equiv r_{i}(u,v) $, $ \eta_{\pm} = \frac{\eta(u) \pm \eta(v)}{2} $, $ \mrm{xn} \equiv \mrm{xn}(u-v,k^{2}) $ are Jacobi elliptic functions with $ (u - v) $ dependence and $ k^{2} $ is an elliptic modulus. It is also important to note that in the limits when $ \eta \in \text{const} $ the 8vB {\Rx} reduces to the regular non-difference elliptic model \cite{Akutsu:1982,Khachatryan:2012wy,Vieira:2017vnw}, whereas in the infinite modulus $ k $ limit it relates to the $ AdS_{2} $ \cite{Hoare:2014kma,Hoare:2015kla,Slavnov:2020xxj} integrable background model.

\paragraph{8vB singular limit behaviour} In addition, it is interesting to check special analytic behaviour of the model above \ref{ND_8vBModel}. Particularly we can note from \eqref{ND_8vB_Sys1} that in $ h_{5} \rightarrow 0 $ limit the model becomes constant and indeed, it is equivalent to $ \eta(s) \rightarrow \dfrac{\pi}{2} $.

Secondly, a special behaviour also occurs when $ h_{3} \rightarrow -h_{4} $, which results in $ h_{7} = c h_{8} $ and $ h_{3} = -h_{4} $ nonzero contributions. One can also find given limit from \ref{ND_8vBModel}, by following applying off-diagonal twist with appropriately fixed entries and parametrisation. Consequently, one can apply diagonal LBT to suppress some entries along with further fixed parameters, where eventually the off-diagonal form is recovered, more detailed in \ref{ND_Apx_8vBLim}.

\paragraph{Comment on Hermiticity}\label{ND_Hermiticity} Completely generic Ansatz in the for non-difference setup is an important questions, which would resolve full classification $ \bb{C}^{2} $-problem with arbitrary spectral dependence. It could be noted already now, that there exists nontrivial integrable sector (an analogue of $ \al{sl}_{2} $ as in \cite{deLeeuw:2019zsi} and beyond), however some of these classes involve highly nontrivial differential systems and other algebraic issues that require further investigation. 

Nevertheless, it is also worth mentioning that there are generators, which respect hermiticity. For instance, one can consider real-imaginary decomposed Ansatz of the form $ \cl{H}_{\text{H}} = \cl{A} + i \cl{B} $ with $ \cl{A}_{ij}, \cl{B}_{ij} \in \bb{R} $ to constitute 16 unfixed spectral functions. Hence we treat $ \Re(\cl{H}) $ and $ \Im(\cl{H}) $ independently and we demand both parts to vanish in order to satisfy commutativity of the integrable hierarchy. We also neglect any generating solution that contains imaginary contribution. Important, that all of these solutions can be transformed into 8-vertex type with \ref{ND_IT}, but the latter does not need to be Hermitian. The opposite is also true, the 8-vertex, which is not Hermitian after specific LBT transform, can be brought into Hermitian form, but will not fall into 8-vertex class.


\newpage

\subsection{Three dimensions}
In this section we develop our approach to address integrable structure that arises from the models with 3-dim local quantum spaces. Some of the relevant works, resolved properties, as well as other investigation methods can be found \cite{Izergin:1981, Jimbo1986, Idzumi:1994kx, Crampe:2013nha,Crampe:2016she, Martins:2013ipa, Martins:2015ufa}. 

Following the symmetry arguments, \textit{i.e.} commutativity of $ \cl{H}/R $ with $ \al{su}(3) $ Cartan subalgebra generators, one can obtain 15-vertex based Ansatz
\begin{equation}\label{ND_H_R_3-dim}
	\cl{O} = 
	\begin{pmatrix}
		\mathrm{x}_{11}  & 0  & 0  & 0  & 0  & 0  & 0  & 0  & 0  \\
		0  & \mathrm{x}_{22} & 0  & \mathrm{x}_{24}& 0  & 0  & 0  & 0  & 0  \\
		0  & 0  & \mathrm{x}_{33} & 0  & 0  & 0  & \mathrm{x}_{37}& 0  & 0  \\
		0  & \mathrm{x}_{42}& 0  & \mathrm{x}_{44} & 0  & 0  & 0  & 0  & 0  \\
		0  & 0  & 0  & 0  & \mathrm{x}_{55} & 0  & 0  & 0  & 0  \\
		0  & 0  & 0  & 0  & 0  & \mathrm{x}_{66} & 0  & \mathrm{x}_{68}& 0  \\
		0  & 0  & \mathrm{x}_{73}& 0  & 0  & 0  & \mathrm{x}_{77} & 0  & 0  \\
		0  & 0  & 0  & 0  & 0  & \mathrm{x}_{86}& 0  & \mathrm{x}_{88} & 0  \\
		0  & 0  & 0  & 0  & 0  & 0  & 0  & 0  & \mathrm{x}_{99}
	\end{pmatrix}
\end{equation}
where we mean equivalent operator configurations for the Hamiltonian and {\Rx}. All entries are functions on the spectral parameters, \textit{e.g.} $ \cl{H}_{ab} \equiv \cl{H}_{ab}(w) $  and $ \cl{R}_{ab} \equiv \cl{R}_{ab}(u,v) $, and given indexing is assumed for both $ \cl{H} $ and $ R $. One can also consider conventional basis with block enumeration $ R_{kl}^{ij} $ \cite{Korepin_1997_QISM_CF,Khachatryan:2014cfj} as 
\begin{equation}\label{key}
	R = 
	\begin{cases}
		R_{kl}^{ij}, \text{ if } i + j = k + l \\
		0, \text{ else}
	\end{cases}
\end{equation}
with the given construction and arguments of \ref{ND_Technique} we shall attempt the search for novel structures in 15-vertex sector \cite{Vieira:2019vog}.

Keeping in mind the steps of \ref{Method} and \ref{ND_IT}, we are able to find ten generating solutions in the first level. However only six of these models appear of true non-difference form. The separated four models do contain full operatorial configuration as of \eqref{ND_H_R_3-dim}, but after applying appropriately adopted LBT, twisting and reparametrisations from \ref{ND_IT} exhibit only difference spectral dependence. On contrary, the models of non-difference form may have some zeros in \eqref{ND_H_R_3-dim}, but appear of novel structure form with arbitrary spectral dependence. Important, that the last do not show hermiticity, which contrasts two dimensional argumentation \ref{ND_Hermiticity}, where it was possible to prove hermitian property to hold for a number of cases by exploiting \ref{ND_IT}. 

\subsubsection{Densities}
We shall provide a description for the six models, which are explicitly of non-difference form. Important that also here we can note the difference with first four models and the remaining two, which demonstrate specific decomposition at the level of the {\Rx}. We can unify the generating densities $ \cl{H} $ according to \eqref{ND_H_R_3-dim} as follows

\begin{table}[h!]
	\centering
	\begin{tabular}{ c | c | c | c | c | c | c | c |}
		Class $ N $ & $ h_{66} $  & $ h_{55} $ & $ h_{99} $ & $ h_{86} $  & $ h_{24} $ & $ h_{73} $ & $ h_{42} $ \\ \hline \hline
		\Rn{1} & $ 1 $ & $ 0 $ & $ 0 $ & $ \fk{c}_{3} $ & $ \fk{c}_{2}\,e^{-w} $ &  $ \fk{c}_{1}\,e^{w} $ & 0 \\ \hline
		\Rn{2} & $ 1 $ & $ 1 $ & $ 0 $ & $ \fk{c}_{3} $ & $ \fk{c}_{2}\,e^{-w} $ &  $ \fk{c}_{1}\,e^{w} $ & 0 \\ \hline
		\Rn{3} & $ 1 $ & $ 0 $ & $ 1 $ & $ \fk{c}_{3} $ & $ \fk{c}_{2}\,e^{-w} $ &  $ \fk{c}_{1}\,e^{w} $ & 0 \\ \hline
		\Rn{4} & $ 1 $ & $ 1 $ & $ 1 $ & $ \fk{c}_{3} $ & $ \fk{c}_{2}\,e^{-w} $ & $ \fk{c}_{1}\,e^{w} $ & 0 \\ \hline
		\Rn{5} & $ 0 $ & $ f_{-} $ & $ 2 \left( f_{+} - f_{0} \right) $ & $ 0 $ & $ 0 $ & $ 0 $ & $ -\frac{f_{-}}{3} \, e^{F} $ \\ \hline
		\Rn{6} & $ 0 $ & $ \fk{f}_{+} $ & $ \fk{f}_{-} $ & $ 0 $ & $ 0 $ & $ -\frac{\fk{f}_{+}}{3}  \, \fk{c}_{1} e^{\fk{F}} $ & $ -\frac{\fk{f}_{+}}{3}  \, e^{\fk{F}} $ \\ \hline
	\end{tabular}	
	\caption{Entries of the generating $ \cl{H} $ classes}
	\label{T:ND_3dim_HSpace}
\end{table}
\vspace{2cm}
\noindent the rest of the entries according to \eqref{ND_H_R_3-dim} are null. The functions involved are defined as follows
\begin{equation}\label{ND_3-dim_DensityFunctions}
	f_{\pm} = 2 \left( f_{2} \pm f_{1} \right), \quad \fk{f}_{+} = 2 \left( f_{3} \pm \partial_{w} \cl{X} \right), \quad \fk{f}_{-} = 2 \left( f_{3} \mp \partial_{w}\cl{X} \right), \quad \fk{F} = 2 \left( \tilde{F} \pm \cl{X} \right)
\end{equation}
\begin{equation}\label{ND_TanhCondition}
	\tanh^{2} \left( 2 \cl{X} \right) = 1 + \fk{c}_{2} e^{4 \tilde{F}}
\end{equation}
where $ \fk{c}_{i} \in \text{const} $, free spectral functions $ f_{k} = f_{k}(w) $ with $ k = \overline{0,3} $ and primitive function
\begin{equation}\label{ND_3-dimPFunction}
	\begin{array}{l}
		F(w) = \int_{0}^{w} \dd s \left[ \varphi_{2}(s) - \varphi_{1}(s) \right] \\ 
		\tilde{F}(w) = \int_{0}^{w} \dd s \varphi_{3}(s)
	\end{array}
\end{equation}
where $ \varphi_{i}(s) $ are free spectral functions. It is important to note, that one need to pay attention on branches \eqref{ND_TanhCondition}, when considering constraints or other specific computations within the class \Rn{6}. We shall now discuss the associated structure of the {\Rxs}.

\subsubsection{$ R $-matrices}
The underlying {\Rxs} for the six models in \ref{T:ND_3dim_HSpace} exhibit specific properties. One common for all six classes, is that can be transformed into XXX type form after appropriately applying \ref{ND_IT}. Namely, one can find Heisenberg type decomposition $ R = \mathds{1} \cdot u + P $, which for classes \Rn{1}-\Rn{6} can be expressed as follows
\begin{equation}\label{key}
	R = \phi \bb{D} + \pi
\end{equation}
where $ \phi \equiv \phi(u, v) $ is a model dependent spectral function, $ \bb{D} $ is a diagonal operator and $ \pi $ is an operator, whose entry configuration coincides with permutation $ P $. The spectral functions involved in $ \bb{D} $ and $ \pi $ both possess $ (u, v) $-dependence.

The models \Rn{1}-\Rn{4} can be characterised by the $ \phi, \bb{D}, \pi $ 
\begin{equation}\label{key}
	\begin{cases}
		\phi = 2 \sinh \left( \dfrac{u-v}{2} \right) \\
		\bb{D} = \fk{c}_{1} e^{U_{+}} E^{33} + \fk{c}_{2} e^{-U_{+}} E^{44} + \fk{c}_{3} e^{U_{-}} E^{66} + \sigma_{1} e^{U_{-}} E^{55} + \sigma_{2} e^{U_{-}} E^{99} \\
		\pi = \left( e^{2 U_{-}} - 1 \right) E^{86} + P
	\end{cases}
	\quad
	U_{\pm} = \dfrac{u \pm v}{2}
\end{equation}
with $ E^{ij} = 1 $ at $ (i,j) $ and $ 0 $ elsewhere, $ P = \sum_{i,j} E_{ij} \otimes E_{ji} $ and model dependent $ \sigma_{i} $ listed in Table \ref{T_Sigma}.
\begin{table}[h!]
	\centering
	\begin{tabular}{ c | c | c |}
		Class $ N $ & $ \sigma_{1} $  & $ \sigma_{2} $ \\ \hline \hline
		\Rn{1} & $ 0 $ & $ 0 $ \\ \hline
		\Rn{2} & $ 1 $ & $ 0 $ \\ \hline
		\Rn{3} & $ 0 $ & $ 1 $ \\ \hline
		\Rn{4} & $ 1 $ & $ 1 $ \\ \hline
	\end{tabular}	
	\caption{Model dependent coefficients $ \sigma $}
	\label{T_Sigma}
\end{table}

In the classes \Rn{5} and \Rn{6} condition $ \pi = P $ holds, hence their description is provided by $ \phi $ and $ \bb{D} $. The class \Rn{5} is given by
\begin{equation}\label{key}
	\begin{cases}
		\phi = 2 \sinh \cl{F}_{-} \\
		\bb{D} = -\dfrac{e^{\cl{F}_{+}}}{3} E^{22} +  e^{\cl{F}_{-}} E^{55} + \dfrac{\sinh \cl{I}_{\varphi}}{\sinh \cl{F}_{-}} e^{\cl{I}_{\varphi}} E^{99}
	\end{cases}
	\quad
	\begin{array}{l}
		\cl{F}_{\pm} = F(u) \pm F(v) \\
		\cl{I}_{\phi} = \int_{v}^{u} \dd s \left[ 2 \varphi_{2}(s) + \varphi_{1}(s) \right]
	\end{array}
\end{equation}
where as above $ \varphi_{i} $ are independent free spectral functions (in analogy to \eqref{ND_3-dimPFunction}). 

The class \Rn{6} is characterised by
\begin{equation}\label{key}
	\begin{cases}
		\phi = -\dfrac{2}{3}\sinh \left[ \tilde{F}_{-} \pm \cl{X}_{-} \right] \\
		\bb{D} = e^{\tilde{F}_{+} \pm \cl{X}_{+}} E^{22} + \fk{c}_{1} e^{\tilde{F}_{+} \pm \cl{X}_{+}} E^{33} - 3 e^{\tilde{F}_{-}} \left( e^{\pm \cl{X}_{-}} E^{55} + e^{\mp \cl{X}_{-}}\dfrac{\sinh \left[ \tilde{F}_{-} \mp \cl{X}_{-} \right]}{\sinh \left[ \tilde{F}_{-} \pm \cl{X}_{-} \right]} E^{99} \right) 
	\end{cases}
\end{equation}
\begin{equation}\label{key}
	\begin{array}{l}
		\tilde{F}_{\pm} = \tilde{F}(u) \pm \tilde{F}(v) \\
		\cl{X}_{\pm} = \cl{X}(u) \pm \cl{X}(v)
	\end{array}
\end{equation}
where the $ \cl{X}(s) $ involves condition \eqref{ND_TanhCondition} on each spectral parameter. Due to the aforementioned argument on signs in \eqref{ND_3-dim_DensityFunctions}-\eqref{ND_TanhCondition} there are two branches in class \Rn{6}, which represent two models. There is no suitable integrable transformation \ref{ND_IT}, which would establish map withing class \Rn{6}. We should also notice that, generic spectral behaviour in the models above admits differences of primitives and other free spectral functions.

\newpage
\subsection{Four dimensions}

In this section we shall discuss the method \ref{ND_Technique} to the models with four dimensional local spaces \ref{D_Hubbard}. Similar to the development in \ref{HubDiff}, we shall create our free spectral Ansatz based $ \al{su}(2) \oplus \al{su}(2) $ symmetry. On the other hand, current setup is exactly the one needed for the corresponding models arising in the context of AdS/CFT integrability. More specifically, based on the symmetry algebra described in \ref{HubDiff}, we can create a framework for studying novel integrable structures with arbitrary spectral dependence, which is relevant for one-dimensional Hubbard chain in {\ads}, Shastry type constructions, the analogues of the classes found in Table \ref{HHubbard_SolTable} \cite{deLeeuw:2019vdb} or novel integrable deformations \cite{deLeeuw:2020xrw,deLeeuw:2020ahe} of the above, which were not possible to find by other means. The analysis will be conducted along the representation scheme provided \ref{D_Hubbard}, \textit{i.e.} 2-dim representations of each copy of $ \al{su}(2) $ (Hubbard), the 4-dim representation of $ \al{so}(4) $ (isomorphism) and distinct integrable generalisations will follow. The investigation of deformations arising in the $ AdS_{\{ 2,3 \}} $ sector will be provided in \ref{p5_Sec_AdS_2_3_Integrable_Deformations}.

\subsubsection{$ \al{su}(2) \oplus \al{su}(2) $}
We consider $ \al{su}(2) \oplus \al{su}(2) $ symmetry with 2-dim representation for each copy of $ \al{su}(2) $, which is also relevant for Shastry type, A/B models and other \cite{Frolov:2011wg}. We shall recapitulate, that we shall start by the 2-state Ansatz of the form
\begin{equation}\label{ND-4-dim_Ansatz}
	\begin{array}{l}
		\cl{O} \ket{\phi_a \phi_b} = \mathrm{x}_1 \ket{\phi_a \phi_b} + \mathrm{x}_2 \ket{\phi_b \phi_a} + \mathrm{x}_3 \epsilon_{ab}\epsilon_{\alpha\beta} \ket{\psi_\alpha \psi_\beta} \\
		\cl{O} \ket{\phi_a \psi_\beta} = \mathrm{x}_4   |\phi_a \psi_\beta \rangle + \mathrm{x}_5  |\psi_\beta \phi_a \rangle \\
		\cl{O} |\psi_\alpha \phi_b \rangle = \mathrm{x}_6  |\psi_\alpha \phi_b \rangle + \mathrm{x}_{7}  |\phi_b \psi_\alpha \rangle \\
		\cl{O}|\psi_\alpha \psi_\beta \rangle = \mathrm{x}_8  |\psi_\alpha \psi_\beta \rangle + \mathrm{x}_9  |\psi_\beta \psi_\alpha \rangle + \mathrm{x}_{10} \epsilon_{ab}\epsilon_{\alpha\beta} |\phi_a \phi_b \rangle
	\end{array}
\end{equation}
where one considers $ \cl{O} = \cl{H} \text{ or } R $\footnote{One can also implement similar action of the scattering matrix $ S $ with appropriate modifications, see 2-state embeddings in \ref{p5_Sec_AdS_2_3_Integrable_Deformations}, \cite{Sfondrini:2014via}} and $ \phi_{i}/\psi_{i} $ are in two fundamental representations of $ \al{su}(2) $.
\begin{equation}\label{ND_4-dim_Ansatz_O}
	\cl{O} = 
	\tiny
	\begin{pmatrix}
		\mathrm{x}_1+\mathrm{x}_2 & 0 & 0 & 0 & 0 & 0 & 0 & 0 & 0 & 0 & 0 & 0 & 0 & 0 & 0 & 0 \\
		0 & \mathrm{x}_1 & 0 & 0 & \mathrm{x}_2 & 0 & 0 & 0 & 0 & 0 & 0 & \mathrm{x}_{10} & 0 & 0 & -\mathrm{x}_{10} & 0 \\
		0 & 0 & \mathrm{x}_4 & 0 & 0 & 0 & 0 & 0 & \mathrm{x}_7 & 0 & 0 & 0 & 0 & 0 & 0 & 0 \\
		0 & 0 & 0 & \mathrm{x}_4 & 0 & 0 & 0 & 0 & 0 & 0 & 0 & 0 & \mathrm{x}_7 & 0 & 0 & 0 \\
		0 & \mathrm{x}_2 & 0 & 0 & \mathrm{x}_1 & 0 & 0 & 0 & 0 & 0 & 0 & -\mathrm{x}_{10} & 0 & 0 & \mathrm{x}_{10} & 0 \\
		0 & 0 & 0 & 0 & 0 & \mathrm{x}_1+\mathrm{x}_2 & 0 & 0 & 0 & 0 & 0 & 0 & 0 & 0 & 0 & 0 \\
		0 & 0 & 0 & 0 & 0 & 0 & \mathrm{x}_4 & 0 & 0 & \mathrm{x}_7 & 0 & 0 & 0 & 0 & 0 & 0 \\
		0 & 0 & 0 & 0 & 0 & 0 & 0 & \mathrm{x}_4 & 0 & 0 & 0 & 0 & 0 & \mathrm{x}_7 & 0 & 0 \\
		0 & 0 & \mathrm{x}_5 & 0 & 0 & 0 & 0 & 0 & \mathrm{x}_6 & 0 & 0 & 0 & 0 & 0 & 0 & 0 \\
		0 & 0 & 0 & 0 & 0 & 0 & \mathrm{x}_5 & 0 & 0 & \mathrm{x}_6 & 0 & 0 & 0 & 0 & 0 & 0 \\
		0 & 0 & 0 & 0 & 0 & 0 & 0 & 0 & 0 & 0 & \mathrm{x}_8+\mathrm{x}_9 & 0 & 0 & 0 & 0 & 0 \\
		0 & \mathrm{x}_3 & 0 & 0 & -\mathrm{x}_3 & 0 & 0 & 0 & 0 & 0 & 0 & \mathrm{x}_8 & 0 & 0 & \mathrm{x}_9 & 0 \\
		0 & 0 & 0 & \mathrm{x}_5 & 0 & 0 & 0 & 0 & 0 & 0 & 0 & 0 & \mathrm{x}_6 & 0 & 0 & 0 \\
		0 & 0 & 0 & 0 & 0 & 0 & 0 & \mathrm{x}_5 & 0 & 0 & 0 & 0 & 0 & \mathrm{x}_6 & 0 & 0 \\
		0 & -\mathrm{x}_3 & 0 & 0 & \mathrm{x}_3 & 0 & 0 & 0 & 0 & 0 & 0 & \mathrm{x}_9 & 0 & 0 & \mathrm{x}_8 & 0 \\
		0 & 0 & 0 & 0 & 0 & 0 & 0 & 0 & 0 & 0 & 0 & 0 & 0 & 0 & 0 & \mathrm{x}_8+\mathrm{x}_9
	\end{pmatrix}
	\normalsize
\end{equation}
where the correspondingly dependent entries $ \mathrm{x}_{i} = h_{i}(w) \text{ or } r_{i}(u, v) $.

\paragraph{Densities and {\Rxs} of integrable classes}\label{ND_4-dim_Densities} In this segment we find eight models with arbitrary spectral dependence by following commutativity constraint and performing required identifications \ref{ND_IT}. We shall unify the first six classes by their common vanishing parameter $ h_{3} = 0 $ and list them in Table \ref{Tab_ND_4-dim_HSpace}.
\begin{table}[ht]
	\centering
	\resizebox{\textwidth}{!}
	{
	\begin{tabular}{c | c | c | c | c | c | c | c | c | c | c |}
		$ N $ & $ h_{1} $ & $ h_{2} $ & $ h_{3} $ & $ h_{4} $ & $ h_{5} $ & $ h_{6} $ & $ h_{7} $ & $ h_{8} $ & $ h_{9} $ & $ h_{10} $ \\ \hline \hline
		\Rn{1} & $\frac{1}{ 2 \mathrm{w}_{+} \mathrm{w}_{-}}$ & $\frac{1}{2}$ & 0 & $ \frac{ 1 - \mathrm{w}_{-} }{\mathrm{w}_{+} \mathrm{w}_{-} }$ & $\frac{\pm 1}{2} \sqrt{\frac{ \mathrm{w}_{+} }{ \mathrm{w}_{-} }}$&$ \frac{ \mathrm{w}_{+} - 1 }{ \mathrm{w}_{+} \mathrm{w}_{-} } $&$ \frac{\pm 1}{2} \sqrt{\frac{ \mathrm{w}_{-} }{ \mathrm{w}_{+} }}$ & $\frac{1}{2 \mathrm{w}_{+} \mathrm{w}_{-} }$ & $ -\frac{1}{2} $ & $ c_{3} $ \\ \hline
		\Rn{2} & $f $& $h $& 0 &$g$& $\frac{c_{3} \, h}{ e^{2 F}}$ & $-g$ &$ \frac{h e^{2 F}}{c_{3} } $&$ -f $& $\pm h$ & 0 \\ \hline
		\Rn{3} & $f$ & $\pm h $& 0&$ g$ & $\frac{c_{3}  h }{e^{2 F}}$ & $-g$ & $\frac{h e^{2 F} }{c_{3}} $ & $h-f $& 0 & 0 \\ \hline
		\Rn{4} & $(c_1+2)f $& 0 & 0 & $c_1(f-g) $& $  \frac{c_1(c_1+2)  g}{c_2 e^{2 F}} $& $(c_1+2)\; (f-g)$ &  $ c_2 e^{2 F}g$ &$ c_1 f  $& 0 & 0\\  \hline
		\Rn{5} & $f$ & 0 & 0 & 0& $g$& 0 & $h$& $-f$ & 0& 0   \\ \hline
		\Rn{6} &$ f-h$ & 0 & 0 &  $f+h  $&$ \frac{2h }{c_{3} \;e^{2F}}$ & $h-f $& $2 c_{3}  h e^{2 F}$&$ h-f$ & $\pm 2 h$ & 0 \\\hline
	\end{tabular}
	}
	\caption{Generating densities $ \cl{H}_{\al{su}(2) \oplus \al{su}(2)} $ following from Ansatz \ref{ND-4-dim_Ansatz}.}
	\label{Tab_ND_4-dim_HSpace}
\end{table}
where $ c_{i} \in const $, $ \mathrm{w}_{\pm} = w \pm 1 $ and $ w $-dependent spectral functions $ f $, $ g $, $ h $, $ F $, $ \dot{F}(w) = f $. We should notice, that all of the models in Table \ref{Tab_ND_4-dim_HSpace} exhibit either bounded or separated fermion dynamics (simultaneous presence suppressed due to vanishing coefficients). We can also note that class \Rn{5} is nothing but a 6vB \ref{ND_6vB} quadro-embedding \cite{deLeeuw:2020xrw,deLeeuw:2020ahe}, since it could immediately accomplished by the appropriate LBT and maps
\begin{equation}\label{key}
	\begin{cases}
		h_{5} \rightarrow h_{4} h_{5}/2, \qquad\text{(spectral redefinition)} \\
		h_{3} \mapsto g, \quad h_{4} \mapsto h, \quad h_{4}h_{5} \mapsto -f
	\end{cases}
\end{equation}

All the classes found as expected of their form, apriori are not Hermitian. For instance, for class \Rn{1} one can demand Hermiticity by $ c_{3} \rightarrow 0 $ and absent spectral dependence. Hermiticity is more involved for classes \Rn{2}-\Rn{6}, whose associated necessary constraints are stated in Table \ref{Tab_ND_4-dim_C1Hermiticity}. 

\begin{table}[ht]
	\centering
		\begin{tabular}{ c | c }
			$ N $ & \textbf{\textit{Hermitian}} constraints\\ \hline\hline
			\Rn{1} & $ w = 0 $, $ c_{3} = 0 $ \\ \hline
			\Rn{2},\Rn{3} & $ e^{4 \Re \left[ F \right]}=|c_{3}|^2,\;\;f,\,h\, \in\, \mathbb{R}$\\ \hline
			\Rn{4} & $\left(e^{4 \Re \left[ F \right]}=\frac{\Re \left[ c_{1} \right] (\Re\left[ c_{1} \right]+2)}{|c_2|^2}\;\;\text{or}\;\;e^{4 \Re \left[ F \right]}=\frac{1}{|c_2|^2}, \Re\left[ c_{1} \right] = -1\right)$,\;\;$c_1,\,f,\,g\, \in\, \mathbb{R}$ \\ \hline
			\Rn{5} & $g=h^*$, $f\,\in\,\mathbb{R}$\\ \hline
			\Rn{6} & $e^{-4 \Re \left[ F \right]}=|c_{3}|^2$, $f,\,h\, \in\, \mathbb{R}$\\ 			
		\end{tabular}
	\caption{Constraints required for classes \Rn{1}-\Rn{6} to be Hermitian}
	\label{Tab_ND_4-dim_C1Hermiticity}
\end{table}
\noindent in Tables \ref{Tab_ND_4-dim_C1Hermiticity} and \ref{Tab_ND_4-dim_C2HermitianMappings}, we denote the real $ \Re $ and imaginary $ \Im $ parts.

In addition, mapping between the classes and their Hermitian conjugates can be shown, for that a form of unitary constraint $ \left[ \cl{H}, \cl{H}^{\dagger} \right] = 0 $ needed to be satisfied. The necessary unitary conditions are grouped for classes \Rn{1}-\Rn{6} in Table \ref{Tab_ND_4-dim_C2HermitianMappings}
\begin{table}[ht]
	\centering
		\begin{tabular}{ c | c }
			$ N $ & \textbf{\textit{Unitary}} constraints \\ \hline\hline
			\Rn{1} & $ \Re \left[ w \right] = 0 $, $ c_{3} = 0 $ \\ \hline
			\Rn{2}, \Rn{3} & $ e^{4 \Re \left[ F \right]}=|c_{3}|^2$\\ \hline
			\Rn{4} & $e^{4 \Re \left[ F \right]}= \frac{{\Im \left[ c_{1} \right]}^2+1}{|{c_2}|^2},\Re \left[ c_{1} \right]=-1$ or $e^{4 \Re \left[ F \right]}= \frac{\Re \left[ c_{1} \right] (\Re \left[ c_{1} \right]+2)}{|c_2|^2}, \Im \left[ c_{1} \right]=0$ or ${c_1}=-1$ \\ \hline
			\Rn{5} & $\forall\; f,g,h$\\ \hline
			\Rn{6} & $e^{-4 \Re \left[ F \right]}=|c_{3}|^2$ \\
		\end{tabular}
	\caption{Constraints for classes \Rn{1}-\Rn{6} to satisy unitarity}
	\label{Tab_ND_4-dim_C2HermitianMappings}
\end{table}\normalsize

We shall now resolve the coupled Sutherland system to derive {\Rxs} for classes \Rn{1}-\Rn{6}. Majority of the systems that follow from \eqref{ND_Technique} and densities obtained above \eqref{ND_4-dim_Densities} can be straightforwardly solved. We recapitulate that as it was conventionally used, also here $ \sigma = \pm 1 $, $ c_{i} \in \text{const} $, $ F_{\pm} = F(u) \pm F(v) $ (analogous $ G_{\pm} $, $ H_{\pm} $). In what follows the {\Rx} structure

\paragraph{Class \Rn{1}}
\begin{equation}\label{key}
	\begin{array}{l}
		r_1=\frac{- r_{10}  \sqrt{1+ v}}{2 c\sqrt{r_{5}}  \sqrt{1 + u}} \\
		r_2=\frac{- r_{1}r_9}{r_8} \\
		r_3=0 \\
		r_4=\pm r_{1} \sqrt{\frac{u+1}{u-1}} \\
		r_5=\frac{\sqrt{1-v^2}}{\sqrt{1-u^2}} \\	
	\end{array}
	\qquad\qquad
	\begin{array}{l}
		r_6=\pm r_1\sqrt{\frac{v-1}{v+1}} \\
		r_7=\frac{1}{r_5} \\
		r_8=-\frac{r_{4} r_{6}}{r_{1}} \\
		r_9=\frac{2  r_{8}}{v-u} \\
		r_{10}=c (v-u)
	\end{array}
\end{equation}

\paragraph{Class \Rn{2}}
\begin{equation}\label{key}
	\begin{array}{l}
		r_1=H_-  e^{F_-}  \\
		r_2=e^{F_-}  \\
		r_3=0 \\
		r_4=c H_- e^{-F_+} \\
		r_5=e^{ G_-} \\
	\end{array}
	\qquad\qquad
	\begin{array}{l}
		r_6=\frac{H_- e^{F_+}}{c} \\
		r_7=e^{-G_-} \\
		r_8=\pm H_- e^{-F_-} \\
		r_9=e^{-F_-} \\
		r_{10}=0 
	\end{array}
\end{equation}

\paragraph{Class \Rn{3}}
\begin{equation}\label{key}
	\begin{array}{l}
		r_1=\pm H_- e^{F_-} \\
		r_2=e^{F_-} \\
		r_3=0 \\
		r_4=\frac{c\; H_-}{ e^{F_+}} \\
		r_5=e^{ G_-} \\
	\end{array}
	\qquad\qquad
	\begin{array}{l}
		r_6=\frac{H_- e^{F_+}}{c} \\
		r_7=e^{-G_-} \\
		r_8=0 \\
		r_9= \frac{(H_-+1)}{e^{F_-}} \\
		r_{10}=0
	\end{array}
\end{equation}

\paragraph{Class \Rn{4}}
\begin{equation}\label{key}
	\begin{array}{l}
		r_1=r_3=r_8=r_{10}=0 \\
		r_2=\frac{\left((c_1+2) e^{2 G_-}-c_1\right)r_7}{2} \\
		r_4= \frac{c_1 (c_1+2) (e^{2 G_-}-1)r_7}{2 c_2 e^{2 F(u)}} \\
		r_5=e^{c_1 (F_--G_-)} \\
	\end{array}
	\qquad\qquad
	\begin{array}{l}
		r_6=\frac{{c_2}^2 e^{2 F_+} r_4}{c_1 (c_1+2)} \\
		r_7=e^{(2+c_1)(F_--G_-)} \\
		r_9=e^{-2 F_-}r_2
	\end{array}
\end{equation}

\paragraph{Class \Rn{5}}
\begin{equation}\label{key}
	\begin{array}{l}
		r_1=0 \\
		r_2=\frac{H_- f(v)}{h(v)}+1 \\
		r_3= 0 \\
		r_4=\frac{1}{H_-}-\frac{r_2 r_9}{H_-} \\
		r_5=1 \\
	\end{array}
	\qquad\qquad
	\begin{array}{l}
		r_6=H_- \\
		r_7=1 \\
		r_8=0 \\
		r_9=1-\frac{H_- f(u)}{h(u)} \\
		r_{10}=0
	\end{array}
\end{equation}

\paragraph{Class \Rn{6}}
\begin{equation}\label{key}
	\begin{array}{l}
		r_1=r_3=r_{10}=0 \\
		r_2=e^{ F_-+H_-} (1-2 H_-) \\
		r_4=\frac{2 H_- e^{H_-}}{c\; e^{F_+}} \\
		r_5=e^{F_-+H_-} \\
	\end{array}
	\qquad\qquad
	\begin{array}{l}
		r_6=2 c H_- e^{ F_++H_-} \\
		r_7=\frac{e^{H_-}}{e^{F_-}} \\
		r_8=\pm2 H_- \frac{e^{H_-}}{e^{F_-}} \\
		r_9=\frac{e^{H_-}}{e^{F_-}}
	\end{array}
\end{equation}
where for \Rn{5} we have introduced $ u \mapsto \varphi (u) = \int ^u\frac{f \dot{h}-h \dot{f}}{h \left( f^2-g h \right)} $ to resolve the corresponding system. Only spectral reparametrisations along with allowed transformations have been used \ref{ND_IT}, so that all the {\Rxs} satisfy YBE and boundary conditions above \ref{ND_Technique}.

As we have stated above the form of the classes \Rn{7} and \Rn{8} appears more general and contains additional set of constraints, in what we shall discuss them separately.

\paragraph{Class \Rn{7}} The seventh class appears as the most general configuration with arbitrary spectral dependence, where as in \Rn{8} the $ h_{3} $ does not vanish. In order to resolve for this generator more efficiently we can set $ h_{\{ 4,6 \}} = 0 $ by LBT \ref{ND_IT} and normalise $ h_{10} = 1 $ (if that cannot be done, then one can use separate scheme with suppressed $ h_{10} $ and and recovered by reverse transformation of the $ P \left( \cl{H} \right)^{T} P $ form). After set of additional algebraic operations this leads us to a differential system
\begin{equation}\label{ND_Class7_HSys}
	\begin{array}{l}
		h_1 + h_8 = h_2+h_9 = 0 \\
		h_3= h_5 h_7-h_9^2 \\ 
		h_8= \frac{(h_5+h_7)^2}{4 h_9}-h_9
	\end{array}
	\qquad
	\begin{array}{l}
		\dot{h}_5= 2 h_7 h_9-\frac{h_5 (h_5+h_7)^2}{2 h_9} \\
		\dot{h}_7= \frac{h_7 (h_5+h_7)^2}{2 h_9}-2 h_5 h_9 \\
		\dot{h}_9= h_7^2-h_5^2		
	\end{array}
\end{equation}
where by dot conventionally denoted the derivative w.r.t. single spectral parameter($ w $). Clearly one can notice the structure of the first three equations \eqref{ND_Class7_HSys} and immediately obtain
\begin{equation}\label{key}
	\begin{cases}
		h_9 = 2 \Gamma_{1} \\
		h_5 = \frac{\sqrt{\gamma_{1}} \left({\gamma_{2}}^2-1\right)}{\sqrt{2}\; \gamma_{2}} \\ 
		h_7 = \frac{\sqrt{\gamma_{1}} \left({\gamma_{2}}^2+1\right)}{\sqrt{2}\; \gamma_{2}} 
	\end{cases}
\end{equation}
where $ \Gamma(w), \gamma(w) $ can be defined as
\begin{equation}\label{key}
	\gamma_{2}= \sigma\frac{\sqrt{\Gamma_{1}} \sqrt{8 \Gamma_{1}+c_1}}{\sqrt{\gamma_{1}}}, \quad \text{with } \Gamma_{k} = \int_{0}^{w_{i}} \dd s \gamma_{k}(s), \,\, \sigma = \pm 1, \,\, c_{i} \in \text{const}
\end{equation}
For the $ \gamma_{1} $ we can obtain a condition from equation on $ h_{7} $ \ref{ND_Class7_HSys} as follows
\begin{equation}\label{ND_C7_Eq1}
	\Gamma_{1} (8 \Gamma_{1}+c_1) \left(2 \dot{\gamma_{1}} -  c_1 \Gamma_{1}(8\Gamma_{1}+ c_1)\right) = \gamma_{1}^2 (16 \Gamma_{1}+c_1)
\end{equation}
From the structure of the equation \eqref{ND_C7_Eq1}, we can find that it can be resolved by an elliptic scheme for arbitrary $ c_{1} $ if we make generic identification
\begin{equation}\label{key}
	\Gamma_{1} \mapsto \phi(w), \quad \gamma_{1} \mapsto \dot{\phi(w)}, \quad \dot{\gamma_{1}} \mapsto \ddot{\phi(w)}
\end{equation}
which results in the solution, which can expressed in the form
\begin{equation}\label{key}
	\gamma_{1}(w) = \dfrac{i}{8} c_{1}^{2} \mrm{cs}(x|m) \mrm{ds}(x|m) \mrm{ns}(x |m)
\end{equation}
where $ x = \dfrac{i}{2} c_{1} \left( c_{2} + w \right) $, $ m = \dfrac{8 c_{3}}{c_{1}^{2}} $, $ c_{i} \in \text{const} $ and we work in the $ \mrm{pq}(u|m) $ notation for the elliptic functions, more detailed in Appendix \ref{Apx_Elliptic_Apparatus}. Eventually we can obtain resolved \eqref{ND_Class7_HSys} as
\begin{equation}\label{key}
	\begin{cases}
		h_5 - h_7 = \frac{i \sigma}{2} c_1  \mrm{ds}(x|m) \\
		h_5+h_7=\frac{\sigma}{2} c_1 \mrm{nc}(x|m) \left(1-\mrm{ns}(x|m)^2\right) \\ 
		h_9=-\frac{1}{4} c_{1} \mrm{ns}(x|m)^{2}
	\end{cases}
\end{equation}
As we it could be noticed there are no explicit \textit{free} functions left, that must turn us to the fact that there might exist a mapping to one of the AdS integrable models. A natural assumption would to analyse the structure of the model with centrally extended $ \al{su(2|2)} $ and compare densities \cite{Beisert:2005tm}}. Indeed, after set of operations with allowed transformations \ref{ND_IT} on the Hamiltonian with $ \al{su(2|2)} $, one can express the entries as
\begin{equation}\label{key}
	\begin{array}{l}
		h_1=-h_8 \\
		h_2=-h_9 \\
		h_3 = -\frac{1}{\alpha ^2} \\
		h_4=h_6 = 0 \\
		h_{10}=1
	\end{array}
	\quad\qquad
	\begin{array}{l}
		h_5=\frac{1-{x^-}^2}{\alpha  (x^--{x^+})}\sqrt{\frac{{x^+}}{x^-}} \\
		h_7=\frac{x^+ \dot x ^-}{\dot x^+ x^-} h_5 \\ 
		h_8= \frac{(h_5+h_7)^2}{4 h_9}-h_9 \\ 
		h_9=\frac{1-x^- {x^+}}{\alpha ( x^--{x^+})}
	\end{array}
\end{equation}
\begin{equation}\label{key}
	x^{\pm} = -\frac{1}{4} i \hbar \left( \mrm{dn}(\zeta|k) + 1 \right) \left(\mrm{cs}(\zeta|k) \pm i \right)
\end{equation}
where Zhukovsky variables are appropriately defined through suitable elliptic functions \cite{Arutyunov:2009ga}, $ \alpha \in \text{const} $, $ k = 16/\hbar^{2} $ ($ \hbar $ can be related to coupling $ g $ as $ \hbar = 2i/g $). Both densities agree when $ \sigma \mapsto 1 $
\begin{equation}\label{key}
	\hbar \mapsto \alpha c_{1} \quad \alpha^{2} \mapsto \dfrac{2}{c_{3}} \quad \zeta \mapsto \dfrac{i}{2} c_{1} \left( c_{2} + w \right)
\end{equation}
it could be seen that for $ \sigma = -1 $ we can relate it to the above by a twist (analogous to \ref{ND_IT})
\begin{equation}\label{key}
	\cl{H} \xRightarrow[]{\fk{W}} \cl{H}'(w,\sigma = 1) = \fk{W} \cl{H}(w,\sigma = -1) \fk{W}^{-1}, \qquad \fk{W} = \cl{V} \otimes \cl{W}
\end{equation}
with $ \cl{V} = \text{diag} (1,1,-1,1) $ and $ \cl{W} = \text{diag} (1,1,1,-1) $. The underlying {\Rx} can be represented by the $ AdS_{5} $ $ R $-/$ S $-matrix, which also contains the Hubbard chain \cite{Beisert:2005tm, Beisert:2006qh}, more detailed on the construction and properties can be found \cite{Arutyunov:2009ga}.

\paragraph{Class \Rn{8}} As it was indicated, the reason the class \Rn{7} was considered first, is that there exists a limit scheme on how recover class \Rn{8} from \Rn{7}. Apriori we can solve \eqref{ND_C7_Eq1} for $ c_{1} \rightarrow 0 $ to obtain resolved system 
\begin{equation}\label{ND_C8_HSolution}
	\begin{array}{l}
		\gamma_{1}(w) = c_{2} e^{c_{3} w} \\ 
		h_{\{ 1, 4, 6, 8 \}} = 0 \\
		h_{10} = 1 \\
		h_3 = -\frac{c_3{}^2}{16} \\
	\end{array}
	\qquad \qquad
	\begin{array}{l}
		h_9=\frac{2 c_2 e^{c_3 w }}{c_3} \\
		h_2 = -\frac{2 c_2 e^{c_3 w }}{c_3} \\
		h_5-h_7 = -\sigma\frac{c_3}{2} \\
		h_5+h_7 = \sigma\frac{4 c_2 e^{c_3 w }}{c_3} \\		
	\end{array}
\end{equation}
It is necessary to remark on the limits of \Rn{7}, since it is forbidden to take directly $ c_{1} \rightarrow 0 $ as singularities become apparent in some of the underlying elliptic functions. Hence appropriate step are required
\begin{itemize}
	\item Implementation of inverse modulus elliptic functions, \textit{e.g.} $ \mrm{ns}(i x| k) = -i \sqrt{k} \, \mrm{cs}\left[ x\sqrt{k} | 1 - \dfrac{1}{k} \right] $ 
	\item $ c_{1} \ll 1 $ perturbation
	\item $ c_{3} \rightarrow c_{1} $
	\item Large $ w $ perturbation
 	\item Redefine $ c_{\{ 1,2 \}} $ to match \eqref{ND_C8_HSolution}
\end{itemize}
An attention needed capture the correct behaviour when performing elliptic perturbation, by this can recover the class \Rn{8}. So that the associated {\Rx} is found to be
\begin{equation}\label{key}
	\begin{array}{l}
		r_2=\frac{1}{\cosh \left(\frac{1}{4} c_3 (u-v)\right)} \\  [1ex]
		r_3=\frac{1}{4} c_3 \tanh \left(\frac{1}{4} c_3 (u-v)\right) \\ [1ex]
		r_5=r_7=\frac{r_2}{r_9}=-\frac{c_3{}^2 r_{10}}{16 r_3} = 1 \\ [1ex]
		r_8 = r_{1} + \sigma \left(r_4+r_6\right) \\ [1ex]
	\end{array}
	\qquad
	\begin{array}{l}
		r_4 = -\frac{e^{-\frac{1}{4} c_3 (u+v)} \left(e^{\frac{c_3 u}{2}}-e^{\frac{c_3 v}{2}}\right) \left(c_3{}^2-8 c_2 e^{\frac{1}{2} c_3 (u+v)}\right)}{2 c_3{}^2 \sigma } \\ [1ex]
		r_6 = \frac{8 c_2 e^{\frac{1}{4} c_3 (u+v)} \left(e^{\frac{c_3 u}{2}}-e^{\frac{c_3 v}{2}}\right)}{c_3{}^2 \sigma }-r_4 \\ [1ex]
		r_1 = \frac{e^{-\frac{1}{4} c_3 (u+v)} \left[ c_3{}^2 \left(e^{\frac{c_3 u}{2}}-e^{\frac{c_3 v}{2}}\right){}^2-16 c_2 e^{c_3 (u+v)} \sinh \left(\frac{1}{2} c_3 (u-v)\right) \right]}{2 c_3{}^2 \left(e^{\frac{c_3 u}{2}}+e^{\frac{c_3 v}{2}}\right)} \\
	\end{array}
\end{equation}

\paragraph{$ \al{so}(4) $ case} As we have discussed in \ref{H_2o2sector}, there is also subsector that possesses an associated isomorphism $ \al{su}(2) \oplus \al{su}(2) \sim \al{so}(4) $, which also contains independent Casimir decomposition. Based on that we can construct generic density Ansatz
\begin{equation}\label{key}
	\cl{H}_{\al{so}(4)} = h_{1} \cdot \mathds{1} + h_{2} \underbrace{E_{i}^{j} \otimes E_{j}^{i}}_{P} + h_{3} \underbrace{E_{i}^{j} \otimes E_{i}^{j}}_{Q} + h_{4}\epsilon_{ijkl} E_{k}^{i} \otimes E_{l}^{j}
\end{equation}
where $ i,j,k,l = \overline{1,4} $, $ \cl{H} \equiv \cl{H}(w), \, h_{k} \equiv h_{k}(w) $. The identity, permutation $ P $, trace $ Q $ are $ 16 \times 16 $ operators with appropriate embedding and $ (E_{i}^{j})^{\alpha}_{\beta} = \delta_{i}^{\alpha}\delta^{j}_{\beta} $. This prescription leads to single class with undetermined spectral dependence
\begin{equation}\label{key}
	\cl{H} = h_{1} \cdot \mathds{1} + h_{2} \left( P - Q \right) + h_{4}\epsilon_{ijkl} E_{k}^{i} \otimes E_{l}^{j}
\end{equation}
with the {\Rx} to appear as
\begin{equation}\label{key}
	R=e^{H_1(u,v)} \left[ \left( H_2(u,v)-\frac{H_4(u,v)^2}{1 + H_2(u,v)} \right) I+ P - \dfrac{H_2(u,v)}{1 + H_2(u,v)} K + \frac{H_4(u,v)}{1 + H_2(u,v)} \epsilon_{ijkl} E_{k}^{i} \otimes E_{l}^{j}  \right] 
\end{equation}
which indeed can be identified with $ \al{so}(4) $ chain with arbitrary spectral dependent $ h_{k} $ and $ H_{k}(u,v) = \int_{v}^{u} \dd s h_{i}(s) $. As we have indicated before there is quasi-difference present, \textit{i.e.} $ u $ and $ v $ separate though generic spectral functions. Even more, it is possible to reabsorb one of the functions via normalisation and another can be spectrally reparametrised, which will single out the only function left. By this we have provided classification in the $ \al{su}(2) \oplus \al{su}(2) $ non-difference sector. We follow now with the discussion on the structure arising for the generalised Hubbard class and comparison of the technique results.

\paragraph{Comment on Generlised Hubbard Class} As we have discussed in \ref{Hub_GHC}, the $ \al{su}(2) \oplus \al{su}(2) $ was shown to be possible to embed in $ \al{su}(2|2) $ centrally extended superalgebra, which is relevant for the worldsheet scattering and integrability of {\ads}, where the latter appeared related to the Shastry {\Rx} \cite{Shastry_1986,Shastry:1988_DSTR_IHM}. So the Hubbard type and Shastry model structure is important in the context of integrability arising in Gauge/Gravity Duality \cite{Maldacena_1997,Witten_1998, Beisert:2005tm, Beisert:2006qh, Arutyunov:2009ga}. In the present setting, as it was indicated \ref{Hub_GHC}
\begin{equation}\label{key}
	\cl{H} = \cl{H}_{\text{Hub}} + \cl{H}_{\text{PH}} + \cl{H}_{\text{Flip}} + V 
\end{equation}
where one can consider an Ansatz with conserved fermion number, but an allowed violation of the spin. As we have mentioned the prescription above contains kinetic Hubbard part, pair hopping, flipping terms and a generic integrable potential (up to quartic order $ \mathbf{n}_{\sigma,k} $). It is important to remark, that within the provided setup, the space of solutions appears to be extremely large and possesses highly nontrivial systems arising in it. To the present moment, it could be stated that among resolvable nontrivial models, we can extract single model with all nonzero entries according to \eqref{ND-4-dim_Ansatz}, but the underlying {\Rx} after transformations and algebra its {\Rx} acquires difference form. As an example, we provide the corresponding $ \cl{H} $ and $ R $
\begin{equation}\label{key}
	\cl{H} = 
	{\tiny{
	\begin{pmatrix}
		-\lambda & 0 & 0 & 0 & 0 & 0 & 0 & 0 & 0 & 0 & 0 & 0 & 0 & 0 & 0 & 0 \\
		0 & \lambda & 0 & 0 & 0 & 0 & 0 & 0 & 0 & 0 & 0 & \rho_2 & 0 & 0 & -\rho_2 & 0 \\
		0 & 0 & 0 & 0 & 0 & 0 & 0 & 0 & \rho_1 & 0 & 0 & 0 & 0 & 0 & 0 & 0 \\
		0 & 0 & 0 & 0 & 0 & 0 & 0 & 0 & 0 & 0 & 0 & 0 & \rho_1 & 0 & 0 & 0 \\
		0 & 0 & 0 & 0 & \lambda & 0 & 0 & 0 & 0 & 0 & 0 & -\rho_2 & 0 & 0 & \rho_2 & 0 \\
		0 & 0 & 0 & 0 & 0 & -\lambda & 0 & 0 & 0 & 0 & 0 & 0 & 0 & 0 & 0 & 0 \\
		0 & 0 & 0 & 0 & 0 & 0 & 0 & 0 & 0 & \rho_1 & 0 & 0 & 0 & 0 & 0 & 0 \\
		0 & 0 & 0 & 0 & 0 & 0 & 0 & 0 & 0 & 0 & 0 & 0 & 0 & \rho_1 & 0 & 0 \\
		0 & 0 & -\rho_1 & 0 & 0 & 0 & 0 & 0 & 0 & 0 & 0 & 0 & 0 & 0 & 0 & 0 \\
		0 & 0 & 0 & 0 & 0 & 0 & -\rho_1 & 0 & 0 & 0 & 0 & 0 & 0 & 0 & 0 & 0 \\
		0 & 0 & 0 & 0 & 0 & 0 & 0 & 0 & 0 & 0 & 0 & 0 & 0 & 0 & 0 & \tau\,\lambda \\
		0 & -\xi & 0 & 0 & \xi & 0 & 0 & 0 & 0 & 0 & 0 & 0 & 0 & 0 & -\lambda & 0 \\
		0 & 0 & 0 & -\rho_1 & 0 & 0 & 0 & 0 & 0 & 0 & 0 & 0 & 0 & 0 & 0 & 0 \\
		0 & 0 & 0 & 0 & 0 & 0 & 0 & -\rho_1 & 0 & 0 & 0 & 0 & 0 & 0 & 0 & 0 \\
		0 & \xi & 0 & 0 & -\xi & 0 & 0 & 0 & 0 & 0 & 0 & -\lambda & 0 & 0 & 0 & 0 \\
		0 & 0 & 0 & 0 & 0 & 0 & 0 & 0 & 0 & 0 & \frac{\lambda}{\tau} & 0 & 0 & 0 & 0 & 0
	\end{pmatrix}
	}}
\end{equation}
where
\begin{equation}\label{key}
	\begin{array}{l}
		\rho_1= i \sqrt{\lambda^2-1} \\ 
		\rho_2=\frac{1-\lambda^2}{\xi}
	\end{array}
	\qquad\qquad 
	\text{with free parameters } \lambda,\tau,\xi
\end{equation}
with the {\Rx}
\begin{equation}\label{key}
	\cl{H} = 
	{\tiny{
			\begin{pmatrix}
				r_1 & 0 & 0 & 0 & 0 & 0 & 0 & 0 & 0 & 0 & 0 & 0 & 0 & 0 & 0 & 0 \\
				0 & r_2 & 0 & 0 & r_{11} & 0 & 0 & 0 & 0 & 0 & 0 & -r_{8} & 0 & 0 & r_{8} & 0 \\
				0 & 0 & r_4 & 0 & 0 & 0 & 0 & 0 & r_{10} & 0 & 0 & 0 & 0 & 0 & 0 & 0 \\
				0 & 0 & 0 & r_4 & 0 & 0 & 0 & 0 & 0 & 0 & 0 & 0 & r_{10} & 0 & 0 & 0 \\
				0 & r_{11} & 0 & 0 & r_2 & 0 & 0 & 0 & 0 & 0 & 0 & r_{8} & 0 & 0 & -r_{8} & 0 \\
				0 & 0 & 0 & 0 & 0 & r_1 & 0 & 0 & 0 & 0 & 0 & 0 & 0 & 0 & 0 & 0 \\
				0 & 0 & 0 & 0 & 0 & 0 & r_4 & 0 & 0 & r_{10} & 0 & 0 & 0 & 0 & 0 & 0 \\
				0 & 0 & 0 & 0 & 0 & 0 & 0 & r_4 & 0 & 0 & 0 & 0 & 0 & r_{10} & 0 & 0 \\
				0 & 0 & r_7 & 0 & 0 & 0 & 0 & 0 & r_3 & 0 & 0 & 0 & 0 & 0 & 0 & 0 \\
				0 & 0 & 0 & 0 & 0 & 0 & r_7 & 0 & 0 & r_3 & 0 & 0 & 0 & 0 & 0 & 0 \\
				0 & 0 & 0 & 0 & 0 & 0 & 0 & 0 & 0 & 0 & r_5 & 0 & 0 & 0 & 0 & r_{13} \\
				0 & -r_9 & 0 & 0 & r_9 & 0 & 0 & 0 & 0 & 0 & 0 & r_6 & 0 & 0 & r_{12} & 0 \\
				0 & 0 & 0 & r_7 & 0 & 0 & 0 & 0 & 0 & 0 & 0 & 0 & r_3 & 0 & 0 & 0 \\
				0 & 0 & 0 & 0 & 0 & 0 & 0 & r_7 & 0 & 0 & 0 & 0 & 0 & r_3 & 0 & 0 \\
				0 & r_9 & 0 & 0 & -r_9 & 0 & 0 & 0 & 0 & 0 & 0 & r_{12} & 0 & 0 & r_6 & 0 \\
				0 & 0 & 0 & 0 & 0 & 0 & 0 & 0 & 0 & 0 & r_{14} & 0 & 0 & 0 & 0 & r_5
			\end{pmatrix}
	}}
\end{equation}
\normalsize
where $ r_{i}(u) $ constitute
\begin{equation}\label{key}
	\begin{array}{l}
		r_1=\cosh u-\lambda\sinh u \\ [1ex]
		r_2=\frac{(1-\lambda^2)\sinh u \tanh u}{1-\lambda\tanh u} \\ [1ex] 
		r_3 = -r_{4} = i \sqrt{\lambda^2-1}\sinh u \\ [1ex]
		r_5 =\cosh u \\ [1ex] 
		r_6=-\frac{\sinh u\left(\lambda-\tanh u\right)}{1-\lambda\tanh u} \\ [1ex] 
		r_{\{ 7,10 \}} = 1 \\ [1ex] 
	\end{array}
	\qquad\qquad
	\begin{array}{l}
		r_8=\frac{(1-\lambda^2\tanh u)}{\xi\left(1-\lambda\tanh u\right)} \\ [1ex] r_9=-\frac{\xi\tanh u}{1-\lambda\tanh u} \\ [1ex] 
		r_{11}=\frac{\text{sech}\, u}{1-\lambda\tanh u} \\ [1ex] r_{12}=\frac{\text{sech}\,u\left(2-\lambda\sinh 2u+2\,\lambda^2\sinh^2 u \right)}{2\left(1-\lambda\tanh u\right)} \\ [1ex]
		r_{13}=\tau\,\lambda\sinh u \\ [1ex] 
		r_{14}=\frac{\lambda\sinh u}{\tau} \\ [1ex]
	\end{array}
\end{equation}
The novel structures beyond {\ads} Ansatz and their deformations is still an open question. It is an ongoing progress to find and classify unifying structures that must be arising in $ AdS_{5} $ \cite{} and would generalise them \cite{Beisert:2005tm,Beisert:2008tw}.

\begin{table}[ht]
	\centering
		\begin{tabular}{ c | c }
			Difference & Arb. spectral dependence \\ \hline\hline
			\Rn{1} & \Rn{3} $ \left( h_{2} \mapsto h \right) $ \\ \hline 
			\Rn{2} & \Rn{6} \\ \hline 
			\Rn{3} & \Rn{7} \\ \hline 
			\Rn{6} & \Rn{4} \\ \hline 
			\Rn{7} & \Rn{4}, \Rn{5} \\ \hline 
			\Rn{8} & \Rn{4} \\ \hline 
			\Rn{9} & \Rn{2} $ \left( h_{9} \mapsto h \right) $ \\ \hline 
			\Rn{10} & \Rn{2} $ \left( h_{9} \mapsto -h \right) $ \\ \hline 
			\Rn{12} & \Rn{7}, \Rn{8} \\ \hline 
		\end{tabular}
	\caption{Non-/Difference class correspondence through reduction limits in four dimensions}
	\label{Tab_D_ND_ReductionCorrespondence}
\end{table}

\paragraph{Non-/Difference comparison in 4-dim} To complete the discussion in four dimensional space, we can establish mappings between classes of difference form \ref{HHubbard_SolTable} and classes with arbitrary spectral dependence (or some quasi-difference) \ref{Tab_ND_4-dim_HSpace}. Namely, to obtain former form the latter by means of allowed transformations \ref{IT} and \ref{ND_IT} (shifts, rescalings, twisting and other) \cite{deLeeuw:2019vdb, deLeeuw:2020ahe}. The established correspondence is reflected in the Table \ref{Tab_D_ND_ReductionCorrespondence}, where core coefficients relabelling is indicated in brackets. 

One can note that non-difference classes \Rn{2}, \Rn{4} and \Rn{7} exhibit degeneracy, since they possess limits, which lead to more than difference form class from Table \ref{HHubbard_SolTable}. On contrary, the classes \Rn{1}, \Rn{3} do not have an independent difference form sub-classes and for difference form classes \Rn{4}\footnote{Specific limit is necessary to difference class \Rn{8}, since it holds for $ c_{1} = -2 $ and $ c_{2} \rightarrow 0 $ is required (singular in one of the $ h $ coefficients. )}, \Rn{5} and \Rn{11} there does not exist a map from the found non-difference classes\footnote{That must be taken into account when considering complete $ \al{su}(2) \oplus \al{su}(2) $ set, since there might more nontrivial limits following from other GHM sector or deformations, or the models that were omitted from the beginning because of their manifest difference dependence}.

\newpage

\section{Free fermions in $ AdS_{2,3} $}\label{FF_Section}

As we have indicated in section \ref{Sec_ND_2-dim}, the corresponding A and B classes possess the configuration suitable to arise in the $ AdS_{\{ 2,3 \}} $ string integrable backgrounds. Here we shall describe consistency of A and B classes with $ AdS $ backgrounds, highlight their special physical properties and derive observables of the related $ AdS_{3} $ system equipped with flux, whereas in the next section, the associated scattering operators, their properties, deformation realisation and algebraic structure is provided \ref{p5_Sec_AdS_2_3_Integrable_Deformations}.

\subsection{$ AdS_{\{ 2,3 \}} $ backgrounds}

An important framework of string integrability arises maximally symmetric {\ads} space \cite{Arutyunov:2009ga,Beisert:2005tm} in the context of $ AdS_{5}/CFT_{4} $ \cite{Maldacena_1997,Witten_1998,Klebanov:2004ya}. Significant progress was achieved in establishing integrability of less symmetric backgrounds {\adsm} and {\adsT}. These lower dimensional string integrable systems \cite{Babichenko_2010,Sax_2011,Borsato:2013qpa,Borsato:2015mma,Hoare:2014kma,Sfondrini:2014via} have also enhanced correlation between massless $ S $-matrices and distinct conformal field theories \cite{Zamolodchikov:1990bu,Zamolodchikov:1992zr,Fendley:1993jh,Bombardelli:2018jkj}. So we shall look at the emerging structures above in the more extended context of B classes above \cite{deLeeuw:2020xrw,deLeeuw:2020ahe}. First, we shall analyse how the free fermion condition is arising in the latter context and illustrate its significance for the underlying transfer matrices. We then perform an algebraic analysis exploiting free fermion condition for the $ AdS_{3} $ case and indicate how the property allows to replace certain steps of the Bethe Ansatz. In what we present the corresponding properties and exact derivations in Appendix \ref{Apx_FF}.

It is known that massless sector of the string integrable setting reflects its structure as a deformed quantum group of the associated 2-dim super-Poincar\'e symmetry. This realisation stems from the five-dimensional prescription \cite{Gomez:2007zr,Young:2007wd}. For the massless $ AdS_{3} $ sector, one identify further symmetry implications allowing one to establish non-relativistic {\Sx} along with its dressing factor to be of difference-form by specific variable transformations \cite{Stromwall:2016dyw,Borsato:2017icj,Fontanella:2016opq}. This difference dependence is equivalent to the one arising in the BMN limit \cite{Berenstein:2002jq}

In general, the $ AdS_{3} $ background admits a mixed-flux, which affects the conventional magnon dispersion relation 
\begin{equation}\label{key}
	E = \sqrt{\Big(m + \frac{k}{2 \pi} p\Big)^2 + 4 h^2 \sin^2\frac{p}{2}}
\end{equation}
with $ k $ being Wess-Zumino-Witten level, $ m $ corresponds to mass and $ h $ is a coupling \cite{OhlssonSax:2018hgc}. Related investigation of relativistic limit of RR- and LL-mode scattering has been addressed in , that lead a class of CFTs, whose exact resolution can be achieved by TBA \cite{Fontanella:2019ury}.

As indicated in \ref{Introduction}, the $ AdS_{2}/CFT_{1} $ duality gives rise to {\adsT} integrable background \cite{Hoare:2014kma,Hoare:2015kla,Fontanella:2017rvu}, which could dualised to superconformal quantum mechanics or a chiral CFT \cite{Maldacena:1998uz,Strominger:1998yg,Chamon:2011xk}. The $ AdS_{2} $ massive are governed by \textit{long} representations, whereas massless are in \textit{short} representations. There is also a clear understanding that massless $ S $-matrices form a separate class, which might correspond to the description of massless Renormalisation Group flows as of \cite{Zamolodchikov:1992zr}.  There is also a nontrivial BMN limit for LL- and RR-modes, which is completely different from $ \cl{N} = 1 $ supersymmetric quantum mechanics \cite{Fendley:1990cy}. It is rather resembling elliptic spin chain behaviour, since the underlying {\Sx} is of 8-vertex form, for which the an absence of reference state (pseudovacuum) is characteristic.

\subsection{$ AdS $ 8v classes and Free fermions}
As it have been shown \cite{deLeeuw:2020xrw,deLeeuw:2020ahe}, the 8-vertex classes with arbitrary spectral dependence exhibit standard form \eqref{p3_8v}. The complete classification of the difference models have been performed before \cite{deLeeuw:2019zsi} and it was established that all the corresponding 8-vertex (and below) models do satisfy certain polynomial constraints and free fermion in particular. 

We know that \eqref{p3_8v} prescription, creates two categories of the models, what we have qualified as the A-class and B-class \ref{Sec_ND_2-dim}. The class A essentially constitutes XXZ and XYZ possessing $ S^{z} \otimes S^{z} $ interaction \ref{Apx_2-dim}. The class B characterises the models that are relevant for holography and would arise in $ AdS_{\{ 2,3 \}} $.

On the other side as we could have noticed from \ref{NA_S_SD} and \cite{deLeeuw:2020xrw,deLeeuw:2020ahe} that the coupled Sutherland system \eqref{SutherlandEq_1}-\eqref{SutherlandEq_2}
\begin{equation}\label{p4_Sutherland_System}
	\begin{cases}
		\left[ R_{13} R_{23}, \cl{H}_{12}(u) \right] = R_{13, v}R_{23} - R_{13}R_{23, v} \\
		\left[ R_{13} R_{12}, \cl{H}_{23}(v) \right] = R_{13}R_{12, u} - R_{13, u}R_{12}
	\end{cases}
\end{equation}
when we plug in the before considered 8-vertex Ansatz for $ \cl{H} $ \eqref{ND_H8v} and {\Rx} \eqref{p3_8v} we obtain an intertwined system. It is possible then to express derivatives and the remaining system to resolve for Hamiltonian entries $ h_{k} $ in terms of $ r_{k} $. Importantly, that in this case one can gather coupled set of constraints, which can combined as follows

\begin{equation}\label{key}
	\begin{array}{l}
		\dfrac{\fk{F}}{r_{2}r_{4}} = \Phi(u) \\ [3ex]
		\dfrac{\fk{F}}{r_{1}r_{3}} = \Gamma(u) \\ [3ex]
	\end{array}
	\qquad
	\begin{array}{l}
		\dfrac{\fk{F}}{r_{1}r_{2}} = \Phi(v) \\ [3ex] 
		\dfrac{\fk{F}}{r_{3}r_{4}} = \Gamma(v) \\ [3ex]
	\end{array}
\end{equation}
where $ \fk{F} = r_{1}r_{4} + r_{2}r_{3} - r_{5}r_{6} - r_{7}r_{8} $, $ \Phi $ and $ \Gamma $ depend on $ h_{k} $. We can immediately see that this leads to the
\begin{equation}\label{key}
	\dfrac{ \left[ r_{1}r_{4} + r_{2}r_{3} - \left( r_{5}r_{6} + r_{7}r_{8} \right) \right]^{2} }{r_{1}r_{2}r_{3}r_{4}} = \Gamma(u)\Phi(u) = \Gamma(v)\Phi(v)
\end{equation}
which for further purposes can be suitably brought into the form
\begin{equation}\label{p4_Baxter_Constraint}
	\dfrac{ \left[ r_{1}r_{4} + r_{2}r_{3} - \left( r_{5}r_{6} + r_{7}r_{8} \right) \right]^{2} }{r_{1}r_{2}r_{3}r_{4}} = \fk{c}_{\tx{B}}
\end{equation}
where $ \fk{c}_{\tx{B}} $ constitutes a characteristic Baxter constant with
\begin{equation}\label{key}
	\begin{cases}
		\fk{c}_{\tx{B}} = 0, \tx{ \textit{Free Fermion} constraint} \\
		\fk{c}_{\tx{B}} \neq 0, \tx{ \textit{Baxter} constraint}
	\end{cases}
\end{equation}
where indeed one can note, that for $ \fk{c}_{\tx{B}} \rightarrow 0 $ \eqref{p4_Baxter_Constraint} reduces to the free fermion polynomial $ r_{1}r_{4} + r_{2}r_{3} = r_{5}r_{6} + r_{7}r_{8} $. Moreover, by means of operations following from \eqref{p4_Sutherland_System}, one can address analogous polynomial constraining also for other symmetry algebras \cite{Korepanov_1993tetrahedral,Korepanov:1994rc,Umeno_1998fermionic}, including the ones associated to {\ads} ($ \al{psu}(2,2|4) $, $ \al{su}(2|2) $, {\Hal} and other), which leads to a more generic algebraic shortening structure and transformations important for multi-layer structure arising in string backgrounds \cite{Mitev:2012vt} and that would find it useful for a construction of TBA.

\paragraph{A Class} We can now see that for 8vA, which represents full elliptic setup (again up to integrable freedoms from \ref{ND_IT}), given in the form
\begin{equation}\label{key}
	R_{\tx{8vA}} = 
	\begin{pmatrix}
		\mrm{sn}(\alpha + u) & 0 & 0 & k\, \mrm{sn}(\alpha) \mrm{sn}(u) \mrm{sn}(\alpha + u) \\
		0 & \mrm{sn}(u) & \mrm{sn}(\alpha) & 0 \\
		0 & \mrm{sn}(\alpha) & \mrm{sn}(u) & 0 \\
		k\, \mrm{sn}(\alpha) \mrm{sn}(u) \mrm{sn}(\alpha + u) & 0 & 0 &  \mrm{sn}(\alpha + u) 
	\end{pmatrix}
\end{equation}
satisfies \eqref{p4_Baxter_Constraint} in generic form, \textit{i.e.}
\begin{equation}\label{key}
	\fk{c}_{\tx{B}} = 4 \, \mrm{cn}^{2}(\alpha,k^{2}) \mrm{dn}^{2}(\alpha,k^{2})
\end{equation}
where Elliptic Jacobi functions $ \mrm{xn} \equiv \mrm{xn}(u,k^{2}) $ (more detailed in \ref{Apx_Elliptic_Apparatus}), elliptic modulus $ \fk{c}_{\tx{B}} $, $ k $, $ \alpha \in \tx{const} $. For the 6vA follows a straightforward check.

\paragraph{B class} We proceed explicitly for the 6- and 8-vertex B class. For the 6vB
\begin{equation}\label{key}
	R_{\tx{6vB}} = 
	\begin{pmatrix}
		1 & 0 & 0 & 0 \\
		0 & 0 & 1 & 0 \\
		0 & 1 & h_{5}(z_{2}) - h_{5}(z_{1}) & 0 \\
		0 & 0 & 0 & 1 \\
	\end{pmatrix}
	+ H_{4}
	\begin{pmatrix}
		h_{5}(z_{1}) & 0 & 0 & 0 \\
		0 & 1 & 0 & 0 \\
		0 & 0 & h_{5}(z_{1})h_{5}(z_{2}) & 0 \\
		0 & 0 & 0 & -h_{5}(z_{2}) \\
	\end{pmatrix}
\end{equation}
with $ H_{4} \equiv H_{4}(z_{1}) - H_{4}(z_{2}) $ and $ h_{5}$ , $ H_{4} $ free spectral functions, we get 
\begin{equation}\label{key}
	r_{1}r_{4} + r_{2}r_{3} = r_{5}r_{6} + r_{7}r_{8} = 1
\end{equation}
which fulfils free fermion condition $ \fk{c}_{\tx{B}} = 0 $. 

\noindent To identify 8vB constraint, we first recast the structure of 8vB into
\begin{equation}\label{p4_R8vB}
	\begin{array}{l}
		r_{1} = \Sigma(u,v) \left[ \sin\eta_+\frac{\mrm{cn}}{\mrm{dn}} 
		-\cos\eta_+ \mrm{sn} \right] \\ [1ex]
		r_{2} = - \Sigma(u,v) \left[ \cos\eta_-\mrm{sn} +\sin\eta_-\frac{\mrm{cn}}{\mrm{dn}} 
		\right] \\ [1ex]
		r_{3} = - \Sigma(u,v) \left[ \cos\eta_-\mrm{sn}  - \sin\eta_-\frac{\mrm{cn}}{\mrm{dn}}  \right] \\ [1ex]
		r_{4} = \Sigma(u,v) \left[ \sin\eta_+\frac{\mrm{cn}}{\mrm{dn}} + \cos\eta_+ \mrm{sn} \right] \\ [1ex]
		r_{5} = r_{6} = 1, \qquad r_{7} = r_{8} = k \, \mrm{sn}\frac{\mrm{cn}}{\mrm{dn}} \\
	\end{array}
\end{equation}
where it is convenient to utilise short Elliptic $ \mrm{xn} $-notation (Appendix \ref{Apx_Elliptic_Apparatus})
\begin{equation}\label{p4_EllipticUnity}
	\mrm{xn} \equiv \mrm{xn}(u-v,k^{2}), \quad\qquad (k \, \mrm{sn})^{2} + \mrm{dn}^{2} = 1
\end{equation}
with $ \mrm{xn} $ to represent any elliptic function involved with elliptic modulus $ k $, and $ \Sigma(u,v) $ and $ \eta_{\pm} $ constitute 
\begin{equation}\label{p5_Eta_Sigma}
	\Sigma(u,v) = \frac{1}{\sqrt{\sin\eta(u) \sin\eta(v)}}, \quad\qquad \eta_{\pm} = \dfrac{\eta(u) \pm \eta(v)}{2}
\end{equation}
We can immediately obtain the corresponding terms of \ref{p4_Baxter_Constraint}
\begin{equation}\label{key}
	r_{1}r_{4} = \Sigma^{2} \left[ (\sin\eta_{+}\dfrac{\mrm{cn}}{\mrm{dn}})^{2} - (\cos\eta_{+} \mrm{sn})^{2} \right]
\end{equation}
along with $ r_{2}r_{3} $
\begin{equation}\label{key}
	r_{2}r_{3} = - \Sigma^{2} \left[ (\sin\eta_{-}\dfrac{\mrm{cn}}{\mrm{dn}})^{2} - (\cos\eta_{-} \mrm{sn})^{2} \right]
\end{equation}
and the rest results in
\begin{equation}\label{key}
	r_{5}r_{6} + r_{7}r_{8} = 1 + \left( k \, \mrm{sn}\dfrac{\mrm{cn}}{dn} \right)^{2}
\end{equation}
In total that will add up to both sides of the equation as follows 
\begin{equation}\label{key}
	\Sigma^{2} \left[ ( \sin^{2}\eta_{+} - \sin^{2}\eta_{-} )\left( \dfrac{\mrm{cn}}{\mrm{dn}} \right)^{2}  + ( \cos^{2}\eta_{-} - \cos^{2}\eta_{+} )\mrm{sn}^{2} \right] = 1 + \left( k \, \mrm{sn}\dfrac{\mrm{cn}}{dn} \right)^{2}
\end{equation}
where on the LHS after appropriate $ \eta $-rescaling
\begin{equation}\label{key}
	\begin{array}{l}
		\left( \dfrac{\mrm{cn}}{\mrm{dn}} \right)^{2}  + \mrm{sn}^{2} = 1 + \left( k \, \mrm{sn}\dfrac{\mrm{cn}}{dn} \right)^{2} \\ [3ex]
		\left[ 1 - \left( k \, \mrm{sn} \right)^{2} \right] \left( \dfrac{\mrm{cn}}{\mrm{dn}} \right)^{2} = 1 - \mrm{sn}^{2} \qquad \quad\tx{\eqref{p4_EllipticUnity} and \ref{Apx_Elliptic_Apparatus}} \\ [3ex]
		\mrm{sn}^{2} \equiv 1 - \mrm{cn}^{2} \\ [3ex]		
	\end{array}
\end{equation}
This indeed, shows that 8vB class does satisfy the free fermion property\footnote{Strictly speaking under rescaling of the free spectral function}.

\subsection{$ AdS_{3} $ with RR-flux: $ R $ and $ t $ analysis}\label{p4_Sec_AdS3RR_R_T_Structure}

In \cite{deLeeuw:2020bgo} we have also shown on how free fermion property can be implemented at the level of ABA to obtain significant reductions in algebraic computations as well as obtain useful relations. We shall demonstrate now on how to derive transfer matrix $ t $ and {\Rx} for the massless $ AdS_{3} $ sector equipped with RR-flux. For this purpose, we shall work in the oscillator formalism and as a spectral parameter analogue we shall use relativistic variable $ \theta $. Important to note, that this prescription is of difference form, \textit{i.e.} one can replace $ \theta = \theta_{1} - \theta_{2} $. At the same time all expression will hold in the non-relativistic case also.
\begin{equation}\label{key}
	\begin{array}{l}
		\begin{cases}
			a(\theta) \equiv \mbox{sech} \frac{\theta}{2} \\
			b(\theta) \equiv \tanh \frac{\theta}{2}
		\end{cases}
	\end{array}
	\qquad
	\begin{array}{l}
		\begin{cases}
			a_{ij} \equiv a(\theta_i - \theta_j) \\
			b_{ij} \equiv b(\theta_i - \theta_j)
		\end{cases}
	\end{array}
\end{equation}
Whereas the corresponding {\Rx} with no specified normalisation (without loss of generality we shall recover upon demand)
\begin{equation}\label{key}
	R_{12}(\theta) = \Big[\cosh \frac{\theta}{2} \, (m_1 m_2 - n_1 n_2) - \sinh \frac{\theta}{2} \, (m_1 n_2 - n_1 m_2) + c^\dagger_1 c_2 - c_1 c^\dagger_2 \Big]
\end{equation}
it can be compared with 6-vertex of \cite{Mitev:2012vt} and obtain mapping
\begin{equation}\label{key}
	a_M = d_M = \cosh \frac{\theta}{2}, \qquad b_M = - c_M = i \sinh \frac{\theta}{2}
\end{equation}
where $ a_{M}, b_{M}, c_{M}, d_{M} $ are the corresponding parameters of \cite{Mitev:2012vt}. We immediately see that FF-condition is satisfied
\begin{equation}\label{key}
	a_M d_M - b_M c_M = 1
\end{equation}
which consistent with free fermion 
\begin{equation}\label{key}
	r_1 r_4 + r_2 r_3 = r_5 r_6 + r_7 r_8
\end{equation}
with $ r_{\{ 7,8 \}} = 0 $ and transformation to \cite{Mitev:2012vt} is governed by
\begin{equation}\label{key}
	a_M = r_4, \quad b_M = i r_2, \quad c_M = i r_3, \quad d_M = r_1, \quad r_6 = r_5 = 1
\end{equation}
Important to note that $ R(0) = P_{g} $ obeys graded permutation.

We shall deviate in this section from the previous sections in the fact, that here it is not critical to get a Hamiltonian. On contrary we would like to demonstrate the effect of free fermion properties on the $ R $- and transfer matrix, as well as acquire useful generating properties that can arise on the latter \ref{Apx_FF}. In fact, the massless $ R_{AdS_{3}} $ is described by nested BA, where pseudovacuum consisting of $ \ket{\phi} $ is level-one pseudovacuum and not the corresponding  BMN vacuum of all $ \ket{Z} $ \cite{sax2013massless}. It necessary then to obtain $ t $
\begin{equation}\label{key}
	T_N = \mbox{str}_0 R_{01} (\theta_0 - \theta_1) ... R_{0N} (\theta_0 - \theta_N)
\end{equation}
with generic inhomogeneities $ \theta_{i} $ \cite{Bombardelli:2018jkj} and satisfy momentum-carrying Bethe equations. We can establish our free fermion mapping by starting from $ t $
\begin{align}
	t_{2} &=&\frac{1 - b_{01} b_{02}}{a_{01}a_{02}} \, (m_1 m_2 - n_1 n_2) + \frac{b_{01} - b_{02}}{a_{01}a_{02}} \, (m_1 n_2 - n_1 m_2) + c_1^\dagger c_2 - c_1 c_2^\dagger\nonumber\\
	&=&\frac{1}{a_{12}} \mathds{1} - e^{-\frac{\theta_{12}}{2}} c_1^\dagger c_1 - e^{\frac{\theta_{12}}{2}} c_2^\dagger c_2 +c_1^\dagger c_2 - c_1 c_2^\dagger
\end{align}
where $ t_{2} $ indicates two quantum spaces and  $ \theta_{ij} \equiv \theta_i - \theta_j $. From here one can get the following transformations 
\begin{equation}\label{key}
	\begin{cases}
		c_1 = \cos \alpha \, \eta_1 - \sin \alpha \, \eta_2 \\ 
		c_2 = \sin \alpha \, \eta_1 + \cos \alpha \, \eta_2 \\ 
		\cot 2\alpha = \sinh \frac{\theta_{12}}{2} \in \bb{R}
	\end{cases}
\end{equation}
where in order to obtain restrictions to physical momenta (frame particles) we demand reality of inhomogeneities. The new canonical creation/annihilation operators arise as follows
\begin{equation}\label{key}
	\{\eta_i,\eta_j^\dagger\}=\delta_{ij}, \qquad \{\eta_i,\eta_j\} = \{\eta_i^\dagger,\eta_j^\dagger\} = 0, \qquad \{i,j\} = 1,2
\end{equation}
that in turn brings the transfer matrix $ t $ into the following form
\begin{equation}\label{p4_t2_1}
	t_{2} = \cosh \frac{\theta_{12}}{2} \left( \mathds{1} - 2 \eta_1^\dagger \eta_1 + \color{red} 0 \cdot \eta_{2}^{\dagger}\eta_{2} \color{black} \right) 
\end{equation}
where term in red is put for the illustrative purpose only, since the established transformation exploits the free fermion and diagonalises $ t $.

It must be obvious now on how to obtain the eigenspectrum due to the additive pattern of free fermions. \textit{i.e.} eigenvalues would come from adding individual energies. Indeed, since $ \eta_{i} $ both annihilate the pseudovacuum, then
\begin{equation}\label{key}
	\ket{\phi} \otimes \ket{\phi} = 0
\end{equation}
with eigenvalue $ \cosh\dfrac{\theta_{12}}{2} $. It is immediate to see how the operators do
\begin{equation}\label{key}
	\eta_1^\dagger |0\rangle = \sin \alpha \Big[ c_2^\dagger |0 \rangle + \cot \alpha \,  c_1^\dagger |0 \rangle\Big]=\sin \alpha \Big[ |\phi\rangle \otimes |\psi\rangle + \cot \alpha \, |\psi\rangle \otimes |\phi\rangle\Big]
\end{equation}
\begin{equation}\label{key}
	\eta_2^\dagger |0\rangle = \cos \alpha \Big[ c_2^\dagger |0 \rangle - \tan \alpha \,  c_1^\dagger |0 \rangle\Big]=\cos \alpha \Big[ |\phi\rangle \otimes |\psi\rangle - \tan \alpha \, |\psi\rangle \otimes |\phi\rangle\Big]
\end{equation}
where eigenvalues accordingly constitute $ \pm \cosh\dfrac{\theta_{12}}{2} $ and creation acquires
\begin{equation}\label{key}
	\eta_1^\dagger \eta_2^\dagger |0\rangle = |\psi \rangle \otimes |\psi\rangle, \qquad \tx{with eigenvalue} \quad \cosh\dfrac{\theta_{12}}{2}.
\end{equation}
After choosing branch from $ \cot \alpha = - \exp\left[ - \frac{\theta_{12}}{2} \right] $ one finds complete agreement with \cite{Bombardelli:2018jkj}. It is also possible to perform explicit checks with ABA and identify the corresponding solutions for operators with evaluation on auxiliary Bethe system. The core structure of that can be given by generic expression for an eigenvalue (with $ N $ spaces)
\begin{equation}\label{key}
	\lambda = {{} \Theta_N}(\theta_0;\theta_1,...,\theta_N) \prod_{m=1}^M \frac{1}{b({{} v}_m - \theta_0)}
\end{equation}
with $ \Theta_{N} $ is a function on rapidities and spectral parameter (to all eigenvalues), whereas $ v_{m} $ provided by $ N $ potential solutions of the auxiliary Bethe system
\begin{equation}\label{key}
	\prod_{i=1}^N b({{} v} - \theta_i) = 1 , \qquad v \in \left[ -i\frac{\pi}{2}, \frac{\pi}{2} \right] 
\end{equation}
where solutions localise around centres $ z_{k} $ on $ v_{k} = z_{k} \pm\frac{\pi}{2} $ \cite{Fendley:1993pi}. It clear that resolution involves distinct configuration of the $ N $ possible solutions with $ M $ excitations. We also know that solution for $ v_{m} $ comes in pairs, where associated pair is obtained from $ v \rightarrow v + i\pi $, hence there always exists $ v = \pm \infty $ for $ N $ even. This becomes consistent with expression for $ t_{N} $
\begin{equation}\label{key}
	t_{N} = e^{i \psi_N \mathds{1} + i \sum_{i=1}^N \omega_i \, \eta_i^\dagger \eta_i}
\end{equation}
where $ \psi_{N}, \omega_{N} = f_{a}(\theta_{0}; \theta_{1} \theta_{2} \dots \theta_{N} ) $ and $ \eta, \eta^{\dagger} = f_{b}(\theta_{1} \theta_{2} \dots \theta_{N}) $. In principle, due to existing $ v = \infty $, the $ \omega = 0 $ and some other $ \omega_{i} = \pi $ and the rest $ (\omega_{k}, -\omega_{k}) $. From these arguments we can see that again for $ N = 2 $
\begin{equation}\label{key}
	t_{2} = e^{i \psi_2} \, e^{i \omega_1 N_1} = e^{i \psi_2} \, \Bigg[1+ N_1 \sum_{n=1}^\infty \frac{(i \omega_1)^n}{n!} \Bigg] = e^{i \psi_2} \Big[1+ N_1 (e^{i \omega_1} - 1) \Big] = e^{i \psi_2} \Big[1 - 2 N_1\Big]
\end{equation}
that reproduces \eqref{p4_t2_1} after $ \psi_2 \rightarrow -i \log \cosh \frac{\theta_{12}}{2} $. For $ N = 3 $ we can obtain \cite{Bombardelli:2018jkj}
\begin{equation}\label{key}
	\begin{array}{c}
		\omega_1 = 0 \\ 
		\omega_2 = i \log \frac{y^2 - \mu^2}{y^2 + \mu^2}\equiv \omega, \qquad \omega_3 = - i \log \frac{y^2 - \mu^2}{y^2 + \mu^2}= -\omega
	\end{array}
\end{equation}
along with 
\begin{equation}\label{key}
	\mu = e^{\frac{\theta_0}{2}}, \qquad y = - e^{- i \frac{\pi}{4}} e^{\frac{\theta_1 + \theta_2 + \theta_3}{4}} (e^{\theta_{1}} + e^{\theta_{2}} + e^{\theta_{3}})^{-\frac{1}{4}}
\end{equation}
The pattern is indeed predictable, but one would anticipate to observe if $ \omega \in \bb{R} $
\begin{equation}\label{key}
	\frac{y^2 - \mu^2}{y^2 + \mu^2} = \frac{e^{-\theta _0}-i e^{-\frac{1}{2} \left(\theta _1+\theta _2+\theta _3\right)} \sqrt{e^{\theta _1}+e^{\theta _2}+e^{\theta _3}}}{e^{-\theta _0}+i e^{-\frac{1}{2} \left(\theta _1+\theta _2+\theta_3\right)} \sqrt{e^{\theta _1}+e^{\theta _2}+e^{\theta _3}}}
\end{equation}
as we can see that is nothing, but a pure phase, which also manifests unitarity of $ t $. One can also confirm the eigenvalue additivity and for $ t_{2} $ that would result in 
\begin{equation}\label{key}
	\{e^0,e^0,e^\omega,e^{-\omega},e^{\omega},e^{-\omega},e^0,e^{0}\}
\end{equation}
that comes from creation of $ \leq 3 $ free fermions with additive energies. It requires further analytic investigation, but one propose under generic commutativity of $ N_{i} $ for $ AdS_{3} $ with RR
\begin{equation}\label{key}
	t_{N} = e^{i \psi_N} \prod_{i=1}^N \left[ 1 + (e^{i \omega_i}-1) N_i \right] .
\end{equation}

Importantly that in the new variables we can also reinstate the {\Rx}
\begin{equation}\label{p4_RRAdS3_R}
	R_{12}(\theta_{1} - \theta_{2}) = t_{2} = \cosh \frac{\theta_{12}}{2} \Big[\mathds{1} - 2 \eta_1^\dagger \eta_1\Big] = \Big[\cosh \frac{\theta_{12}}{2}\Big] \, e^{i \pi N_1}
\end{equation}
it is useful to remark that here it exceptionally considered on two sites with associated canonical creation/annihilation operators $ \eta_{i},\eta_{i}^{\dagger} $ and number operators $ N_{i} $
\begin{equation}\label{key}
	N_{i} \equiv \eta_{i}^{\dagger}\eta_{i}, \qquad N_{1}^{2} \equiv N_{1}
\end{equation}
it can then be also recovered \cite{Fontanella:2016opq,Stromwall:2016dyw,Bombardelli:2018jkj} on hermiticity and unitarity of 
\begin{equation}\label{key}
	\tilde{R}^2 = \tilde{R}^{\dagger} \tilde{R} =  (\mathds{1} - 2 N_1)(\mathds{1} - 2 N_1) = \mathds{1}, \qquad \tilde{R} = \dfrac{R}{\cosh\frac{\theta_{12}}{2}}
\end{equation}
under $ \theta_{i} \in \bb{R} $. So For the massless $ R_{AdS_{3}} $ one can derive
\begin{equation}\label{key}
	\tilde{R} = e^{i \pi N_1}, \qquad N_1 = \eta_1^\dagger \eta_1
\end{equation}
After all one can see on how the free fermion property is reflected in the emergent supersymmetry charges
\begin{equation}\label{p4_RRAdS3_Supercharges}
	\begin{cases}
		Q \equiv e^{\frac{\theta_1}{2}} c_1 +e^{\frac{\theta_2}{2}} c_2 \\ 
		Q^\dagger \equiv e^{\frac{\theta_1}{2}} c_1^\dagger +e^{\frac{\theta_2}{2}} c_2^\dagger \\ 
		\left[ Q, R \right] = \left[ Q^{\dagger}, R \right] = 0
	\end{cases}
\end{equation}
which can obtained by virtue of free fermion anti-commutativity.  Along with supergenerators we also have an associated Hopf algebra \ref{Apx1_Hopf_Algebra} structure
\begin{equation}\label{key}
	\begin{array}{l}
		\Delta(q) = q_1 \otimes \mathds{1} + \mathds{1} \otimes q_2 \\ [2ex]
		\Delta(q^\dagger) = q^\dagger_1 \otimes \mathds{1} + \mathds{1} \otimes q^\dagger_2 \\ [2ex]
	\end{array}
	\qquad
	\begin{array}{l}
		q_i = \sqrt{p_i} \, E_{12} \\ [2ex]
		q^\dagger_i = \sqrt{p_i} \, E_{12} \\ [2ex]
	\end{array}
	\qquad
	\tx{with} \, \, p_i = {{} e^{\theta_i}} \\ [2ex]
\end{equation}
where $ p_{i\{ 1,2 \}} $ are momenta. Part of the cocommutativity is reflected by the last commutative condition in \eqref{p4_RRAdS3_Supercharges}
\begin{equation}\label{key}
	\Delta^{op}(q) R = R \Delta(q), \qquad \Delta^{op}(q^\dagger) R = R \Delta(q^\dagger), 
\end{equation}
this supersymmetry has a graded structure encoded in it and on contrary respects $ R(\theta_{1},\theta_{2}) = R(\theta_{1} - \theta_{2}) $.

\newpage

\section{$ AdS_{\{ 2,3 \}} $ integrable deformations}\label{p5_Sec_AdS_2_3_Integrable_Deformations}

The discussion that we have initiated in \ref{ND_Technique} led us to the existence of a novel deformed structure in $ AdS_{2} $ and $ AdS_{3} $ integrable backgrounds. We shall now reinstate these structures and provide investigation of the associated 6- and 8-vertex sector in the AdS/CFT integrable context, \textit{i.e.} how $ AdS_{\{ 2, 3 \}} $ integrability is emergent and related to our deformed classes. More specifically, it will be shown how our deformed $ R $-/$ S $-matrices can describe equivalent/opposite particle scattering formalism, and what is the right mapping language to use. It will be also demonstrated that known $ AdS_{3} $ $ R $-/$ S $-matrices and some of their $ q $-deformations can be obtained from our classes \cite{deLeeuw:2020ahe,deLeeuw:2020xrw,deLeeuw:2020bgo}, as well as their relevance for studying sigma models. In addition, we shall discuss a generic algebra prescription of the aforementioned classes \cite{deLeeuw:2021ufg} and implications of further symmetry constraining.

\subsection{$ AdS $ string integrable structure}

One of the core properties of \textbf{I}ntegrable \textbf{Q}uantum \textbf{F}ield \textbf{T}heories in $ 1+1 $-dimensions is complete factorisation of the scattering process \cite{Zamolodchikov:1978xm,Zamolodchikov:1990bu,Zamolodchikov_1979z}. Indeed, the scattering can be viewed as an elastic one, where in this elementary intermediate processes a creation/annihilation is not happening. It can be shown from many perspectives \cite{Bombardelli:2016scq}, but a fundamental argument comes from the fact of dimensionality and an amount of conserved symmetries. In the case of IQFT one can find and infinite tower of symmetries corresponding to an infinite number of conserved charges (obtained from conserved currents). In this respect a three body simultaneous scattering is not present, but instead a two-body consequential scattering is involved, which is governed by the braided structure of YBE
\begin{equation}\label{key}
	S_{12}S_{23}S_{12} = S_{23}S_{12}S_{23}
\end{equation}
which can be recast in many suitable ways \cite{Drinfeld:1985rx,Perk2006Yang,Jimbo:1989,Jimbo:1985vd}. Moreover, 2-dim IQFT define an integrable structure in holographic context \cite{Maldacena_1997,Witten_1998,Gubser_1998,Polchinski:2010hw}, as well as in a variety of string worldsheet sigma models \cite{Zamolodchikov:1992zr,Fateev:1996ea,Metsaev:1998it,Fateev:2018yos}. Specifically, one can exploit light-cone gauge \cite{Arutyunov:2009ga} and in the decompactifying limit map a worldsheet into a plane, where an asymptotic formalism is emergent. With the help of the latter one can define {\Sx} structure and attempt integrable algebraic approaches and perturbation to investigate.

On the other hand, as we have discussed in \ref{NA_S_SD}, one needs to account for symmetry considerations, where perturbation is only needed for identifications of the last in certain cases. It has been shown that light-cone symmetry breaks certain superisometries of the string backgrounds, but provides dramatic simplifications and restrictions on the characteristic observable computations. In this respect, here we shall look at {\adsm} \cite{Borsato:2013qpa,Borsato:2014hja,Borsato:2015mma,Sfondrini:2014via,Fontanella:2019ury,Torrielli:2021hnd} and {\adsT} \cite{Hoare:2014kma,Hoare:2014oua,Hoare:2015kla,Fontanella:2017rvu,Fontanella:2019baq} string backgrounds and integrable structure that could arise in this geometries. Dual super-CFTs in this sector are still not completely determined, but a number of proposals exist form worldsheet orbifold perspective. Nevertheless there are associated string sigma models known \cite{Metsaev:1998it}, described by super-quotients, also known as supercoset models \cite{Zarembo:2010sg,Hoare:2014kma,Hoare:2015kla,Delduc:2018xug,Lukyanov:2012zt}. There is also a class of integrable deformations on various backgrounds that makes it connection with an underlying classical Yang-Baxter algebra \cite{Klimcik:2008eq,Klimcik_2014,Hoare:2014oua}.

Another relevant direction that correlates with the present discussion is a definition of integrable theories on various deformed backgrounds, \textit{e.g.} $ \eta $-/$ \lambda $-deformed backgrounds and non-abelian $ T $-duality symmetry arising in this theories \cite{Borsato:2017qsx,Hoare:2016wsk,Arutyunov:2015mqj,Sfetsos:2013wia,Sfetsos:2014cea,Delduc:2013qra,Hollowood:2014qma,Arutyunov:2013ega}. In a number of cases above can study a $ q $-deformed symmetry algebra \cite{Delduc:2013qra,Delduc:2014kha,Delduc_2013}, as of universal enveloping algebra $ U_{q} \left[ \al{psu}(2 \vert 2)_{\tx{ce}} \right]  $ in {\ads} space. The last prescription have been also resolved also for the $ AdS_{3} $ case \cite{Delduc_2013,Hoare:2014oua,Seibold:2019dvf,Bocconcello:2020qkt,Garcia:2021iox,Seibold:2021lju}.

Conventionally, the deformation of the string (sigma) models above is implemented in the functional of the corresponding theory. However it is not always explicit or not obvious at all on how such deformations could arise at the level of an integrable {\Sx} or a corresponding symmetry algebra. In what, we shall address here the description of 6- and 8-vertex B models that followed from the results of technique \ref{Sec_ND_2-dim}. In fact, these models admit mapping and embedding of known string models from above in $ AdS_{\{ 2,3 \}} $ at the level of {\Sx} under appropriate constraining of free spectral functions of the B class. Moreover, parameters present in 6vB/8vB source deformations under their flow (these parameters provide different type of deformations). Hence our prescription forms a class of deformed models \cite{deLeeuw:2020ahe,deLeeuw:2020xrw,deLeeuw:2020bgo,deLeeuw:2021ufg} that admit string background deformed integrable models.

\begin{figure}[!h]
	\centering
	\includegraphics[width=0.80\textwidth]{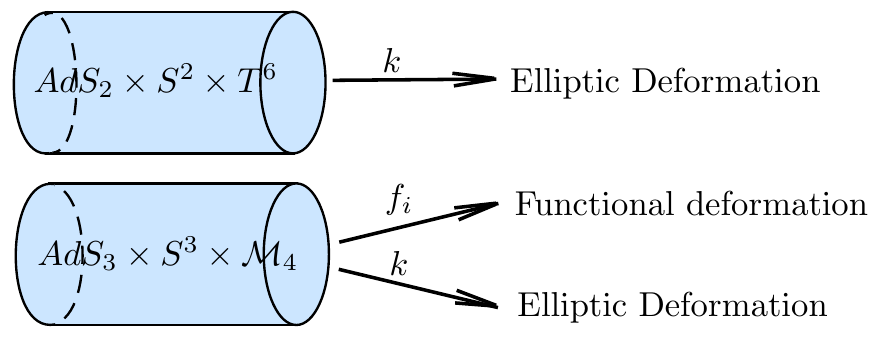}
	\caption{The $ AdS_{\{ 2,3 \}} $ deformations based on 8vB and 6vB classes}
	\label{p5_AdS_Deformation_Scheme}
\end{figure}

In this regard, the {\adsm} string background will allow for two deformations, which will be realised through 6vB and 8vB models. These deformations can be mapped to $ q $-deformed string sigma models of \cite{Hoare:2014oua,Seibold:2021lju}. One advantage that this is realised through finite amount of unconstrained functions and the remainder allows for further functional deformations. Concretely, the last ar not restricted by any physical constraint, \textit{e.g.} mass-energy-momentum relations, reality conditions, representation parametrisations and more. The 8vB deformation here, is a novel elliptic deformation and respects all integrable constraints of the corresponding {\Sx}, \textit{e.g.} existence of special spectral points, braiding-unitarity and crossing symmetry. Interesting, that for certain special points of deformation parameter a symmetry breaking occurs.

It is then only 8vB class that gives rise to {\adsT} deformation, which appears to be of elliptic form. This model is parametrised by the elliptic modulus and demonstrates non-trivial limits when the $ k $ parameter flows. There is another specific fact, that this deformation possesses analogous algebra to that of \cite{Hoare:2014kma}, however is manifestly of distinct representation. Peculiar, that deformation does not affect higher level Yangian generators. It is also of interest to resolve for the underlying quantum algebra of the $ AdS_{2} $ deformed class. The deformation realisations are reflected in the diagram \ref{p5_AdS_Deformation_Scheme}. As we have noticed above \ref{FF_Section} all these models satisfy free fermion property, which provides remarkable relations that have use during computations and symmetry proofs \cite{deLeeuw:2020bgo}.

\subsection{Blocks: Structure of vertex models}
The novel 2-dim classes \cite{deLeeuw:2020xrw,deLeeuw:2020ahe} with unspecified spectral dependence (non-difference) turn to serve as a generalised deformation, which by their generic structure can consistently arise in $ AdS_{2} \times S^{2} \times T^{6} $ and $ AdS_{3} \times S^{3} \times \cl{M}^{4} $ backgrounds. For string backgrounds in spin chain picture, we assume scattering that respects fermion number is an invariant. The building block of this prescription is an elliptic form {\Rx}
\begin{equation}\label{p5_8vBase}
	R_{\text{8v}} = 
	\begin{pmatrix}
		r_{1} & 0 & 0 & r_{8} \\
		0 & r_{2} & r_{6} & 0 \\
		0 & r_{5} & r_{3} & 0 \\
		r_{7} & 0 & 0 & r_{4} \\
	\end{pmatrix}
\end{equation}
which later will be exploited in the $ 16 \times 16 $ embedding, that in turn, is a unified basis for studying the structure, scattering and other properties of $ AdS_{n} $ integrability. In the two-particle representation we consider the following pair combinations
\begin{equation}\label{p5_ScatteringAnsatz_4-dim}
	\begin{array}{l}
		\cl{O} \ket{\phi_a \phi_b} \rightarrow \mrm{x}_{1} \ket{\phi_a \phi_b} + \mrm{x}_{2} \ket{\psi_\alpha \psi_\beta} \\
		\cl{O} \ket{\psi_{\alpha} \psi_\beta} \rightarrow \mrm{x}_{3} \ket{\psi_\alpha \psi_\beta} \mrm{x}_{4} \ket{\phi_a \phi_b} \\
		\cl{O} \ket{\phi_a \psi_{\beta}} \rightarrow \mrm{y}_{1} \ket{\phi_a \psi_{\beta}} + \mrm{y}_{2} \ket{\psi_{\beta} \phi_{a}} \\
		\cl{O} \ket{\psi_{\alpha} \phi_b} \rightarrow \mrm{y}_{3} \ket{\psi_{\alpha} \phi_{b}} + \mrm{y}_{4} \ket{\phi_{b} \psi_{\alpha}} \\
	\end{array}
\end{equation}
where under $ \cl{O} $ one assumes an operator with appropriate parametric dependence, \textit{e.g.} $ (u,v) $ for $ S $-/$ R $-matrix. As we have noted before, when the spin conservation is manifest, the $ r_{\{ 7,8 \}} \equiv 0 $ since boson pair does not create the fermionic one and the opposite is true. For our current analysis, only B subclass \ref{ND_6vB} is of importance as it produces generic deformed construction consistent with $ AdS $ sector \cite{deLeeuw:2020ahe}, we shall refer to them as 6vB and 8vB. These classes are special from several perspectives, \textit{e.g.} deformation and functional freedoms, they possess physical constraint in the form of free fermion analogue \cite{deLeeuw:2020bgo} and additional described below.

\paragraph{6vB}\label{p5_6vB} We shall bring equivalent, but suitably modified version of 6vB, where extra functional parametrisations are introduced. With respect to \eqref{p5_8vBase} we can define the form of $ r_{k} \equiv r_{k}(u,v) $
\begin{equation}\label{key}
	\begin{array}{l}
		r_1(u,v)=\frac{h_2(v)-h_1(u)}{h_2(u)-h_1(u)} \\  [1ex]
		r_5(u,v) =\frac{\fk{Y}(u)}{\fk{Y}(v)} \\ [1ex]
		r_6(u,v) =\frac{\fk{X}(u)}{\fk{X}(v)} \\ [1ex]
	\end{array}
	\qquad
	\begin{array}{l}
		r_2(u,v) =(h_2(u)-h_2(v))\fk{X}(u)\fk{Y}(u) \\ [1ex]
		r_3(u,v) =\frac{h_1(u)-h_1(v)}{(h_2(u)-h_1(u))(h_2(v)-h_1(v))}\frac{1}{\fk{X}(v)\fk{Y}(v)} \\ [1ex]
		r_4(u,v) =\frac{h_2(u)-h_1(v)}{h_2(v)-h_1(v)}\frac{\fk{X}(u)\fk{Y}(u)}{\fk{X}(v)\fk{Y}(v)} \\ [1ex]
	\end{array}
\end{equation}
where obviously for 6v $ r_{\{ 7,8 \}} = 0 $, $ h_{k} $, $ \fk{X} $, and $ \fk{Y} $ are undetermined spectral functions. The role of the last two functions was not highlighted in the previous discussion, because they were involved in the set of acting integrable transforms. However here $ \fk{X}/\fk{Y} $ appear unaffected explicitly as a part of the twist transform, which will become useful when analysing the embedding of $ AdS_{3} $ models. Undoubtedly, all the transformations and spectral freedoms that are involved form a consistent integrable block, that satisfies all integrable constraints according to \ref{Method} and \ref{ND_Technique}.

\paragraph{8vB}\label{p5_8vB} As we have indicated before the content of $ R_{\text{8vB}}(u,v) $ is given by \eqref{p4_R8vB}
\begin{equation}\label{key}
	\begin{array}{l}
		r_{1} = \Sigma(u,v) \left[ \sin\eta_+\frac{\mrm{cn}}{\mrm{dn}} 
		-\cos\eta_+ \mrm{sn} \right] \\ [1ex]
		r_{2} = - \Sigma(u,v) \left[ \cos\eta_-\mrm{sn} +\sin\eta_-\frac{\mrm{cn}}{\mrm{dn}} 
		 \right] \\ [1ex]
		r_{3} = - \Sigma(u,v) \left[ \cos\eta_-\mrm{sn}  - \sin\eta_-\frac{\mrm{cn}}{\mrm{dn}}  \right] \\ [1ex]
		r_{4} = \Sigma(u,v) \left[ \sin\eta_+\frac{\mrm{cn}}{\mrm{dn}} + \cos\eta_+ \mrm{sn} \right] \\ [1ex]
		r_{5} = r_{6} = 1, \qquad r_{7} = r_{8} = k \, \mrm{sn}\frac{\mrm{cn}}{\mrm{dn}} \\
	\end{array}
\end{equation}
with $ \Sigma(u,v) = \left[ \sin\eta(u) \sin\eta(v) \right]^{-\frac{1}{2}}  $ and $ \eta_{\pm} = (\eta(u) \pm \eta(v))/2 $.

\subsection{$ AdS $ Deformations}
In this section we further develop the understanding of the models described in \ref{Sec_ND_2-dim} \cite{deLeeuw:2020xrw,deLeeuw:2020ahe} by investigating the $ AdS $ sectors they could arising in and how these models are relevant for potential holographic duals. In particular, we shall ask for possible embeddings of the known $ AdS_{\{ 2,3 \}} $ models into the last ones \cite{deLeeuw:2021ufg}.

\subsubsection{Prescription}\label{p5_R_Prescription}
Before comparing the structures of $ AdS $ and our models, we need to make manifest the decomposition into chiral blocks of our classes, which will create the consistent framework for the desired analysis. Namely, we shall start with the regular {\Rx}, which constitutes a building block for $ 16 \times 16 $ embedding
\begin{equation}\label{p5_8vBlock}
	R^{\cl{X}}(u,v) = \sigma
	\begin{pmatrix}
		r_{1} & 0 & 0 & r_{8} \\
		0 & r_{2} & r_{6} & 0 \\
		0 & r_{5} & r_{3} & 0 \\
		r_{7} & 0 & 0 & r_{4} \\
	\end{pmatrix}
\end{equation}
where in comparison to \eqref{p5_8vBase} we allow for extra free scalar factor $ \sigma $\footnote{The block dressing factor in general is identified such that (corresponding) blocks and the full {\Rx} obey braiding unitarity and crossing symmetric relations.} (\textit{dressing}) and $ \cl{X} $ indicates the corresponding chirality of the block. We shall have four blocks -- two in the pure sector (\textbf{L}eft-\textbf{L}eft, \textbf{R}ight-\textbf{R}ight) and two in the mixed one (\textbf{LR}, \textbf{RL}). In this context, one also has regularity of the pure blocks, \textit{i.e.}
\begin{equation}\label{p5_P-Regularity}
	R^{\tx{LL}}(u,u) \tx{ and } R^{\tx{RR}}(u,u) \sim P
\end{equation}
where $ P $ conventionally is appropriate dimension permutation operator. As it will be shown for the full $ 16 \times 16 $ embedding, all entries will acquire chiral sector indication coming from chirality of the corresponding block \eqref{p5_8vBlock}. The four block implementation as of \cite{Sfondrini:2014via} will result in the $ R $ operator of the form
\begin{equation}\label{p5_R_Full}
	R(u,v) = {\tiny{
	\begin{pmatrix}
		r_1^{\tx{LL}} & 0 & 0 & 0 & 0 & r_8^{\tx{LL}} & 0 & 0 & 0 & 0 & 0 & 0 & 0 & 0 & 0 & 0 \\ 
		0 & r_2^{\tx{LL}} & 0 & 0 & r_6^{\tx{LL}} & 0 & 0 & 0 & 0 & 0 & 0 & 0 & 0 & 0 & 0 & 0 \\ 
		0 & 0 & r_1^{\tx{LR}} & 0 & 0 & 0 & 0 & r_8^{\tx{LR}} & 0 & 0 & 0 & 0 & 0 & 0 & 0 & 0 \\ 
		0 & 0 & 0 & r_2^{\tx{LR}} & 0 & 0 & r_6^{\tx{LR}} & 0 & 0 & 0 & 0 & 0 & 0 & 0 & 0 & 0 \\ 
		0 & r_5^{\tx{LL}} & 0 & 0 & r_3^{\tx{LL}} & 0 & 0 & 0 & 0 & 0 & 0 & 0 & 0 & 0 & 0 & 0 \\ 
		r_7^{\tx{LL}} & 0 & 0 & 0 & 0 & r_4^{\tx{LL}} & 0 & 0 & 0 & 0 & 0 & 0 & 0 & 0 & 0 & 0 \\ 
		0 & 0 & 0 & r_5^{\tx{LR}} & 0 & 0 & r_3^{\tx{LR}} & 0 & 0 & 0 & 0 & 0 & 0 & 0 & 0 & 0 \\ 
		0 & 0 & r_7^{\tx{LR}} & 0 & 0 & 0 & 0 & r_4^{\tx{LR}} & 0 & 0 & 0 & 0 & 0 & 0 & 0 & 0 \\ 
		0 & 0 & 0 & 0 & 0 & 0 & 0 & 0 & r_1^{\tx{RL}} & 0 & 0 & 0 & 0 & r_8^{\tx{RL}} & 0 & 0 \\ 
		0 & 0 & 0 & 0 & 0 & 0 & 0 & 0 & 0 & r_2^{\tx{RL}} & 0 & 0 & r_6^{\tx{RL}} & 0 & 0 & 0 \\ 
		0 & 0 & 0 & 0 & 0 & 0 & 0 & 0 & 0 & 0 & r_1^{\tx{RR}} & 0 & 0 & 0 & 0 & r_8^{\tx{RR}} \\ 
		0 & 0 & 0 & 0 & 0 & 0 & 0 & 0 & 0 & 0 & 0 & r_2^{\tx{RR}} & 0 & 0 & r_6^{\tx{RR}} & 0 \\ 
		0 & 0 & 0 & 0 & 0 & 0 & 0 & 0 & 0 & r_5^{\tx{RL}} & 0 & 0 & r_3^{\tx{RL}} & 0 & 0 & 0 \\ 
		0 & 0 & 0 & 0 & 0 & 0 & 0 & 0 & r_7^{\tx{RL}} & 0 & 0 & 0 & 0 & r_4^{\tx{RL}} & 0 & 0 \\ 
		0 & 0 & 0 & 0 & 0 & 0 & 0 & 0 & 0 & 0 & 0 & r_5^{\tx{RR}} & 0 & 0 & r_3^{\tx{RR}} & 0 \\ 
		0 & 0 & 0 & 0 & 0 & 0 & 0 & 0 & 0 & 0 & r_7^{\tx{RR}} & 0 & 0 & 0 & 0 & r_4^{\tx{RR}} \\ 
	\end{pmatrix}}}
\end{equation}
where $ r_{k}^{\cl{X}} \equiv r_{k}^{\cl{X}}(u,v) $, $ \cl{X} \in \{ \tx{LL, RR, LR, RL} \} $ and $ R $ will correspond to the full $ 16 \times 16 $ {\Rx} (if not stated otherwise). The latter must satisfy qYBE \ref{YBE_ArbDep} according with action $ R_{ij} \in \text{End}\left( \bb{V} \otimes \bb{V} \otimes \bb{V} \right) $, which implies that underlying combinations of the chiral blocks do as well
\begin{equation}\label{p5_Chiral_YBE}
	R_{12}^{\mrm{x}_{1}\mrm{x}_{2}}\left( u, v \right) R_{13}^{\mrm{x}_{1}\mrm{x}_{3}}\left( u,w \right) R_{23}^{\mrm{x}_{2}\mrm{x}_{3}}\left( v, w \right) = R_{23}^{\mrm{x}_{2}\mrm{x}_{3}}\left( v, w \right) R_{13}^{\mrm{x}_{1}\mrm{x}_{3}}\left( z_{1}, z_{3} \right) R_{12}^{\mrm{x}_{1}\mrm{x}_{2}}\left( z_{1}, z_{2} \right) 
\end{equation}
where we span each $ \bb{V} $ by the chirality $ \mrm{x}_{i} \in \{ \tx{L, R} \} $, which results in eight independent qYBE on $ R_{ij}^{\mrm{x}_{i}\mrm{x}_{j}} $.

\paragraph{Constructing R} \label{p5_Com_R} In order to construct \eqref{p5_R_Full}, we need to resolve $ \forall r_{k}^{\cl{X}} $, or more appropriately for the mixed sector \cite{deLeeuw:2021ufg}. The reason for that is the two-dimensional classes \cite{deLeeuw:2020ahe} obtained in \ref{Sec_ND_2-dim} obey regularity \eqref{p5_P-Regularity}, hence they will serve as an input for the pure blocks $ R^{\{ \tx{LL, RR} \}} $. However one must distinguish between them as they can be deformed independently.

We can note that we are left with six independent qYBE, since for $ \mrm{x}_{i} = \tx{L or R} $ the underlying YB equations are automatically satisfied. It is natural then to consider the chiral combinations $ \mrm{x}_{1} = \mrm{x}_{2} = \tx{L} $, $ \mrm{x}_{3} = \tx{R}$ and $ \tx{L} \leftrightarrow \tx{R} $, where we shall solve the associated YBEs for $ r_{k}^{\tx{LR}} $ and $ r_{k}^{\tx{RL}} $ accordingly.

Now the remainder of YB equations is used to identify some functions on one parameter. By this resolved, the corresponding $ r_{k}^{\cl{X}} $ allow to assemble \ref{p5_R_Full} and perform a qYBE check for itself. The required braiding unitarity and crossing properties will be discussed below.


\subsubsection{6vB as $ AdS_{3} $ deformation}
First we shall do an analysis on how deformations of massive $ AdS_{3} $ \cite{Borsato:2013qpa,Borsato:2014hja,Borsato:2015mma} integrable models appear as 6vB. Such behaviour is accomplished due to present freedoms and allowed transformation of the structure, which is exactly one of the goals and might be important in many context, \textit{e.g.} supercoset models, $ AdS $ deformations and their classification, quantum analogues of sigma models \cite{Delduc:2013qra,Delduc_2013,Delduc:2018xug} and other \ref{Conclusions and remarks}.

\paragraph{The {\Rx}}\label{p5_R6vB_Full} As we have discussed above \ref{p5_Com_R}, we shall attempt the independent twisting of the LL and RR blocks, which apriori does not guarantee getting consistent \ref{p5_R_Full}, hence we need to work out the rest sectors. So the LL-block could be provided as
\begin{equation}\label{key}
	\begin{array}{l}
		r_1^{\tx{LL}}=\frac{h_2^{\tx{L}}(v)-h_1^{\tx{L}}(u)}{h_2^{\tx{L}}(u)-h_1^{\tx{L}}(u)} \\ [2ex]
		r_2^{\tx{LL}}=(h_2^{\tx{L}}(u)-h_2^{\tx{L}}(v))\fk{X}^{\tx{L}}(u)\fk{Y}^{\tx{L}}(u) \\ [2ex]
		r_3^{\tx{LL}}=\frac{h_1^{\tx{L}}(u)-h_1^{\tx{L}}(v)}{\left(h_2^{\tx{L}}(u)-h_1^{\tx{L}}(u)\right)\left(h_2^{\tx{L}}(v)-h_1^{\tx{L}}(v)\right)}\frac{1}{\fk{X}^{\tx{L}}(v)\fk{Y}^{\tx{L}}(v)} \\ [2ex]
	\end{array}
	\quad
	\begin{array}{l}
		r_4^{\tx{LL}}=\frac{h_2^{\tx{L}}(u)-h_1^{\tx{L}}(v)}{h_2^{\tx{L}}(v)-h_1^{\tx{L}}(v)}\frac{\fk{X}^{\tx{L}}(u)\fk{Y}^{\tx{L}}(u)}{\fk{X}^{\tx{L}}(v)\fk{Y}^{\tx{L}}(v)} \\ [2ex]
		r_5^{\tx{LL}}=\frac{\fk{Y}^{\tx{L}}(u)}{\fk{Y}^{\tx{L}}(v)}, \quad 	r_6^{\tx{LL}}=\frac{\fk{X}^{\tx{L}}(u)}{\fk{X}^{\tx{L}}(v)}	
	\end{array}
\end{equation}
one can note distinction with \ref{Sec_ND_2-dim} because of independent twist transformation (suitable for fulfilling crossing constraints and comparison analysis). Accordingly for the RR-block we also assume independent twisting, which becomes
\begin{equation}\label{key}
	\begin{array}{l}
		r_1^{{\tx{RR}}}=\frac{h_2^{\tx{R}}(v)-h_1^{\tx{R}}(u)}{h_2^{\tx{R}}(u)-h_1^{\tx{R}}(u)} \\ [2ex]
		r_2^{{\tx{RR}}}=\frac{h_2^{\tx{R}}(u)-h_2^{\tx{R}}(v)}{h_2^{\tx{R}}(u)^2}\fk{X}^{\tx{R}}(u)\fk{Y}^{\tx{R}}(u) \\ [2ex] 
		r_3^{{\tx{RR}}}=\frac{h_2^{\tx{R}}(v)^2\left(h_1^{\tx{R}}(u)-h_1^{\tx{R}}(v)\right)}{\left(h_2^{\tx{R}}(u)-h_1^{\tx{R}}(u)\right)\left(h_2^{\tx{R}}(v)-h_1^{\tx{R}}(v)\right)}\frac{1}{\fk{X}^{\tx{R}}(v)\fk{Y}^{\tx{R}}(v)} \\ [2ex]
	\end{array}
	\quad
	\begin{array}{l}
		r_4^{{\tx{RR}}}=\frac{h_2^{\tx{R}}(v)^2}{h_2^{\tx{R}}(u)^2}\frac{h_2^{\tx{R}}(u)-h_1^{\tx{R}}(v)}{h_2^{\tx{R}}(v)-h_1^{\tx{R}}(v)}\frac{\fk{X}^{\tx{R}}(u)\fk{Y}^{\tx{R}}(u)}{\fk{X}^{\tx{R}}(v)\fk{Y}^{\tx{R}}(v)} \\ [2ex] 
		r_5^{{\tx{RR}}}=\frac{h_2^{\tx{R}}(v)}{h_2^{\tx{R}}(u)}\frac{\fk{X}^{\tx{R}}(u)}{\fk{X}^{\tx{R}}(v)},\quad r_6^{{\tx{RR}}}=\frac{h_2^{\tx{R}}(v)}{h_2^{\tx{R}}(u)}\frac{\fk{Y}^{\tx{R}}(u)}{\fk{Y}^{\tx{R}}(v)}
	\end{array}
\end{equation}
where for both LL and RR $ r_{\{ 7,8 \}}^{\tx{LL, RR}} = 0 $ and LLL/RRR YBE are autonomously satisfied. Whereas the LR- and RL-block resolve into
\begin{equation}\label{key}
	\begin{array}{l}
		r_1^{{\tx{LR}}}=1 \\ [2ex]
		r_2^{{\tx{LR}}}=-\frac{1}{h_2^{\tx{L}}(u)-h_1^{\tx{L}}(u)}\frac{1+h_1^{\tx{L}}(u)h_1^{\tx{R}}(v)}{1+h_2^{\tx{L}}(u)h_1^{\tx{R}}(v)}\frac{1}{\fk{X}^{\tx{L}}(u)\fk{Y}^{\tx{L}}(u)} \\ [2ex] 
		r_3^{{\tx{LR}}}=-\frac{h_2^{\tx{R}}(v)-h_1^{\tx{R}}(v)}{h_2^{\tx{R}}(v)^2}\frac{1+h_2^{\tx{L}}(u)h_2^{\tx{R}}(v)}{1+h_2^{\tx{L}}(u)h_1^{\tx{R}}(v)}\fk{X}^{\tx{R}}(v)\fk{Y}^{\tx{R}}(v) \\ [2ex] 
	\end{array}
	\quad
	\begin{array}{l}
		r_4^{{\tx{LR}}}=-\frac{1}{h_2^{\tx{R}}(v)^2}\frac{h_2^{\tx{R}}(v)-h_1^{\tx{R}}(v)}{h_2^{\tx{L}}(u)-h_1^{\tx{L}}(u)}\frac{1+h_1^{\tx{L}}(u)h_2^{\tx{R}}(v)}{1+h_2^{\tx{L}}(u)h_1^{\tx{R}}(v))}\frac{\fk{X}^{\tx{R}}(u)\fk{Y}^{\tx{R}}(u)}{\fk{X}^{\tx{L}}(u)\fk{Y}^{\tx{L}}(u)} \\ [2ex] 
		r_7^{{\tx{LR}}}=\frac{i}{h_2^{\tx{R}}(v)}\frac{h_2^{\tx{R}}(v)-h_1^{\tx{R}}(v)}{1+h_2^{\tx{L}}(u)h_1^{\tx{R}}(v)}\frac{\fk{X}^{\tx{R}}(v)}{\fk{X}^{\tx{L}}(u)} \\ [2ex] 
		r_8^{{\tx{LR}}}=-\frac{i}{h_2^{\tx{R}}(v)}\frac{h_2^{\tx{R}}(v)-h_1^{\tx{R}}(v)}{1+h_2^{\tx{L}}(u)h_1^{\tx{R}}(v)}\frac{\fk{Y}^{\tx{R}}(v)}{\fk{Y}^{\tx{L}}(u)}
	\end{array}
\end{equation}
and
\begin{equation}\label{key}
	\begin{array}{l}
		r_1^{{\tx{RL}}} = 1 \\ [2ex]
		r_2^{{\tx{RL}}}=-\frac{h_2^{\tx{R}}(u)^2}{h_2^{\tx{R}}(u)-h_1^{\tx{R}}(u)}\frac{1+h_1^{\tx{L}}(v)h_1^{\tx{R}}(u)}{1+h_1^{\tx{L}}(v)h_2^{\tx{R}}(u)}\frac{1}{\fk{X}^{\tx{R}}(u)\fk{Y}^{\tx{R}}(u)} \\ [2ex] 
		r_3^{{\tx{RL}}}=-\left(h_2^{\tx{L}}(v)-h_1^{\tx{L}}(v)\right)\frac{1+h_2^{\tx{L}}(v)h_2^{\tx{R}}(u)}{1+h_1^{\tx{L}}(v)h_2^{\tx{R}}(u)}\fk{X}^{\tx{L}}(v)\fk{Y}^{\tx{L}}(v) \\ [2ex] 
		r_4^{{\tx{RL}}}=-h_2^{\tx{R}}(u)^2\frac{h_2^{\tx{L}}(v)-h_1^{\tx{L}}(v)}{h_2^{\tx{R}}(u)-h_1^{\tx{R}}(u)}\frac{1+h_2^{\tx{L}}(v)h_1^{\tx{R}}(u)}{1+h_1^{\tx{L}}(v)h_2^{\tx{R}}(u))}\frac{\fk{X}^{\tx{L}}(v)\fk{Y}^{\tx{L}}(v)}{\fk{X}^{\tx{R}}(u)\fk{Y}^{\tx{R}}(u)} \\ [2ex]
	\end{array}
	\quad
	\begin{array}{l} 
		r_7^{{\tx{RL}}}=i\,h_2^{\tx{R}}(u)\frac{h_2^{\tx{L}}(v)-h_1^{\tx{L}}(v)}{1+h_1^{\tx{L}}(v)h_2^{\tx{R}}(u)}\frac{\fk{Y}^{\tx{L}}(v)}{\fk{Y}^{\tx{R}}(u)} \\ [2ex]  r_8^{{\tx{RL}}}=-i\,h_2^{\tx{R}}(u)\frac{h_2^{\tx{L}}(v)-h_1^{\tx{L}}(v)}{1+h_1^{\tx{L}}(v)h_2^{\tx{R}}(u)}\frac{\fk{X}^{\tx{L}}(v)}{\fk{X}^{\tx{R}}(u)}
	\end{array}
\end{equation}
where for LR and RL blocks $ r_{\{ 5,6 \}}^{\tx{LR, RL}} = 0 $. As we have indicated all of the blocks satisfy the corresponding YB equations, as well as the full {\Rx} \eqref{p5_R_Full} does. In what one can confirm that such construction based on 6vB class unifies into both deformations of {\adsm}, \textit{i.e.} $ \cl{M}^{4} = T^{4} \tx{ or } S^{3} \times S^{1} $ \cite{Borsato:2013qpa, Borsato:2015mma}. 

\paragraph{Braiding unitarity}\label{p5_Braiding} The aforementioned braiding properties, can be derived based on the the corresponding braiding unitarity constraint
\begin{equation}\label{key}
	R^{\cl{X}} P \bar{R}^{\bar{\cl{X}}} P = \fk{B}^{\cl{X}} \mathds{1}
\end{equation}
where $ R \equiv R(u,v) $, $ \fk{B} \equiv \fk{B}(u,v) $, the chiral sector $ \cl{X} $ and bar implies swap of spectral  parameters and chiralities (only mixed sectors affected). That results in four conditions according to \{LL, RR, LR, RL\}, and $ \fk{B}^{\cl{X}}(u,v) $ finds
\begin{equation}\label{key}
	\begin{array}{l}
		\fk{B}^{\tx{LL}} = \frac{h_2^{\tx{L}}(u)-h_1^{\tx{L}}(v)}{h_2^{\tx{L}}(u)-h_1^{\tx{L}}(u)}\frac{h_2^{\tx{L}}(v)-h_1^{\tx{L}}(u)}{h_2^{\tx{L}}(v)-h_1^{\tx{L}}(v)}\sigma^{\tx{LL}}(u,v)\sigma^{\tx{LL}}(v,u) \\ [2ex] 
		\fk{B}^{\tx{RR}} = \frac{h_2^{\tx{R}}(u)-h_1^{\tx{R}}(v)}{h_2^{\tx{R}}(u)-h_1^{\tx{R}}(u)}\frac{h_2^{\tx{R}}(v)-h_1^{\tx{R}}(u)}{h_2^{\tx{R}}(v)-h_1^{\tx{R}}(v)}\sigma^{\tx{RR}}(u,v)\sigma^{\tx{RR}}(v,u) \\ [2ex] 
		\fk{B}^{\tx{LR}} = \frac{1+h_2^{\tx{L}}(u)h_2^{\tx{R}}(v)}{1+h_1^{\tx{L}}(u)h_2^{\tx{R}}(v)}\frac{1+h_1^{\tx{R}}(v)h_1^{\tx{L}}(u)}{1+h_1^{\tx{R}}(v)h_2^{\tx{L}}(u)}\sigma^{\tx{LR}}(u,v)\sigma^{\tx{RL}}(v,u) \\ [2ex] 
		\fk{B}^{\tx{RL}} = \frac{1+h_2^{\tx{L}}(v)h_2^{\tx{R}}(u)}{1+h_1^{\tx{L}}(v)h_2^{\tx{R}}(u)}\frac{1+h_1^{\tx{R}}(v)h_1^{\tx{L}}(u)}{1+h_1^{\tx{R}}(u)h_2^{\tx{L}}(v)}\sigma^{\tx{RL}}(u,v)\sigma^{\tx{LR}}(v,u)
	\end{array}
\end{equation}
however the full {\Rx} \ref{p5_R_Full} will satisfy
\begin{equation}\label{p5_R_Full_BU}
	R(u,v) P R(v,u) P = B(u,v) \mathds{1}
\end{equation}
\begin{equation}\label{p5_R_Full_BUCondition}
	\tx{iff} \quad \fk{B}^{\{ \tx{LL,RR,LR,RL} \}} = B
\end{equation}
Namely, when the braiding factors $ \fk{B}(u,v) $ coincide $ \forall \cl{X} $ sectors. 

\paragraph{6vB $ \mapsto \mathbf{ AdS_{3} \times  S^{3} \times S^{3} \times S^{1} } $} It is important to note, that one can establish maps also to other models from 6vB. In particular it can be shown on how {\adst} (more detailed in \cite{Borsato:2015mma} and Appendix \ref{Apx_AdS_Backgrounds}) can be emerging from 6vB, the matching conditions would include
\begin{equation}\label{p5_6vB_S3S1_1}
	\begin{array}{l}
		h_{1}^{\tx{R}}(u) = -\beta^{-1} x_{\tx{R}}^-(u)  \\ [2ex] 
		h_1^{\tx{L}}(u)=\beta\, x_{\tx{L}}^-(u), \\ [2ex] 
		h_2^{\tx{R}}(u) = - \beta^{-1}x_{\tx{R}}^+(u)  \\ [2ex]  
		h_2^{\tx{L}}(u)=\beta\, x_{\tx{L}}^+(u)  \\ [2ex] 
	\end{array}
	\qquad
	\begin{array}{l}
		\sigma^{\tx{LL}}(u,v)=\frac{x_{\tx{L}}^+(u)-x_{\tx{L}}^-(u)}{x_{\tx{L}}^+(v)-x_{\tx{L}}^-(u)} \\ [2ex]  
		\sigma^{\tx{RR}}(u,v)=\frac{x_{\tx{R}}^+(u)-x_{\tx{R}}^-(u)}{x_{\tx{R}}^+(v)-x_{\tx{R}}^-(u)} \\ [2ex]  
		\sigma^{\tx{LR}}(u,v)=\sqrt{\frac{x_{\tx{L}}^-(u)}{x_{\tx{L}}^+(u)}}\frac{1-x_{\tx{L}}^+(u)x_{\tx{R}}^-(v)}{1-x_{\tx{L}}^-(u)x_{\tx{R}}^-(v)}\zeta^{\tx{LR}}(u,v) \\ [2ex] 
		\sigma^{\tx{RL}}(u,v)=\sqrt{\frac{x_{\tx{R}}^-(u)}{x_{\tx{R}}^+(u)}}\frac{1-x_{\tx{R}}^+(u)x_{\tx{L}}^-(v)}{1-x_{\tx{R}}^-(u)x_{\tx{L}}^-(v)}\zeta^{\tx{RL}}(u,v)  \\ [2ex] 
	\end{array}
\end{equation}
with $ \beta,\rho \in \tx{const} $, also on contrary to $ \cl{M}^{4} = T^{4} $, Zhukovsky pair splits for the case of two spheres. Finally, the $ \fk{X}^{\mrm{x}}(u) $ and $ \fk{Y}^{\mrm{x}}(u) $ functions map to
\begin{equation}\label{p5_6vB_S3S1_2}
	\begin{array}{l}
		\fk{X}^{\tx{L}}=\frac{\rho}{\gamma^{\tx{L}}(u)} \\ [2ex] 
		\fk{X}^{\tx{R}} = -i \, \rho \, \frac{ x_{\tx{R}}^+(u)}{\gamma^{\tx{R}}(u)} \\ [2ex]  
	\end{array}
	\qquad
	\begin{array}{l}
		\fk{Y}^{\tx{L}}=\frac{\gamma^{\tx{L}}(u)}{\beta\rho\left(x_{\tx{L}}^-(u)-x_{\tx{L}}^+(u)\right)}\sqrt{\frac{x_{\tx{L}}^-(u)}{x_{\tx{L}}^+(u)}} \\ [2ex]
		\fk{Y}^{\tx{R}}=\frac{-i\,\gamma^{\tx{L}}(u)}{\beta\,\rho}\frac{\sqrt{x_{\tx{R}}^-(u)x_{\tx{R}}^+(u)}}{x_{\tx{R}}^-(u)-x_{\tx{R}}^+(u)} \\ [2ex]
	\end{array}
\end{equation}
by completing mapping above, we can reconstruct {\adst} from the full $ R_{\tx{6vB}} $.

\paragraph{6vB $ \mapsto $ double $ \mathbf{q} $-deformed $ \mathbf{AdS_{3}} $}\label{p5_q_deformed_AdS3} Moreover, it can be shown that for the (double) q-deformed models \cite{Klimcik:2008eq,Klimcik_2014,Delduc_2013,Hoare:2014oua} there also exists a map from 6vB, that provides a basis for further functional deformations. The matching scheme follows 
\begin{equation}\label{p5_6vB_qAdS3_1}
	\begin{array}{l}
		h_1^{\tx{R}}(u) = -\beta^{-1} x_{\tx{R}}^-(u) \\ [2ex] 
		h_1^{\tx{L}}(u)=\beta\,x_{\tx{L}}^-(u) \\ [2ex] 
		h_2^{\tx{R}}(u) = -\beta^{-1} x_{\tx{R}}^+(u) \\ [2ex] 
		h_2^{\tx{L}}(u)=\beta\,x_{\tx{L}}^+(u) \\ [2ex] 
	\end{array}
	\qquad
	\begin{array}{l}
		\sigma^{\tx{LL}} = - \frac{U_L(u)V_L(u)W_L(u)}{U_L(v)V_L(v)W_L(v)}\frac{x_{\tx{L}}^-(u)-x_{\tx{L}}^+(u)}{x_{\tx{L}}^-(v)-x_{\tx{L}}^+(u)} \\ [2ex] 
		\sigma^{\tx{RR}} = - \frac{U_R(u)V_R(u)W_R(u)}{U_R(v)V_R(v)W_R(v)}\frac{x_{\tx{R}}^-(u)-x_{\tx{R}}^+(u)}{x_{\tx{R}}^-(v)-x_{\tx{R}}^+(u)} \\ [2ex] 
		\sigma^{\tx{LR}} = U_R(v)V_R(v)W_R(v)\frac{(1-x_{\tx{L}}^+(u)x_{\tx{R}}^-(v))}{(1-x_{\tx{L}}^+(u)x_{\tx{R}}^+(v))} \\ [2ex] 
		\sigma^{\tx{RL}} = U_L(v)V_L(v)W_L(v)\frac{(1-x_{\tx{R}}^+(u)x_{\tx{L}}^-(v))}{(1-x_{\tx{R}}^+(u)x_{\tx{L}}^+(v))}
	\end{array}
\end{equation}
where $ \sigma^{\cl{X}} \equiv \sigma^{\cl{X}}(u,v)  $, $ \beta,\rho \in \tx{const} $ and 
\begin{equation}\label{p5_6vB_qAdS3_2 }
	\begin{array}{l}
		\fk{X}^{\tx{L}} = - \frac{\rho}{\gamma_L(u)} \\ [2ex]
		\fk{X}^{\tx{R}} = - \frac{\rho\,\gamma_R(u)}{U_R(u)V_R(u)W_R(u)}\frac{x_{\tx{R}}^+(u)}{x_{\tx{R}}^-(u)-x_{\tx{R}}^+(u)}  \\ [2ex]
	\end{array}
	\qquad
	\begin{array}{l}
		\fk{Y}^{\tx{L}} = \frac{1}{\beta\,\rho}\frac{\gamma_L(u)}{U_L(u)V_L(u)W_L(u)}\frac{1}{x_{\tx{L}}^-(u)-x_{\tx{L}}^+(u)} \\ [2ex] 
		\fk{Y}^{\tx{R}} = \frac{1}{\beta\,\rho}\frac{x_{\tx{R}}^+(u)}{\gamma_R(u)}  \\ [2ex] 
	\end{array}
\end{equation}
Worth mentioning that due to possible representation parameter redefinitions present in \cite{Hoare:2014oua} it appears as a single $ q $-deformation. However the remaining unfixed functions could serve as an additional freedom for deformations, which could be crucial for associated (supercoset) sigma model construction.

\subsubsection{6vB sector: Symmetries of $ AdS_{3} $ deformation}\label{p5_6vB_DS}

\paragraph{Universal algebraic structure} In order to be able to judge on some properties of the 6vB and 8vB models, it is useful to address its underlying symmetry structure. This could be achieved by several methods, although some of them could be not as useful as others, since we aim to keep as many freedoms in our models as possible. One of the consistent approaches is based on the Algebraic Ansatz \cite{Faddeev:1996iy} arguments, namely, to identify the structure of the underlying algebra from \textbf{F}undamental \textbf{C}ommutation \textbf{R}elations \ref{Introduction} that follows from the $ RTT $-relations
\begin{equation}\label{p5_RTT}
	R_{mn}(u,v) T_{m}(u) T_{n}(v) = T_{n}(v) T_{m}(u) R_{mn}(u,v)
\end{equation}
On the other hand the associated generators $ \fk{G} $ must also follow Hopf structure \cite{Drinfeld:1985rx,Drinfeld1986quantum,Drinfeld_1986DAHAY}, and in particular, satisfy co-commutativity
\begin{equation}\label{p5_R_Cocommutativity}
	R \Delta(\fk{G}) = \Delta^{\mrm{op}}(\fk{G}) R
\end{equation}
Such prescriptions are used to identify enveloping, affine and Yangian symmetry algebras of the corresponding Lie groups $ \cl{G} $. For the case {\ads} it was implemented in \cite{Beisert:2006qh, Beisert:2008tw, Beisert:2014hya}, whereas the $ S $-matrix and symmetry algebra of {\adsm} and {\adst} first appeared in \cite{Borsato:2014hja,Borsato:2015mma,Hoare:2014kma,Hoare:2015kla}.

From the form of $ RTT $, which provide FCR, it is clear that one should be able to obtain a constraining set of commutation relations that in turn will define the corresponding symmetry algebra on the generators. If we bring the constituents of \eqref{p5_RTT} in the component form
\begin{equation}\label{key}
	R(u,v) = R_{i_{1} i_{2}}^{j_{1} j_{2}} E_{j_{1}}^{i_{1}} \otimes E_{j_{2}}^{i_{2}}, 
	\quad
	\begin{cases}
		T_{\rho}(u) = E_{\, j}^{i} \otimes 1 \otimes T_{i}^{\, j}(u) \\
		T_{\sigma}(u) = E_{\, j}^{i} \otimes T_{i}^{\, j}(u) \otimes 1 
	\end{cases}
\end{equation}
where from monodromic perturbation, it is also known that
\begin{equation}\label{key}
	T_{\rho}^{\, \sigma} = T_{\rho, \, -1}^{\, \sigma} + \sum_{k = 0}^{+\infty} T_{\rho, \, k}^{\, \sigma} u^{-k -1}
\end{equation}
where $ T_{\rho, \, -1}^{\, \sigma} $ is nothing but the coproduct \textit{braiding} element $ \delta_{\rho \sigma}\fk{U} $ and the associated algebra is produced by the commutation relations on $ T_{\rho, \, 0 }^{\, \sigma} $. Important, that along with the YB constraint, the {\Rx} components identify a representation\footnote{In the present and other settings, the fusion always allows to construct the corresponding derived representations} of the algebra \cite{Beisert:2005tm,Beisert:2015msa}. In fact, the $ S $-matrices, which arise in string integrable backgrounds are obtained from the symmetry algebra of the associated systems. Here we attempt a bottom-up approach, based on the identified representation of the corresponding {\Rx} and the involved co-commutative structure \eqref{p5_R_Cocommutativity}.

Another important remark, is that the technique mentioned above, requires existence of the well-defined special point of $ R $ in the expansion (\textit{e.g.} identity vicinity, regularity or other model specific). For the example in the case 8v type deformation of $ AdS_{3} $, it is an obstacle to identify such a diagonal reduction point, whereas it is possible for the $ AdS_{2} $ case. In this respect, we can propose an algebraic way, in which we can exploit universal Hopf structure, co-commutativity, the generic configuration of the $ R_{\tx{8vB}} $, from which also the underlying representation follows. More specifically, we can impose \eqref{p5_R_Cocommutativity} in the form
\begin{equation}\label{p5_Cocommutativity_2}
	\check{R} \fk{G}_{ij} = \bar{\fk{G}}_{ij} \check{R}
\end{equation}
with
\begin{equation}\label{key}
	\check{R} = P R, \quad \bar{\fk{G}}_{ij} = \fk{G}_{ij}(v,u), \quad \fk{G}_{ij}(u,v) = \fk{G}(u) \otimes 1 + \fk{U}(u) \otimes \fk{G}(v)
\end{equation}
the $ \fk{G}_{ij}(u,v) $ is unspecified symmetry generator. If we would exploit $ \partial_{v} \check{R}(u,v)|_{v \mapsto u} \equiv = \cl{H}(u) $ on \eqref{p5_Cocommutativity_2}, we can obtain a differential constraint on generators intertwined with Hamiltonian data
\begin{equation}\label{p5_Differential_Cocommutativity}
	\dot{\fk{U}} \otimes \fk{G} - \fk{U} \otimes \dot{\fk{G}} + \dot{\fk{G}} \otimes 1 = \left[ \fk{G} \otimes 1 + \fk{U} \otimes \fk{G}, \, \cl{H} \right]  
\end{equation}
where the dots indicate derivative. It indeed resembles Sutherland structure developed in \ref{ND_Technique} and also eliminates some obstacles that could arise due to the form and limits of the 6vB \ref{p5_R6vB_Full} and 8vB \ref{p5_R8vB_Full} classes. By solving \eqref{p5_Differential_Cocommutativity} and taking into account \eqref{p5_Cocommutativity_2} one can constrain the generator functions.

\paragraph{Algebraic structure} In order to gain an algebraic understanding of the $ AdS_{3} $ deformation of the 6-vertex type \ref{p5_R6vB_Full}, we can first implement the two block-diagonal supercharges of the form
\begin{equation}\label{p5_6vB_Supercharge}
	\fk{G}_{\pm} = 
	\begin{pmatrix}
		\fk{G}_{\pm}^{\tx{L}} & 0 \\
		0 & \fk{G}_{\pm}^{\tx{R}} 
	\end{pmatrix}
	, \quad \tx{with} \qquad
	\begin{array}{l}
		\fk{G}_{+}^{\tx{L}} = 
		\begin{pmatrix}
			0 & 0 \\
			a_{+} & 0 
		\end{pmatrix} \\ [4ex]
		\fk{G}_{-}^{\tx{L}} = 
		\begin{pmatrix}
			0 & a_{-} \\
			0 & 0 
		\end{pmatrix}
	\end{array}
	\quad
	\begin{array}{l}
		\fk{G}_{+}^{\tx{R}} = 
		\begin{pmatrix}
			0 & b_{+} \\
			0 & 0 
		\end{pmatrix} \\ [4ex]
		\fk{G}_{-}^{\tx{R}} = 
		\begin{pmatrix}
			0 & 0 \\
			b_{-} & 0 
		\end{pmatrix} 
	\end{array}
\end{equation}
and the corresponding comultiplication arise as
\begin{equation}\label{key}
	\begin{cases}
		\Delta\left( \fk{G}_{+} \right) = \fk{G}_{+} \otimes 1 + \fk{U}^{-1} \otimes \fk{G}_{+} \\
		\Delta\left( \fk{G}_{-} \right) = \fk{G}_{-} \otimes 1 + \fk{V}^{-1} \otimes \fk{G}_{-}
	\end{cases}
\end{equation}
where braidings $ \cl{X} = \fk{U}, \, \fk{V} $ constitute
\begin{equation}\label{key}
	\cl{X} = 
	\begin{pmatrix}
		\cl{X}^{\tx{L}} & 0 \\
		0 & \cl{X}^{\tx{R}} 
	\end{pmatrix}
\end{equation}
with $ \cl{X}^{\mrm{x}} \sim c_{k} \cdot \id $, $ c_{k} \in \tx{const} $. The structure of the $ R_{\tx{6vB}} $ and co-commutativity immediately provides
\begin{equation}\label{key}
	\begin{array}{l}
		a_{+} = \dfrac{1}{(\xi - h_{2}^{\tx{L}})} \dfrac{1}{\fk{X}^{\tx{L}}} \\ [2ex] 
		a_{-} = \dfrac{h_{1}^{\tx{L}} - h_{2}^{\tx{L}}}{\zeta - h_{1}^{\tx{L}}} \fk{X}^{\tx{L}}
	\end{array}
	\qquad
	\begin{array}{l}
		b_{+} = i \dfrac{h_{1}^{\tx{R}} - h_{2}^{\tx{R}} }{h_{2}^{\tx{R}} (1 + \xi h_{1}^{\tx{R}})} \fk{Y}^{\tx{R}} \\ [2ex] 
		b_{-} = i \dfrac{h_{2}^{\tx{R}}}{1 + \zeta h_{2}^{\tx{R}}} \dfrac{1}{\fk{Y}^{\tx{R}}}
	\end{array}
\end{equation}
along with $ \fk{U}^{\mrm{x}} $ and $ \fk{V}^{\mrm{x}} $
\begin{equation}\label{key}
	\begin{array}{l}
		\fk{U}^{\tx{L}} = \left( h_{1}^{\tx{L}} - h_{2}^{\tx{L}} \right) \dfrac{\xi - h_{2}^{\tx{L}}}{\xi - h_{1}^{\tx{L}}} \fk{X}^{\tx{L}}\fk{Y}^{\tx{L}} \\ [2ex] 
		\fk{U}^{\tx{R}} = \dfrac{(h_{2}^{\tx{R}})^{2}}{h_{1}^{\tx{R}} - h_{2}^{\tx{R}}} \dfrac{1 + \xi h_{1}^{\tx{R}}}{1 + \xi h_{2}^{\tx{R}}} \dfrac{1}{\fk{X}^{\tx{R}}\fk{Y}^{\tx{R}}} \\ [2ex] 
	\end{array}
	\qquad
	\begin{array}{l}
		\fk{V}^{\tx{L}} = \dfrac{1}{h_{1}^{\tx{L}} - h_{2}^{\tx{L}}} \dfrac{\zeta - h_{1}^{\tx{L}}}{\zeta - h_{3}^{\tx{L}}} \dfrac{1}{\fk{X}^{\tx{L}}\fk{Y}^{\tx{L}}} \\ [2ex] 
		\fk{V}^{\tx{R}} = \dfrac{h_{1}^{\tx{R}} - h_{2}^{\tx{R}}}{(h_{2}^{\tx{R}})^{2}} \dfrac{1 + \zeta h_{2}^{\tx{R}}}{1 + \zeta h_{1}^{\tx{R}}} \fk{X}^{\tx{R}}\fk{Y}^{\tx{R}} \\ [2ex] 
	\end{array}
\end{equation}
The supercharge structure \eqref{p5_6vB_Supercharge} shows that these would constitute one-parameter freedom for each sector, which in fact due to generic combination $ \fk{G}_{+}(X) = c_{1} \fk{G}_{+}(\xi_{1}) + c_{2}\fk{G}_{+}(\xi_{2}) $, for distinct $ \xi_{i} $ brings to four linearly independent supercharge combination $ \{ \fk{G}_{+}(\xi_{i}), \fk{G}_{-}(\zeta_{j}) \} $, $ i,j = 1,2 $. The cocommutativity along with these supercharges will fix the {\Rx} for all blocks up to normalisation. From this perspective, one can bring the following generator algebra to hold
\begin{equation}\label{key}
	\begin{array}{l}
		\left[ \fk{S}, \fk{G}_{\pm} \right] = \pm \fk{G}_{\pm} \\ [2ex]
		\{ \fk{G}_{-}(\zeta_{i}), \fk{G}_{+}(\xi_{j}) \} = \fk{Q}(\zeta_{i},\xi_{j}) \\ [2ex]
		\{ \fk{G}_{+}(\xi_{1}), \fk{G}_{+}(\xi_{2}) \} = \{ \fk{G}_{-}(\zeta_{1}), \fk{G}_{-}(\zeta_{2}) \} = 0 \\ [2ex]
		\Delta(\fk{S}) = \fk{S} \otimes 1 + (\fk{U} \fk{V})^{-1} \otimes \fk{S}, \quad \Delta(\cl{X}) = \cl{X} \otimes \cl{X}
	\end{array}
\end{equation}
where $ \fk{S} $ complementary extension for $ \fk{G}_{\pm} $, four two-parametric supercharges $ \fk{Q} $ in the centre of the algebra and braidings $ \cl{X} = \fk{U}, \fk{V} $.

\subsubsection{$ AdS_{\{ 2, 3 \}} $ deformations in 8vB}
We shall now derive the {\Rx} based on the 8vB $ 4 \times 4 $ blocks \ref{ND_8vBModel} according to the construction scheme \ref{p5_R_Prescription} \cite{deLeeuw:2021ufg}.

\paragraph{The {\Rx}}\label{p5_R8vB_Full} We start resolution by incorporating the 8vB block for the corresponding LL- and RR-block with allowed freedoms, then the form acquired
\begin{equation}\label{key}
	\begin{array}{l}
		r_1^{\tx{LL}} = \Sigma(u,v) \left[ \frac{\mrm{cn}^{\tx{LL}}_-}{\mrm{dn}^{\tx{LL}}_-}\sin\eta_+ - \cos\eta_+\mrm{sn}^{\tx{LL}}_- \right] \\ [2ex] 
		r_2^{\tx{LL}} = \Sigma(u,v) \left[ \frac{\mrm{cn}^{\tx{LL}}_-}{\mrm{dn}^{\tx{LL}}_-}\sin\eta_- - \cos\eta_-\mrm{sn}^{\tx{LL}}_-\right]  \\ [2ex] 
		r_3^{\tx{LL}} = \Sigma(u,v) \left[ \frac{\mrm{cn}^{\tx{LL}}_-}{\mrm{dn}^{\tx{LL}}_-}\sin\eta_- - \cos\eta_-\mrm{sn}^{\tx{LL}}_-\right]  \\ [2ex] 
		r_4^{\tx{LL}}=\Sigma(u,v) \left[ \frac{\mrm{cn}^{\tx{LL}}_-}{\mrm{dn}^{\tx{LL}}_-}\sin\eta_+ + \cos \eta_+ \mrm{sn}^{\tx{LL}}_- \right] \\ [2ex] 
	\end{array}
	\quad
	\begin{array}{l}
		r_{5}^{\tx{LL}} = \left( r_{6}^{\tx{LL}} \right)^{-1} = \sqrt{\frac{\fk{f}_{\tx{L}}(v)}{\fk{f}_{\tx{L}}(u)}} \\ [2ex] 
		r_7^{\tx{LL}}=\frac{k \alpha}{\sqrt{\fk{f}_{\tx{L}}(u)\fk{f}_{\tx{L}}(v)}}\frac{\mrm{cn}^{\tx{LL}}_-\mrm{sn}^{\tx{LL}}_-}{\mrm{dn}^{\tx{LL}}_-} \\ [2ex]
		r_8^{\tx{LL}}=\frac{k \sqrt{\fk{f}_{\tx{L}}(u)\fk{f}_{\tx{L}}(v)}}{\alpha}\frac{\mrm{cn}^{\tx{LL}}_-\mrm{sn}^{\tx{LL}}_-}{\mrm{dn}^{\tx{LL}}_-}
	\end{array}
\end{equation}
and
\begin{equation}\label{key}
	\begin{array}{l}
		r_1^{\tx{RR}} = \Sigma(u,v) \left[ \frac{\mrm{cn}^{\tx{RR}}_-}{\mrm{dn}^{\tx{RR}}_-}\sin\eta_+ - \cos\eta_+\mrm{sn}^{\tx{RR}}_- \right] \\ [2ex] 
		r_2^{\tx{RR}}=-\Sigma(u,v) \left[ \cos\eta_-\mrm{sn}^{\tx{RR}}_-+\frac{\mrm{cn}^{\tx{RR}}_-}{\mrm{dn}^{\tx{RR}}_-}\sin\eta_- \right] \\ [2ex] 
		r_3^{\tx{RR}}=-\Sigma(u,v) \left[ \cos\eta_-\mrm{sn}^{\tx{RR}}_--\frac{\mrm{cn}^{\tx{RR}}_-}{\mrm{dn}^{\tx{RR}}_-}\sin\eta_- \right]  \\ [2ex] 
		r_4^{\tx{RR}}=\Sigma(u,v) \left[ \frac{\mrm{cn}^{\tx{RR}}_-}{\mrm{dn}^{\tx{RR}}_-}\sin\eta_+ + \cos\eta_+\mrm{sn}^{\tx{RR}}_- \right]  \\ [2ex] 
	\end{array}
	\quad
	\begin{array}{l}
		r_{5}^{\tx{RR}} = \left( r_{6}^{\tx{RR}} \right)^{-1} = \sqrt{ \frac{\fk{f}_{\tx{R}}(u)}{\fk{f}_{\tx{R}}(v)} } \\ [2ex] 
		r_7^{\tx{RR}}=\frac{k \sqrt{\fk{f}_{\tx{R}}(u)\fk{f}_{\tx{R}}(v)}}{\alpha}\frac{\mrm{cn}^{\tx{RR}}_-\mrm{sn}^{\tx{RR}}_-}{\mrm{dn}^{\tx{RR}}_-} \\ [2ex]
		r_8^{\tx{RR}}=\frac{k \alpha}{\sqrt{\fk{f}_{\tx{R}}(u)\fk{f}_{\tx{R}}(v)}}\frac{\mrm{cn}^{\tx{RR}}_-\mrm{sn}^{\tx{RR}}_-}{\mrm{dn}^{\tx{RR}}_-}
	\end{array}
\end{equation}
where $ \eta_{\pm} $, $ \Sigma(u,v) $ defined in \ref{p5_Eta_Sigma} and the modified elliptic notation is
\begin{equation}\label{key}
	\mrm{xn}_{\pm}^{\mrm{x}_{1}\mrm{x}_{2}} = \mrm{xn} \left( \cl{F}^{\mrm{x}_{1}}(u) - \cl{F}^{\mrm{x}_{2}}(v), k^{2} \right)
\end{equation}
and $ \fk{f}_{\mrm{x}} $ is the freedom that comes the LBT on the initial blocks with $ \mrm{x} $ indicating chirality. Following \ref{p5_R_Prescription}, we can now also resolve for the mixed sector
\begin{equation}\label{p5_r_LR}
	\begin{array}{l}
		r_1^{\tx{LR}}=\Sigma(u,v)\sqrt{-r_{7}^{\tx{LR}}(v,u)}\left(\cos\eta_-\mrm{sn}^{\tx{LR}}_++\frac{\mrm{cn}^{\tx{LR}}_+}{\mrm{dn}^{\tx{LR}}_+}\sin\eta_-\right) \\ [2ex] 
		r_2^{\tx{LR}}=\Sigma(u,v)\sqrt{-r_{7}^{\tx{LR}}(v,u)}\left(\cos\eta_+\mrm{sn}^{\tx{LR}}_+-\frac{\mrm{cn}^{\tx{LR}}_+}{\mrm{dn}^{\tx{LR}}_+}\sin\eta_+\right) \\ [2ex] 
		r_3^{\tx{LR}}=\Sigma(u,v)\sqrt{-r_{7}^{\tx{LR}}(v,u)}\left(\cos\eta_+\mrm{sn}^{\tx{LR}}_++\frac{\mrm{cn}^{\tx{LR}}_+}{\mrm{dn}^{\tx{LR}}_+}\sin\eta_+\right) \\ [2ex] 
		r_4^{\tx{LR}}=\Sigma(u,v)\sqrt{-r_{7}^{\tx{LR}}(v,u)}\left(-\cos\eta_-\mrm{sn}^{\tx{LR}}_++\frac{\mrm{cn}^{\tx{LR}}_+}{\mrm{dn}^{\tx{LR}}_+}\sin\eta_-\right) \\ [2ex] 
	\end{array}
	\quad
	\begin{array}{l}
		r_5^{\tx{LR}}=-\frac{k \alpha}{\fk{f}_{\tx{L}}(u)}\frac{\mrm{cn}^{\tx{LR}}_+\mrm{sn}^{\tx{LR}}_+}{\mrm{dn}^{\tx{LR}}_+} \\ [2ex] 
		r_6^{\tx{LR}}=\frac{k \fk{f}_{\tx{R}}(v)}{\alpha}\frac{\mrm{cn}^{\tx{LR}}_+\mrm{sn}^{\tx{LR}}_+}{\mrm{dn}^{\tx{LR}}_+} \\ [2ex]
		r_7^{\tx{LR}} = - \frac{\fk{f}_{\tx{R}}(u)}{\fk{f}_{\tx{L}}(v)}, \quad r_8^{\tx{LR}} = 1
	\end{array}
\end{equation}
where by $ r_{7}^{\tx{LR}}(v,u) $ we mean a swap of $ \left( u,v \right)  $ parameters in $ r_{7} $ \eqref{p5_r_LR}. For RL we obtain
\begin{equation}\label{key}
	\begin{array}{l}
		r_1^{\tx{RL}}=\Sigma(u,v)\sqrt{-r_{7}^{\tx{RL}}(u,v)}\left(\cos\eta_-\mrm{sn}^{\tx{RL}}_++\frac{\mrm{cn}^{\tx{RL}}_+}{\mrm{dn}^{\tx{RL}}_+}\sin\eta_-\right) \\ [2ex] 
		r_2^{\tx{RL}}=\Sigma(u,v)\sqrt{-r_{7}^{\tx{RL}}(u,v)}\left(\cos\eta_+\mrm{sn}^{\tx{RL}}_+-\frac{\mrm{cn}^{\tx{RL}}_+}{\mrm{dn}^{\tx{RL}}_+}\sin\eta_+\right) \\ [2ex] 
		r_3^{\tx{RL}}=\Sigma(u,v)\sqrt{-r_{7}^{\tx{RL}}(u,v)}\left(\cos\eta_+\mrm{sn}^{\tx{RL}}_++\frac{\mrm{cn}^{\tx{RL}}_+}{\mrm{dn}^{\tx{RL}}_+}\sin\eta_+\right) \\ [2ex] 
		r_4^{\tx{RL}}=\Sigma(u,v)\sqrt{-r_{7}^{\tx{RL}}(u,v)}\left(-\cos\eta_-\mrm{sn}^{\tx{RL}}_++\frac{\mrm{cn}^{\tx{RL}}_+}{\mrm{dn}^{\tx{RL}}_+}\sin\eta_-\right) \\ [2ex] 
	\end{array}
	\quad
	\begin{array}{l}
		r_5^{\tx{RL}}=-\frac{k \fk{f}_{\tx{R}}(u)}{\alpha}\frac{\mrm{cn}^{\tx{RL}}_+\mrm{sn}^{\tx{RL}}_+}{\mrm{dn}^{\tx{RL}}_+} \\ [2ex] 
		r_6^{\tx{RL}}=\frac{k \alpha}{\fk{f}_{\tx{L}}(v)}\frac{\mrm{cn}^{\tx{RL}}_+\mrm{sn}^{\tx{RL}}_+}{\mrm{dn}^{\tx{RL}}_+} \\ [2ex] 
		r_7^{\tx{RL}}=-\frac{\fk{f}_{\tx{R}}(u)}{\fk{f}_{\tx{L}}(v)},\quad r_8^{\tx{RL}} = 1 \\ [2ex]
	\end{array}
\end{equation}
One can ask if starting from the pure blocks (LL/RR) an arbitrary elliptic moduli $ k_{\mrm{x}} $ and spectral functions $ \eta_{\mrm{x}} $ must be allowed independently for each of the sectors. The answer appears to be negative, since mixed YB equations constrain them, and indeed $ \eta_{\tx{L}} $ and $ \eta_{\tx{R}} $ can be set to add up to $ \pi $. Eventually the associated full {\Rx} does satisfy the corresponding qYBE.

\paragraph{Braiding unitarity} In similar manner \ref{p5_Braiding} we can demand the braiding unitarity also for the 8vB construction
\begin{equation}\label{key}
	R^{\cl{X}} P \bar{R}^{\bar{\cl{X}}} P = \fk{B}^{\cl{X}} \mathds{1}
\end{equation}
which imposes four constraints and fixes the braiding factors. In what conditions on $ \fk{B}^{\cl{X}} $ follow as 
\begin{equation}\label{key}
	\begin{array}{l}
		\frac{\fk{B}^{\tx{LL}}(u,v)}{\sigma^{\tx{LL}}(u,v)\sigma^{\tx{LL}}(v,u)} = \frac{\mrm{cn}_{\text{L,L},-}^2}{\mrm{dn}_{\text{L,L},-}^2} \left[ \left( k \, \mrm{sn}_{\text{L,L},-} \right)^{2} - (\Sigma(u,v) \sin \eta_{+})^{2} \right]  - \mrm{sn}_{\text{L,L},-}^{2} \frac{\cos ^2\eta_+}{\sin \eta(u) \sin \eta(v)} \\ [2ex] 
		\frac{\fk{B}^{\tx{RR}}(u,v)}{ \sigma^{\tx{RR}}(u,v)\sigma^{\tx{RR}}(v,u)} = \frac{\mrm{cn}_{\text{R,R},-}^2}{\mrm{dn}_{\text{R,R},-}^2} \left[ \left( k \, \mrm{sn}_{\text{R,R},-} \right)^{2} - (\Sigma(u,v) \sin \eta_{+})^{2} \right] - \mrm{sn}_{\text{R,R},-}^{2} \frac{\cos ^2\eta_+}{\sin \eta(u) \sin \eta(v)} \\ [2ex] 
		\frac{\fk{f}^{\tx{R}}(v)}{\fk{f}^{\tx{L}}(u)}\frac{\fk{B}^{\tx{LR}}(u,v)}{\sigma^{\tx{LR}}(u,v)\sigma^{\tx{RL}}(v,u)} = \Sigma(u,v)^{2} \left[ \mrm{sn}_{\text{L,R},+}^2 \cos^2 \eta_{-} - \frac{\mrm{cn}_{\text{L,R},+}^2}{\mrm{dn}_{\text{L,R},+}^2} \sin^2 \eta_{-} \right]  - 1 \\ [2ex] 
		\frac{\fk{f}^{\tx{R}}(u)}{\fk{f}^{\tx{L}}(v)}\frac{\fk{B}^{\tx{RL}}(u,v)}{\sigma^{\tx{RL}}(u,v)\sigma^{\tx{LR}}(v,u)} = \Sigma(u,v)^{2} \left[ \mrm{sn}_{\text{R,L},+}^2 \cos^2 \eta_{-} - \frac{\mrm{cn}_{\text{R,L},+}^2}{\mrm{dn}_{\text{R,L},+}^2} \sin^2 \eta_{-} \right] - 1\\ [2ex] 
	\end{array}
\end{equation}
Analogously in order for the full $ 16 \times 16 $ embedding to satisfy braided unitarity \ref{p5_R_Full_BU}, one must require braiding factors of all sectors to agree $ \fk{B}^{\{ \tx{LL,RR,LR,RL} \}} = B $.

\paragraph{8vB as $ \mathbf{ AdS_{2} \times S^{2} \times T^{6} } $ deformation}\label{p5_R8vB_AdS3_1} As it can be noted \cite{deLeeuw:2020xrw}, a natural proposal following from the structure of the $ 16 \times 16 $ 8vB is to investigate if there matching limits to an integrable models of the massive $ AdS_{2} $ string \cite{Hoare:2014kma,Hoare:2015kla}. Before initiating such an analysis, one needs to act by certain transformations \ref{ND_IT} in addition to the construction above \ref{p5_R8vB_Full}, \textit{e.g.} we need to allow for distinguishable spectral dependence $ (u,v) \mapsto (\cl{F}(u), \cl{F}(v)) $ (twist type and other allowed freedom implementation is performed at the {\Rx} \ref{p5_R8vB_Full}, unless stated otherwise).

We can first attempt the characteristic off-diagonal entries of the 8-vertex, \textit{i.e.} let us analyse entries $ (1,4) $ of the $ AdS_{2} $ and 8vB
	\begin{numcases}{}
		r_{14}^{AdS_{2}} = \dfrac{\sqrt{\dfrac{x_{u}^{+}x_{v}^{+}}{x_{u}^{-}x_{v}^{-}}} \left(x_{u}^{-} - \dfrac{1}{x_{u}^{+}}\right) \left( x_{v}^{-} - \dfrac{1}{x_{v}^{+}} \right) }{\sqrt{x_{u}^{-}x_{u}^{+}x_{v}^{-}x_{v}^{+}}\left( 1 - \dfrac{1}{x_{u}^{-}x_{u}^{+}x_{v}^{-}x_{v}^{+}} \right) } \label{p5_AdS2_r7} \\
		r_{14}^{\tx{8vB}} = k \, \mrm{sn}\left[ \Delta \cl{F}, k^{2} \right] \dfrac{\mrm{cn}\left[ \Delta \cl{F}, k^{2} \right] }{\mrm{dn}\left[ \Delta \cl{F}, k^{2} \right] } \label{p5_8vB_r7}
	\end{numcases}
by the subscript we denote spectral dependence of the Zhukovsky pair $ x^{\pm} $ and elliptic functions $ \mrm{xn} = \mrm{xn}\left[  \cl{F}(u) - \cl{F}(v) , k^{2} \right]  $ introduced as above \ref{p5_8vB}. From the perturbation of $ r_{14}^{AdS_{2}} $ for $ u $ in the vicinity of $ v $, one can get
\begin{equation}\label{key}
	\underbrace{\dfrac{ \dd_{\,\, u}\left( x^{-} x^{-} \right) }{2 \sqrt{x^{-} x^{-}} \left( x^{-} x^{-} - 1 \right) } }_{\fk{c}_{x}} \left(u - v\right) + \cl{O}\left[ \left( u - v \right)^{2} \right]
\end{equation}
and taking into account the form of the $ r_{14}^{\tx{8vB}} $, the coefficient $ \fk{c}_{x} $ acquires constant value. We can now resolve above and incorporate generic dependence $ \cl{F} $, in what obtained
\begin{equation}\label{p5_xplus_res}
	x_{v}^{+} = \dfrac{\tanh \left[ \fk{c}_{x} \cl{F}(v) + c_{0}/2 \right] }{x_{v}^{-}}
\end{equation}
after substituting \eqref{p5_xplus_res} into \eqref{p5_AdS2_r7}
\begin{equation}\label{key}
	r_{14}^{AdS_{2}} = - \tanh \left[ \fk{c}_{x} \left( \cl{F}(u) - \cl{F}(v) \right)  \right] 
\end{equation}
now it can be seen that $ k \rightarrow \infty $ will produce $ \tanh $ in \eqref{p5_8vB_r7}. Moreover for the complete agreement between \eqref{p5_AdS2_r7} and \eqref{p5_8vB_r7}, we are required to have $ \fk{c}_{x} = - i $ and $ c_{0} $ can be set to 0, so that it results in 
\begin{equation}\label{key}
	x_{v}^{+} = -\dfrac{\tan^{2} \left[ \cl{F}(v) \right] }{x_{v}^{-}}
\end{equation}
Eventually to recover the $ R_{AdS_{2}} $ from 8vB, one needs to take $ k \rightarrow \infty $ limit with the followed perturbation in $ u $ around $ v $ and identify for $ \eta(u) $ to take the form
\begin{equation}\label{key}
	\eta(u) = \mrm{arccot} \left[ -\dfrac{k}{2} \sec\left[ \cl{F} \right] \csc\left[ \cl{F} \right] \dfrac{\cot\left[ \cl{F} \right] x^{-} + i}{ \cot\left[ \cl{F} \right] x^{-} - i } \right] 
\end{equation}
Important, that there are no further constraining needed to fix functions and for two models to completely agree. Still that is rather deformation and limits under which one can reconstruct the $ AdS_{2} $ model from above. One can also note, that if one considers a flow of $ k $, then for finite $ k $ values it brings deformation freedom back. In addition one can also consider functional shifts of the form $ \eta(u) \rightarrow \eta(u) + \fk{q} \rho(u) $, with $ \fk{q} $ to be additional deformation freedom. By looking at the shortening condition $ 	x^{+} + (x^{+})^{-1} - \left( x^{-} + (x^{-})^{-1} \right) = 2 i m/g $ arising in this context for Zhukovsky space, we can notice that such deformations could lead to the observables dependent on the spectral parameter (like mass $ m(u) $).

\paragraph{8vB $ \mapsto \mathbf{ AdS_{3} \times  S^{3} \times S^{3} \times S^{1} } $} Interestingly, one can find that a mapping limit from 8vB to $ \cl{M}^{4} = S^{3} \times S^{1} $ exists under $ k \rightarrow 0 $. Hereafter we analyse both structures \cite{Borsato:2015mma,deLeeuw:2020ahe} and resolve for mapping

\begin{equation}\label{p5_8vB_S3S1_2}
	\begin{array}{l}
		\sigma^{\text{LL}} = \frac{\sqrt{\fk{f}_{\tx{L}}(v)}}{\sqrt{\fk{f}_{\tx{L}}(u)}}\frac{x_{\tx{L}}^-(u)-x_{\tx{L}}^+(u)}{x_{\tx{L}}^-(u)-x_{\tx{L}}^+(v)}\frac{\gamma^L(v)}{\gamma^L(u)} \\ [2ex] 
		\sigma^{\text{RR}} = \frac{\sqrt{\fk{f}_{\tx{R}}(v)}}{\sqrt{\fk{f}_{\tx{R}}(u)}}\frac{x_{\tx{R}}^-(u)-x_{\tx{R}}^+(u)}{x_{\tx{R}}^-(u)-x_{\tx{R}}^+(v)}\frac{\gamma^R(v)}{\gamma^R(u)} \\ [2ex] 
	\end{array}
	\qquad
	\begin{array}{l}
		\sigma^{\tx{LR}}=\sqrt{\frac{x_{\tx{R}}^-(v)}{x_{\tx{R}}^+(v)}}\frac{x_{\tx{L}}^-(u)-x_{\tx{L}}^+(u)}{1-x_{\tx{L}}^-(u)x_{\tx{R}}^-(v)}\frac{\gamma^R(v)}{\gamma^L(u)} \\ [2ex] 
		\sigma^{\tx{RL}}=\sqrt{\frac{x_{\tx{R}}^-(u)}{x_{\tx{R}}^+(u)}}\frac{x_{\tx{L}}^-(u)-x_{\tx{L}}^+(u)}{1-x_{\tx{L}}^-(v)x_{\tx{R}}^-(u)}\frac{\gamma^R(u)}{\gamma^L(v)} \\ [2ex] 
	\end{array}
\end{equation}
with $ \sigma^{\cl{X}} \equiv \sigma^{\cl{X}}(u,v) $ and
\begin{equation}\label{p5_8vB_S3S1_1}
	\begin{array}{l}
		\fk{f}_{\tx{L}} = \fk{s} \frac{x_{\tx{L}}^-(u)-x_{\tx{L}}^+(u)}{\gamma^L(u)^2}\sqrt{\frac{x_{\tx{L}}^+(u)}{x_{\tx{L}}^-(u)}}  \\ [2ex] 
		\fk{f}_{\tx{R}} = \fk{s}\frac{x_{\tx{R}}^-(u)-x_{\tx{R}}^+(u)}{\gamma^R(u)^2}\sqrt{\frac{x_{\tx{R}}^+(u)}{x_{\tx{R}}^-(u)}} \\ [2ex] 
		\eta = -\frac{i}{2} \log\left(\sqrt{\frac{x_{\tx{L}}^+(u)}{x_{\tx{L}}^-(u)}}\right)  \\ [2ex] 
	\end{array}
	\qquad
	\begin{array}{l}
		\cl{F}^L = \pi -\frac{i}{4}\log\left(\sqrt{x_{\tx{L}}^-(u)x_{\tx{L}}^+(u)}\right)  \\ [2ex] 
		\cl{F}^R = \pi -\frac{i}{4}\log\left(\sqrt{x_{\tx{R}}^-(u)x_{\tx{R}}^+(u)}\right)  \\ [2ex] 
	\end{array}
\end{equation}
where $ \fk{f}_{\mrm{x}} $, $ \eta $, $ \cl{F}^{\mrm{x}} $ depend on single spectral parameter, $ \fk{s} = \pm 1 $ and one can identify ratios of sectors $ x_{\tx{R}}^{-}/x_{\tx{L}}^{-} = x_{\tx{R}}^{+}/x_{\tx{L}}^{+} $. More detailed description on the structure, limits and parametrisations of {\adsm} is provided in Appendix \ref{Apx_AdS_Backgrounds}.

\subsubsection{8vB sector: Symmetries of $ AdS_{\{ 2,3 \}} $ deformations} 

\paragraph{8vB $ \mathbf{AdS_{2}} $ deformation} In analogy with \ref{p5_6vB_DS} one can propose consistent Ansatz for the supercharge and the associated algebra structure
\begin{equation}\label{p5_8vB_G}
	\fk{G} = 
	\begin{pmatrix}
		0 & b \\
		a & 0
	\end{pmatrix}
	\qquad 
	\begin{array}{l}
		\Delta(\fk{G}) = \fk{G} \otimes 1 + \fk{U} \otimes \fk{G} \\[2ex] 
		R \Delta(\fk{G}) = \Delta^{\mrm{op}}(\fk{G}) R
	\end{array}
\end{equation}
so that after resolving cocommutativity for the supercharge, we obtain
\begin{equation}\label{key}
	\begin{array}{l}
			\dfrac{a}{b} = \dfrac{\dn{2 v} - 1}{\sigma_{1} \mrm{sn}(2 v)} \\ [2ex] 
			\fk{U} = \dfrac{\sigma_{1} \sigma_{2}\cos \eta \, \sn{2 v} + \sigma_{1} \sin \eta \, \cn{2 v}}{\sin \eta + \sigma_{1}\sigma_{2} \, \sn{2v}}
	\end{array}
\end{equation}
with $ \sigma_{i} = \pm 1 $ and one can get a differential constraint on either $ a $ or $ b $ similar to \eqref{p5_Differential_Cocommutativity}, but its explicit form does not appear convenient for use. Instead one can use similar double parametric central element of the form $ \fk{Q} \sim \{ \fk{G},\fk{G} \} $, which implies
\begin{equation}\label{p5_G_Constraint_1}
	\Delta(\fk{Q}) = \Delta^{\mrm{op}}(\fk{Q}) \qquad \fk{Q} = \fk{c} ( 1 - \fk{U}^{2} )
\end{equation}
solving the last completely fixes \eqref{p5_8vB_G}, but sign ambiguity affects only the supercharge normalisation, hence constant $ \fk{c} \equiv 1 $ without loss of generality. 

In addition there are two more constraining clarifications needed for the supercharges and $ \sigma_{i} $ constants. First of all it can be shown that there are only two generators \eqref{p5_8vB_G} in the current setting, which are distinguished by $ \{ \sigma_{1}, \sigma_{2} \} $ and we can define 
\begin{equation}\label{key}
	\fk{G}_{\pm} = \fk{G} \big|_{\sigma_{1} = -1, \, \sigma_{2} = \pm 1 }
\end{equation}
and 
\begin{equation}\label{key}
	\begin{array}{l}
		\fk{U} \xrightarrow[]{\sigma_{2} \rightarrow -\sigma_{2}} \fk{U}^{-1} \\ [2ex] 
		\Delta(\fk{G}_{\pm}) = \fk{G}_{\pm} \otimes 1 + \fk{U}^{\left( \pm 1 \right)} \otimes \fk{G}_{\pm}
	\end{array}
\end{equation}
Unrestricted form of the $ \fk{G}_{\pm} $ involves nontrivial ratios and branches of the elliptic functions, but for $ k \rightarrow 1 $ they could characterised by $ a $, $ b $ and $ \fk{U} $ as follows
\begin{equation}\label{key}
	\begin{array}{l}
		a = \dfrac{\sigma_{2} \sin(\eta/2) \tanh(v) - \cos(\eta/2)}{\sigma_{2} - \sin \eta \coth(2 v)} \coth^{\frac{1 - \sigma_{2}}{2}}(v) \\ [3ex] 
		b = \dfrac{\cosh(v) \tanh^{\frac{1 - \sigma_{2}}{2}}(v)}{\sin(\eta/2)\cosh(v) - \sigma_{2} \cos(\eta/2) \sinh(v)} \\ [3ex] 
		\fk{U} = \dfrac{\sin \eta \, \tx{sech}(2 v) + \cos \eta \tanh(2 v)}{\tanh(2 v) - \sin \eta} \\ [3ex] 
	\end{array}
\end{equation}

\paragraph{Higher orders in $ \mathbf{\textit{T}} $} It is instructive to analyse higher orders in $ RTT $, since the structure of the both $ AdS_{2} $ \cite{Hoare:2014kma,Hoare:2015kla} and our emerging $ AdS_{2} $-deformation exhibit analogous algebra. To achieve that, we must perturb $ R $ and $ T $, which for the second order results in
\begin{equation}\label{key}
	T(u) =
	\begin{pmatrix}
		1 & 0 \\
		0 & \fk{U}
	\end{pmatrix}
	+
	\begin{pmatrix}
		\frac{B + H}{2} & \fk{U} \fk{G}_{-} \\ 
		\fk{G}_{+} & \fk{U}\frac{B - H}{2}
	\end{pmatrix}
	u + \cl{O}(u^{2})
\end{equation}
and $ \eta(u) $ in the $ R_{\tx{8vB}} $
\begin{equation}\label{key}
	\eta(u) = c_{0} + c_{1} u + c_{2} u^{2} + \cl{O}(u^{3})
\end{equation}
where the $ R_{\tx{8vB}} $ would become diagonal iff $ c_{\{ 0,1 \}} = 0 $ (assumption in the vicinity $ u = 0 $) and we identify FCR from expanded $ RTT $. Up to the firs order $ \fk{U} $ is in the centre (commutes with $ T(u) $). Order two provides $ \al{psu}(1|1)_{ce} $ along with additional element $ B $ in the centre. Since there no additional deformation freedoms appear in the first two orders, as well as $ k $-dependence is absent, it implies that the corresponding level one Yangians $ \cl{Y}_{1} $ remain equivalent and algebra of \cite{Hoare:2015kla} is reconstructed (including proposed secret symmetry).

It must be noted, that the deformation we consider is not just deformation at the level of representations or their characteristic parameters. Already in the second order the dependence on $ k $ arises in the structure constants and apparently not removable by means of transformations or redefinitions at any level of the symmetry. Also algebra that is centrally extended by $ \cl{Q} $, which is $ \{ \fk{G},\fk{G} \} = 2 \fk{Q} $ along with its level one Yangian structure $ \{ ^{(1)}\fk{G}, \fk{G} \} = 2 \, ^{(1)}\fk{Q} $ not to be $ k $-dependent, but level two is $ \{ ^{(1)}\fk{G}, ^{(1)}\fk{G} \} = 2 \, ^{(2)}\fk{Q} $. On the other hand it is known that $ \cl{Y} $ fundamental representation does not allow an evaluation one. In the present context the $ k $-dependent second level forbids the evaluation representation\footnote{One needs to take care of $ k \rightarrow \infty $ limit, since in this case it becomes possible.}, which is indeed, also the case for $ AdS_{2} $. The associated Yangian structure if present in certain form, would arise nontrivially (more remarks in \ref{Conclusions and remarks}).

There is another remark that given deformation cannot be recasted, for instance at the level of the physical characteristics like $ e $, $ m $ or $ p $. The initial $ AdS_{2} $ model exhibits linear constraint $ r_{1} = r_{2} + r_{3} + r_{4} $, whereas 8vB apriori does not (if not $ k \rightarrow \infty $ limit considered (similar to the discussion in Section \ref{FF_Section}).

\paragraph{8vB $ \mathbf{AdS_{3}} $ deformation} We have already noticed, that pure blocks (LL/RR) will take the form of the block that characterised the 8vB $ AdS_{2} $ case \ref{p5_R8vB_Full}. As we have mentioned, the corresponding pure sector scattering operators satisfy $ \al{psu}(1|1)_{ce} $ (with $ k $ dependent representations). Hence if one would start with the charge of the form
\begin{equation}\label{key}
	\begin{pmatrix}
		\fk{G}^{\tx{L}} & 0 \\ 
		0 & \fk{G}^{\tx{R}} 
	\end{pmatrix}
\end{equation}
where the $ \fk{G}^{\tx{L}} $ corresponds to the one as in 8vB $ AdS_{2} $ type deformation, with accordingly replaced $ v \rightarrow \cl{F}^{\tx{L}}(v) $. On the other hand, the $ \fk{G}^{\tx{R}} $ can be obtained from part of the cocommutativity conditions. Moreover from the latter, one can find that conditions on distinct blocks do not intertwine and conditions on pure blocks map to each other, \textit{i.e.} $ a^{\tx{L}} \mapsto b^{\tx{R}} $, $ b^{\tx{L}} \mapsto a^{\tx{R}} $, $ \fk{U}^{\tx{L}} \mapsto \fk{U}^{\tx{R}} $ and $ \cl{F}^{\tx{L}} \mapsto \cl{F}^{\tx{R}} $, the normalisation is set for the $ \fk{G}^{\tx{R}} $. Commutativity with the associated full {\Rx} restricts to the existence of two charges of the form (constitutes one copy of the $ \al{psu}(1|1)_{ce} $)
\begin{equation}\label{key}
	\begin{pmatrix}
		\fk{G}_{-}^{\tx{L}} & 0 \\ 
		0 & \fk{G}_{+}^{\tx{R}} 
	\end{pmatrix}
	\qquad
	\begin{pmatrix}
		\fk{G}_{+}^{\tx{L}} & 0 \\ 
		0 & \fk{G}_{-}^{\tx{R}} 
	\end{pmatrix}
\end{equation}
The previous arguments on $ k $-dependence of the higher level deformed $ ^{(n)}\cl{Y} $ absence of the evaluation representation become apparent also for 8vB $ AdS_{3} $ deformation.


\newpage
\section{Conclusions and Remarks}\label{Conclusions and remarks}

In the present work we have constructed a new approach for deriving novel integrable structures, solve classification problem and study their properties. It was developed on universal grounds and is applicable to quantum and string integrability. This method demonstrated its significance in a number of subfields, \textit{e.g.} new integrable structure in system with local space dimensions $ D = 2, 3, 4 $, novel type of the emerging symmetries and scattering operator properties, identification of the completeness and sufficiency constraints and many more.

We have started our analysis from the automorphic and recursive structure \cite{deLeeuw:2019zsi} that arise in integrable systems of dimension two. We have addressed classification problem, on the example of two dimensional integrable spin chains with the nearest-neighbour interactions and periodic boundary conditions. We have shown that by proposing a full $ 16 $-parametric Ansatz for the Hamiltonian density $ \cl{H} $, we can in general, construct an infinite hierarchy of conserved charges by virtue of the automorphic nature of the so called $ \cl{B} $oost operator. Such recursive generating structures have been first discussed in the context of master symmetries arising in quantum integrability \cite{Tetelman}. The proposed approach exploited the commutative property of the aforementioned tower of conserved charges $ \bb{Q}_{r} $, where as an initial constraint is taken the commutativity of the second $ \bb{Q}_{2} \equiv \cl{H} $ and third $ \bb{Q}_{3} $ charges, \textit{i.e.} $ \left[ \bb{Q}_{2}, \bb{Q}_{3} \right] = 0  $. We call this condition as the leading integrability constraint, as it generates a polynomial system necessary to resolve and appears to provide completely saturating construction in majority of cases \cite{Grabowski}, meaning that it is necessary and sufficient to generate only the first commutator \cite{deLeeuw:2019zsi}.

The next step was to demonstrate not only the integrable models that have not appeared before, but incorporate certain scheme that would allow for a compact classification in terms of generating solutions (\textit{generators}). For that reason we brought generic integrable transforms/identifications, which created a class structure on the full solution space \ref{IT}. As a consequence a question of finding the corresponding {\Rx} for each class came next. In this respect, an inverse or bottom-up construction appeared helpful, by that said, meaning finding the {\Rx} from the associated $ \cl{H} $, which is contrary to conventional approaches. It turned out that one could have found the {\Rxs} from the first three orders of the $ \cl{H} $-perturbation. However, it can be efficiently accomplished only for the two-dimensional setups and {\Rxs} with difference spectral dependence, in what later, a universal differential procedure was developed, that does not contain issues of the perturbative {\Rx} reconstruction. We have also cross-checked all the {\Rx} to satisfy qYBE, hence completing quantum integrable consistency of the found classes.

In addition, the classification of two-dimensional chains with graded properties was addressed, which appears relevant for the supersymmetric spin chains, \textit{e.g.} $ \al{gl}(n|m) $ program. The graded/non-graded bijection mapping was independently proven, whose general form took elliptic type (as expected for the even sector, first discovered in \cite{Kulish1982}). One could also address classification of the systems with supersymmetric symmetry algebra and higher dimensionality of the local spaces.

One of the several interesting directions in this framework, is the quantum algebras and Yangians $ \cl{Y} $ of the new structure that arose for the non-hermitian sector. In particular one of the models generalises \cite{Stolin1997} to multi-parameter deformation, which in turn, has limiting cases to non-equilibrium transport system and criticality \cite{Alcaraz_1994}. It proposed that these multi-parametric deformations would correspond non-trivial deformation of the $ \cl{Y}^{\ast}\left[ \al{sl}_{2} \right]  $. It would be also useful to develop Bethe Ansatz type techniques to solve eigenproblems of such models. In this context, it is interesting to see if this sector has relations to other non-unitary (field-theoretic) models in spin chain picture, \textit{e.g.} fishnet class models or other conformal integrable deformations \cite{Caetano:2016ydc,Gromov:2017cja,Ipsen:2018fmu}. One can also ask for the application to open spin chain systems or the existence of continuum framework of the boost structure.

At the next level, we have asked for the novel systems, which could exhibit fermion pair formation and in general correspond to models with dynamics in the conduction zone. In particular, our interest would span integrable systems with four dimensional local space, \textit{e.g.} that could belong to the Hubbard model class or related system on the lattice \cite{deLeeuw:2019vdb}. Since this sector apriori is characterised by $ \al{su}(2) \oplus \al{su}(2) $ symmetry algebra, it was taken as a building block. However it was noted that Shastry type construction was not present in the investigation of 4-dim difference form setting, because of its manifestly distinct spectral dependence and properties.

In the first stage, for the search of new structures with $ \fk{h}^{4} $ local Hilbert space, we have brought in appropriately modified notion of boost operator and identifying symmetry transformations \ref{ND_IT}. By going along the lines of this classifying scheme, we have found novel models with the property of the pair formation, in certain sense corresponding to superconductive modes in the system. Undoubtedly, our setup reproduced the known structures as well, as it should, in order to be universal and complete (\textit{e.g.} $ \al{su}(4) $, $ \al{sp}(4) $, decomposed anti-ferromagnets and other).

Specifically, we have extracted 18 classes, including modified and generalised one. The classes obeying $ \al{su}(2) \oplus \al{su}(2) $ symmetry algebra in according representations were given in Table \ref{HHubbard_SolTable}. Thereby, classes \Rn{1}-\Rn{5} represent the new structure, where \Rn{1} and \Rn{2} share properties reminiscent of $ \al{su}(4) $ and $ \al{su}(2|2) $ integrable models. The latter shows rational form of the {\Rxs}, it is of interest to identify the corresponding Yangians and if such construction can universally appear $ \al{su}(m) \oplus \al{su}(n) $. The rest classes \Rn{3}-\Rn{5} demonstrate the system where all individual modes are suppressed and only pair dynamics is allowed (bosonised constraining).

An important direction would be to address the completely generic ansatz, which would result in over 30 free parameters, where we could impose certain restrictions, \textit{e.g.} identifications, physical and other symmetry constraints. Remarkable fact, that our prescription can handle this challenging query through the techniques already developed by the current moment. The sufficiency and uniqueness of the first commutator still holds, but it would interesting to find in generic ansatz a class for which the higher commutativity is required for constraining (analogous to ZSM model \cite{Korepin_1997_QISM_CF}). The other problem that we have come up with is the solution of the eigenproblem via the Bethe Ans\"atze. It appeared that CBA, or even more generic approach as inhomogeneous nested CBA do not appear to work for the novel sector (level two break). The problem also persists for the Generalised Hubbard Class, and it is currently under investigation if Generalised Bethe Ansatz \cite{Slavnov:2020xxj} or Quantum Spectral Curve \cite{Gromov_2014, Gromov:2015wca, Kazakov_2018} have a realisation for this sector.

It would be interesting to further study physical properties of the new classes. For instance it would be interesting to see phase diagrams of the first five classes or of the class \Rn{18}, where we have free parameters that can be treated as two coupling (resembles A and B models \cite{Frolov:2011wg}). It would be also interesting to investigate relation to the other generalised constructions \cite{Drummond:2007gt,Drummond:2007sa,Maassarani_1998_su_n_HM,Maassarani_1998_su_n_XX} and analyse further deformations or novel classes also in these sector. 



We have then followed our discussion with the generalization of both the automorphism technique and higher dimension $ D $ of the models \cite{deLeeuw:2020ahe}. It is known that for a number of integrable models, the spectral dependence cannot be brought into difference form and is arbitrary in general. In particular, it appears to be a property of the Shastry class or Hubbard chain in {\ads} and integrable structures in string $ AdS_n $ backgrounds. In what we have obtained non-difference boost construction and generic integrable transformation set. It was then possible to address various general Ans\"atze at the level of $ \cl{H} $ in $ D = 2,3,4 $. For two dimensions we have analysed 8-vertex integrable structure (and subsequent lower vertex classes), for three dimensions it was 15-vertex prescription and in four-dimensional case Shastry type, Generalised Hubbard Model and other classes have been studied in arbitrary spectral form, meaning Hamiltonians contain undetermined functions on a parameter. In number of the subcases above we found a novel structure. We have also developed unified construction in the form of coupled Sutherland system to resolve for the {\Rxs}, which generated all the $ R $ and confirmed their quantum consistency.

For two dimensions we were able to identify four novel structures, which lie in the 6- and 8-vertex sectors and which exhibit arbitrary functional dependence. On contrary, by suppression of the differential term in the boost expression and recovering constant parameters, we reproduce difference form technique. Later it became apparent that 6vB and 8vB contain deformation parameters and can be realised as $ AdS_{2} $ and $ AdS_{3} $ deformations. Finding physical interpretation of these parameters, their potential relation to the ones in $ AdS_{\{ 2,3 \}} $ backgrounds with RR and NSNS fluxes, as well as opposite chirality mapping questions remain open.

The three dimensional setting showed reduced 15-vertex construction of the non-difference form. In addition, we have found that Zhiber-Shabat-Mikhalov type Ansatz when put on the lattice appears to be unique, as even with generalised prescription it have shown only ZSM or subcases to exist. In this respect, there is space for further investigation in higher vertex sector (\textit{e.g.} 19-, 22-, 33-vertex \cite{Makoto:1994JPhy1,Fonseca:2014mqa,Crampe:2016she}), it would be interesting if these constructions possess any implications in field theoretic or worldsheet string framework.

In four dimensions, it turned that there are no interesting non-difference classes, which at the same time would respect $ \al{su}(2) \oplus \al{su}(2) $ symmetry algebra. However the $ \cl{H}^{(0)} + \epsilon \cl{H}^{(0)} + \epsilon^{2}\cl{H}^{(2)} + \cl{O}(\epsilon^{3}) $ perturbation shows that the new structures must exists in the $ q $-deformed \cite{Beisert:2008tw} sector of $ AdS_{5}/CFT_{4} $, where quantum affine algebras are emergent \cite{Beisert_2012,Beisert:2011wq}. An analysis on the potential relation of difference and non-difference have been provided, showing that for some models there are no maps from either sector. For instance, we have proven that classes \Rn{4}, \Rn{5} and \Rn{11} do not have any non-difference counterpart, hence it is a valid question if the latter can be recovered from some broken version of the $ \al{su}(2) \oplus \al{su}(2) $, $ \al{su}(4) $ or a centrally extended $ \fk{g} $. It would also interesting to develop a boundary formalism \cite{Sklyanin:1987bi,Sklyanin:1988yz} and construct open spin chains associated to classes and some of the deformations above.

Another important point yet to be proven for both approaches, \textit{i.e.} non- and difference form, how exactly the data that encoded in polynomial systems $ \left[ \bb{Q}_{2}, \bb{Q}_{3} \right] = \dots = \left[ \bb{Q}_{r}, \bb{Q}_{s} \right] = 0 $ and differential conditions on $ \cl{H}/R $ is equivalent to the one provided by the YB constraint and existing limits of the associated {\Rx}. An analytic closed expressions of objects on both sides could provide an insight into intertwining structure, that would important in the algebraic techniques for studying models where standard methods are not applicable (QISM, $ RTT $, QSC, nested BA and other). In the given context, it can be also found that scheme \ref{p2_BoostScheme}, could brought into diagram with commutative edges and investigation of multiple steps is an important question itself. Moreover, one could address an algebraic structure of the commutative hierarchies associated to different symmetry algebras, hence study properties of the varieties and characteristic (genus $ c $) curves, which could provide an answer to some of the intermediate steps above, unclassified classes or bring more useful algebraic structure.

In the last segment, we proceeded further on the non-additive 6v/8v models and their deformations \cite{deLeeuw:2021ufg} relevant for $ AdS_{2} $ and $ AdS_{3} $ string backgrounds. We have started by noting, that these models satisfy special constraint, that is also characteristic for a number of statistical models -- \textit{free fermion} \cite{deLeeuw:2020bgo}. We then have shown the usefulness for the scattering operators arising in lower $ D $ cases of the AdS/CFT. In fact, one can gain significantly more compact formalism for the {\Rxs} in terms of free fermionic creation/annihilation operators via Bogoliubov transformations.

Beyond showing relations of the 6vB and 8vB class relation to integrable models of the $ AdS $ space, we have demonstrated {\adsT} and {\adsm} deformations based on B class and established mappings \cite{deLeeuw:2021ufg}. Detailed analysis of the latter has indicated that {\adsm} admits both parametric and functional deformation, whereas {\adsT} case only the parametric one. In both sectors, the braiding unitarity and crossing properties have been confirmed, the associated conjugation matrix and braiding factor have been found. It would be useful to study their structure in more details. The algebra of deformations exhibits significant correlation with $ AdS_{\{ 2,3 \}} $ models, but at the same time appears to constitute nontrivial deformation, especially at the level of Yangian structure. Further understanding of the models and their algebra would include specific constraining and identification of characteristic observables, which would provide more building blocks at the quantum algebraic level and associated sigma model class.

In this context, it would very important to work out the corresponding current and algebra for the sigma model associated to these new deformations. It is clear that a more general approach might be needed, based on the screening charge formalism \cite{Fateev:2018yos,Alfimov_2020_OSPSigma} to find a complete structure of the sigma or its deformation. In such prescription it would also suitable to investigate limiting cases, as all controlling parameters are contained in screening charges. It would also interesting if there are specific classical limits to recover some of the Yang-Baxter class sigma model or higher parametric deformations \cite{Fateev:1996ea, Delduc:2018xug, Bocconcello:2020qkt, Lukyanov:2012zt}. It would very important to understand if due to specific structure of the scattering operators and symmetry algebra of the 6v/8v classes above, they could have a discrete classification on quiver space or if their worldsheet counterparts exhibit geometrical reduction limits \cite{Affleck:2021jls,Bykov:2021dbk,Bykov:2022aka}. It is of current progress, an investigation if the algebraically constrained or extended B class can develop novel (super)coset structure or if it would lead to a new integrable class on the worldsheet.

\vspace{4cm}


\newpage

\section{Acknowledgments}
I especially would like to thank Marius de Leeuw, Sergey A. Frolov, Tristan McLoughlin, Samson L. Shatashvili, Vladimir A. Kazakov, Vladimir E. Korepin, Leon A. Takhtadzhan, Dmitry I. Gurevich, Alessandro Sfondrini, Ben Hoare, Fiona Seibold, Alessandro Torrielli, Mikhail A. Vasiliev, Konstantin L. Zarembo, Vladimir V. Bazhanov. Current work progress is supported by SFI and Royal Society RGF$\backslash$EA$\backslash$180167.


\newpage
\begin{appendix}
	
		\section{Spin chain operators and algebras}
		\subsection{$ \bb{V} = \bb{C}^{2} $}\label{Apx_2-dim}
		The spin operators $ \mathbb{S} $ in the context of spin chains provide a description of internal angular characteristic associated to each site. Their components $ S_{\mu} $ obey angular momentum algebra in analogy to quantum mechanical systems
		\begin{equation}\label{key}
			\left[ S^{\mu} , S^{\nu} \right] = i \hbar \varepsilon_{\mu \nu \rho} S^{\rho} \qquad \qquad \qquad \mu, \nu, \rho = 1,2,3 
		\end{equation}
		with totally antisymmetric tensor $ \varepsilon_{123} = -\varepsilon_{213} = 1 $ and $ S^{2} = (S^{1})^{2} + (S^{2})^{2} + (S^{3})^{2} $ that commutes with the corresponding spin-components. In the conventional setting, for the spin-$ s $ quantum particle we have (anti)symmetrisation condition on the corresponding wave function
		\begin{equation}\label{key}
			\Psi\left( \, \dots \, ; \, \textbf{r}_{i}, \mathfrak{s}_{i} \, ; \, \dots \, ; \, \textbf{r}_{j}, \mathfrak{s}_{j} \, ; \, \dots \, \right) = (-1)^{2 s}  \Psi\left( \, \dots \, ; \, \textbf{r}_{j}, \mathfrak{s}_{j} \, ; \, \dots \, ; \, \textbf{r}_{i}, \mathfrak{s}_{i} \, ; \, \dots \, \right)
		\end{equation}
		where $ \Psi\left( \textbf{r}, \mathfrak{s} \right) $ characterises particle spatial position and spin state, with $ s \in \mathbb{Z}_{+} $ for bosons and $ s \in \mathbb{Z}_{+} + \dfrac{1}{2} $ for fermions, so that spin state is spanned by $ \mathfrak{s} \in \left( -s, \dots s \right) $. From the prefactor one can also identify (anti)commutation nature of the corresponding statistics. For the spin-$ \dfrac{1}{2} $ particles the spin operators are defined
		\begin{equation}\label{key}
			\mathbb{S} = \dfrac{1}{2} \left( \sigma_{1}, \sigma_{2}, \sigma_{3}  \right) \qquad
			\sigma_{1} = \begin{pmatrix}
				0 & 1 \\
				1 & 0
			\end{pmatrix} \text{, } \quad
			\sigma_{2} = \begin{pmatrix}
				0 & -i \\
				i & 0
			\end{pmatrix} \text{, }
			\quad
			\sigma_{3} = \begin{pmatrix}
				1 & 0 \\
				0 & -1
			\end{pmatrix}
		\end{equation}
		with $ \sigma_{1,2,3} \equiv \sigma_{x,y,z} $\footnote{or in extended algebraic setting, $ 2 $-dim identity is can be implemented on the equivalent footing $ \sigma_{0} = \mathds{1}_{2} $} and we deal with the Pauli algebra (natural units)
		\begin{equation}\label{key}
			\left( \sigma^{\mu} \right)^{2} = \mathds{1} \qquad \text{and} \qquad \sigma^{\mu}\sigma^{\nu} = i \varepsilon_{\mu\nu\rho}\sigma^{\rho}  \quad\text{with } \mu \neq \nu
		\end{equation}
	
		For quantum spin chain computations it is useful to define $ S^{z} $-basis with the suitable spin-up/down eigenvectors $ \{ \mathfrak{e}_{+}, \mathfrak{e}_{-} \} $
		\begin{equation}\label{key}
			\mathfrak{e}_{+} = 
			\begin{pmatrix}
				1 \\ 0
			\end{pmatrix}
			\qquad \qquad
			\mathfrak{e}_{-} =
			\begin{pmatrix}
				0 \\ 1
			\end{pmatrix}
		\end{equation}
		where raising/lowering operators defined accordingly
		\begin{equation}\label{key}
			S^{+} = S^{1} + i S^{2} =
			\begin{pmatrix}
				0 & 1 \\
				0 & 0
			\end{pmatrix}
			\qquad
			S^{-} = S^{1} - i S^{2} =
			\begin{pmatrix}
				0 & 0 \\
				1 & 0
			\end{pmatrix}
		\end{equation}
		with $ S^{z,+,-} $ action
		\begin{equation}\label{key}
			S^{+} \mathfrak{e}_{+} = S^{-} \mathfrak{e}_{-} = 0 \quad S^{+} \mathfrak{e}_{-} = \mathfrak{e}_{+} \quad S^{-} \mathfrak{e}_{+} = \mathfrak{e}_{-} \quad S^{z} \mathfrak{e}_{+} = \dfrac{1}{2} \mathfrak{e}_{+} \qquad S^{z} \mathfrak{e}_{-} = - \dfrac{1}{2} \mathfrak{e}_{-}
		\end{equation}
		so one obtains 2-dimensional irreducible $ \al{sl}(2) $ representation
		\begin{equation}\label{key}
			S^{+} =
			\begin{pmatrix}
				0 & 1 \\
				0 & 0
			\end{pmatrix}
			\qquad
			S^{-} =
			\begin{pmatrix}
				0 & 0 \\
				1 & 0
			\end{pmatrix}
			\qquad
			S^{z} = \dfrac{1}{2}
			\begin{pmatrix}
				1 & 0 \\
				0 & -1
			\end{pmatrix}
		\end{equation}
		and the correspondingly closing algebra
		\begin{equation}\label{key}
			\left[ S^{z}, S^{\pm} \right] = \pm S^{\pm} \qquad \left[ S^{+}, S^{-} \right] = 2 S^{z}
		\end{equation}
		
		\newpage
		\subsection{Graded spaces and algebra}\label{Apx1_Graded_Spaces_Algebra}
		In order to be able to implement the bosons and fermions on a unified ground, one is required to build the corresponding structure on the vector spaces. As it is known such structure can be created by notion of grading, hence defining graded vector spaces. In the integrability that admits fermionic degrees along with bosonic ones in various methods \cite{Essler2005_HB} (Bethe Ans\"atze, QISM, $ RTT $ and other). Specifically, for a vector space $ \bb{V} $ it implies decomposition
		\begin{equation}\label{key}
			\bb{V} = \bb{V}_{0} \oplus \bb{V}_{1}
		\end{equation}
		where $ \dim \bb{V}_{0} = d_{0} $, $ \dim \bb{V}_{1} = d_{1} $, $ v_{0} \in \bb{V}_{0} $ are even, $ v_{1} \in \bb{V}_{1} $ are odd and subspaces $ \bb{V}_{i} $ are homogeneous. The function $ p(\cdot) $ is taken to identify \textit{parity}
		\begin{equation}\label{key}
			p(v_{i}) = i, \qquad i = 0,1 \tx{ and } v_{i} \in \bb{V}_{i}
		\end{equation}
		hence $ \bb{V} $ constitutes \textit{graded vector space} or a \textit{super space}. In the quantum $ (2|2) $-graded vector space that is considered in four dimensions, one could establish the following grading
		\begin{equation}\label{key}
			p(x) = 
			\begin{cases}
				0, \quad x = 1,2 \\
				1, \quad x = 3,4
			\end{cases}
		\end{equation}
		For a basis $ \{ e_{1}, e_{2}, \dots, e_{d_{0}+d_{1}} \} $ with prescribed parity one can define $ p(\alpha) = p(e_{\alpha}) $. In addition we must have an extension of $ e_{\alpha}^{\beta} $
		\begin{equation}\label{key}
			e_{\alpha}^{\beta} e_{\rho} = \delta_{\rho}^{\beta}e_{\alpha}, \quad\qquad \{e_{\alpha}^{\beta} \in \tx{End}(\bb{V}) \,|\, \alpha,\beta = 1, 2, \dots, d_{0}+d_{1} \}
		\end{equation}
		and associated parity,
		\begin{equation}\label{Apx1_parity_1}
			p(e_{\alpha}^{\beta}) = p(\alpha) + p(\beta)
		\end{equation}
		which defines grading on $ \tx{End}(\bb{V}) $ and an element $ a = A_{\alpha}^{\beta} e_{\beta}^{\alpha} \in \tx{End}(\bb{V}) $ is homogeneous on parity $ p(a) $ iff
		\begin{equation}\label{key}
			(-1)^{p(a)} A_{\alpha\beta} = (-1)^{p(\alpha) + p(\beta)} A_{\alpha\beta}
		\end{equation}
		We can now consider $ L $-folded generalisation $ \tx{End} (\bb{V})^{\otimes L} $ of \eqref{Apx1_parity_1}
		\begin{equation}\label{Apx1_parity_2}
			p(e_{\alpha_{1}}^{\beta_{1}} \otimes \cdots \otimes e_{\alpha_{L}}^{\beta_{L}}) = p(\alpha_{1}) + p(\beta_{1}) + \cdots + p(\alpha_{L}) + p(\beta_{L})
		\end{equation}
		hence the homogeneity can lifted
		\begin{equation}\label{Apx1_Homogeneity_2}
			(-1)^{ \sum_{k = 1}^{L} (p(\alpha_{k}) + p(\beta_{k})) } A_{\beta_{1}\dots\beta_{L}}^{\alpha_{1}\dots\alpha_{L}} = (-1)^{p(a)} A_{\beta_{1}\dots\beta_{L}}^{\alpha_{1}\dots\alpha_{L}}
		\end{equation}
		One can notice that \eqref{Apx1_parity_2}-\eqref{Apx1_Homogeneity_2} impose graded associative algebra, whose superbracket is defined
		\begin{equation}\label{key}
			\left[ \cl{X}, \cl{Y} \right] = \cl{X}\cl{Y} -(-1)^{p(\cl{X})p(\cl{Y})}\cl{Y}\cl{X}
		\end{equation}
		where $ \cl{X},\cl{Y} \in \tx{End}(\bb{V}) $. Under linear extension for both arguments of $ \tx{End}(\bb{V}) $ we acquire a superbracket on Lie superalgebra $ \al{gl}(d_{0}|d_{1}) $ \cite{Essler2005_HB}.
		
		One can now address the fermionic sector and identify the corresponding integrability constraint. It can be shown that one can obtain graded (fermionic) solutions from the non-graded YBE \cite{Kulish1982,Gohmann_2000}, \textit{i.e.} for a fixed grading and $ R(\lambda,\mu) $ from
		\begin{equation}\label{Apx1_NG_qYBE}
			R_{\alpha'\beta'}^{\alpha\beta}(\lambda,\mu)R_{\alpha''\gamma'}^{\alpha'\gamma}(\lambda,\nu)R_{\beta''\gamma''}^{\beta'\gamma'}(\mu,\nu) = R_{\beta'\gamma'}^{\beta\gamma}(\mu,\nu)R_{\alpha'\gamma''}^{\alpha\gamma'}(\lambda,\nu)R_{\alpha''\beta''}^{\alpha'\beta'}(\lambda,\mu)
		\end{equation}
		that is compatible with the given grading, one can obtain a solution of graded qYBE
		\begin{equation}\label{key}
			R_{12}^{g_{f}}(\lambda,\mu)R_{13}^{g_{f}}(\lambda,\nu)R_{23}^{g_{f}}(\mu,\nu) = R_{23}^{g_{f}}(\mu,\nu)R_{13}^{g_{f}}(\lambda,\nu)R_{12}^{g_{f}}(\lambda,\mu)
		\end{equation}
		by
		\begin{equation}\label{Apx1_RG_RNG}
			R_{ij}^{g_{f}}(\lambda,\mu) = (-1)^{p(\gamma) + p(\alpha)\left[ p(\beta) + p(\gamma) \right] } R_{\alpha\beta}^{\gamma\delta} e_{i,\,\gamma}^{\alpha} e_{j,\,\delta}^{\beta} \qquad p\left[ R_{12}^{g_{f}}(\lambda,\mu) \right] = 0
		\end{equation}
		along with $ R_{\alpha\beta}^{\gamma\delta} $ from \eqref{Apx1_NG_qYBE} to satisfy consistency
		\begin{equation}\label{key}
			R_{\alpha\beta}^{\gamma\delta}(\lambda,\mu) = (-1)^{ p(\alpha) + p(\beta) + p(\gamma) + p(\delta) } R_{\alpha\beta}^{\gamma\delta}(\lambda,\mu)
		\end{equation}
		where above we have defined projectors on graded space as follows 
		\begin{equation}\label{key}
			e_{i, \alpha}^{\beta} = (-1)^{( p(\alpha) + p(\beta) ) \sum_{j = i+1}^{L} p(\delta_{j})} \mathds{1}_{(d_{0}+d_{1})}^{\otimes (i - 1)} \otimes e_{\alpha}^{\beta} \otimes e_{\delta_{i + 1}}^{\delta_{i + 1}} \otimes e_{\delta_{L}}^{\delta_{L}} \qquad e_{i, \alpha}^{\beta}e_{i, \epsilon}^{\gamma} = \delta_{\gamma}^{\beta} e_{i, \alpha}^{\epsilon}
		\end{equation}
		All the characteristic {\Rx} constraints arise as
		\begin{equation}\label{key}
			\begin{array}{l}
				R^{g_{f}}(\lambda*,\mu*) = P_{ij} \quad \tx{(Regularity)}  \\ [2ex]
				P_{ij} = (-1)^{p(\alpha)}e_{i,\,\beta}^{\alpha} e_{j,\,\alpha}^{\beta} \quad (\tx{graded permutation}) \\ [2ex]
				R_{ij}^{g_{f}}(\lambda^{*},\mu^{*}) R_{ji}^{g_{f}}(\mu*,\lambda*) = \mathds{1} \quad \tx{(Graded unitarity)} \\ [2ex]
			\end{array}
		\end{equation}
		and we can reconstruct the local Hamiltonian from
		\begin{equation}\label{key}
			\cl{H}_{i, i+1} = (-1)^{p(\gamma) ( p(\alpha) + p(\gamma) ) } \partial_{\lambda} R_{\gamma\delta}^{\alpha\beta}(\lambda,\mu^{*}) \big|_{\lambda = \lambda^{*}} e_{i, \, \beta}^{\gamma} e_{i+1, \, \alpha}^{\delta}
		\end{equation}
		One can also work out the algebra by the associated $ RTT $ relations
		\begin{equation}\label{key}
				R_{ab}^{g_{f}}(\lambda,\mu)T_{a}^{g_{f}}(\lambda)T_{b}^{g_{f}}(\mu) = T_{b}^{g_{f}}(\mu)T_{a}^{g_{f}}(\lambda)R_{ab}^{g_{f}}(\lambda,\mu)
		\end{equation}
		with monodromy $ T_{a}^{g_{f}} $ appearing as
		\begin{equation}\label{key}
			T_{a}^{g_{f}}(\lambda) = R_{a,L}^{g_{f}}(\lambda,\rho_{L}) \dots R_{a,1}^{g_{f}}(\lambda,\rho_{1})
		\end{equation}
		where we have analogously two fermionic auxiliary spaces $ a,b $ and as of \eqref{Apx1_RG_RNG}
		\begin{equation}\label{key}
			T_{a}^{g_{f}}(\lambda) = (-1)^{p(\alpha) + p(\alpha)p(\beta)} e_{a, \, \beta}^{\alpha} T_{\alpha}^{\beta}(\lambda)
		\end{equation}
	
		\newpage
		\subsection{Hopf algebras and tensor algebraic properties}\label{Apx1_Hopf_Algebra}

		\paragraph{Hopf algebras} Here we describe fundamental structure and properties of the Hopf algebras \cite{Reshetikhin:1990ep, Artamonov_2015_SemisimpleHA}, whose structure is relevant for the discussion of the symmetry algebras in Sec. \ref{p1} and \ref{p5_Sec_AdS_2_3_Integrable_Deformations}. The Hopf algebra forms a core structure for the existence of a generic solution of the qYBE \cite{Drinfeld:1985rx} and its perturbative analysis \ref{Introduction}.
		
		An associative algebra equipped with a unity and multiplication $ \mu: \, \fk{H}^{2} \rightarrow \fk{H} $  is called the \textit{Hopf algebra} $ \fk{H} $, where the following is defined
		\begin{itemize}
			\item $ \exists $ Algebra homomorphism with unity $ \Delta: \, \fk{H} \rightarrow \fk{H}^{\otimes 2} $, which called \textit{co-multiplication} (or a coproduct)
			\item $ \exists $ Algebra homomorphism with unity $ \varepsilon: \, \fk{H} \rightarrow k $, which called \textit{co-unity}
			\item $ \exists $ Algebra anti-homomorphism with unity $ S: \, \fk{H} \rightarrow \fk{H} $, which called an \textit{antipode}
		\end{itemize}
		and commutativity of the appropriate diagrams is required to hold.

		At first, an important class of the Hopf algebras are the \textit{group algebras} $ k \cl{G} $ of the arbitrary finite group $ \cl{G} $. In this case the coproduct, unity and antipode from $ g \in \cl{G} $ are define according to
		\begin{equation}\label{key}
			\Delta(g) = g \otimes g, \qquad \varepsilon(g) = 1, \qquad S(g) = g^{-1}
		\end{equation}

		For the second class, one can consider universal enveloping algebra $ \cl{U} $ for the Lie algebra $ \cl{G}_{L} $. The coproduct, unity and antipode from $ g \in \cl{G}_{L} $ appear as
		\begin{equation}\label{key}
			\Delta(g) = g \otimes 1 + 1 \otimes g, \qquad \varepsilon(g) = 0, \qquad S(g) = -g
		\end{equation}
		Which appears in a number of integrable deformed symmetry algebras, \textit{e.g.} deformed {\Rx} in \cite{Hoare:2014oua}.
		
		The third class constitutes the algebras of regular functions $ \cl{F}(\cl{G}) = k \left[ \cl{G} \right] $ for the algebraic groups $ \cl{G} $ over field $ k $. For that $ \Delta, \, \varepsilon, \, S $ are defined for the function $ f \in \cl{F}(\cl{G}) $
		\begin{equation}\label{key}
			\Delta(f)(x,y) = f(xy), \qquad \varepsilon(f) = f(1), \qquad S(f)(g) = f(g^{-1})
		\end{equation}
		here $ \cl{F}(G) \otimes \cl{F}(G) \simeq \cl{F}(G \times G) $ and coproduct $ \Delta: \, \cl{F}(G) \rightarrow \cl{F}(G \times G) $ is dual to $ G \times G \rightarrow G $ mapping.


		\paragraph{Tensor algebra properties}
		The tensor product algebra involved in \ref{AA} can be reflected on the 3-site product
		\begin{equation}\label{Apx1_TA_3}
			\left( \rho_{i} \otimes \rho_{j} \otimes \rho_{k} \right).\left( \sigma_{i} \otimes \sigma_{j} \otimes \sigma_{k} \right).\left( \tau_{i} \otimes \tau_{j} \otimes \tau_{k} \right) = \left( \rho_{i} . \sigma_{i} . \tau_{i} \right) \otimes \left( \rho_{j} . \sigma_{j} . \tau_{j} \right) \otimes \left( \rho_{k} . \sigma_{k} . \tau_{k} \right)
		\end{equation}
		where the corresponding combinations of the Pauli operators $ \sigma^{k} $ arising in the higher charges $ \cl{Q}_{r} $ and their commutator, will be grouped according to \eqref{Apx1_TA_3} for the local charges of interaction range 3. It can then be immediately extended to the $ L $-fold tensor embedding, which would generalise the above to $ M $-folded $ L $-site tensor multiplication with product commutativity
		\begin{equation}\label{key}
			\prod_{j=1}^{M} \left( \bigotimes_{i=1}^{L} T_{i,j} \right) = \bigotimes_{i=1}^{L} \left( \prod_{j=1}^{M} T_{i,j} \right)
		\end{equation}
		As we have noticed, the computation in \ref{p1_Graded_Mapping} involved the graded mapping of the distinct models \cite{Kulish1982}, which have been also shown to be generated by our method \ref{Method}. In this respect, if one generates the hierarchy from Ansatz
		\begin{equation}\label{key}
			\cl{Q} = \sum \mathcal{A}_{ijkl}E_{ij}\otimes_{g}E_{kl}
		\end{equation}
		where $ E_{i}^{j} $ constitute unities at $ (i,j) $, but tensor product are graded. Taking into account the structure and properties defined \ref{Apx1_Graded_Spaces_Algebra} the operator graded tensor blocks as of \eqref{Apx1_TA_3}, will result in generalisation of $ L $-folded products to the corresponding graded property
		\begin{equation}\label{key}
			\mathcal{O}_{\Rn{1}}\mathcal{O}_{\Rn{2}} = \bigotimes_{i=1}^{L}\mathfrak{e}_{\Rn{1},i} \bigotimes_{j=1}^{L}\mathfrak{e}_{\Rn{2},j} = \prod_{i=1}^{L-1} (-1)^{p(\mathfrak{e}_{\Rn{2},i}) \sum\limits_{j=i+1}^{L} p(\mathfrak{e}_{\Rn{1},j})} \bigotimes_{k=1}^{L} \mathfrak{e}_{\Rn{1},k}\mathfrak{e}_{\Rn{2},k}
		\end{equation}
		which fixes the graded properties and allows to find a complete graded solution space, which extends \cite{Kulish1982} result by nontrivial multi-parameter deformations in the $ \al{sl}_{2} $-sector.

	\newpage
	\section{Hubbard type: $ \al{su}(2) \oplus \al{su}(2) $ 4-dim representations}\label{App:4dim_rep_su2}

	We shall here discuss the representations involved in the Hubbard type models, that exhibit the {\Hal} symmetry algebra. In this respect we would need to consider a four-dimensional representations as our local Hilbert spaces are 4-dim, where on Hubbard semi-simple Lie algebra one acquires a 4-dim representation for each $ \al{su}(2) $ copy. Specifically, there are five possible 4-dim representations can be assigned to $ \al{su}(2) $
	\begin{equation}\label{key}
		4,  \quad 3 \oplus 1, \quad 2 \oplus 2, \quad 2 \oplus 1 \oplus 1, \quad 1 \oplus 1 \oplus 1 \oplus 1
	\end{equation}
	where the last decomposition can be omitted as we interested in nontrivial representations only. One can then form three generic $ 4 \times 4 $ operators that must close under $ \al{su}(2) $ symmetry algebra and demand they commute with representation of the first copy, hence this will fix the representation of the second copy of $ \al{su}(2) $ (up to similarity transform). From here we can span the representations of {\Hal} by the above identified representation of one $ \al{su}(2) $ copy. For the {\Hal} algebra one can consider the algebra generators associated to each copy
	\begin{equation}\label{key}
		\left[ t_{i}^{\mrm{x}}, t_{j}^{\mrm{x}} \right] = \epsilon_{ijk} t_{k}^{\mrm{x}}
	\end{equation}
	where single copy can be identified from $ \mrm{x} = \{ L, R\} $ algebra generators and the corresponding irreducible $ n $-dim representation is provided by $ \fk{r}_{n}(t_{i}) $. 
	
	\paragraph{$ \mb{\fk{r}_{2 \oplus 1 \oplus 1}} $} It is possible to show that in this case embeddings are conventional spin-charge $ \al{su}(2) $ symmetry. For the $ 2 \oplus 1 \oplus 1 $ representation the first part can be associated to upper left $ 2 \times 2 $ block
	\begin{equation}\label{key}
		\fk{r}_{2 \oplus 1 \oplus 1}(t_{i}^{\tx{L}}) = \fk{r}_{2}(t_{i}) \oplus 0 = 
		\begin{pmatrix}
			\fk{r_{2}}(t_{i}) & 0 \\
			0 & 0
		\end{pmatrix}
	\end{equation}
	\begin{equation}\label{key}
		\fk{r}_{2 \oplus 1 \oplus 1}(t_{i}^{\tx{R}}) = 0 \oplus \fk{r}_{2}(t_{i}) = 
		\begin{pmatrix}
			0 & 0 \\
			0 & \fk{r_{2}}(t_{i})
		\end{pmatrix}
	\end{equation}
	where according to the algebraic argument above, the representation correlates with the first one, hence the second corresponds to the lower right $ 2 \times 2 $ block.

	\paragraph{$ \mb{\fk{r}_{2 \oplus 2}} $} The $ 2 \oplus 2 $ decomposition is block-diagonal 2-dim $ \al{su}(2) $ representation
	\begin{equation}\label{key}
		\fk{r}_{2 \oplus 2}(t_{i}^{\tx{L}}) = 1 \otimes \fk{r}_{2}(t_{i}) = 
		\begin{pmatrix}
			\fk{r_{2}}(t_{i}) & 0 \\
			0 & \fk{r_{2}}(t_{i})
		\end{pmatrix}
	\end{equation}
	it can be immediately seen that it induces
	\begin{equation}\label{key}
		\fk{r}_{2 \oplus 2}(t_{i}^{\tx{R}}) = \fk{r}_{2}(t_{i}) \otimes 1
	\end{equation}
	which can be checked to be isomorphic to the $ \al{so}(4) $ decomposed realisation.
	
	The decomposed $ 3 \oplus 1 $ and $ 4 $ dimensional irreducible representations, will not be considered, since it can be straightforwardly concluded that they induce no nontrivial representations on the second copy. Namely, such prescription will not admit consistent representations for {\Hal} and will result in the embedded $ \al{su}(2) $ spin chain, which is not of interest for the 4-dim setting.
	
	\paragraph{Oscillator representation} By the above said, it is useful to identify the symmetry algebra generators $ t_{i} $ in the oscillator representation as follows
	\begin{equation}\label{Apx_Hubbard_OR}
		\begin{array}{l}
			\fk{r}_{osc} (t^L_1) = \frac{1}{2}\left(\cdd_\uparrow \cdd_\downarrow  + \mb{c}_\uparrow \mb{c}_\downarrow\right) \\ [2ex]
			\fk{r}_{osc} (t^L_2) = \frac{i}{2}\left( \cdd_\uparrow \cdd_\downarrow  - \mb{c}_\uparrow \mb{c}_\downarrow\right) \\ [2ex]
			\fk{r}_{osc} (t^L_3) = \frac{i}{2}\left(\mb{n}_\uparrow + \mb{n}_\downarrow -1\right) \\ [2ex]
		\end{array}
		\qquad
		\begin{array}{l}
			\fk{r}_{osc} (t^R_1) = \frac{1}{2}\left(\cdd_\uparrow \mb{c}_\downarrow  + \mb{c}_\uparrow \cdd_\downarrow\right) \\ [2ex]
			\fk{r}_{osc} (t^R_2) = \frac{i}{2}\left(\cdd_\uparrow \mb{c}_\downarrow - \mb{c}_\uparrow \cdd_\downarrow\right) \\ [2ex]
			\fk{r}_{osc} (t^R_3) = -\frac{i}{2}\left(\mb{n}_\downarrow - \mb{n}_\uparrow .\right) \\ [2ex]
		\end{array}
	\end{equation}	
	where $ \mb{c}_{\sigma}^{\dagger} $/$ \mb{c}_{\sigma} $ are conventionally creation and annihilation operators, along with number operators $ \mb{n}_{\sigma} $.

	\newpage
	\section{(Non-)difference apparatus and computations} 
	In this Appendix we shall elaborate on the algebraic structure and analytic properties of the boost automorphic symmetry on the lattice and its counterpart arising in the technique for integrable models, whose scattering operators exhibit arbitrary spectral dependence (including pseudo-difference). A computation in 6-vertex sector that possesses conventional reduction to difference form will demonstrated. Also necessary details of elliptic apparatus relevant for \ref{NA_S_SD},\ref{FF_Section} and \ref{p5_Sec_AdS_2_3_Integrable_Deformations} are provided, along with additional nontrivial limits of the 8vB class.

	\subsection{Difference and Non-difference $ \cl{B} $oost construction}\label{Apx_A_NA_Boost_Automorphic_Symmetry}
	
	\paragraph{Boost automorphism on the lattice} We analyse here the structure of the boost operator on the lattice systems and see on how the automorphic condition is emergent \cite{Tetelman,Loebbert:2016cdm}. As it was indicated for $ \cl{B} $ automorphism on constant models Sec. \ref{p1} its generating property is based on the integrable master symmetries \ref{Introduction}. One can note that it is useful to consider $ R \cl{L}\cl{L} $ relation along the lines of QISM \cite{Sklyanin_1979quantum,Sklyanin_1982quantum,Faddeev:1996iy}, which can be written in the form
	\begin{equation}\label{key}
		R_{12}(v) \cl{L}_{10}(u + v) \cl{L}_{20}(u) = \cl{L}_{20}(u) \cl{L}_{10}(u + v) R_{12}(v)
	\end{equation}
	it is instructive now to study the regular existing points, in order to achieve, we can take derivative w.r.t. one of the spectral parameters ($ v $ in this case, and subsequently set it to 0), which after basic algebra will bring
	\begin{equation}\label{key}
		i \left[ \cl{L}_{10} \cl{L}_{20}, \cl{H}_{12} \right] = \cl{L}_{10}\cl{L}'_{20} - \cl{L}_{10}'\cl{L}_{20}
	\end{equation}
	where $ \cl{L}_{i0} \equiv \cl{L}_{i0}(u) $ and the following was implemented
	\begin{equation}\label{key}
		R_{12}(0) = P_{12}, \quad\qquad R'_{j,j+1}(0) = \dd_{\,\, u} R_{j,j+1}(u) \big|_{u = 0}
	\end{equation}
	along with 
	\begin{equation}\label{key}
		\cl{H}_{j,j+1} = i P_{j,j+1} R'_{j,j+1}(0)
	\end{equation}
	Now we unify the formulation from spaces $ 1 $ and and $ 2 $ to some $ k $ and $ k + 1 $.
	\begin{equation}\label{Apx1_HLL_1}
		i \left[ \cl{L}_{k,0} \cl{L}_{k+1,0}, \cl{H}_{k,k+1} \right] = \cl{L}_{k,0}\cl{L}'_{k+1,0} - \cl{L}_{k,0}'\cl{L}_{k+1,0}
	\end{equation}
	and multiply by monodromic complements from left and right, so that it becomes
	\begin{align}\label{key}
		& i \prod_{j=1}^{k-1}\cl{L}_{0,j} \left[ \cl{L}_{0,k} \cl{L}_{0,k+1}, \cl{H}_{k,k+1} \right] \prod_{n=k+2}^{L}\cl{L}_{0,n} = \\ 
		& \prod_{j=1}^{k-1}\cl{L}_{0,j}\cl{L}_{0,k}\cl{L}'_{0,k+1} \prod_{n=k+2}^{L}\cl{L}_{0,n} - \prod_{j=1}^{k-1}\cl{L}_{0,j}\cl{L}_{0,k}'\cl{L}_{0,k+1} \prod_{n=k+2}^{L}\cl{L}_{0,n}
	\end{align}
	which after proceeding with monodromy arrangement
	\begin{equation}\label{Apx1_B_RLL_1}
		i \left[ \prod_{j=1}^{L}\cl{L}_{0,j}, \cl{H}_{k,k+1} \right] = \prod_{j=1	}^{k}\cl{L}_{0,j} \cl{L}'_{0,k+1} \prod_{n=k+2}^{L}\cl{L}_{0,n} - \prod_{j=1}^{k-1}\cl{L}_{0,j}\cl{L}_{0,k}'\prod_{n=k+1}^{L}\cl{L}_{0,n}
	\end{equation}
	if \eqref{Apx1_B_RLL_1} is now multiplied by $ k $ and summed as $ \sum_{k=1}^{L} $, one obtains
	\begin{equation}\label{Apx1_B_RLL_2}
		i \left[ \prod_{j=1}^{L}\cl{L}_{0,j}, \sum_{k=1}^{L}k \cl{H}_{k,k+1} \right] = \sum_{k=1}^{L}k \prod_{j=1}^{k}\cl{L}_{0,j} \cl{L}'_{0,k+1} \prod_{n=k+2}^{L}\cl{L}_{0,n} - \sum_{k=1}^{L}k \prod_{j=1}^{k-1}\cl{L}_{0,j}\cl{L}_{0,k}'\prod_{n=k+1}^{L}\cl{L}_{0,n}
	\end{equation}
	Recalling $ \cl{L} $-monodromy definition and analysing intertwined structure on the right hand side of \eqref{Apx1_B_RLL_2}
	\begin{align}\label{Apx1_B_RLL_3}
		i \left[ \sum_{k=1}^{L}k \cl{H}_{k,k+1}, T(u) \right] & = \sum_{k=1}^{L}k \prod_{j=1}^{k-1}\cl{L}_{0,j}\cl{L}_{0,k}'\prod_{n=k+1}^{L}\cl{L}_{0,n} - \sum_{k=1}^{L}k \prod_{j=1}^{k}\cl{L}_{0,j} \cl{L}'_{0,k+1} \prod_{n=k+2}^{L}\cl{L}_{0,n} \\ 
		& = \dd_{\,\,u}T(u) + 0 \cdot \cl{L}_{0,1}'\prod_{j=2}^{L}\cl{L}_{0,j} - L \cdot \prod_{j=1}^{L}\cl{L}_{0,j} \cl{L}'_{0,L+1} \label{Apx1_RLL_2_2}
	\end{align}
	with monodromy $ T(u) $ and the boundary terms present in the last line \eqref{Apx1_RLL_2_2}, are subject to our choice of initial ($ 1 $) and final ($ L $) site of the spin chain. By considering consistent infinite limit, \textit{i.e.} shifting boundaries to $ -\infty \leftarrow 1 \dots j \dots L \rightarrow +\infty $, one obtains
	\begin{equation}\label{Apx1_B_RLL_4}
		i \left[ \sum_{k= -\infty}^{+ \infty}k \cl{H}_{k,k+1}, T(u) \right] = \dd_{\,\,u}T(u) + 0_{\rightarrow -\infty} \cdot \cancelto{0}{\cl{L}_{0,1}'\prod_{j=2}^{L}\cl{L}_{0,j}} - L_{\rightarrow +\infty} \cdot \cancelto{0}{\prod_{j=1}^{L}\cl{L}_{0,j} \cl{L}'_{0,L+1}}
	\end{equation}
	as expected the boundary terms vanish and we left with boost recursion formula.
	\begin{equation}\label{Apx1_B_Recursion_T}
		i \left[ \cl{B}, T \right] = \dot{T} \qquad \cl{B} \equiv \sum_{k= -\infty}^{+ \infty}k \cl{H}_{k,k+1}
	\end{equation}
	or
	\begin{equation}\label{Apx1_B_Recursion_t}
		i \left[ \cl{B}, t \right] = \dot{t}
	\end{equation}
	where  $ \dot{T} \equiv \dd_{\,\, u} T(u) $ and \eqref{Apx1_B_Recursion_t} corresponds to a transfer matrix $ t(u) $ analogue. It can be noted, that \eqref{Apx1_B_Recursion_T} constitutes nothing but a discrete form\footnote{Importantly, that the boost for the systems on lattice is associated to the Drinfeld boost automorphism of $ \cl{Y}\left[ \fk{g} \right] $ \cite{Drinfeld:1985rx,Drinfeld:1987sy} } \cite{Loebbert:2016cdm} of the field theoretic boost symmetry \cite{Tetelman} with a standard discretisation scheme $ \int x \dd \, x \mapsto \sum_{k} k $.
	
	\paragraph{Non-difference generalisation} One can consider a generalisation of the boost symmetry \ref{Apx1_B_Recursion_t} to the case, when scattering describing operators $ S $-/$ R $-matrix possess an arbitrary spectral dependence form \ref{NA_S_SD}. In order to achieve that, one can perform analysis with steps close to the above and deriving the corresponding boost expression. More specifically, we need to recall the coupled differential structure \eqref{SutherlandEq_1}, which in turn was derived by exploiting differentiation of the $ RTT $ and qYBE
	\begin{equation}\label{Apx1_R_DS}
			\left[ R_{13} R_{12}, \cl{H}_{23}(v) \right] = R_{13}R_{12, v} - R_{13, v}R_{12} \qquad R_{ij} \equiv R_{ij}(u,v)
	\end{equation}
	which can now be proposed to generic expression
	\begin{equation}\label{key}
		\left[ R_{a,k+1} R_{a,k}, \cl{H}_{k,k+1} \right] = R_{a,k+1}R'_{a,k} - R'_{a,k+1}R_{a,k}
	\end{equation}
	by the shifts $ 1 \rightarrow a $ (auxiliary), $ 2 \rightarrow k $, $ 3 \rightarrow k + 1 $. We then can analyse $ R $-monodromies with appropriate product extensions
	\begin{align}\label{key}
		& \prod_{j=L}^{k+2}R_{a,j} \left[ R_{a,k+1} R_{a,k}, \cl{H}_{k,k+1} \right] \prod_{n=k-1}^{1}R_{a,n} = \\ & \prod_{j=L}^{k+2}R_{a,j} R_{a,k+1}R'_{a,k}\prod_{n=k-1}^{1}R_{a,n} - \prod_{j=L}^{k+2}R_{a,j} R'_{a,k+1}R_{a,k}\prod_{n=k-1}^{1}R_{a,n}
	\end{align}
	which analogously provides after arranging products
	\begin{equation}\label{key}
		\left[ \prod_{j=L}^{1} R_{a,j}, \cl{H}_{k,k+1} \right] = \prod_{j=L}^{k+1}R_{a,j} R'_{a,k} \prod_{n=k-1}^{1}R_{a,n} - \prod_{j=L}^{k+2}R_{a,j} R'_{a,k+1} \prod_{n=k}^{1}R_{a,n}
	\end{equation}
	It can be immediately noticed, that there are internal mutual cancellations analogous to \eqref{Apx1_B_RLL_3}. Let us multiply by $ k $ and sum towards the boost
	\begin{equation}\label{key}
		\left[ \prod_{j=L}^{1} R_{a,j}, \sum_{k} k \cl{H}_{k,k+1} \right] = \sum_{k} k \prod_{j=L}^{k+1}R_{a,j} R'_{a,k} \prod_{n=k-1}^{1}R_{a,n} - \sum_{k} k \prod_{j=L}^{k+2}R_{a,j} R'_{a,k+1} \prod_{n=k}^{1}R_{a,n}
	\end{equation}
	where now one can take the infinite limit $ -\infty \leftarrow 1,L \rightarrow +\infty $ and accordingly assuming sums, we obtain
	\begin{equation}\label{key}
		\left[ T_{a}, \sum_{k=-\infty}^{+\infty} k \cl{H}_{k,k+1} \right] = \dd_{\,\, v} T_{a} + \underbrace{ 1 \cdot \prod_{j=L}^{2} R_{a,j} R'_{a,1} - L \cdot R'_{a,L+1} \prod_{n=L}^{1} R_{a,n} }_{ \lim \limits_{\{ 1,L \} \rightarrow \mp \infty}  R'_{a,L+1} \prod_{n=L}^{1}R_{a,n} \rightarrow L \cdot \prod_{j=L}^{2}R_{a,j} R'_{a,1} \,\, \Rightarrow 0}
	\end{equation}
	One can observe that due to closed spin chain periodicity there will be boundary terms that will be become commutative (boundary factors can swap, \textit{i.e.} $ L+1 \mapsto 1 $), however strictly speaking their absence is also guaranteed at infinity. For the study of finite spin chains and local short-range charges it establishes consistent unified boost automorphism in the form
	\begin{equation}\label{key}
		\left[ T_{a}, \cl{B} \right] = \dot{T}_{a} \quad\qquad T_{a} \equiv T_{a}(u,v), \quad \cl{H}_{k,k+1} \equiv \cl{H}_{k,k+1}(v)
	\end{equation}
	where for transfer matrix $ t $ we obtain
	\begin{equation}\label{Apx1_B_ND_1}
		\left[ t, \sum_{k=-\infty}^{+\infty} k \cl{H}_{k,k+1} \right] = \dot{t}
	\end{equation}
	where $ \dot{t} \equiv \dd_{\,\, v} t(u,v) $. By recalling perturbation of the $ t $ \eqref{I_Qr}
	\begin{equation}\label{Apx1_ConjectureE_Exp}
		\log t(u,v) = \bb{Q}_{1} + \bb{Q}_{2} (u-v) + \dfrac{\bb{Q}_{3}}{2} (u-v)^{2} + \cl{O}\left[ (u-v)^{3} \right]
	\end{equation}
	with $ \bb{Q}_{i} \equiv \bb{Q}_{i}(v) $. One is required to take into account the existence of special spectral points and expansion. From the automorphic structure of the boost, one can derive from \eqref{Apx1_B_ND_1} and \eqref{Apx1_ConjectureE_Exp} that at each level of the recursion the following contributions arise
	\begin{equation}\label{Apx1_B_ND_2}
		 \bb{Q}_{r+1} = \left[ \cl{B}, \bb{Q}_{r}(v) \right] + \alpha_{1} \partial_{v}\bb{Q}(v) \qquad r \geq 2
	\end{equation}
	where $ r=1 $ is fixed, but \eqref{Apx1_B_ND_2} is applicable for it as well. Important to note, that it is possible to construct a long-range counter with the notion of bi-local charges \cite{Loebbert:2016cdm}, which allow to generate dressing contributions to operators (\textit{e.g.} dilatation of $ \cl{N} = 4 $ SYM \cite{Bargheer:2008jt, Bargheer:2009xy}) and derive long-range deformation of the charges of the associated spin chains, including Hamiltonians $ \cl{Q}_{2} = \cl{H} $.
	
	\newpage
	\subsection{Elliptic Functions and properties}\label{Apx_Elliptic_Apparatus}
	We describe here the apparatus of \textbf{J}acobi \textbf{E}lliptic \textbf{F}unctions \cite{Abramowitz_1988_Handbook}, their structure and properties, that have used in the derivations for 6vB/8vB $ AdS_{\{ 2,3 \}} $ deformations. In Sec. \ref{NA_S_SD}, where coupled differential systems involved substitutions and derivation of the solution in terms of JEF, Sec. \ref{FF_Section} it was necessary for the proof of free fermionic property and other, or transformation and limits in the proofs of Sec. \ref{p5_Sec_AdS_2_3_Integrable_Deformations}.
	
	\paragraph{Elliptic integrals} A meromorphic function that is double periodic is called an \textit{elliptic function}. An elliptic integral is any integral of the class
	\begin{equation}\label{key}
		\int\dfrac{A + B}{C + D\sqrt{S}} \dd \, x
	\end{equation}
	where $ A,B,C,D $ are polynomial in x, but $ S $ is a polynomial of x of degree three or four. Class of elliptic integrals can be taken as a generalisation of the inverse trigonometric functions.
	
	The first kind elliptic integral can be formulated as an incomplete Legendre elliptic integral
	\begin{equation}\label{key}
		F(\phi, k) = \int_{0}^{\phi} \dd \, \omega \dfrac{1}{\sqrt{1 - k^{2} \sin^{2}\omega}} \qquad
		\begin{array}{l}
			0 \leq k^{2} \leq 1 \\ [2ex]
			0 \leq \phi \leq \frac{\pi}{2}
		\end{array}
	\end{equation}
	where $ k $ is \textit{elliptic modulus} and at $ \phi = \pi/2 $ it is called complete. Analogously the second kind elliptic integral
		\begin{equation}\label{key}
		E(\phi, k) = \int_{0}^{\phi} \dd \, \omega \sqrt{1 - k^{2} \sin^{2}\omega}
	\end{equation}

	In fact it useful to define the quarter periods \cite{Abramowitz_1988_Handbook} through the 1st elliptic integral
	\begin{equation}\label{key}
		\begin{cases}
			K(k) = \int_{0}^{\pi/2} \dd \, \omega \left(1 - k^{2} \sin^{2} \omega\right)^{-\frac{1}{2}}  \\
			i K'(k) = i \int_{0}^{\pi/2} \dd \, \omega \left( 1 - k'^{2} \sin^{2} \omega \right)^{-\frac{1}{2}}
		\end{cases}
	\end{equation}
	\begin{equation}\label{key}
		\tx{along with} \qquad k + k' = 1, \quad K(k) = K'(k') = K'(1 - k)
	\end{equation}
	where $ K,K' \in \bb{R} $ are complete integrals, which correspondingly form real $ K $ and $ i K' $ imaginary periods.

	\paragraph{JEF in $ pq $-notation} One can identify the vertices of the rectangle as $ 0 $, $ K $, $ K + i K' $ and $ i K' $ by s, c, d, n (Argand diagram). The corresponding translations within the rectangle by $ \alpha K $ and $ i \beta K' $ (with $ \alpha, \beta \lessgtr 0 $) will induce
	\begin{table}[h]
		\begin{center}
			\begin{tabular}{c | c | c | c }
				s & c & s & c \\ \hline
				n & d & n & d \\ \hline
				s & c & s & c \\ \hline
				n & d & n & d \\
			\end{tabular}
		\end{center}
		\label{JEF_Argand_Diagram}
	\end{table}
	where any $ pq \in \{ \tx{s, c, d, n} \} $ define a Jacobi elliptic function $ pq \, u $ with the following properties
	\begin{itemize}
		\item $ pq \, u $ possesses a simple zero $ p $ and simple pole at $ q $
		\item $ p \rightarrow q $-step is half period of $ pq \, u $.
		\item The leading coefficient of $ pq \, u $ expansion in $ u $ in the vicinity of zero is unity (the leading term will appear to be $ u $, $ u^{-1} $ or $ 1 $ iff $ u = 0 $ corresponds to a zero, a pole or an ordinary point).
	\end{itemize}
	The JEF w.r.t to the 1st kind integral appear as 
	\begin{equation}\label{key}
		u = \int_{0}^{\phi} \dd \, \omega \left( 1 - k^{2} \sin^{2} \omega \right)^{-\frac{1}{2}}
	\end{equation}
	where the $ \phi = \tx{am} u $ is an \textit{amplitude}, then conventional definition includes
	\begin{equation}\label{key}
		\begin{array}{l}
			\sn{u} = \sin \phi \\ [2ex]
			\cn{u} = \cos \phi \\ [2ex]
			\dn{u} = (1 - k^{2} \sin^{2}\phi)^{\frac{1}{2}} \\ [2ex]
		\end{array}
	\end{equation}
	in analogy to the above all JEF can expressed through $ \phi $. Moreover the rest of the $ pq \, u $ functions can be defined
	\begin{equation}\label{key}
		pq \, u = \dfrac{pr \, u}{qr u}
	\end{equation}
	with $ pqr \in \{ \tx{s, c, d, n} \} $ and for coinciding letters function is unity. The properties of the squares can be given by
	\begin{equation}\label{key}
		\begin{array}{l}
		\tx{sn}^{2} + \tx{cn}^{2} = 1 \\ [2ex] 
		k^{2} \tx{sn}^{2} + \tx{dn}^{2} = 1 \\ [2ex] 
		\end{array}
		\qquad
		\begin{array}{l}
			k^{2} \tx{cn}^{2} + \tx{k'}^{2} = \tx{dn}^{2} \\ [2ex] 
			\tx{cn}^{2} + \tx{k'}^{2} \tx{sn}^{2} = \tx{dn}^{2} \\ [2ex] 
		\end{array}
	\end{equation}
	where $ \tx{xn} \equiv \tx{xn} u $	The properties of the shifted periods
	\begin{equation}\label{key}
		\begin{array}{l}
			\tx{sn}(u+K) = \dfrac{\cn{u}}{\dn{u}} \\ [2ex] 
			\tx{cn}(u+K) = - \dfrac{k' \sn{u}}{\dn{u}} \\ [2ex] 
			\tx{dn}(u+K) = \dfrac{k'}{\dn{u}} \\ [2ex] 
		\end{array}
		\qquad
		\begin{array}{l}
			\tx{sn}(u+2K) = -\sn{u} \\ [2ex] 
			\tx{cn}(u+2K) = -\cn{u} \\ [2ex] 
			\tx{dn}(u+2K) = \dn{u} \\ [2ex] 
		\end{array}
	\end{equation}
	along with half and double period expressions
	\begin{equation}\label{key}
		\begin{array}{l}
			\sn{\frac{1}{2} K} = \left( 1+k' \right)^{-\frac{1}{2}} \\ [2ex] 
			\cn{\frac{1}{2} K} = \left( \dfrac{k'}{1+k'} \right)^{\frac{1}{2}} \\ [2ex] 
			\dn{\frac{1}{2} K} = k'^{\frac{1}{2}} \\ [2ex] 
		\end{array}
		\qquad
		\begin{array}{l}
			\tx{sn}(2u) = \dfrac{2 \, \sn{u}\cn{u}\dn{u}}{1- k^{2}\sn(u)^{4}} \\ [2ex] 
			\tx{cn}(2u) = \dfrac{1 - 2 \, \sn{u}^{2} + k^{2}\sn(u)^{4}}{1- k^{2}\sn(u)^{4}} \\ [2ex] 
			\tx{dn}(2u) = \dfrac{1 - 2 \, k^{2} \sn{u}^{2} + k^{2}\sn(u)^{4}}{1- k^{2}\sn(u)^{4}} \\ [2ex] 
		\end{array}
	\end{equation}

	\newpage
	\subsection{6v computation}\label{ND_6vComputation}
	\paragraph{Differential system}In order for the resolution to appear more clear, here we can bring an example of a computation carried in the 6-vertex sector with some spectral dependence. For the illustrative reasons, one can take the Hamiltonian density $ \cl{H}_{\tx{6v}} $ of the form
	\begin{equation}\label{Apx_H6v_Ansatz}
		\cl{H}_{\tx{6v}} = 
		\begin{pmatrix}
			0 & 0 & 0 & 0 \\
			0 & h_1 & h_3 & 0 \\
			0 & h_4 & h_2 & 0 \\
			0 & 0 & 0 & 0 
		\end{pmatrix}
	\end{equation}
	whose $ \cl{B} $oost-generated local charge density associated to length 3 charge $ \bb{Q}_{3} $ appears as follows 
	\begin{equation}\label{key}
		\cl{Q}_{ijk} = 
		\begin{pmatrix}
			0 & 0 & 0 & 0 & 0 & 0 & 0 & 0 \\
			0 & 0 & -h_1 h_3 & 0 & -h_3^2 & 0 & 0 & 0 \\
			0 & h_1 h_4 & \dot{h}_1 & 0 & \dot{h}_3-h_2 h_3 & 0 & 0 & 0 \\
			0 & 0 & 0 & \dot{h}_1 & 0 & \dot{h}_3+h_1 h_3 & h_3^2 & 0 \\
			0 & h_4^2 & \dot{h}_4+h_2 h_4 & 0 & \dot{h}_2 & 0 & 0 & 0 \\
			0 & 0 & 0 & \dot{h}_4-h_1 h_4 & 0 & \dot{h}_2 & h_2 h_3 & 0 \\
			0 & 0 & 0 & -h_4^2 & 0 & -h_2 h_4 & 0 & 0 \\
			0 & 0 & 0 & 0 & 0 & 0 & 0 & 0 
		\end{pmatrix}
	\end{equation}
	where from now on the following denotions will be useful to implement
	\begin{equation}\label{key}
		\begin{array}{l}
			h_{i} \equiv h_{i}(w) \in \cl{H} \equiv \cl{Q}_{2}, \quad \dot{h}_{i} = \partial_{w} h_{i}(w) \\ [2ex] 
			r_{i} \equiv r_{i}(u,v) \in R_{ij}, \quad \dot{r}_{i} = \partial_{u} r_{i}(u,v), \quad r'_{i} = \partial_{v} r_{i}(u,v) \\ [2ex] 
			H_{i}(u) \equiv \int_{0}^{u} \dd \, s h_{i}(s), \quad H_{i}(u,v) \equiv H_{i}(u) - H_{i}(v) = \int_{v}^{u} \dd \, s h_{i}(s)
		\end{array}
	\end{equation}
	One can now demand commutativity $ \left[ \bb{Q}_{2}, \bb{Q_{3}} \right] = 0 $, which establishes decomposed ODE
	\begin{equation}\label{key}
		\begin{cases}
			\dot{h}_3 (h_1+h_2) = (\dot{h}_1+\dot{h}_2) h_3 \\
			\dot{h}_4 (h_1+h_2) = (\dot{h}_1+\dot{h}_2) h_4
		\end{cases}
	\end{equation}
	from where one straightforwardly obtains
	\begin{equation}\label{key}
		\begin{cases}
			h_3 = \frac{c_{3}}{2} (h_1+h_2) \\
			h_4 = \frac{c_{4}}{2} (h_1+h_2)
		\end{cases}
	\end{equation}
	with $ c_{i} \in \tx{const} $ and fixes the density $ \cl{H} $ to
	\begin{equation}\label{Apx_H6v}
		\cl{H}_{\tx{6v}} = 
		\begin{pmatrix}
			0 & 0 & 0 & 0 \\
			0 & h_1 & \frac{c_3}{2} (h_1+h_2) & 0 \\
			0 & \frac{c_4}{2} (h_1+h_2) & h_2 & 0 \\
			0 & 0 & 0 & 0
		\end{pmatrix}
	\end{equation}
	Consequently the Ansatz for the {\Rx} can be established as
	\begin{equation}\label{key}
		R_{\tx{6v}} = 
		\begin{pmatrix}
			r_1 & 0 & 0 & 0 \\
			0 & r_2 & r_3 & 0 \\
			0 & r_4 & r_5 & 0 \\
			0 & 0 & 0 & r_6
		\end{pmatrix}
	\end{equation}
	According coupled differential system developed in \ref{NA_S_SD}, which involves Sutherland type structure, leads to the system to be resolved for $ r_{i} $
	\begin{equation}\label{Apx_RSystem_6v}
		\begin{array}{l}
			c_3 r_2 r_6 = c_4 r_1r_5 \\ [2ex] 
			\dfrac{\dot{r}_1}{r_1} = \dfrac{\dot{r}_6}{r_6} \\ [2ex] 
			\dfrac{\dot{r}_2}{r_2} = \dfrac{\dot{r}_5}{r_5} \\ [2ex] 
		\end{array}
		\qquad\qquad
		\begin{array}{l}
			\dfrac{\dot{r}_2}{r_2} - \dfrac{\dot{r}_3}{r_3} = h_1+h_3\dfrac{r_6}{r_5}\\ [2ex] 
			\dfrac{\dot{r}_4}{r_4} - \dfrac{\dot{r}_3}{r_3} = h_1-h_2\\ [2ex] 
			\dfrac{\dot{r}_3}{r_3} - \dfrac{\dot{r}_1}{r_1} = h_2+h_3\dfrac{r_2}{r_1}\\ [2ex] 
			\dfrac{r_4 r_3}{r_1r_5}- \dfrac{r_6}{r_5}-\dfrac{r_2}{r_1} = \dfrac{2}{c_{3}} \\ [2ex]
		\end{array}
	\end{equation}
	We shall then apply identification transforms, which will make the comparison and embedding transparent. From the \eqref{Apx_RSystem_6v}, one can immediately fix
	\begin{equation}\label{key}
		r_6 = A r_1, \quad r_5 = B r_2, \quad A c_3 = B c_4 
	\end{equation}
	where the first two induced first condition of \eqref{Apx_RSystem_6v}. The regularity of {\Rx}, sets
	\begin{equation}\label{key}
		A = 1, \qquad B = \dfrac{c_{3}}{c_{4}}, \qquad \tx{under } R(u,u) \equiv \mathds{1}
	\end{equation}
	next one can resolve for $ r_{4} $
	\begin{equation}\label{key}
		r_{4} = r_{3} \exp\left[ H_{1}(u,v) - H_{2}(u,v) \right] 
	\end{equation}
	where $ H_{i}(u,v) $ indicated above. The remaining three equations follow
	\begin{equation}\label{Apx_6v_R2}
		\begin{array}{l}
			\dfrac{\dot{r}_2}{r_2} - \dfrac{\dot{r}_3}{r_3} = h_1+h_3\dfrac{r_6}{r_5} \\ [2ex] 
			\dfrac{\dot{r}_3}{r_3} - \dfrac{\dot{r}_1}{r_1} = h_2+h_3\dfrac{r_2}{r_1} \\ [2ex] 
			\dfrac{r_4 r_3}{r_1r_5} - \dfrac{r_1}{r_5}-\dfrac{r_2}{r_1} = \dfrac{2}{c_{3}} \\ [2ex] 
		\end{array}
	\end{equation}
	it is then useful to remap $ r_{i} $ to
	\begin{equation}\label{key}
		r_{1} \mapsto r_3 \left( \tilde{r}_{1} - \frac{\tilde{r}_2}{c_4} \right), \qquad
		r_{2} \mapsto r_3 \tilde{r}_{2}, \qquad
		r_3 \mapsto r_3 
	\end{equation}
	so that the algebraic equation of \eqref{Apx_6v_R2} becomes
	\begin{equation}\label{key}
		c_{4}^{2} e^{H_1-H_2}=c^2_4 \tilde{r}_1^2 +\omega^2\tilde{r}_2^2
	\end{equation}
	with $ H_{i} = H_{i}(u,v) $ and $ \omega^{2} = c_{3}c_{4} - 1 $, which can be further resolved in ellipsoidal coordinates by
	\begin{equation}\label{Apx_6v_R3}
		\begin{cases}
			\tilde{r}_1 =  e^{\frac{H_1-H_2}{2}} \cos \phi \\
			\tilde{r}_2 =  e^{\frac{H_1-H_2}{2}} \frac{c_4}{\omega} \sin \phi
		\end{cases}
	\end{equation}
	where the emerging $ \phi $ can be resolved from the two differential equations (as conventionally established, the system appears to be overdetermined). So substituting \eqref{Apx_6v_R3} into the first two equations of \eqref{Apx_6v_R2}, one derives
	\begin{equation}\label{Apx_6v_R4}
		\begin{cases}
			\phi(u,v)\big|_{v \rightarrow u} = 0 \\ 
			\dot{\phi} = \dfrac{\omega}{2} \left( h_{1} + h_{2} \right) 
		\end{cases}
	\end{equation}
	where the first constraint of \eqref{Apx_6v_R4} constitutes an appropriate initial condition. Hence resolving the last we can perform all differential consistency checks and acquire for the {\Rx}
	\begin{equation}\label{Apx_R6v}
		R_{\tx{6v}} = e^{H_{+}}
		\begin{pmatrix}
			\cos \omega H_+ -  \frac{\sin \omega H_+}{\omega} & 0 & 0 \\
			0 & c_4\frac{\sin \omega H_+}{\omega}  &  e^{-H_-} & 0 \\
			0 &  e^{H_-} & c_3 \frac{\sin \omega H_+}{\omega} & 0 \\
			0 & 0 & 0 & \cos \omega H_+ - \frac{\sin \omega H_+}{\omega}
		\end{pmatrix}
	\end{equation}
	where we have defined $ H_{\pm}(u,v) \equiv \frac{H_{1}(u,v) \pm H_{2}(u,v)}{2}$ and fixed normalisation of $ R $ based on $ r_{3} $. We can notice that due to the spectral functions $ H_{\pm} $ it is obvious that the spectral dependence is decomposed, however functional form is undetermined, which qualifies it to a separate class. It can be immediately checked that \eqref{Apx_R6v} satisfies qYBE, all quantum consistencies and reproduces the Hamiltonian density \eqref{Apx_H6v}.
	
	\paragraph{Resolution by identification transformations}
	It can be demonstrated that the case resolved above by means of coupled Sutherland type system, can be obtained faster by implementing identification transformations \ref{ND_IT}. More specifically, first one can apply LBT to \eqref{Apx_H6v_Ansatz} and identify $ h_{1} = h_{2} $, it is then $ \cl{V} $ defined through
	\begin{equation}\label{key}
		\cl{V}(w) = \exp \left[ \dfrac{\sigma_{z}}{2} H_{-}(w) \right]
	\end{equation}
	along with $ H_{\pm}(w) = \frac{1}{2}\left[ H_{1}(w) \pm H_{2}(w) \right] $. The reparametrisation freedom will allow $ h_{1} = h_{2} = 1 $ and the corresponding Hamiltonian density prescription will reduce to 
	\begin{equation}\label{key}
		\cl{H}_{\tx{6v'}} =
		\begin{pmatrix}
			0 & 0 & 0 & 0 \\
			0 & 1 & c_3 & 0 \\
			0 & c_4 & 1 & 0 \\
			0 & 0 & 0 & 0 
		\end{pmatrix}
	\end{equation}
	We can further proceed with twisting and identify the constants $ c_{3} = c_{4} = \fk{c} $, in what one can show that the appropriate twist transformation \eqref{ND_HTwist} can be accomplished by 
	\begin{equation}\label{key}
		\cl{W} = \tx{Diag}\left[ \sqrt{c_{3}}, \sqrt{c_{4}} \right] 
	\end{equation}
	which results in
	\begin{equation}\label{key}
		\cl{H}' = 
		\begin{pmatrix}
			0 & 0 & 0 & 0 \\
			0 & 1 & c & 0 \\
			0 & c & 1 & 0 \\
			0 & 0 & 0 & 0 
		\end{pmatrix}
		\qquad
		\cl{H} \xrightarrow[]{\cl{W}} \cl{H}' = \cl{W} \cl{H} \cl{W}^{-1}
	\end{equation}
	Eventually, one can notice that resulting coupled differential system is dramatically reduced and the associated {\Rx} can eb written in difference form, \textit{i.e.} due to $ \cl{H} $ becomes constant the $ R $ is an XXZ solution 
	\begin{equation}\label{Apx_Rp6v}
		R' = e^{u}
		\begin{pmatrix}
			\cos \omega u -\frac{\sin \omega u}{\omega}  & 0 & 0 \\
			0 & c \frac{\sin \omega u}{\omega} &  1 & 0 \\
			0 & 1  & c \frac{\sin \omega u}{\omega} & 0 \\
			0 & 0 & 0 &  \cos \omega u -\frac{\sin \omega u}{\omega}
		\end{pmatrix}
	\end{equation}
	with $ \omega^{2} = \fk{c}^{2} - 1 $. One can immediately ask, on how \eqref{Apx_Rp6v} relates to \eqref{Apx_R6v}. First, one is required to backtrack the integrable identification transformations performed above. Namely, one needs to apply counter-twist to apply
	\begin{equation}\label{key}
		\cl{H} \xrightarrow[]{\cl{W}} \cl{H}' = \cl{W}^{-1} \cl{H} \cl{V} \qquad \cl{W} \simeq \cl{V}
	\end{equation}
	for \eqref{Apx_Rp6v} and set back constants $ \fk{c} \mapsto \sqrt{c_{3} c_{4}} $, which establishes the right configuration for the {\Rx}. It is then due to the spectral generalisation one must allow for arbitrary spectral functions $ u \mapsto H_{+}(u) $. The inverse local basis transformation will directly recover the form of \eqref{Apx_R6v}.
	
	\paragraph{(Non-) difference remark} Indeed we see that this unconstrained spectral function dependence is located in the LBT, reparametrisation and twisting \underline{for the given example}, which means one can dress other true difference form of models by this freedoms. However it must be noted that this example of 6-vertex is rather a compact demonstration, since initially it was proven that difference form reduction followed from a non-difference form prescription, but this is \underline{only} a restricted class, and the rest of the models do not possess such reductions and exhibit true arbitrary spectral dependence (such classes can be immediately identified at the Hamiltonian $ \cl{H} $ level, \textit{e.g.} by looking at \eqref{Apx_H6v} it can be concluded that it is of difference form due to a number of freedoms versus integrable transforms). In fact, our method provides unified setting as it stratifies the solution space to contain generic and spectrally reduced dependence. It can be also seen for a number of classes, \textit{e.g.} 8-vertex or in higher dimensions that such reductions are impossible and after combination of transformations \ref{ND_IT} there are left generic spectral functions that cannot be removed. Hence a number of these spectral functions is identified by the corresponding setup, but if one considers further manual constraining of the last one can obtain reductions to known models and many novel interesting limits.


	\newpage
	\subsection{8vB limits}\label{ND_Apx_8vBLim}
	
	Additional nontrivial behaviour could analysed on recovering the \textit{side-diagonal} model from 8vB model \ref{ND_8vBModel}. In this case, it could be noticed that when $ h_{3} \rightarrow -h_{4} $, one obtains from \eqref{ND_8vB_Sys1} that $ h_{\{ 1,2,5,6 \}} = 0 $, $ h_{3} = -h_{4} $ and $ h_{7} = c \, h_{8} $, where $ c $ is some constant.
	
	The recovery of such reduction from 8vB model can be achieved by the following operations
	
	First, one needs to consider a twist, which would potentially bring certain necessary freedom into 8vB side diagonal. In this respect, we can apply accordingly adopted \eqref{ND_RTwist}
	\begin{equation}\label{ND_SDTwist}
		\cl{H} \xRightarrow[]{\{ \fk{W} \}} \cl{H}' = \fk{W} \cl{H} \fk{W}^{-1} \qquad \mathfrak{W} = \cl{W} \otimes \cl{W}
	\end{equation}
	where $ \cl{W} $ could be chosen to be constant
	\begin{equation}\label{key}
		\cl{W} = 
		\begin{pmatrix}
			0 & \fk{b} \\
			\fk{a} & 0
		\end{pmatrix}
	\end{equation}
	it turns out that we need to map these entries in order to satisfy commutation hierarchy, \textit{i.e.} $ \fk{b} = \sigma_{0} \sqrt{c} \, \fk{a} $ with sign freedom $ \sigma_{0} = \{ \pm 1, \pm i \} $, which we shall identify below. It also leads to
	\begin{equation}\label{key}
		\begin{cases}
			h_{3} - h_{4} = \dfrac{\dot{\kappa} \sinh \kappa}{\sqrt{1 + \cosh_{2}\dot{\kappa}}}\\
			\dot{\kappa} = h_{3} + h_{4}
		\end{cases}
	\end{equation}
	where $ \kappa $ is some spectral function and dot indicates derivative w.r.t to spectral parameter.
	
	Secondly we need to apply diagonal LBT with $ \cl{V}(u) $ according to \eqref{ND_LBT} and fix $ (\partial_{s}\cl{V})\cl{V}^{-1} $, so that entries $ \cl{H}_{22} $ and $ \cl{H}_{33} $ are suppressed. One fixes $ \cl{V} $ from differential system on it.
	
	Finally, we obtain the demanded side-diagonal $ \cl{H} $, where $ \cl{H}_{14}/\cl{H}_{41} \sim const $, also $ \cl{H}_{23} $ and $ \cl{H}_{32} $ add up to 0 iff $ \sigma_{0} = \{ \pm i \} $. We can resolve for the {\Rx} from the obtained Hamiltonian and Sutherland system, which results in
	\begin{equation}\label{key}
		\tilde{R}_{sdiag} = 
		\begin{pmatrix}
			\cosh H_3(u,v) & 0 & 0 & \sin H_7(u,v) \\
			0 & -\sinh H_3(u,v)  & \cos  H_7(u,v)  & 0 \\
			0 & \cos H_7(u,v)  & \sinh  H_3(u,v)  & 0 \\
			\sin H_7(u,v) & 0 & 0 & \cosh H_3(u,v) 
		\end{pmatrix}
	\end{equation}
	where spectral pseudo-difference is encoded in $ H_{i}(u,v) = H_{i}(u) - H_{i}(v) $. We can also note that the twist \eqref{ND_SDTwist} is of non-conventional form and it is implicit on how it could implemented directly for $ R_{\text{8vB}} $ along with subsequent algebraic operations. It illustrates one of the instances of the Sutherland system advantage in the setting provided.

	\newpage 
	
	\section{AdS integrable backgrounds}\label{Apx_AdS_Backgrounds}

	\subsection{Sigma models and $ q $-deformed $ R $-matrix}
	
	\subsubsection{$ \al{psu}(1,1\vert2)_{\tx{c.e.}} $ superalgebra}\label{Apx_PSU112_Superalgebra}
	
	We shall discuss the structure of the $ \al{psu}(1,1 \vert 2)_{\tx{c.e.}} $ superalgebra that arises as double copy in {\adsm} and a single for {\adsT} background \cite{Hoare:2014kma}. So the associated $ \al{psu}(1 \vert 1) \times \bb{R}^{3} $ is produced by the 
	\begin{equation}\label{key}
		\{ \fk{O}, \fk{O} \} = 2 \fk{P}, \qquad \{ \fk{S}, \fk{S} \} = 2 \fk{K}, \qquad \{ \fk{O}, \fk{S} \} = 2 \fk{C}
	\end{equation}
	where the bosonic $ \phi $ and fermionic $\psi$ fields transform in representations below
	\begin{equation}\label{Apx_psu112_Representation}
		\begin{array}{l}
			\fk{O}\ket{\phi} = a \ket{\psi} \\ [2ex]
			\fk{O}\ket{\psi} = b \ket{\phi} \\ [2ex]
			\fk{S}\ket{\phi} = c \ket{\psi} \\ [2ex]
			\fk{S}\ket{\psi} = d\ket{\phi} \\ [2ex]
		\end{array}
		\qquad\qquad
		\begin{array}{l}
			\fk{C}\ket{\Phi} = C \ket{\Phi} \\ [2ex]
			\fk{P}\ket{\Phi} = P \ket{\Phi} \\ [2ex]
			\fk{K}\ket{\Phi} = K\ket{\Phi} \\ [2ex]
		\end{array}
	\end{equation}
	with the associated coproducts
	\begin{equation}\label{key}
		\begin{array}{l}
			\Delta(\fk{O}) = \fk{O} \otimes \mathds{1} + \fk{U} \otimes \fk{O}\\ [2ex]
			\Delta(\fk{S}) = \fk{S} \otimes \mathds{1} + \fk{U}^{-1} \otimes \fk{S}\\ [2ex]
		\end{array}
		\qquad\qquad
		\begin{array}{l}
			\Delta(\fk{P}) = \fk{P} \otimes \mathds{1} + \fk{U}^{2} \otimes \fk{P} \\ [2ex]
			\Delta(\fk{C}) = \fk{C} \otimes \mathds{1} + \mathds{1} \otimes \fk{C}\\ [2ex]
			\Delta(\fk{K}) = \fk{K} \otimes \mathds{1} + \fk{U}^{-2} \otimes \fk{K}\\ [2ex]
		\end{array}
	\end{equation}
	where opposite coproduct is $ \Delta^{\tx{op}}(\fk{J}) = P_{g} \Delta(\fk{J}) $, with $ P_{g} $ to be graded permutation operator and $ \{ a,b,c,d,C,P,K \} $ constitute the representation characterising parameters that later become functions on the energy and momentum of the arising states. The superalgebra closure constraint can be unified in
	\begin{equation}\label{Apx_psu112_Closure_Constraints}
		ab = P, \qquad cd = K, \qquad ad + bc = 2C
	\end{equation}
	The provided representation \eqref{Apx_psu112_Representation} constitutes the \textit{long} Kac module. The conjugation on the supercharges
	\begin{equation}\label{key}
		\fk{O}\dagger = \fk{S}, \qquad \fk{P}\dagger = \fk{K}, \qquad \fk{C}\dagger = \fk{C}
	\end{equation}
	induces further restriction of the representation parameters to
	\begin{equation}\label{key}
		a^{*} = d, \qquad b^{*} = c , \qquad C^{*} = C , \qquad P* = K
	\end{equation}
	then the closure \eqref{Apx_psu112_Closure_Constraints} implies
	\begin{equation}\label{key}
		C^{2} = \dfrac{(ad - bc)^{2}}{4} + PK . 
	\end{equation}
	With long representations for scattering for {\adsT} there is no shortening condition and one can identify arbitrarily identify $ m \equiv ad - bc $. After imposing reality conditions and introducing Zhukovsky variables we can derive the dispersion relation in the form
	\begin{equation}\label{key}
		e^{2} = m^{2} + 4 h^{2} \sin^{2}\dfrac{p}{2}
	\end{equation}
	and accordingly rewritten representation parameters appear as
	\begin{equation}\label{key}
		\begin{array}{l}
			a = \dfrac{\alpha \exp\left[ \frac{i p}{4} - \frac{i \pi}{4} \right] }{\sqrt{2}} \sqrt{e + m} \\  [2ex] 
			b = \dfrac{\alpha^{-1} \exp\left[ -\frac{i p}{4} + \frac{i \pi}{4} \right] }{\sqrt{2}} \dfrac{h(1 - \exp\left[i p\right])}{\sqrt{e + m}} \\  [2ex]  
		\end{array}
		\qquad\qquad
		\begin{array}{l}
			c = \dfrac{\alpha \exp\left[ \frac{i p}{4} - \frac{i \pi}{4} \right] }{\sqrt{2}} \dfrac{h(1 - \exp\left[i p\right])}{\sqrt{e + m}} \\  [2ex] 
			d = \dfrac{\alpha^{-1} \exp\left[ -\frac{i p}{4} + \frac{i \pi}{4} \right] }{\sqrt{2}} \sqrt{e + m}
		\end{array}
	\end{equation}
	the $ \alpha $ parameter can be set to $ 1 $ for light-cone gauge-fixed string backgrounds {\ads}, {\adsm} and {\adsT}.
	
	\subsubsection{$ q $-deformed {\adsm}} 
	There are known various deformations and extensions of sigma models and their derived supercoset couterparts. We shall bring core structure of the $ q $-deformed {\Rx} arising in \cite{Hoare:2014oua} and relevant for mapping in Sec. \ref{p5_q_deformed_AdS3}. It was found that deformed $ AdS_{3} $ sigma model 
	\begin{equation}\label{key}
		AdS_{3} \times S^{3} \;: \hspace{10mm} \dfrac{PSU(1,1\vert2) \times PSU(1,1\vert2)}{SO(1,2) \times SO(3)}
	\end{equation}
	studied in \cite{Hoare:2014oua} admits further deformation of the worldsheet sigma model \cite{Klimcik:2008eq,Klimcik_2014,Delduc_2013}. This prescription exploits Metsaev-Tseytlin model \cite{Metsaev:1998it}
	\begin{equation}
		S _{MT} = \int d^{2}x \; \text{STr}\left[ \left( \mathcal{P}_{+}\mathcal{J}_{+} \right) \mathcal{J}_{-} \right]
	\end{equation}
	which comes from the $ S^{3} $ sigma model ($ \al{su}(2) $ PCM), $ \mathcal{P} $ to be projector combinations and $ \mathcal{J} $ \textit{left-invariant} supergroup-valued currents. From here based on Klimcik's bi-Yang-Baxter proposal \cite{Klimcik_2014} one can define 
	\begin{equation}\label{key}
		S = \int d^{2}x \; \text{STr} \left[ \mathcal{J}_{+} \left( \mathcal{P}_{-}^{\eta_{L,R}} \dfrac{1}{1-I_{\eta_{L,R}} R_{f} \mathcal{P}_{-}^{\eta_{L,R}}} \mathcal{J}_{-} \right) \right]
	\end{equation}
	\begin{align}\label{key}
		& R_{f} = \text{Ad}_{f}^{-1} R \text{Ad}_{f}, \qquad I_{\eta_{L,R}} = \dfrac{2}{\sqrt{\left( 1-\eta_{L}^{2} \right) \left( 1-\eta_{R}^{2}	 \right)}}
		\begin{pmatrix}
			\eta_{L}\mathds{1} & 0 \\
			0 & \eta_{R}\mathds{1}
		\end{pmatrix}\\
		& \mathcal{P}_{\pm}^{\eta_{L,R}} = P_{2} \mp \dfrac{\sqrt{\left( 1-\eta_{L}^{2} \right) \left( 1-\eta_{R}^{2}	 \right)}}{2} \left( \mathcal{P}_{1}- \mathcal{P}_{3} \right)
	\end{align}
	where $ \eta_{L,R} $ are deformation parameters. For $ \eta_{L,R} = \eta $ and $ \eta_{L,R} \rightarrow 0 $ these define single-parameter \cite{Delduc_2013} and undeformed cases accordingly (which also will be the source a double deformation). As it was noticed above such realisation leads to $ AdS_{3} $ independent deformation of each copy of the superalgebra central element, i.e.
	\begin{align}
		\{ \mathfrak{O}_{\alpha}, \mathfrak{S}_{\beta} \} = \left[ \mathfrak{C}_{I} \right]_{q_{I}} =\dfrac{\mathfrak{V}_{I}-\mathfrak{V}_{I}^{-1}}{q_{I}-q_{I}^{-1}}, \quad \mathfrak{V}_{I} = q_{I}^{\mathfrak{C}_{I}}
		\qquad
		\begin{cases}
			\alpha = +, \beta = -, I = L \\
			\alpha = -, \beta = +, I = R
		\end{cases}
	\end{align}
	along with action of supergenerators
	\begin{equation}\label{key}
		\begin{array}{l}
			\{ \mathfrak{O}_{+}, \mathfrak{O}_{-} \} = \mathfrak{P} \\ [2ex] 
			\{ \mathfrak{S}_{+}, \mathfrak{S}_{-} \} = \mathfrak{K} \\ [2ex] 
		\end{array}
		\qquad
		\begin{array}{l}
			\{ \mathfrak{O}_{+}, \mathfrak{S}_{-} \} = \mathfrak{C} + \mathfrak{M} = \mathfrak{C}_{L} \\ [2ex] 
			\{ \mathfrak{O}_{-}, \mathfrak{S}_{+} \} = \mathfrak{C} - \mathfrak{M} = \mathfrak{C}_{R}  \\ [2ex] 
		\end{array}
		\qquad
		\begin{array}{l}
			[\mathfrak{B}, \mathfrak{O}_{\pm}] = \pm 2 \i \mathfrak{O}_{\pm} \\ [2ex] 
			[\mathfrak{B}, \mathfrak{S}_{\pm}] = \pm 2 \i \mathfrak{S}_{\pm} \\ [2ex] 
		\end{array}
	\end{equation}
	and coproduct algebra
	\begin{equation}\label{key}
		\begin{array}{l}
			\Delta(\mathfrak{O}_{+}) = \mathfrak{O}_{+} \otimes \mathds{1} + \mathfrak{U} \mathfrak{V}_{L} \otimes \mathfrak{O}_{+} \\ [2ex] 
			\Delta(\mathfrak{O}_{-}) = \mathfrak{O}_{-} \otimes \mathds{1} + \mathfrak{U} \mathfrak{V}_{R} \otimes \mathfrak{O}_{-} \\ [2ex]			
		\end{array}
		\qquad
		\begin{array}{l}
			\Delta(\mathfrak{S}_{+}) = \mathfrak{S}_{\pm} \otimes \mathfrak{V}_{R}^{-1} + \mathfrak{U}^{-1} \otimes \mathfrak{S}_{\pm} \\ [2ex] 
			\Delta(\mathfrak{S}_{-}) = \mathfrak{S}_{\pm} \otimes \mathfrak{V}_{L}^{-1} + \mathfrak{U}^{-1} \otimes \mathfrak{S}_{\pm} \\ [2ex] 
		\end{array}
	\end{equation}
	\begin{equation}\label{key}
		\begin{array}{l}
			\Delta(\mathfrak{P}) = \mathfrak{P} \otimes \mathds{1} + \mathfrak{U}^{2} \mathfrak{V}_{L} \mathfrak{V}_{R} \otimes \mathfrak{P} \\ [2ex] 
			\Delta(\mathfrak{K}) = \mathfrak{K} \otimes \mathfrak{V}_{L}^{-1} \mathfrak{V}_{R}^{-1} + \mathfrak{U}^{-2}  \otimes \mathfrak{K} \\ [2ex] 
		\end{array}
	\end{equation}
	
	The Fundamental R-matrix \cite{Delduc_2013,Hoare:2014oua} defined on universal enveloping algebra
	\begin{equation}\label{key}
		\mathcal{U}_{q} \left( \alg{u}(1) \in \alg{psu}(1\vert1)^{2} \ltimes \alg{u}(1) \ltimes \mathbb{R}^{3} \right)
	\end{equation}
	can be fixed by co-commutativity with the coproduct
	\begin{equation}
		\Delta^{op}(\mathfrak{J}) R = R \Delta(\mathfrak{J}) \quad \Delta^{op}(\mathfrak{J}) = \mathcal{P} \Delta(\mathfrak{J})
	\end{equation}
	It defines its structure through the action in two-particle representation 
	\small
	\begin{align}
		R^{e} \ket{\phi_{\pm}\phi'_{\pm}} & = s_{1}^{e} \ket{\phi_{\pm}\phi'_{\pm}} + q_{1}^{e} \ket{\psi_{\pm}\psi'_{\pm}} \hspace{2mm}
		R^{e} \ket{\psi_{\pm}\psi'_{\pm}} = s_{2}^{e} \ket{\psi_{\pm}\psi'_{\pm}} + q_{2}^{e} \ket{\phi_{\pm}\phi'_{\pm}} \\
		R^{e} \ket{\phi_{\pm}\psi'_{\pm}} & = t_{1}^{e} \ket{\phi_{\pm}\psi'_{\pm}} + r_{1}^{e} \ket{\psi_{\pm}\phi'_{\pm}} \hspace{2mm}
		R^{e} \ket{\psi_{\pm}\phi'_{\pm}} = t_{2}^{e} \ket{\psi_{\pm}\phi'_{\pm}} + r_{2}^{e} \ket{\phi_{\pm}\psi'_{\pm}} \\
		R^{i} \ket{\phi_{\pm}\phi'_{\mp}} & = s_{1}^{i} \ket{\phi_{\pm}\phi'_{\mp}} + q_{1}^{i} \ket{\psi_{\pm}\psi'_{\mp}} \hspace{2mm}
		R^{i} \ket{\psi_{\pm}\psi'_{\mp}} = s_{2}^{i} \ket{\psi_{\pm}\psi'_{\mp}} + q_{2}^{i} \ket{\phi_{\pm}\phi'_{\mp}} \\
		R^{i} \ket{\phi_{\pm}\psi'_{\mp}} & = t_{1}^{i} \ket{\phi_{\pm}\psi'_{\mp}} + r_{1}^{i} \ket{\psi_{\pm}\phi'_{\mp}} \hspace{2mm}
		R^{i} \ket{\psi_{\pm}\phi'_{\mp}} = t_{2}^{i} \ket{\psi_{\pm}\phi'_{\mp}} + r_{2}^{i} \ket{\phi_{\pm}\psi'_{\mp}}
	\end{align}
	where $ e $, $ i $ correspond to equivalent and opposite representations and {\Rx} itself satisfies braiding unitarity, YBE for each sector and crossing constraints.

	\newpage
	\subsection{Free fermion properties: Recursion for $ AdS_{3} $ $ T $-matrix}\label{Apx_FF}
	
	We provide here recursive formulas on transfer matrix $ t_{N} $ that do not exploit auxiliary Bethe system (generic site number). One can notice the special form of $ R_{AdS_{3}} $ with Ramond-Ramond flux 
	\begin{equation}\label{key}
		R_{0i}(\theta_0 - \theta_i) = \cosh \frac{\theta_{0i}}{2} \Big[1-2 \eta_{0i}^\dagger \eta_{0i}\Big] \equiv \cosh \frac{\theta_{0i}}{2} \Big[1-2 N_{0i}\Big]
	\end{equation}
	where we again emphasize that $ \eta_{1}(2), \eta_{1}U\dagger(2) $ are site associated operators \eqref{p4_RRAdS3_R} and $ \eta_{0,i}, \eta_{0,i}^{\dagger} $ indicate that inhomogeneities are associated to sites $ 0 $ and $ i $. One can notice that this fact brings a nontrivial structure of the {\Rx}. The for $ t_{N} $ one can derive
	\begin{align}\label{Apx_FF_tN}
		& t_{N} = \Bigg[\prod_{i=1}^N \cosh \frac{\theta_{0i}}{2} \Bigg] \mbox{str}_0 \Big[1 - 2 N_{01}\Big]... \Big[1 - 2 N_{0N}\Big] = \\ 
		& \Bigg[\prod_{i=1}^N \cosh \frac{\theta_{0i}}{2}\Bigg] \Bigg(\langle 0_0| \Big[1 - 2 N_{01}\Big]... \Big[1 - 2 N_{0N}\Big] |0_0\rangle - \langle 0_0| c_0 \Big[1 - 2 N_{01}\Big]... \Big[1 - 2 N_{0N}\Big] c^\dagger_0|0_0\rangle \Bigg)
	\end{align}
	where $ \ket{0_{0}} $ means a state $ \ket{\phi} $ in auxiliary space $ 0 $. Clearly in order to compute \eqref{Apx_FF_tN} we are required to establish resolution scheme of 
	\begin{equation}\label{Apx_FF_Elements}
		\langle 0_0 | N_{01} ... N_{0m}|0_0\rangle \quad \tx{and} \quad \langle 0_0 | c_0 N_{01} ... N_{0m} c^\dagger_0 |0_0\rangle
	\end{equation}
	with $ m = \overline{1,N} $. In the first step, it can be shown that one derive recursions on the elements above. Hence for the first one in \ref{Apx_FF_Elements} we need to define
	\begin{equation}\label{key}
		X_m \equiv \langle 0_0 | N_{01} ... N_{0m}|0_0\rangle, \qquad Y_m \equiv \langle 0_0 | N_{01} ... N_{0m} c_0^\dagger|0_0\rangle
	\end{equation}
	one can also observe
	\begin{align}
		& \langle 0_0 | N_{01} ... N_{0m}|0_0\rangle = \langle 0_0 | N_{01} ... N_{0,m-1} \eta^\dagger_m \eta_m |0_0\rangle = \\
		& \langle 0_0 | N_{01} ... N_{0,m-1}\eta^\dagger_m (\alpha_m c_0 + \beta_m c_m)|0_0\rangle = \\ 
		& \beta_m \langle 0_0 | N_{01} ... N_{0,m-1}(\alpha_m c_0^\dagger + \beta_m c_m^\dagger)|0_0\rangle c_m= \\
		& \alpha_m \beta_m \langle 0_0 | N_{01} ... N_{0,m-1}c_0^\dagger|0_0\rangle c_m + \beta_m^2 \langle 0_0 | N_{01} ... N_{0,m-1}|0_0\rangle n_m
	\end{align}
	from where one can define
	\begin{equation}\label{key}
		\alpha_i \equiv \cos \alpha_{0i}, \qquad \beta_i \equiv \sin \alpha_{0i}, \qquad \cot 2 \alpha_{0i} = \sinh \frac{\theta_{0i}}{2}
	\end{equation}
	In what a recursion on $ X_{m} $ follows
	\begin{equation}\label{Apx_FF_XR}
		X_m = \alpha_m \beta_m Y_{m-1} c_m + \beta_m^2 X_{m-1} n_m,
	\end{equation}
	similarly by the observations above, but now for $ \langle 0_0 | c_0 N_{01} ... N_{0m} c^\dagger_0 |0_0\rangle $, we can derive intertwined recursion on $ Y_{m} $
	\begin{equation}\label{Apx_FF_YR}
		Y_m = \alpha_m^2 Y_{m-1} + \alpha_m \beta_m X_{m-1} c^\dagger_m + \beta_m^2 Y_{m-1} n_m
	\end{equation}
	It is possible to write resolved recursion in the form
	\begin{equation}\label{key}
		X_m = x_{m-1} \nu_m + \sum_{i=1}^{m-1} X_{i-1} \psi^\dagger_i \xi_{im},\qquad m \geq 1, \qquad X_0=1
	\end{equation} 
	and proceed with \textit{block-molecules} $ \Gamma_{ij} = \psi^\dagger_i \xi_{ij} $ for a given $ m $. However a universal expression can be provided. So that after redefinitions on operators and index algebra we can resolve the coupled recursion system \eqref{Apx_FF_XR}-\eqref{Apx_FF_YR}, which decouples into the following form 
	\begin{equation}\label{Apx_FF_XY_Recursion}
		\begin{aligned}
			X_m =& \prod _{i=0}^{m-1} \Omega _{2,i+1} + \sum _{i=0}^{m-1} \Omega _{1,i+1} \left(\prod _{j=i}^{m-2} \Omega _{2,j+2}\right) \sum _{\mu =0}^{i-1}  \Omega _{3,\mu +1} \left(\prod _{\nu =\mu }^{i-2} \left(\Omega _{4,\nu +2}+\Omega_{2,\nu +2}\right)\right) X_{\mu }, \\
			Y_m =& \sum _{i=0}^{m-1} \Omega _{3,i+1} \left(\prod _{j=i}^{m-2} \left(\Omega _{4,j+2}+\Omega _{2,j+2}\right)\right) \left(\prod _{\alpha =0}^{i-1} \Omega _{2,\alpha +1}+\sum _{\alpha =0}^{i-1}  \Omega _{1,\alpha +1} \left(\prod _{\beta =\alpha }^{i-2} \Omega _{2,\beta +2}\right)	Y_\alpha\right) ,\\
			& \text{for } m \geq 0, \qquad \Omega _{1,q} = \alpha _q \beta _q c_q, \quad \Omega _{2,q} = \beta _q^2 n_q, \quad\Omega _{3,q}=\alpha _q \beta _q c_q{}^{\dagger }, \quad\Omega _{4,q}=\alpha _q^2,
		\end{aligned}
	\end{equation}
	subject to the initial constraints provided in the last line.
	
	We now recall that \eqref{Apx_FF_tN} computation one is left with the elements of the form
	\begin{equation}\label{key}
		Z_m \equiv \langle 0_0 | c_0 N_{01} ... N_{0m} c_0^\dagger|0_0\rangle, \qquad {{} L_m} \equiv \langle 0_0 |c_0 N_{01} ... N_{0m} c_0 c_0^\dagger|0_0\rangle , 
	\end{equation}
	which exhibits analogous structure with extra operators acting on both sides. By analogous prescription we can define closed recursion system on combinations of operators through $ Z_{m} $ and $ L_{m} $
	\begin{equation}\label{key}
		\begin{cases}
			L_{m} = \alpha_m \beta_m Z_{m-1}c_m + \beta_m^2 {{} L_{m-1}} n_m \\ 
			Z_{m} = \alpha_m^2 Z_{m-1} + \alpha_m \beta_m {{} L_{m-1}} c_m^\dagger + \beta_m^2 Z_{m-1}n_m
		\end{cases},
	\end{equation}
	in what now we can solve this system via operatorial set $  \Xi_{i,s} $ with $ i = \overline{1,4} $
	\begin{equation}\label{Apx_FF_ZL_Recursion}
		\begin{aligned}
			{{} L_m} =& \sum _{i=0}^{m-1} \Xi_{4,i+1} \prod _{j=i}^{m-2} \Xi_{3,j+2} \left(\prod _{\alpha =0}^{i-1} \left(\Xi_{1,\alpha
				+1}+\Xi_{3,\alpha +1}\right)+\sum _{\beta =1}^{i-1} \Xi_{2,\beta +1} \prod _{\alpha =\beta }^{i-2} \left(\Xi_{1,\alpha +2}+\Xi_{3,\alpha +2}\right) U_{\beta } \right)\\
			Z_m =& \prod _{i=0}^{m-1} \left(\Xi_{1,i+1}+\Xi_{3,i+1}\right) + \sum _{j=1}^{m-1} \Xi_{2,j+1} \left(\prod _{i=j}^{m-2} \left(\Xi_{1,i+2}+\Xi_{3,i+2}\right)\right) \sum _{\mu =0}^{j-1} \Xi_{4,\mu +1} \left(\prod _{\nu =\mu }^{j-2} \Xi_{3,\nu +2}\right) Z_{\mu},\\
			&  \text{for } m\geq 0 \text{, } \quad \Xi _{1,s}=\alpha _s^2, \quad \Xi _{2,s}=\alpha _s \beta _s c_s{}^{\dagger }, \quad \Xi _{3,s} = \beta _s^2 n_s, \quad \Xi_{4,s} = \alpha _s \beta _s c_s.
		\end{aligned}
	\end{equation}
	The block-molecule form recursion is also applicable here for generic $ m $.
	
	In fact, it appears a closed efficient way of generating $ t_{N} $. Clearly it is still necessary to arrange this operator combinations into certain compact expansion type expression for \eqref{Apx_FF_tN}, which apriori is not obvious. Already at initial values of $ N $ an operator structure develops nontrivially. Hence for instance $ N = 2 $ case provides 
	\begin{equation}\label{Apx_FF_t2}
		t_{2} = \cosh \frac{\theta_{01}}{2} \cosh \frac{\theta_{02}}{2} \, \Big[1 - 2 X_1 - 2 \tilde{X}_1 + 4 X_2 - (1 - 2 Z_1 - 2 \tilde{Z}_1 + 4 Z_2) \Big]
	\end{equation}
	where $ \tilde{X}_{i} = \bra{0_{0}} N_{0i} \ket{0_{0}} $, which from initial operatorial input and \eqref{Apx_FF_XR}-\eqref{Apx_FF_YR}. The $ t_{3} $ develops further
	\begin{align}
		& t_{3} \propto 2 \sum_{i=1}^3 \alpha_i^2 + 4 \Big[\alpha_1 \beta_1 \alpha_2 \beta_2\Big(c_1^\dagger c_2 (1 - 2 \beta_3^2 n_3) - c_1 c_2^\dagger  (1 - 2 \alpha_3^2 - 2 \beta_3^2 n_3)\Big) \\ 
		& + \alpha_1 \beta_1 \alpha_3 \beta_3\Big(c_1^\dagger c_3 (1 - 2 \alpha_2^2 - 2 \beta_2^2 n_2) - c_1 c_3^\dagger  (1 - 2 \beta_2^2 n_2)\Big) \\ 
		& + \alpha_2 \beta_2 \alpha_3 \beta_3\Big(c_2^\dagger c_3 (1 - 2 \beta_1^2 n_1) - c_2 c_3^\dagger  (1 - 2 \alpha_1^2 - 2 \beta_1^2 n_1)\Big)\Big] \\ 
		& -8\Big[\beta_1^2 \beta_2^2 \beta_3^2 n_1 n_2 n_3 - (\alpha_1^2 + \beta_1^2 n_1)(\alpha_2^2 + \beta_2^2 n_2)(\alpha_3^2 + \beta_3^2 n_3)\Big] \\ 
		& - 4\Big[(\alpha_1^2 + \beta_1^2 n_1)(\alpha_2^2 + \beta_2^2 n_2) + (\alpha_1^2 + \beta_1^2 n_1)(\alpha_3^2 + \beta_3^2 n_3) +(\alpha_2^2 + \beta_2^2 n_2)(\alpha_3^2 + \beta_3^2 n_3)\Big]
	\end{align}
	and there is no universal identification in terms of developing $ X $ type combinations as of \eqref{Apx_FF_t2}. Such structure requires either expansion grouping scheme or which is more efficient recursion analytic resolution for the $ 
	t_{N} $ case.

	\newpage
	\subsection{$ AdS_{\{ 2,3 \}} $ crossing symmetry}
	
		\subsubsection{6v $ AdS_{3} $} The crossing symmetry of the models studied in \ref{p5_Sec_AdS_2_3_Integrable_Deformations} can be obtained based on the base crossing condition, where one needs to identify the associated crossing structure including the conjugation operator. It can be proven that for the aforementioned deformations of $ AdS_{3} $ and $ AdS_{2} $ coming from 6vB and 8vB bases do obey the crossing relations. More specifically, the 6vB $ AdS_{3} $ deformation {\Rx} \eqref{p5_R6vB_Full} satisfies crossing through
		\begin{equation}\label{Apx_Crossing_Conditions_AdS3}
			\bb{C}_{i} R(u + \Delta_{\omega, 1}, v + \Delta_{\omega, 2})^{t_{i}} \bb{C}_{i}^{-1} = R(u,v)^{-1}
			\qquad
			\begin{cases}
				i = 1, \quad \Delta_{\omega, 1} = \omega, \, \Delta_{\omega, 2} = 0 \\
				i = 2, \quad \Delta_{\omega, 1} = 0, \, \Delta_{\omega, 2} = -\omega
			\end{cases}
		\end{equation}
		\begin{equation}\label{Apx_COperator_AdS3_1}
			\bb{C}_{AdS_{3}}^{\, \tx{6vB}} = 
			\begin{pmatrix}
				0 & 0 & 1 & 0\\
				0 & 0 & 0 & i\\
				1 & 0 & 0 & 0\\
				0 & i & 0 & 0
			\end{pmatrix}		
		\end{equation}
		where by $ i $ we identify the corresponding vector space, $ t_{i} $ are transpositions in space $ i $ and $ \omega $ is a crossing parameter. The above appears to hold under
		\begin{itemize}
			\item Constraint on $ h_{i}^{\tx{L/R}}(u \pm \omega) $ with $ i = 1,2 $
			\begin{equation}\label{key}
				h_i^{\tx{R}}(u\pm\omega)=-\frac{1}{h_i^{\tx{L}}(u)}, \qquad\qquad h_i^{\tx{L}}(u\pm\omega)=-\frac{1}{h_i^{\tx{R}}(u)}
			\end{equation}
			which implies
			\begin{equation}\label{key}
				h_{i}^{\mrm{x}}(u) = h_{i}^{\mrm{x}}(u \pm 2 \omega)
			\end{equation}
			
			\item Constraint on $ \fk{X} $ and $ \fk{Y} $
			\begin{equation}\label{key}
				\begin{array}{l}
					\fk{X}^{\mrm{x}_{1}}(u\pm 2\omega)=-\fk{X}^{\mrm{x}_{1}}(u) \\ [2ex] 
					\fk{Y}^{\mrm{x}_{1}}(u\pm 2\omega)=-\fk{Y}^{\mrm{x}_{1}}(u) \\ [2ex] 
				\end{array}
				\qquad
				\begin{array}{l}
					\fk{X}^{\tx{R}}(u)=\fk{X}^{\tx{L}}(u+\omega) \\ [2ex] 
					\fk{Y}^{\tx{R}}(u)=\fk{Y}^{\tx{L}}(u+\omega) \\ [2ex] 
				\end{array}
			\end{equation}
		
			\item Constraining the scalar $ \sigma $-factors
			\begin{equation}\label{key}
				\begin{array}{l}
					\sigma^{\mrm{x}_{1}\mrm{x}_{2}}(u+\omega,v)\sigma^{\mrm{x}_{2}\mrm{x}_{2}}(u,v)=\frac{h_2^{\mrm{x}_{2}}(u)-h_1^{\mrm{x}_{2}}(u)}{h_2^{\mrm{x}_{2}}(v)-h_1^{\mrm{x}_{2}}(u)} \\ [2ex] 
					\sigma^{\mrm{x}_{1}\mrm{x}_{1}}(u+\omega,v)\sigma^{\mrm{x}_{2}\mrm{x}_{1}}(u,v)=\frac{h_2^{\mrm{x}_{2}}(u)-h_1^{\mrm{x}_{2}}(u)}{h_2^{\mrm{x}_{2}}(u)}\frac{1+h_1^{\mrm{x}_{1}}(v)h_2^{\mrm{x}_{2}}(u)}{\left(1+h_1^{\mrm{x}_{2}}(u)h_1^{\mrm{x}_{1}}(v)\right)\left(1+h_2^{\mrm{x}_{2}}(u)h_2^{\mrm{x}_{1}}(v)\right)} \\ [2ex] 
					\sigma^{\mrm{x}_{1}\mrm{x}_{1}}(u,v-\omega)=-h_2^{\mrm{x}_{1}}(u)h_2^{\mrm{x}_{2}}(v)\sigma^{\mrm{x}_{2}\mrm{x}_{2}}(u+\omega,v) \\ [2ex] 
					\sigma^{\mrm{x}_{2}\mrm{x}_{1}}(u,v-\omega)=\sigma^{\mrm{x}_{1}\mrm{x}_{2}}(u+\omega,v)
				\end{array}
			\end{equation}
		\end{itemize}
		where $ \mrm{x}_{k} = \{ \tx{L, R} \} $ denotes appropriate chirality with $ k = 1,2 $ and $ \mrm{x}_{1} \neq \mrm{x}_{2} $. According to conventions used, one can establish that \eqref{Apx_COperator_AdS3_1} agrees with \cite{Borsato:2015mma} and is conjugate of \cite{Sfondrini:2014via}, it can be obtained by LBT from the latter.

		\subsubsection{8vB $ \mathbf{AdS_{3}} $} Taking into account the above crossing conditions, it is possible to establish that the last are satisfied when
		\begin{equation}\label{Apx_COperator_AdS3_2}
			\bb{C}_{AdS_{3}}^{\, \tx{8vB}} = 
			\begin{pmatrix}
				0 & 0 & 1 & 0\\
				0 & 0 & 0 & i\\
				1 & 0 & 0 & 0\\
				0 & i & 0 & 0
			\end{pmatrix}
		\end{equation}
		along with the internal functions of \eqref{p5_R8vB_AdS3_1} to satisfy
		\begin{equation}\label{key}
			\begin{array}{l}
				\cl{F}^{\tx{L}}(u+\omega) = \cl{F}^{\tx{L}}(u-\omega) = 2K - \cl{F}^{\tx{R}}(u) \\ [2ex] 
				\cl{F}^{\tx{R}}(u+\omega) = \cl{F}^{\tx{R}}(u-\omega) = 2K - \cl{F}^{\tx{L}}(u) \\ [2ex] 
				\eta(u+\omega) = \eta(u-\omega) = 2\pi-\eta(u) \\ [2ex] 
				\sqrt{\fk{f}_{\tx{R/L}}(u+\omega)}= -\sqrt{\fk{f}_{\tx{R/L}}(u-\omega)} = \sqrt{\fk{f}_{\tx{L/R}}(u)} \\ [2ex] 
			\end{array}
		\end{equation}
		with the 1st kind Elliptic Integral $ K $, whereas for the $ \sigma^{\mrm{x}_{1}\mrm{x}_{2}} $ factors must acquire the following form
		\begin{equation}\label{key}
			\begin{array}{l}
				\sigma^{\tx{LL}}(u+\omega,v)\sigma^{\tx{RL}}(u,v) = i \sqrt{\frac{\fk{f}_L(v)}{\fk{f}_R(u)}} \left[ \text{sn}_{\text{R,L},+}^2 \cos ^2\eta_{-}\Sigma^{2} -\frac{\text{cn}_{\text{R,L},+}^2}{\text{dn}_{\text{R,L},+}^2} \sin^2 \eta_{-}\Sigma^{2} - 1 \right]^{-1}  \\ [2ex]
				\sigma^{\tx{RR}}(u+\omega,v)\sigma^{\tx{LR}}(u,v) = i \sqrt{\frac{\fk{f}_L(u)}{\fk{f}_R(v)}}\left[ \text{sn}_{\text{L,R},+}^2 \cos ^2\eta_{-}\Sigma^{2} -  \frac{\text{cn}_{\text{L,R},+}^2}{\text{dn}_{\text{L,R},+}^2} \sin^2 \eta_{-}\Sigma^{2} - 1 \right]^{-1}  \\ [2ex]
				\sigma^{\tx{LR}}(u+\omega,v)\sigma^{\tx{RR}}(u,v) = i \sqrt{\frac{\fk{f}_R(u)}{\fk{f}_R(v)}} \left[ \text{sn}_{\text{R,R},-}^2 \cos ^2\eta_{-}\Sigma^{2} -  \frac{\text{cn}_{\text{R,R},-}^2}{\text{dn}_{\text{R,R},-}^2} \sin^2 \eta_{-}\Sigma^{2} - 1 \right]^{-1}  \\ [2ex]
				\sigma^{\tx{RL}}(u+\omega,v)\sigma^{\tx{LL}}(u,v) = i \sqrt{\frac{\fk{f}_L(v)}{\fk{f}_L(u)}} \left[ \text{sn}_{\text{L,L},-}^2 \cos ^2\eta_{-}\Sigma^{2} -  \frac{\text{cn}_{\text{L,L},-}^2}{\text{dn}_{\text{L,L},-}^2} \sin^2 \eta_{-}\Sigma^{2} - 1 \right]^{-1} \\ [2ex]
			\end{array}
		\end{equation}
		in parallel with
		\begin{equation}\label{key}
			\begin{array}{l}
				\sigma^{\tx{LL}}(u,v-\omega)=-\sigma^{\tx{RR}}(u+\omega,v) \\ [2ex]
				\sigma^{\tx{RR}}(u,v-\omega)=-\sigma^{\tx{LL}}(u+\omega,v) \\ [2ex]
			\end{array}
			\qquad
			\begin{array}{l}
				\sigma^{\tx{RL}}(u,v-\omega)=\sigma^{\tx{LR}}(u+\omega,v)\frac{\fk{f}_R(v)}{\fk{f}_R(u)} \\ [2ex]
				\sigma^{\tx{LR}}(u,v-\omega)=\sigma^{\tx{RL}}(u+\omega,v)\frac{\fk{f}_L(u)}{\fk{f}_L(v)} \\ [2ex]
			\end{array}
		\end{equation}
		in what one can also confirm that \eqref{Apx_COperator_AdS3_2} is in agreement with \cite{Borsato:2015mma}.

		\subsubsection{8vB $ \mathbf{AdS_{2}} $} For the 8vB as {\adsT} deformation, it is also possible to show under which constraints the crossing symmetry holds. In this case, it requires special attention due to the present deformation parameter $ k $ \ref{p5_8vB}. In contrast to the arguments for $ AdS_{3} $ crossing symmetry, it is useful to consider bosofermion {\Rx} and do not restrict $ k $, in which case we need to appropriately modify \eqref{Apx_Crossing_Conditions_AdS3} to 
		\begin{equation}\label{Apx_Crossing_Conditions_AdS2}
			\bb{C}_{i} R(u + \Delta_{\omega, 1}, v + \Delta_{\omega, 2})^{st_{i}} \bb{C}_{i}^{-1} = R(u,v)^{-1}
			\qquad
			\begin{cases}
				i = 1, \quad \Delta_{\omega, 1} = \omega, \, \Delta_{\omega, 2} = 0 \\
				i = 2, \quad \Delta_{\omega, 1} = 0, \, \Delta_{\omega, 2} = -\omega
			\end{cases}
		\end{equation}
		where in analogy to above $ i = 1,2 $ spans two vector spaces, but with supertransposition $ st_{i} $ accordingly and $ \omega $ crossing parameter. As we have indicated, now the $ R $ appears to be the bosofermion form of \eqref{p5_R8vB_Full}, which can be obtained by remapping $ r_{4} \mapsto -r_{4} $, $ r_{7} \mapsto i r_{7} $ and $ r_{8} \mapsto i r_{8} $ ($ r_{k} \equiv r_{k}(u,v) $). The conjugation operator then takes the form 
		\begin{equation}\label{key}
			\bb{C}_{AdS_{2}}^{\tx{8vB}} = 
			\begin{pmatrix}
				0 & 1 \\
				-i & 0 
			\end{pmatrix}
		\end{equation}
		Important to note that based on conjugation, there also exists a solution with $ i \mapsto -i $. The internal functions must satisfy
		\begin{equation}\label{key}
			\begin{array}{l}
				\eta(u+\omega)=-\eta(u) + 2 \pi n \\ [2ex] 
				\eta(u-\omega)=-\eta(u) + 2\pi m \\ [2ex] 
			\end{array}
			\qquad
			\begin{array}{l}
				\cl{F}(u+\omega)=\cl{F}(u)+2\, n\,K \\ [2ex] 
				\cl{F}(u-\omega)=\cl{F}(u)+2\, m\,K \\ [2ex] 
			\end{array}
		\end{equation}
		where $ m,n \in \bb{Z} $ and the 1st kind elliptic integral $ K(k^{2}) $. For the dressing factors one obtains 
		\begin{equation}\label{Apx_AdS2_Dressing_Factors}
			\sigma(u + \omega, v)\sigma(u, v) =  \sigma(u, v - \omega)\sigma(u, v) = i \left[ \left( \tx{sn} \, \Sigma \cos \eta_{-} \right)^{2} - \left( \dfrac{\tx{cn}}{\tx{dn}} \Sigma \sin \eta_{-} \right)^{2} - 1 \right]^{-1}
		\end{equation}
		It can be noted, that nevertheless one deals with the $ AdS_{2} $ deformation the conjugation operator is distinct from the one found in \cite{Hoare:2014kma}, where it appears to be diagonal. It implies that boso-fermionic sector is brought into the antiparticle counterpart, whereas in our case boson $ \leftrightarrow $ fermion transformation occurs. In addition, both non-deformed \cite{Hoare:2014kma} and deformed (for $ k \rightarrow \infty $) conjugation operators satisfy crossing constraints \eqref{Apx_Crossing_Conditions_AdS2}.
		
		For the $ AdS_{2} $ deformation the boson-boson {\Rx} does satisfy crossing symmetry $ \forall k \textbackslash \{ k \rightarrow 1 \} $, with conditionals of the form
		\begin{equation}\label{key}
			\begin{pmatrix}
				0 & 1 \\
				-1 & 0
			\end{pmatrix}
		\end{equation}
		along with 
		\begin{equation}\label{key}
			\begin{array}{l}
				\eta(u+\omega)=-\eta(u)+\pi n \\ [2ex] 
				\eta(u-\omega)=-\eta(u)-\pi m \\ [2ex] 
			\end{array}
			\qquad
			\begin{array}{l}
				\cl{F}(u+\omega)=\cl{F}(u)+\, n\,K \\ [2ex] 
				\cl{F}(u-\omega)=\cl{F}(u)+\, m\,K \\ [2ex] 
			\end{array}
		\end{equation}
		where $ m $, $ n $ hold to be odd. The dressing factors differ from \eqref{Apx_AdS2_Dressing_Factors} by the common replacement $ i \mapsto 1 $. One can also immediately notice, that in the case of bosofermionic {\Rx} one can suppress the elliptic integral by the admissible $ m = n = 0 $, which is not possible in the boson-boson prescription. Moreover, it would be important to derive an underlying current algebra and corresponding sigma model for the bosofermion {\Rx} associated with $ AdS_{2} $ deformation.
		
	\newpage
	\subsection{$ AdS_{3} \times S^{3} \times \cl{M}^{4} $ string sector}
	Here we explicitly show on how {\adso} \cite{Sfondrini:2014via,Borsato:2013qpa,Borsato:2014hja} and {\adst} \cite{Borsato:2015mma} are required to be rewritten in order to provide compatible comparison analysis with the 6vB/8vB stemming deformations. In the former case, we bring the $ S $- to $ R $-matrices, where all the quantum constraints are explicitly satisfied for both ($ \gamma(p) $ corresponds to $ \eta(p) $ in \cite{Sfondrini:2014via,Borsato:2013qpa,Borsato:2014hja,Borsato:2014exa,Borsato:2015mma}).
	
	\subsubsection{$ AdS_{3} \times S^{3} \times T^{4} $}

	From the $ S $-/$ R $-matrix in \cite{Borsato:2014hja,Sfondrini:2014via} one can work out the block to take the form
	\begin{equation}\label{Apx_R8v_Chi}
		R^{\chi} = \zeta^{\chi}
		\begin{pmatrix}
			r_{1}^{\chi} & 0 & 0 & r_{8}^{\chi} \\
			0 & r_{2}^{\chi} & r_{6}^{\chi} & 0 \\
			0 & r_{5}^{\chi} & r_{3}^{\chi} & 0 \\
			r_{7}^{\chi} & 0 & 0 & r_{4}^{\chi} \\
		\end{pmatrix}
		\qquad
		\begin{array}{l}
			r_{i}^{\chi} \equiv r_{i}^{\chi}(p,q) \\
			\zeta^{\chi} \equiv \zeta^{\chi}(p,q)
		\end{array}
	\end{equation}
	with $\chi$ to correspond to chiral sectors $ \chi = \{ \tx{LL}, \tx{RR}, \tx{LR}, \tx{RL} \} $. We shall now present all the four blocks that assemble the full {\Rx}. The pure blocks (LL ,RR) constitute
	\begin{equation}\label{key}
		\begin{array}{l}
			r_3^{{\rm LL}}=\sqrt{\frac{x^+(p)}{x^-(p)}}\frac{x^-(p)-x^-(q)}{x^+(p)-x^-(q)} \\ [2ex]
			r_4^{{\rm LL}}=-1 \\ [2ex] 
			r_7^{{\rm LL}}=0 \\ [2ex] 
			r_8^{{\rm LL}}=0 \\ [2ex]
		\end{array}
		\quad\qquad
		\begin{array}{l}
			r_1^{{\rm LL}}=-\sqrt{\frac{x^-(q)}{x^+(q)}\frac{x^+(p)}{x^-(p)}} \frac{x^-(p)-x^+(q)}{x^-(q)-x^+(p)} \\ [2ex]
			r_2^{{\rm LL}}=\sqrt{\frac{x^-(q)}{x^+(q)}}\frac{x^+(p)-x^+(q)}{x^+(p)-x^-(q)} \\ [2ex]
			r_5^{{\rm LL}}=-\left(\frac{x^-(q)}{x^+(q)}\frac{x^+(p)}{x^-(p)}\right)^{1/4}\frac{x^+(q)-x^-(q)}{x^+(p)-x^-(q)}\frac{\gamma(p)}{\gamma(q)} \\ [2ex]
			r_6^{{\rm LL}}=-\left(\frac{x^-(q)}{x^+(q)}\frac{x^+(p)}{x^-(p)}\right)^{1/4}\frac{x^+(p)-x^-(p)}{x^+(p)-x^-(q)}\frac{\gamma(q)}{\gamma(p)} \\ [2ex]
		\end{array}
	\end{equation}
	\begin{equation}\label{key}
		\begin{array}{l}			
			r_2^{{\rm RR}}=\sqrt{\frac{x^-(q)}{x^+(q)}}\frac{x^+(p)-x^+(q)}{x^+(p)-x^-(q)} \\ [2ex]
			r_4^{{\rm RR}}=-1 \\ [2ex]
			r_7^{{\rm RR}}=0 \\ [2ex]
			r_8^{{\rm RR}}=0 \\ [2ex]
		\end{array}
		\quad\qquad
		\begin{array}{l}
			r_1^{{\rm RR}}=-\sqrt{\frac{x^-(q)}{x^+(q)}\frac{x^+(p)}{x^-(p)}} \frac{x^-(p)-x^+(q)}{x^-(q)-x^+(p)} \\ [2ex]
			r_3^{{\rm RR}}=\sqrt{\frac{x^+(p)}{x^-(p)}}\frac{x^-(p)-x^-(q)}{x^+(p)-x^-(q)} \\ [2ex]
			r_5^{{\rm RR}}=-\left(\frac{x^-(q)}{x^+(q)}\frac{x^+(p)}{x^-(p)}\right)^{3/4}\frac{x^+(p)-x^-(p)}{x^+(p)-x^-(q)}\frac{\gamma(q)}{\gamma(p)} \\ [2ex]
			r_6^{{\rm RR}}=-\left(\frac{x^-(p)}{x^+(p)}\frac{x^+(q)}{x^-(q)}\right)^{1/4}\frac{x^+(q)-x^-(q)}{x^+(p)-x^-(q)}\frac{\gamma(p)}{\gamma(q)} \\ [2ex]
		\end{array}
	\end{equation}
	For the mixed blocks (LR, RL) one has 
	\begin{equation}\label{key}
		\begin{array}{l}
			r_3^{{\rm LR}}=\sqrt{\frac{x^+(p)}{x^-(p)}} \\ [2ex] 
			r_4^{{\rm LR}}=-\frac{x^+(p)}{x^-(p)}\frac{1-x^+(q)x^-(p)}{1-x^+(p)x^+(q)} \\ [2ex] 
			r_5^{{\rm LR}}=0 \\ [2ex] 
			r_6^{{\rm LR}}=0 \\ [2ex] 
		\end{array}
		\quad\qquad
		\begin{array}{l}
			r_1^{{\rm LR}}=\sqrt{\frac{x^+(p)}{x^-(p)}\frac{x^+(q)}{x^-(q)}}\frac{1-x^+(p)x^-(q)}{1-x^+(p)x^+(q)} \\ [2ex] 
			r_2^{{\rm LR}}=\frac{x^+(p)}{x^-(p)}\sqrt{\frac{x^+(q)}{x^-(q)}}\frac{1-x^-(p)x^-(q)}{1-x^+(p)x^+(q)} \\ [2ex] 
			r_7^{{\rm LR}}=\left(\frac{x^+(p)}{x^-(p)}\frac{x^+(q)}{x^-(q)}\right)^{3/4}\frac{x^+(q)-x^-(q)}{1-x^+(p)x^+(q)}\frac{\gamma(p)}{\gamma(q)} \\ [2ex] 
			r_8^{{\rm LR}}=-\left(\frac{x^-(q)}{x^+(q)}\right)^{1/4}\left(\frac{x^+(p)}{x^-(p)}\right)^{3/4}\frac{x^+(p)-x^-(p)}{1-x^+(p)x^+(q)}\frac{\gamma(q)}{\gamma(p)} \\ [2ex] 
		\end{array}
	\end{equation}
	\begin{equation}\label{key}
		\begin{array}{l}
			r_3^{{\rm RL}}=\sqrt{\frac{x^-(q)}{x^+(q)}} \\ [2ex] 
			r_4^{{\rm RL}}=-\frac{x^-(q)}{x^+(q)}\frac{1-x^+(q)x^-(p)}{1-x^-(p)x^-(q)} \\ [2ex] 
			r_5^{{\rm RL}}=0 \\ [2ex] 
			r_6^{{\rm RL}}=0 \\ [2ex] 
		\end{array}
		\quad\qquad
		\begin{array}{l}
			r_1^{{\rm RL}}=\sqrt{\frac{x^-(p)}{x^+(p)}\frac{x^-(q)}{x^+(q)}}\frac{1-x^+(p)x^-(q)}{1-x^-(p)x^-(q)} \\ [2ex] 
			r_2^{{\rm RL}}=\frac{x^-(q)}{x^+(q)}\sqrt{\frac{x^-(p)}{x^+(p)}}\frac{1-x^+(p)x^+(q)}{1-x^-(p)x^-(q)} \\ [2ex] 
			r_7^{{\rm RL}}=\left(\frac{x^-(q)}{x^+(q)}\right)^{3/4}\left(\frac{x^+(p)}{x^-(p)}\right)^{1/4}\frac{x^+(p)-x^-(p)}{1-x^-(p)x^-(q)}\frac{\gamma(q)}{\gamma(p)} \\ [2ex] 
			r_8^{{\rm RL}}=-\left(\frac{x^-(p)}{x^+(p)}\frac{x^-(q)}{x^+(q)}\right)^{3/4}\frac{x^+(q)-x^-(q)}{1-x^-(p)x^-(q)}\frac{\gamma(p)}{\gamma(q)} \\ [2ex] 
		\end{array}
	\end{equation}
	It is important to note that w.r.t \cite{Borsato:2014hja,Sfondrini:2014via} the modification was done. Specifically, in order for the blocks above to be completely in Zhukovski space, the identification
	\begin{equation}\label{key}
		e^{i p} = \dfrac{x^{+}(p)}{x^{-}(p)}
	\end{equation}
	was performed and the blocks \eqref{Apx_R8v_Chi} satisfy qYBE for any $ \gamma(p) $, so that with
	\begin{equation}\label{key}
		\gamma(p) = e^{\frac{i p}{4}} \sqrt{ \frac{i h}{2} \left( x^{-}(p) - x^{+}(p) \right) }
	\end{equation}
	the \cite{Sfondrini:2014via} {\Rx} is reconstructed.

	\subsubsection{$ AdS_{3} \times S^{3} \times S^{3} \times S^{1} $}
	For the $ \cl{M}^{4} = S^{3} \times S^{1} $ case,  which is along the lines of \cite{Borsato:2015mma} one again is required to proceed with blocks of the form 
	\begin{equation}\label{Apx_R8v_Chi}
		R^{\chi} = \rho^{\chi}\upsilon^{\chi}
		\begin{pmatrix}
			r_{1}^{\chi} & 0 & 0 & r_{8}^{\chi} \\
			0 & r_{2}^{\chi} & r_{6}^{\chi} & 0 \\
			0 & r_{5}^{\chi} & r_{3}^{\chi} & 0 \\
			r_{7}^{\chi} & 0 & 0 & r_{4}^{\chi} \\
		\end{pmatrix}
		\qquad
		\begin{array}{l}
			r_{i}^{\chi} \equiv r_{i}^{\chi}(p,q) \\
			\rho^{\chi} \equiv \rho^{\chi}(p,q) \\
			\upsilon^{\chi} \equiv \upsilon^{\chi}(p,q) \\
		\end{array}
	\end{equation}
	where $ \chi $ conventionally captures the chiral sectors $ \chi = \{ \tx{LL}, \tx{RR}, \tx{LR}, \tx{RL} \} $. So the LL and RR sectors constitute
	\begin{equation}\label{key}
		\begin{array}{l}
			r_1^{{\rm LL}}=1 \\ [2ex] 
			r_2^{{\rm LL}}=\sqrt{\frac{x^-_{{\rm L}}(p)}{x^+_{{\rm L}}(p)}}\frac{x^+_{{\rm L}}(p)-x^+_{{\rm L}}(q)}{x^-_{{\rm L}}(p)-x^-_{{\rm L}}(q)} \\ [2ex] 
			r_7^{{\rm LL}}=0 \\ [2ex] 
			r_8^{{\rm LL}}=0 \\ [2ex] 
		\end{array}
		\quad\qquad
		\begin{array}{l}
			r_3^{{\rm LL}}=\sqrt{\frac{x^+_{{\rm L}}(q)}{x^-_{{\rm L}}(q)}}\frac{x^-_{{\rm L}}(p)-x^-_{{\rm L}}(q)}{x^-_{{\rm L}}(p)-x^+_{{\rm L}}(q)} \\ [2ex] 
			r_4^{{\rm LL}}=\sqrt{\frac{x^-_{{\rm L}}(p)}{x^+_{{\rm L}}(p)}\frac{x^+_{{\rm L}}(q)}{x^-_{{\rm L}}(q)}}\frac{x^-_{{\rm L}}(q)-x^+_{{\rm L}}(p)}{x^-_{{\rm L}}(p)-x^+_{{\rm L}}(q)} \\ [2ex] 
			r_5^{{\rm LL}}=\sqrt{\frac{x^-_{{\rm L}}(p)}{x^+_{{\rm L}}(p)}\frac{x^+_{{\rm L}}(q)}{x^-_{{\rm L}}(q)}}\frac{x^-_{{\rm L}}(q)-x^+_{{\rm L}}(q)}{x^-_{{\rm L}}(p)-x^+_{{\rm L}}(q)}\frac{\gamma^{{\rm L}}(p)}{\gamma^{{\rm L}}(q)} \\ [2ex] 
			r_6^{{\rm LL}}= \frac{x^-_{{\rm L}}(p)-x^+_{{\rm L}}(p)}{x^-_{{\rm L}}(p)-x^+_{{\rm L}}(q)}\frac{\gamma^{{\rm L}}(q)}{\gamma^{{\rm L}}(p)} \\ [2ex] 
		\end{array}
	\end{equation}

	\begin{equation}\label{key}
		\begin{array}{l}
			r_1^{{\rm RR}}=1 \\ [2ex] 
			r_2^{{\rm RR}}=\sqrt{\frac{x^-_{{\rm R}}(p)}{x^+_{{\rm R}}(p)}}\frac{x^+_{{\rm R}}(p)-x^+_{{\rm R}}(q)}{x^-_{{\rm R}}(p)-x^-_{{\rm R}}(q)} \\ [2ex] 
			r_7^{{\rm RR}}=0 \\ [2ex] 
			r_8^{{\rm RR}}=0 \\ [2ex] 
		\end{array}
		\quad\qquad
		\begin{array}{l}
			r_3^{{\rm RR}}=\sqrt{\frac{x^+_{{\rm R}}(q)}{x^-_{{\rm R}}(q)}}\frac{x^-_{{\rm R}}(p)-x^-_{{\rm R}}(q)}{x^-_{{\rm R}}(p)-x^+_{{\rm R}}(q)} \\ [2ex] 
			r_4^{{\rm RR}}=\sqrt{\frac{x^-_{{\rm R}}(p)}{x^+_{{\rm R}}(p)}\frac{x^+_{{\rm R}}(q)}{x^-_{{\rm R}}(q)}}\frac{x^-_{{\rm R}}(q)-x^+_{{\rm R}}(p)}{x^-_{{\rm R}}(p)-x^+_{{\rm R}}(q)} \\ [2ex] 
			r_5^{{\rm RR}}=\frac{x^-_{{\rm R}}(p)-x^+_{{\rm R}}(p)}{x^-_{{\rm R}}(p)-x^+_{{\rm R}}(q)}\frac{\gamma^{{\rm R}}(q)}{\gamma^{{\rm R}}(p)} \\ [2ex] 
			r_6^{{\rm RR}}= \sqrt{\frac{x^-_{{\rm R}}(p)}{x^+_{{\rm R}}(p)}\frac{x^+_{{\rm R}}(q)}{x^-_{{\rm R}}(q)}}\frac{x^-_{{\rm R}}(q)-x^+_{{\rm R}}(q)}{x^-_{{\rm R}}(p)-x^+_{{\rm R}}(q)}\frac{\gamma^{{\rm R}}(p)}{\gamma^{{\rm R}}(q)} \\ [2ex]
		\end{array}
	\end{equation}
	Whereas the LR and RL sectors arise accordingly
	\begin{equation}\label{key}
		\begin{array}{l}
			r_2^{{\rm LR}}=1 \\ [2ex] 
			r_4^{{\rm LR}}=-\sqrt{\frac{x^-_{{\rm R}}(q)}{x^+_{{\rm R}}(q)}}\frac{1-x^-_{{\rm L}}(p)x^+_{{\rm R}}(q)}{1-x^-_{{\rm L}}(p)x^-_{{\rm R}}(q)} \\ [2ex] 
			r_5^{{\rm LR}}=0 \\ [2ex] 
			r_6^{{\rm LR}}=0 \\ [2ex] 
		\end{array}
		\quad\qquad
		\begin{array}{l}
			r_1^{{\rm LR}}=\sqrt{\frac{x^-_{{\rm L}}(p)}{x^+_{{\rm L}}(p)}}\frac{1-x^+_{{\rm L}}(p)x^-_{{\rm R}}(q)}{1-x^-_{{\rm L}}(p)x^-_{{\rm R}}(q)} \\ [2ex] 
			r_3^{{\rm LR}}=\sqrt{\frac{x^-_{{\rm L}}(p)}{x^+_{{\rm L}}(p)}\frac{x^-_{{\rm R}}(q)}{x^+_{{\rm R}}(q)}}\frac{1-x^+_{{\rm L}}(p)x^+_{{\rm R}}(q)}{1-x^-_{{\rm L}}(p)x^-_{{\rm R}}(q)} \\ [2ex] 
			r_7^{{\rm LR}}=\sqrt{\frac{x^-_{{\rm L}}(p)}{x^+_{{\rm L}}(p)}}\frac{x_{{\rm R}}^+(q)-x_{{\rm R}}^-(q)}{1-x_{{\rm L}}^-(p)x_{{\rm R}}^-(q)}\frac{\gamma^{{\rm L}}(p)}{\gamma^{{\rm R}}(q)} \\ [2ex] 
			r_8^{{\rm LR}}=-\sqrt{\frac{x^-_{{\rm R}}(q)}{x^+_{{\rm R}}(q)}}\frac{x_{{\rm L}}^+(p)-x_{{\rm L}}^-(p)}{1-x_{{\rm L}}^-(p)x_{{\rm R}}^-(q)}\frac{\gamma^{{\rm R}}(q)}{\gamma^{{\rm L}}(p)} \\ [2ex] 
		\end{array}
	\end{equation}
	\begin{equation}\label{key}
		\begin{array}{l}
			r_1^{{\rm RL}}=\sqrt{\frac{x^-_{{\rm R}}(p)}{x^+_{{\rm R}}(p)}}\frac{1-x^+_{{\rm R}}(p)x^-_{{\rm L}}(q)}{1-x^-_{{\rm R}}(p)x^-_{{\rm L}}(q)} \\ [2ex] 
			r_2^{{\rm RL}}=1 \\ [2ex] 
			r_5^{{\rm RL}}=0 \\ [2ex] 
			r_6^{{\rm RL}}=0 \\ [2ex] 
		\end{array}
		\quad\qquad
		\begin{array}{l}
			r_3^{{\rm RL}}=\sqrt{\frac{x^-_{{\rm R}}(p)}{x^+_{{\rm R}}(p)}\frac{x^-_{{\rm L}}(q)}{x^+_{{\rm L}}(q)}}\frac{1-x^+_{{\rm R}}(p)x^+_{{\rm L}}(q)}{1-x^-_{{\rm R}}(p)x^-_{{\rm L}}(q)} \\ [2ex] 
			r_4^{{\rm RL}}=-\sqrt{\frac{x^-_{{\rm L}}(q)}{x^+_{{\rm L}}(q)}}\frac{1-x^-_{{\rm R}}(p)x^+_{{\rm L}}(q)}{1-x^-_{{\rm R}}(p)x^-_{{\rm L}}(q)} \\ [2ex] 
			r_7^{{\rm RL}}=\sqrt{\frac{x^-_{{\rm L}}(q)}{x^+_{{\rm L}}(q)}}\frac{x_{{\rm R}}^+(p)-x_{{\rm R}}^-(p)}{1-x_{{\rm R}}^-(p)x_{{\rm L}}^-(q)}\frac{\gamma^{{\rm L}}(q)}{\gamma^{{\rm R}}(p)} \\ [2ex] 
			r_8^{{\rm RL}}=-\sqrt{\frac{x^-_{{\rm R}}(p)}{x^+_{{\rm R}}(p)}}\frac{x_{{\rm L}}^+(q)-x_{{\rm L}}^-(q)}{1-x_{{\rm R}}^-(p)x_{{\rm L}}^-(q)}\frac{\gamma^{{\rm R}}(p)}{\gamma^{{\rm L}}(q)} \\ [2ex] 
		\end{array}
	\end{equation}
	The $ \upsilon(p,q)^{\chi} $ can be specified as
	\begin{equation}\label{key}
		\upsilon(p,q)^{\tx{LR}} = \left( \dfrac{x_{\tx{L}}^{+}(p)}{x_{\tx{L}}^{-}(p)} \right)^{-\frac{1}{4}} \left( \dfrac{x_{\tx{R}}^{+}(q)}{x_{\tx{R}}^{-}(q)} \right)^{-\frac{1}{4}} \left( \dfrac{1 - \frac{1}{x_{\tx{L}}^{-}(p)x_{\tx{R}}^{-}(q)}}{1 - \frac{1}{x_{\tx{L}}^{+}(p)x_{\tx{R}}^{+}(q)}} \right)
	\end{equation}
	and RL is obtained due to symmetry $ \tx{L} \leftrightarrow \tx{R} $, whereas for the pure factors are fixed
	\begin{equation}\label{key}
		\upsilon^{\tx{LL}} = \upsilon^{\tx{RR}} = 1
	\end{equation}
	Also the corresponding sectors of $ \gamma $ and Zhukovski variables impose can be found to be related
	\begin{equation}\label{key}
		\begin{array}{l}
			\gamma^{\tx{L}}(p + \omega) = i \dfrac{\gamma^{\tx{R}}(p)}{x_{\tx{R}}^{+}(p)} \\ [2ex]
			x_{\tx{L}}^{\pm}(p + \omega) = \dfrac{1}{x_{\tx{R}}^{\pm}(p)}
		\end{array}
	\end{equation}
	where $ p + \omega \rightarrow \bar{p} $ of \cite{Borsato:2015mma} and the right sector obtained through $ L \leftrightarrow R $.

	\subsubsection{$ AdS_{3} \times S^{3} \times S^{3} \times S^{1} \mapsto AdS_{3} \times S^{3} \times T^{4} $}
	It can be shown that by the set of restrictions and mappings one can acquire {\adso} {\Rx} from the one of {\adst}. First one is required to remap Zhukovski variables
	\begin{equation}\label{key}
		\begin{array}{l}
			x_{\tx{L}}^{-} \mapsto x^{-} \\
			x_{\tx{L}}^{+} \mapsto x^{+} \\
		\end{array}
		\qquad\qquad
		\begin{array}{l}
			x_{\tx{R}}^{-} \mapsto x^{-} \\
			x_{\tx{R}}^{+} \mapsto x^{+} \\
		\end{array}
	\end{equation}
	 and $ \gamma^{\mrm{x}} $ ($ \mrm{x} = \{ \tx{L, R} \} $)
	\begin{equation}\label{key}
		\begin{array}{l}
	 		\gamma^{\tx{L}}(p)=a \left(\frac{x^+(p)}{x^-(p)}\right)^{1/4}\gamma(p) \\ [2ex]
			\gamma^{\tx{R}}(p)=a \left(\frac{x^-(p)}{x^+(p)}\right)^{1/4}\gamma(p) \\ [2ex]
		\end{array}
	\end{equation}
	where $ a \in \tx{const} $ and $ \rho^{\chi} $ related to $ \zeta^{\chi} $ by
	\begin{equation}\label{key}
		\begin{array}{l}
			\rho^{{\rm LL}}(p,q)=-\sqrt{\frac{x^+(p)}{x^-(p)}\frac{x^-(q)}{x^+(q)}}\frac{x^-(p)-x^+(q)}{x^-(q)-x^+(p)}\zeta^{{\rm LL}}(p,q) \\ [2ex] 
			\rho^{{\rm RR}}(p,q) = -\sqrt{\frac{x^+(p)}{x^-(p)}\frac{x^-(q)}{x^+(q)}}\frac{x^-(p)-x^+(q)}{x^-(q)-x^+(p)}\zeta^{{\rm RR}}(p,q) \\ [2ex] 
		\end{array}
		\qquad
		\begin{array}{l}
			\rho^{{\rm LR}}(p,q) =\frac{x^+(p)}{x^-(p)}\sqrt{\frac{x^+(q)}{x^-(q)}}\frac{1-x^-(p)x^-(q)}{1-x^+(p)x^+(q)}\frac{\zeta^{{\rm LR}}(p,q)}{\upsilon^{{\rm LR}}(p,q)} \\ [2ex] 
			\rho^{{\rm RL}}(p,q)=\sqrt{\frac{x^-(q)}{x^+(q)}}\frac{\zeta^{{\rm RL}}(p,q)}{\upsilon^{{\rm RL}}(p,q)} \\ [2ex] 
		\end{array}
	\end{equation}

\end{appendix}

\newpage
\bibliographystyle{JHEP}
\bibliography{References}

\providecommand{\href}[2]{#2}\begingroup\raggedright\begin{thebibliography}{100}

\bibitem{zakharov1971korteweg}
V.E.~Zakharov and L.D.~Faddeev, \emph{Korteweg--de vries equation: A completely
  integrable hamiltonian system}, {\emph{Funktsional'nyi Analiz i ego
  Prilozheniya} {\bfseries 5} (1971) 18}.

\bibitem{Korepin_1997_QISM_CF}
V.E.~Korepin, N.M.~Bogoliubov and A.G.~Izergin, \emph{Quantum inverse
  scattering method and correlation functions}, vol.~3, Cambridge university
  press (1997).

\bibitem{Sklyanin_1979quantum}
E.K.~Sklyanin, L.A.~Takhtadzhyan and L.D.~Faddeev, \emph{Quantum inverse
  problem method. i}, {\emph{Teoreticheskaya i Matematicheskaya Fizika}
  {\bfseries 40} (1979) 194}.

\bibitem{Sklyanin_1982quantum}
E.K.~Sklyanin, \emph{Quantum version of the method of inverse scattering
  problem}, {\emph{Journal of Soviet Mathematics} {\bfseries 19} (1982) 1546}.

\bibitem{Zamolodchikov:1978xm}
A.B.~Zamolodchikov and A.B.~Zamolodchikov, \emph{{Factorized S Matrices in
  Two-Dimensions as the Exact Solutions of Certain Relativistic Quantum Field
  Models}}, \href{https://doi.org/10.1016/0003-4916(79)90391-9}{\emph{Annals
  Phys.} {\bfseries 120} (1979) 253}.

\bibitem{Zamolodchikov_1979z}
A.B.~Zamolodchikov, \emph{Z$ _{4} $-symmetric factorized s-matrix in two
  space-time dimensions}, {\emph{Communications in Mathematical Physics}
  {\bfseries 69} (1979) 165}.

\bibitem{Baxter_1980hardhexagons}
R.J.~Baxter, \emph{Hard hexagons: exact solution}, {\emph{Journal of Physics A:
  Mathematical and General} {\bfseries 13} (1980) L61}.

\bibitem{Stroganov_1979new}
Y.G.~Stroganov, \emph{A new calculation method for partition functions in some
  lattice models}, {\emph{Physics Letters A} {\bfseries 74} (1979) 116}.

\bibitem{Zamolodchikov:1990bu}
A.B.~Zamolodchikov, \emph{{Factorized S-matrices and Lattice Statistical
  systems}}, .

\bibitem{Faddeev:1996iy}
L.D.~Faddeev, \emph{{How algebraic Bethe ansatz works for integrable model}},
  in \emph{{Les Houches School of Physics: Astrophysical Sources of
  Gravitational Radiation}}, pp.~pp. 149--219, 5, 1996
  [\href{https://arxiv.org/abs/hep-th/9605187}{{\ttfamily hep-th/9605187}}].

\bibitem{Tetelman}
M.~Tetelman, \emph{{Lorentz group for two-dimensional integrable lattice
  systems.}}, {\emph{Sov. Phys. JETP} {\bfseries 55(2)} (1982) 306}.

\bibitem{Loebbert:2016cdm}
F.~Loebbert, \emph{{Lectures on Yangian Symmetry}},
  \href{https://doi.org/10.1088/1751-8113/49/32/323002}{\emph{J. Phys.}
  {\bfseries A49} (2016) 323002}
  [\href{https://arxiv.org/abs/1606.02947}{{\ttfamily 1606.02947}}].

\bibitem{Curtright_1993}
T.~Curtright and C.~Zachos, \emph{Supersymmetry and the nonlocal yangian
  deformation symmetry},
  \href{https://doi.org/10.1016/0550-3213(93)90120-e}{\emph{Nuclear Physics B}
  {\bfseries 402} (1993) 604}.

\bibitem{Zamolodchikov:2004ce}
A.B.~Zamolodchikov, \emph{{Expectation value of composite field T anti-T in
  two-dimensional quantum field theory}},
  \href{https://arxiv.org/abs/hep-th/0401146}{{\ttfamily hep-th/0401146}}.

\bibitem{Smirnov:2016lqw}
F.A.~Smirnov and A.B.~Zamolodchikov, \emph{{On space of integrable quantum
  field theories}},
  \href{https://doi.org/10.1016/j.nuclphysb.2016.12.014}{\emph{Nucl. Phys. B}
  {\bfseries 915} (2017) 363}
  [\href{https://arxiv.org/abs/1608.05499}{{\ttfamily 1608.05499}}].

\bibitem{Cavaglia:2016oda}
A.~Cavagli\`a, S.~Negro, I.M.~Sz\'ecs\'enyi and R.~Tateo, \emph{{$T
  \bar{T}$-deformed 2D Quantum Field Theories}},
  \href{https://doi.org/10.1007/JHEP10(2016)112}{\emph{JHEP} {\bfseries 10}
  (2016) 112} [\href{https://arxiv.org/abs/1608.05534}{{\ttfamily
  1608.05534}}].

\bibitem{Frolov:2019nrr}
S.~Frolov, \emph{{$T\overline{T}$ Deformation and the Light-Cone Gauge}},
  \href{https://doi.org/10.1134/S0081543820030098}{\emph{Proc. Steklov Inst.
  Math.} {\bfseries 309} (2020) 107}
  [\href{https://arxiv.org/abs/1905.07946}{{\ttfamily 1905.07946}}].

\bibitem{Jiang:2019epa}
Y.~Jiang, \emph{{A pedagogical review on solvable irrelevant deformations of 2D
  quantum field theory}},
  \href{https://doi.org/10.1088/1572-9494/abe4c9}{\emph{Commun. Theor. Phys.}
  {\bfseries 73} (2021) 057201}
  [\href{https://arxiv.org/abs/1904.13376}{{\ttfamily 1904.13376}}].

\bibitem{Drinfeld:1985rx}
V.G.~Drinfeld, \emph{{Hopf algebras and the quantum Yang-Baxter equation}},
  {\emph{Sov. Math. Dokl.} {\bfseries 32} (1985) 254}.

\bibitem{Drinfeld1986quantum}
V.G.~Drinfeld, \emph{Quantum groups}, {\emph{Zapiski Nauchnykh Seminarov POMI}
  {\bfseries 155} (1986) 18}.

\bibitem{Drinfeld_1986DAHAY}
V.G.~Drinfeld, \emph{Degenerate affine hecke algebras and yangians},
  {\emph{Functional Analysis and Its Applications} {\bfseries 20} (1986) 58}.

\bibitem{Kulish1982}
P.P.~Kulish and E.K.~Sklyanin, \emph{Solutions of the yang-baxter equation},
  {\emph{Journal of Soviet Mathematics} {\bfseries 19} (1982) 1596}.

\bibitem{Akutsu:1982}
K.~Sogo, M.~Uchinami, Y.~Akutsu and M.~Wadati, \emph{{Classification of Exactly
  Solvable Two-Component Models: }},
  \href{https://doi.org/10.1143/PTP.68.508}{\emph{Progress of Theoretical
  Physics} {\bfseries 68} (1982) 508}.

\bibitem{Khachatryan:2012wy}
S.~Khachatryan and A.~Sedrakyan, \emph{{On the solutions of the Yang-Baxter
  equations with general inhomogeneous eight-vertex $R$-matrix: Relations with
  Zamolodchikov's tetrahedral algebra}},
  \href{https://doi.org/10.1007/s10955-012-0666-8}{\emph{J. Statist. Phys.}
  {\bfseries 150} (2013) 130}
  [\href{https://arxiv.org/abs/1208.4339}{{\ttfamily 1208.4339}}].

\bibitem{Hietarinta:1992ix}
J.~Hietarinta, \emph{{Solving the two-dimensional constant quantum Yang-Baxter
  equation}}, \href{https://doi.org/10.1063/1.530185}{\emph{J. Math. Phys.}
  {\bfseries 34} (1993) 1725}.

\bibitem{Pourkia:2018}
A.~{Pourkia}, \emph{{Solutions to the constant Yang-Baxter equation in all
  dimensions}}, {\emph{arXiv e-prints} (2018) arXiv:1806.08400}
  [\href{https://arxiv.org/abs/1806.08400}{{\ttfamily 1806.08400}}].

\bibitem{Vieira:2017vnw}
R.S.~Vieira, \emph{{Solving and classifying the solutions of the Yang-Baxter
  equation through a differential approach. Two-state systems}},
  \href{https://doi.org/10.1007/JHEP10(2018)110}{\emph{JHEP} {\bfseries 10}
  (2018) 110} [\href{https://arxiv.org/abs/1712.02341}{{\ttfamily
  1712.02341}}].

\bibitem{Vieira:2020lem}
R.S.~Vieira and A.~Lima-Santos, \emph{{Solutions of the Yang\textendash{}Baxter
  equation for ($n$ + 1) (2$n$ + 1)-vertex models using a differential
  approach}}, \href{https://doi.org/10.1088/1742-5468/abf7be}{\emph{J. Stat.
  Mech.} {\bfseries 2105} (2021) 053103}
  [\href{https://arxiv.org/abs/2012.02543}{{\ttfamily 2012.02543}}].

\bibitem{FOKAS_1981_341}
A.~Fokas and B.~Fuchssteiner, \emph{The hierarchy of the benjamin-ono
  equation},
  \href{https://doi.org/https://doi.org/10.1016/0375-9601(81)90551-X}{\emph{Physics
  Letters A} {\bfseries 86} (1981) 341}.

\bibitem{Fuchssteiner_1983}
B.~Fuchssteiner, \emph{{Mastersymmetries, Higher Order Time-Dependent
  Symmetries and Conserved Densities of Nonlinear Evolution Equations}},
  \href{https://doi.org/10.1143/PTP.70.1508}{\emph{Progress of Theoretical
  Physics} {\bfseries 70} (1983) 1508}
  [\href{https://arxiv.org/abs/https://academic.oup.com/ptp/article-pdf/70/6/1508/5208894/70-6-1508.pdf}{{\ttfamily
  https://academic.oup.com/ptp/article-pdf/70/6/1508/5208894/70-6-1508.pdf}}].

\bibitem{Bajnok_2011}
Z.~Bajnok, \emph{Review of {AdS}/{CFT} integrability, chapter {III}.6:
  Thermodynamic bethe ansatz},
  \href{https://doi.org/10.1007/s11005-011-0512-y}{\emph{Letters in
  Mathematical Physics} {\bfseries 99} (2011) 299}.

\bibitem{Tongeren_2016}
S.J.~van Tongeren, \emph{Introduction to the thermodynamic bethe ansatz},
  \href{https://doi.org/10.1088/1751-8113/49/32/323005}{\emph{Journal of
  Physics A: Mathematical and Theoretical} {\bfseries 49} (2016) 323005}.

\bibitem{Arutyunov:2009ga}
G.~Arutyunov and S.~Frolov, \emph{{Foundations of the AdS$_{5} \times S^{5}$
  Superstring. Part I}},
  \href{https://doi.org/10.1088/1751-8113/42/25/254003}{\emph{J. Phys. A}
  {\bfseries 42} (2009) 254003}
  [\href{https://arxiv.org/abs/0901.4937}{{\ttfamily 0901.4937}}].

\bibitem{Frolov:2021fmj}
S.~Frolov and A.~Sfondrini, \emph{{New dressing factors for AdS3/CFT2}},
  \href{https://doi.org/10.1007/JHEP04(2022)162}{\emph{JHEP} {\bfseries 04}
  (2022) 162} [\href{https://arxiv.org/abs/2112.08896}{{\ttfamily
  2112.08896}}].

\bibitem{Frolov:2021bwp}
S.~Frolov and A.~Sfondrini, \emph{{Mirror thermodynamic Bethe ansatz for
  AdS3/CFT2}}, \href{https://doi.org/10.1007/JHEP03(2022)138}{\emph{JHEP}
  {\bfseries 03} (2022) 138}
  [\href{https://arxiv.org/abs/2112.08898}{{\ttfamily 2112.08898}}].

\bibitem{Frolov:2021zyc}
S.~Frolov and A.~Sfondrini, \emph{{Massless S matrices for AdS3/CFT2}},
  \href{https://doi.org/10.1007/JHEP04(2022)067}{\emph{JHEP} {\bfseries 04}
  (2022) 067} [\href{https://arxiv.org/abs/2112.08895}{{\ttfamily
  2112.08895}}].

\bibitem{Minahan_2003}
J.A.~Minahan and K.~Zarembo, \emph{The bethe-ansatz for script n = 4 super
  yang-mills},
  \href{https://doi.org/10.1088/1126-6708/2003/03/013}{\emph{Journal of High
  Energy Physics} {\bfseries 2003} (2003) 013}.

\bibitem{Beisert:2005tm}
N.~Beisert, \emph{{The SU(2|2) dynamic S-matrix}},
  \href{https://doi.org/10.4310/ATMP.2008.v12.n5.a1}{\emph{Adv. Theor. Math.
  Phys.} {\bfseries 12} (2008) 945}
  [\href{https://arxiv.org/abs/hep-th/0511082}{{\ttfamily hep-th/0511082}}].

\bibitem{Beisert:2006qh}
N.~Beisert, \emph{{The Analytic Bethe Ansatz for a Chain with Centrally
  Extended su(2|2) Symmetry}},
  \href{https://doi.org/10.1088/1742-5468/2007/01/P01017}{\emph{J. Stat. Mech.}
  {\bfseries 0701} (2007) P01017}
  [\href{https://arxiv.org/abs/nlin/0610017}{{\ttfamily nlin/0610017}}].

\bibitem{Arutyunov:2006yd}
G.~Arutyunov, S.~Frolov and M.~Zamaklar, \emph{{The Zamolodchikov-Faddeev
  algebra for $\text{AdS}_5\times \text{S\;}^5$ superstring}},
  \href{https://doi.org/10.1088/1126-6708/2007/04/002}{\emph{JHEP} {\bfseries
  04} (2007) 002} [\href{https://arxiv.org/abs/hep-th/0612229}{{\ttfamily
  hep-th/0612229}}].

\bibitem{Maldacena_1997}
J.M.~Maldacena, \emph{{The Large N limit of superconformal field theories and
  supergravity}}, \href{https://doi.org/10.1023/A:1026654312961}{\emph{Int. J.
  Theor. Phys.} {\bfseries 38} (1999) 1113}
  [\href{https://arxiv.org/abs/hep-th/9711200}{{\ttfamily hep-th/9711200}}].

\bibitem{Witten_1998}
E.~Witten, \emph{{Anti-de Sitter space and holography}},
  \href{https://doi.org/10.4310/ATMP.1998.v2.n2.a2}{\emph{Adv. Theor. Math.
  Phys.} {\bfseries 2} (1998) 253}
  [\href{https://arxiv.org/abs/hep-th/9802150}{{\ttfamily hep-th/9802150}}].

\bibitem{Aharony:1999ti}
O.~Aharony, S.S.~Gubser, J.M.~Maldacena, H.~Ooguri and Y.~Oz, \emph{{Large N
  field theories, string theory and gravity}},
  \href{https://doi.org/10.1016/S0370-1573(99)00083-6}{\emph{Phys. Rept.}
  {\bfseries 323} (2000) 183}
  [\href{https://arxiv.org/abs/hep-th/9905111}{{\ttfamily hep-th/9905111}}].

\bibitem{DHoker:2002nbb}
E.~D'Hoker and D.Z.~Freedman, \emph{{Supersymmetric gauge theories and the AdS
  / CFT correspondence}},  in \emph{{Theoretical Advanced Study Institute in
  Elementary Particle Physics (TASI 2001): Strings, Branes and EXTRA
  Dimensions}}, pp.~3--158, 1, 2002
  [\href{https://arxiv.org/abs/hep-th/0201253}{{\ttfamily hep-th/0201253}}].

\bibitem{Polchinski:2010hw}
J.~Polchinski, \emph{{Introduction to Gauge/Gravity Duality}},  in
  \emph{{Theoretical Advanced Study Institute in Elementary Particle Physics}:
  {String theory and its Applications: From meV to the Planck Scale}},
  pp.~3--46, 10, 2010, \href{https://doi.org/10.1142/9789814350525_0001}{DOI}
  [\href{https://arxiv.org/abs/1010.6134}{{\ttfamily 1010.6134}}].

\bibitem{Maldacena:2003nj}
J.M.~Maldacena, \emph{{TASI 2003 lectures on AdS / CFT}},  in
  \emph{{Theoretical Advanced Study Institute in Elementary Particle Physics
  (TASI 2003): Recent Trends in String Theory}}, pp.~155--203, 9, 2003
  [\href{https://arxiv.org/abs/hep-th/0309246}{{\ttfamily hep-th/0309246}}].

\bibitem{Sorokin_2011_SAdS_2}
D.~Sorokin, A.~Tseytlin, L.~Wulff and K.~Zarembo, \emph{{Superstrings in $
  AdS2\times S2 \times T6 $}},
  \href{https://doi.org/10.1088/1751-8113/44/27/275401}{\emph{J.Phys.A}
  {\bfseries 44} (2011) 275}.

\bibitem{Beisert:2010jr}
N.~Beisert et~al., \emph{{Review of AdS/CFT Integrability: An Overview}},
  \href{https://doi.org/10.1007/s11005-011-0529-2}{\emph{Lett. Math. Phys.}
  {\bfseries 99} (2012) 3} [\href{https://arxiv.org/abs/1012.3982}{{\ttfamily
  1012.3982}}].

\bibitem{Arutyunov:2013ega}
G.~Arutyunov, R.~Borsato and S.~Frolov, \emph{{S-matrix for strings on
  $\eta$-deformed $AdS_{5} \times S^{5} $}},
  \href{https://doi.org/10.1007/JHEP04(2014)002}{\emph{JHEP} {\bfseries 04}
  (2014) 002} [\href{https://arxiv.org/abs/1312.3542}{{\ttfamily 1312.3542}}].

\bibitem{Borsato:2013qpa}
R.~Borsato, O.~Ohlsson~Sax, A.~Sfondrini, B.~Stefanski and A.~Torrielli,
  \emph{{The all-loop integrable spin-chain for strings on AdS$_3 \times S^3
  \times T^4$: the massive sector}},
  \href{https://doi.org/10.1007/JHEP08(2013)043}{\emph{JHEP} {\bfseries 08}
  (2013) 043} [\href{https://arxiv.org/abs/1303.5995}{{\ttfamily 1303.5995}}].

\bibitem{Arutyunov:2009ax}
G.~Arutyunov, S.~Frolov and R.~Suzuki, \emph{{Exploring the mirror TBA}},
  \href{https://doi.org/10.1007/JHEP05(2010)031}{\emph{JHEP} {\bfseries 05}
  (2010) 031} [\href{https://arxiv.org/abs/0911.2224}{{\ttfamily 0911.2224}}].

\bibitem{Arutyunov:2011uz}
G.~Arutyunov and S.~Frolov, \emph{{Comments on the Mirror TBA}},
  \href{https://doi.org/10.1007/JHEP05(2011)082}{\emph{JHEP} {\bfseries 05}
  (2011) 082} [\href{https://arxiv.org/abs/1103.2708}{{\ttfamily 1103.2708}}].

\bibitem{Arutyunov:2012zt}
G.~Arutyunov, M.~de~Leeuw and S.J.~van Tongeren, \emph{{The Quantum Deformed
  Mirror TBA I}}, \href{https://doi.org/10.1007/JHEP10(2012)090}{\emph{JHEP}
  {\bfseries 10} (2012) 090} [\href{https://arxiv.org/abs/1208.3478}{{\ttfamily
  1208.3478}}].

\bibitem{Arutyunov:2012ai}
G.~Arutyunov, M.~de~Leeuw and S.J.~van Tongeren, \emph{{The Quantum Deformed
  Mirror TBA II}}, \href{https://doi.org/10.1007/JHEP02(2013)012}{\emph{JHEP}
  {\bfseries 02} (2013) 012} [\href{https://arxiv.org/abs/1210.8185}{{\ttfamily
  1210.8185}}].

\bibitem{Bombardelli:2009xz}
D.~Bombardelli, D.~Fioravanti and R.~Tateo, \emph{{TBA and Y-system for planar
  AdS(4)/CFT(3)}},
  \href{https://doi.org/10.1016/j.nuclphysb.2010.04.005}{\emph{Nucl. Phys. B}
  {\bfseries 834} (2010) 543}
  [\href{https://arxiv.org/abs/0912.4715}{{\ttfamily 0912.4715}}].

\bibitem{Alday:2010vh}
L.F.~Alday, J.~Maldacena, A.~Sever and P.~Vieira, \emph{{Y-system for
  Scattering Amplitudes}},
  \href{https://doi.org/10.1088/1751-8113/43/48/485401}{\emph{J. Phys. A}
  {\bfseries 43} (2010) 485401}
  [\href{https://arxiv.org/abs/1002.2459}{{\ttfamily 1002.2459}}].

\bibitem{Gromov_2014}
N.~Gromov, V.~Kazakov, S.~Leurent and D.~Volin, \emph{Quantum spectral curve
  for planar n= 4 super-yang-mills theory}, {\emph{Physical Review Letters}
  {\bfseries 112} (2014) 011602}.

\bibitem{Gromov:2015wca}
N.~Gromov, F.~Levkovich-Maslyuk and G.~Sizov, \emph{{Quantum Spectral Curve and
  the Numerical Solution of the Spectral Problem in AdS5/CFT4}},
  \href{https://doi.org/10.1007/JHEP06(2016)036}{\emph{JHEP} {\bfseries 06}
  (2016) 036} [\href{https://arxiv.org/abs/1504.06640}{{\ttfamily
  1504.06640}}].

\bibitem{Kazakov_2018}
V.~Kazakov, \emph{Quantum spectral curve of $\gamma$-twisted n= 4 sym theory
  and fishnet cft},  in \emph{Ludwig Faddeev Memorial Volume: A Life in
  Mathematical Physics}, pp.~293--342, World Scientific (2018).

\bibitem{Klose:2010ki}
T.~Klose, \emph{{Review of AdS/CFT Integrability, Chapter IV.3: N=6
  Chern-Simons and Strings on AdS4xCP3}},
  \href{https://doi.org/10.1007/s11005-011-0520-y}{\emph{Lett. Math. Phys.}
  {\bfseries 99} (2012) 401} [\href{https://arxiv.org/abs/1012.3999}{{\ttfamily
  1012.3999}}].

\bibitem{Bykov:2010tv}
D.~Bykov, \emph{{The worldsheet low-energy limit of the AdS$_4$ x
  $\mathbb{C}P^{3}$ superstring}},
  \href{https://doi.org/10.1016/j.nuclphysb.2010.05.013}{\emph{Nucl. Phys. B}
  {\bfseries 838} (2010) 47} [\href{https://arxiv.org/abs/1003.2199}{{\ttfamily
  1003.2199}}].

\bibitem{Borsato:2014exa}
R.~Borsato, O.~Ohlsson~Sax, A.~Sfondrini and B.~Stefanski, \emph{{Towards the
  All-Loop Worldsheet S Matrix for $AdS_3\times S^3\times T^4$}},
  \href{https://doi.org/10.1103/PhysRevLett.113.131601}{\emph{Phys. Rev. Lett.}
  {\bfseries 113} (2014) 131601}
  [\href{https://arxiv.org/abs/1403.4543}{{\ttfamily 1403.4543}}].

\bibitem{Borsato:2014hja}
R.~Borsato, O.~Ohlsson~Sax, A.~Sfondrini and B.~Stefanski, \emph{{The complete
  AdS$_{3} \times$ S$^3 \times$ T$^4$ worldsheet S matrix}},
  \href{https://doi.org/10.1007/JHEP10(2014)066}{\emph{JHEP} {\bfseries 10}
  (2014) 066} [\href{https://arxiv.org/abs/1406.0453}{{\ttfamily 1406.0453}}].

\bibitem{Borsato:2015mma}
R.~Borsato, O.~Ohlsson~Sax, A.~Sfondrini and B.~Stefa\'nski, \emph{{The
  $\mathrm{AdS}_3\times \mathrm{S}^3\times \mathrm{S}^3\times\mathrm{S}^1$
  worldsheet S matrix}},
  \href{https://doi.org/10.1088/1751-8113/48/41/415401}{\emph{J. Phys. A}
  {\bfseries 48} (2015) 415401}
  [\href{https://arxiv.org/abs/1506.00218}{{\ttfamily 1506.00218}}].

\bibitem{Hoare:2014kma}
B.~Hoare, A.~Pittelli and A.~Torrielli, \emph{{Integrable S-matrices, massive
  and massless modes and the AdS$_{2}\times$S$^{2}$ superstring}},
  \href{https://doi.org/10.1007/JHEP11(2014)051}{\emph{JHEP} {\bfseries 11}
  (2014) 051} [\href{https://arxiv.org/abs/1407.0303}{{\ttfamily 1407.0303}}].

\bibitem{Hoare:2015kla}
B.~Hoare, A.~Pittelli and A.~Torrielli, \emph{{S-matrix algebra of the
  AdS$_{2}\times$S$^{2}$ superstring}},
  \href{https://doi.org/10.1103/PhysRevD.93.066006}{\emph{Phys.Rev.D}
  {\bfseries 93} (2016) 066006}
  [\href{https://arxiv.org/abs/1509.07587}{{\ttfamily 1509.07587}}].

\bibitem{Fontanella:2017rvu}
A.~Fontanella and A.~Torrielli, \emph{{Massless $AdS_2$ scattering and Bethe
  ansatz}}, \href{https://doi.org/10.1007/JHEP09(2017)075}{\emph{JHEP}
  {\bfseries 09} (2017) 075}
  [\href{https://arxiv.org/abs/1706.02634}{{\ttfamily 1706.02634}}].

\bibitem{Sfondrini:2014via}
A.~Sfondrini, \emph{{Towards integrability for ${\rm Ad}{{{\rm S}}_{{\bf
  3}}}/{\rm CF}{{{\rm T}}_{{\bf 2}}}$}},
  \href{https://doi.org/10.1088/1751-8113/48/2/023001}{\emph{J. Phys.}
  {\bfseries A48} (2015) 023001}
  [\href{https://arxiv.org/abs/1406.2971}{{\ttfamily 1406.2971}}].

\bibitem{Zarembo:2017muf}
K.~Zarembo, \emph{{Integrability in Sigma-Models}},
  \href{https://arxiv.org/abs/1712.07725}{{\ttfamily 1712.07725}}.

\bibitem{Babichenko_2010}
A.~Babichenko, B.~Stefa{\'{n}}ski and K.~Zarembo, \emph{Integrability and the
  {AdS} 3/{CFT} 2 correspondence},
  \href{https://doi.org/10.1007/jhep03(2010)058}{\emph{Journal of High Energy
  Physics} {\bfseries 2010} (2010) }.

\bibitem{Green:1983wt}
M.B.~Green and J.H.~Schwarz, \emph{{Covariant Description of Superstrings}},
  \href{https://doi.org/10.1016/0370-2693(84)92021-5}{\emph{Phys. Lett. B}
  {\bfseries 136} (1984) 367}.

\bibitem{Grisaru:1985fv}
M.T.~Grisaru, P.S.~Howe, L.~Mezincescu, B.~Nilsson and P.K.~Townsend,
  \emph{{N=2 Superstrings in a Supergravity Background}},
  \href{https://doi.org/10.1016/0370-2693(85)91071-8}{\emph{Phys. Lett. B}
  {\bfseries 162} (1985) 116}.

\bibitem{Berenstein:2002jq}
D.E.~Berenstein, J.M.~Maldacena and H.S.~Nastase, \emph{{Strings in flat space
  and pp waves from N=4 superYang-Mills}},
  \href{https://doi.org/10.1088/1126-6708/2002/04/013}{\emph{JHEP} {\bfseries
  04} (2002) 013} [\href{https://arxiv.org/abs/hep-th/0202021}{{\ttfamily
  hep-th/0202021}}].

\bibitem{Metsaev:1998it}
R.R.~Metsaev and A.A.~Tseytlin, \emph{{Type IIB superstring action in $ AdS_{5}
  \times S^{5} $ background}},
  \href{https://doi.org/10.1016/S0550-3213(98)00570-7}{\emph{Nucl. Phys. B}
  {\bfseries 533} (1998) 109}
  [\href{https://arxiv.org/abs/hep-th/9805028}{{\ttfamily hep-th/9805028}}].

\bibitem{Bena:2003wd}
I.~Bena, J.~Polchinski and R.~Roiban, \emph{{Hidden symmetries of the $ AdS_{5}
  \times S^{5} $ superstring}},
  \href{https://doi.org/10.1103/PhysRevD.69.046002}{\emph{Phys. Rev. D}
  {\bfseries 69} (2004) 046002}
  [\href{https://arxiv.org/abs/hep-th/0305116}{{\ttfamily hep-th/0305116}}].

\bibitem{Kulish:1982a}
P.P.~Kulish and E.K.~Sklyanin, \emph{Quantum spectral transform method recent
  developments},  in \emph{Integrable Quantum Field Theories}, J.~Hietarinta
  and C.~Montonen, eds., (Berlin, Heidelberg), pp.~61--119, Springer Berlin
  Heidelberg, 1982.

\bibitem{Grabowski}
M.P.~Grabowski and P.~Mathieu, \emph{Integrability test for spin chains},
  {\emph{Journal of Physics A: Mathematical and General} {\bfseries 28} (1995)
  4777}.

\bibitem{deLeeuw:2019zsi}
M.~de~Leeuw, A.~Pribytok and P.~Ryan, \emph{{Classifying two-dimensional
  integrable spin chains}},  \href{https://arxiv.org/abs/1904.12005}{{\ttfamily
  1904.12005}}.

\bibitem{Beisert_2013_idXXZ}
N.~Beisert, L.~Fi{\'e}vet, M.~de~Leeuw and F.~Loebbert, \emph{Integrable
  deformations of the xxz spin chain}, {\emph{Journal of Statistical Mechanics:
  Theory and Experiment} {\bfseries 2013} (2013) P09028}.

\bibitem{Alcaraz_1994}
F.~Alcaraz, M.~Droz, M.~Henkel and V.~Rittenberg, \emph{Reaction-diffusion
  processes, critical dynamics, and quantum chains},
  \href{https://doi.org/10.1006/aphy.1994.1026}{\emph{Annals of Physics}
  {\bfseries 230} (1994) 250}.

\bibitem{Kulish_1997}
P.P.~Kulish and A.A.~Stolin, \emph{Deformed yangians and integrable models},
  \href{https://doi.org/10.1023/a:1022869414679}{\emph{Czechoslovak Journal of
  Physics} {\bfseries 47} (1997) 1207}.

\bibitem{Kirillov1990representations}
A.~Kirillov and N.Y.~Reshetikhin, \emph{Representations of the algebra u$ _{q}
  $(sl(2)), q-orthogonal polynomials and invariants of links},  in \emph{New
  developments in the theory of knots}, pp.~202--256, World Scientific (1990).

\bibitem{Bytsko:2009a}
A.G.~Bytsko, \emph{Non-hermitian spin chains with inhomogeneous coupling}, .

\bibitem{Gromov:2017cja}
N.~Gromov, V.~Kazakov, G.~Korchemsky, S.~Negro and G.~Sizov,
  \emph{{Integrability of Conformal Fishnet Theory}},
  \href{https://doi.org/10.1007/JHEP01(2018)095}{\emph{JHEP} {\bfseries 01}
  (2018) 095} [\href{https://arxiv.org/abs/1706.04167}{{\ttfamily
  1706.04167}}].

\bibitem{Caetano:2016ydc}
J.~Caetano, O.~Gurdogan and V.~Kazakov, \emph{{Chiral limit of $ \mathcal{N} $
  = 4 SYM and ABJM and integrable Feynman graphs}},
  \href{https://doi.org/10.1007/JHEP03(2018)077}{\emph{JHEP} {\bfseries 03}
  (2018) 077} [\href{https://arxiv.org/abs/1612.05895}{{\ttfamily
  1612.05895}}].

\bibitem{Gainutdinov:2016pxy}
A.M.~Gainutdinov and R.I.~Nepomechie, \emph{{Algebraic Bethe ansatz for the
  quantum group invariant open XXZ chain at roots of unity}},
  \href{https://doi.org/10.1016/j.nuclphysb.2016.06.007}{\emph{Nucl. Phys.}
  {\bfseries B909} (2016) 796}
  [\href{https://arxiv.org/abs/1603.09249}{{\ttfamily 1603.09249}}].

\bibitem{Morin-Duchesne:2011a}
A.~{Morin-Duchesne} and Y.~{Saint-Aubin}, \emph{{The Jordan structure of
  two-dimensional loop models}},
  \href{https://doi.org/10.1088/1742-5468/2011/04/P04007}{\emph{Journal of
  Statistical Mechanics: Theory and Experiment} {\bfseries 2011} (2011) 04007}
  [\href{https://arxiv.org/abs/1101.2885}{{\ttfamily 1101.2885}}].

\bibitem{Kulish:1985bj}
P.P.~Kulish, \emph{{Integrable graded magnets}},
  \href{https://doi.org/10.1007/BF01083770}{\emph{J. Sov. Math.} {\bfseries 35}
  (1986) 2648}.

\bibitem{Bracken:1994hz}
A.J.~Bracken, M.D.~Gould, Y.-Z.~Zhang and G.W.~Delius, \emph{{Solutions of the
  quantum Yang-Baxter equation with extra nonadditive parameters}},
  \href{https://doi.org/10.1088/0305-4470/27/19/025}{\emph{J. Phys.} {\bfseries
  A27} (1994) 6551} [\href{https://arxiv.org/abs/hep-th/9405138}{{\ttfamily
  hep-th/9405138}}].

\bibitem{batchelor2008quantum}
M.~Batchekor, A.~Foerster, X.-W.~Guan, J.~Links and H.-Q.~Zhou, \emph{The
  quantum inverse scattering method with anyonic grading}, {\emph{Journal of
  physics. A, Mathematical and theoretical} {\bfseries 41} (2008) }.

\bibitem{Hubbard_1965RSPSA}
J.~{Hubbard}, \emph{{Electron Correlations in Narrow Energy Bands. IV. The
  Atomic Representation}},
  \href{https://doi.org/10.1098/rspa.1965.0124}{\emph{Proceedings of the Royal
  Society of London Series A} {\bfseries 285} (1965) 542}.

\bibitem{Essler2005_HB}
F.H.~Essler, H.~Frahm, F.~G{\"o}hmann, A.~Kl{\"u}mper and V.E.~Korepin,
  \emph{The one-dimensional Hubbard model}, Cambridge University Press (2005).

\bibitem{Lieb:1968zza}
E.H.~Lieb and F.Y.~Wu, \emph{{Absence of Mott transition in an exact solution
  of the short-range, one-band model in one dimension}},
  \href{https://doi.org/10.1103/PhysRevLett.21.192.2,
  10.1103/PhysRevLett.20.1445}{\emph{Phys. Rev. Lett.} {\bfseries 20} (1968)
  1445}.

\bibitem{Shastry:1988_DSTR_IHM}
B.~Sriram~Shastry, \emph{Decorated star-triangle relations and exact
  integrability of the one-dimensional hubbard model},
  \href{https://doi.org/10.1007/BF01022987}{\emph{Journal of Statistical
  Physics} {\bfseries 50} (1988) 57}.

\bibitem{Dye_2002}
H.A.~Dye, \emph{Unitary solutions to the yang--baxter equation in dimension
  four}, \href{https://doi.org/10.1023/A:1025843426102}{\emph{Quantum
  Information Processing} {\bfseries 2} (2003) 117}.

\bibitem{Makoto:1994JPhy1}
M.~{Idzumi}, T.~{Tokihiro} and M.~{Arai}, \emph{{Solvable nineteen-vertex
  models and quantum spin chains of spin one}},
  \href{https://doi.org/10.1051/jp1:1994245}{\emph{Journal de Physique I}
  {\bfseries 4} (1994) 1151}.

\bibitem{Crampe:2013nha}
N.~Crampé, L.~Frappat and E.~Ragoucy, \emph{{Classification of three-state
  Hamiltonians solvable by the coordinate Bethe ansatz}},
  \href{https://doi.org/10.1088/1751-8113/46/40/405001}{\emph{J. Phys. A}
  {\bfseries 46} (2013) 405001}
  [\href{https://arxiv.org/abs/1306.6303}{{\ttfamily 1306.6303}}].

\bibitem{Fonseca:2014mqa}
T.~Fonseca, L.~Frappat and E.~Ragoucy, \emph{{R-matrices of three-state
  Hamiltonians solvable by Coordinate Bethe Ansatz}},
  \href{https://doi.org/10.1063/1.4905893}{\emph{J. Math. Phys.} {\bfseries 56}
  (2015) 013503} [\href{https://arxiv.org/abs/1406.3197}{{\ttfamily
  1406.3197}}].

\bibitem{Crampe:2016she}
N.~Crampé, L.~Frappat, E.~Ragoucy and M.~Vanicat, \emph{{3-state Hamiltonians
  associated to solvable 33-vertex models}},
  \href{https://doi.org/10.1063/1.4962920}{\emph{J. Math. Phys.} {\bfseries 57}
  (2016) 093504} [\href{https://arxiv.org/abs/1509.07589}{{\ttfamily
  1509.07589}}].

\bibitem{Heilmann:2006}
O.~Heilmann and E.~Lieb, \emph{Violation of the noncrossing rule: The hubbard
  hamiltonian for benzene},
  \href{https://doi.org/10.1111/j.1749-6632.1971.tb34956.x}{\emph{Annals of the
  New York Academy of Sciences} {\bfseries 172} (2006) 584 }.

\bibitem{Essler:1991wg}
F.H.L.~Essler, V.E.~Korepin and K.~Schoutens, \emph{{Completeness of the SO(4)
  extended Bethe ansatz for the one-dimensional Hubbard model}},
  \href{https://doi.org/10.1016/0550-3213(92)90575-V}{\emph{Nucl. Phys.}
  {\bfseries B384} (1992) 431}
  [\href{https://arxiv.org/abs/cond-mat/9209012}{{\ttfamily
  cond-mat/9209012}}].

\bibitem{Frolov:2011wg}
S.~Frolov and E.~Quinn, \emph{{Hubbard-Shastry lattice models}},
  \href{https://doi.org/10.1088/1751-8113/45/9/095004}{\emph{J. Phys.}
  {\bfseries A45} (2012) 095004}
  [\href{https://arxiv.org/abs/1111.5304}{{\ttfamily 1111.5304}}].

\bibitem{deLeeuw:2015ula}
M.~de~Leeuw and V.~Regelskis, \emph{{An algebraic approach to the Hubbard
  model}}, \href{https://doi.org/10.1016/j.physleta.2015.12.013}{\emph{Phys.
  Lett.} {\bfseries A380} (2016) 645}
  [\href{https://arxiv.org/abs/1509.06205}{{\ttfamily 1509.06205}}].

\bibitem{deLeeuw:2019vdb}
M.~De~Leeuw, A.~Pribytok, A.L.~Retore and P.~Ryan, \emph{{New integrable 1D
  models of superconductivity}},
  \href{https://doi.org/10.1088/1751-8121/aba860}{\emph{J. Phys. A} {\bfseries
  53} (2020) 385201} [\href{https://arxiv.org/abs/1911.01439}{{\ttfamily
  1911.01439}}].

\bibitem{Slavnov:2020xxj}
N.~Slavnov, A.~Zabrodin and A.~Zotov, \emph{{Scalar products of Bethe vectors
  in the 8-vertex model}},
  \href{https://doi.org/10.1007/JHEP06(2020)123}{\emph{JHEP} {\bfseries 06}
  (2020) 123} [\href{https://arxiv.org/abs/2005.11224}{{\ttfamily
  2005.11224}}].

\bibitem{BERG1978125}
B.~Berg, M.~Karowski, P.~Weisz and V.~Kurak, \emph{Factorized u(n) symmetric
  s-matrices in two dimensions},
  \href{https://doi.org/https://doi.org/10.1016/0550-3213(78)90489-3}{\emph{Nuclear
  Physics B} {\bfseries 134} (1978) 125 }.

\bibitem{KAROWSKI1979244}
M.~Karowski, \emph{On the bound state problem in 1+1 dimensional field
  theories},
  \href{https://doi.org/https://doi.org/10.1016/0550-3213(79)90600-X}{\emph{Nuclear
  Physics B} {\bfseries 153} (1979) 244 }.

\bibitem{Kulish:1979cr}
P.P.~Kulish and N.Y.~Reshetikhin, \emph{{Generalised Heisenberg Ferromagnet and
  the Gross-Neveu model}}, {\emph{Sov. Phys. JETP} {\bfseries 53} (1981) 108}.

\bibitem{Reshetikhin:1986vd}
N.Y.~Reshetikhin, \emph{{Integrable Models of Quantum One-dimensional Magnets
  With O($N$) and Sp(2k) Symmetry}},
  \href{https://doi.org/10.1007/BF01017501}{\emph{Theor. Math. Phys.}
  {\bfseries 63} (1985) 555}.

\bibitem{Li:2018xrb}
G.-L.~Li, J.~Cao, P.~Xue, Z.-R.~Xin, K.~Hao, W.-L.~Yang et~al., \emph{{Exact
  solution of the $sp(4)$ integrable spin chain with generic boundaries}},
  \href{https://doi.org/10.1007/JHEP05(2019)067}{\emph{JHEP} {\bfseries 05}
  (2019) 067} [\href{https://arxiv.org/abs/1812.03618}{{\ttfamily
  1812.03618}}].

\bibitem{Tu_2008}
H.-H.~Tu, G.-M.~Zhang and T.~Xiang, \emph{Class of exactly solvable so(n)
  symmetric spin chains with matrix product ground states},
  \href{https://doi.org/10.1103/physrevb.78.094404}{\emph{Physical Review B}
  {\bfseries 78} (2008) }.

\bibitem{baxter1972partition}
R.J.~Baxter, \emph{Partition function of the eight-vertex lattice model},
  {\emph{Annals of Physics} {\bfseries 70} (1972) 193}.

\bibitem{jimbo1990yang}
M.~Jimbo, \emph{Yang-Baxter Equation in Integrable Systems}, Advanced series in
  mathematical physics, World Scientific (1990).

\bibitem{Perk2006Yang}
J.~Perk and H.~Au-Yang, \emph{Yang-baxter equations}, {\emph{Encyclopedia of
  Mathematical Physics, Vol. 5, (Elsevier Science, Oxford, 2006), pp. 465-473}
  (2006) }.

\bibitem{Beisert_2004}
N.~Beisert, V.~Dippel and M.~Staudacher, \emph{A novel long range spin chain
  and planar $ n=4 $ super yang-mills},
  \href{https://doi.org/10.1088/1126-6708/2004/07/075}{\emph{Journal of High
  Energy Physics} {\bfseries 2004} (2004) 075}.

\bibitem{Gubser_1998}
S.S.~Gubser, I.R.~Klebanov and A.M.~Polyakov, \emph{Gauge theory correlators
  from non-critical string theory}, {\emph{Physics Letters B} {\bfseries 428}
  (1998) 105}.

\bibitem{Kulish:1981gi}
P.~Kulish, N.~Reshetikhin and E.~Sklyanin, \emph{{Yang-Baxter Equation and
  Representation Theory. 1.}},
  \href{https://doi.org/10.1007/BF02285311}{\emph{Lett. Math. Phys.} {\bfseries
  5} (1981) 393}.

\bibitem{Jimbo1986}
M.~Jimbo, \emph{Quantum r matrix for the generalized toda system},
  \href{https://doi.org/10.1007/BF01221646}{\emph{Communications in
  Mathematical Physics} {\bfseries 102} (1986) 537}.

\bibitem{Drinfeld:1987sy}
V.G.~Drinfeld, \emph{{A New realization of Yangians and quantized affine
  algebras}}, {\emph{Sov. Math. Dokl.} {\bfseries 36} (1988) 212}.

\bibitem{Jimbo:1985vd}
M.~Jimbo, \emph{{A q Analog of u (Gl (n+1)), Hecke Algebra and the Yang-Baxter
  Equation}}, \href{https://doi.org/10.1007/BF00400222}{\emph{Lett. Math.
  Phys.} {\bfseries 11} (1986) 247}.

\bibitem{cheng1991yang}
Y.~Cheng, M.~Ge and K.~Xue, \emph{Yang-baxterization of braid group
  representations}, {\emph{Communications in mathematical physics} {\bfseries
  136} (1991) 195}.

\bibitem{zhang1991representations}
R.~Zhang, M.~Gould and A.~Bracken, \emph{From representations of the braid
  group to solutions of the yang-baxter equation}, {\emph{Nuclear Physics B}
  {\bfseries 354} (1991) 625}.

\bibitem{li1993yang}
Y.-Q.~Li, \emph{Yang baxterization}, {\emph{Journal of mathematical physics}
  {\bfseries 34} (1993) 757}.

\bibitem{Arnaudon:2002tu}
D.~Arnaudon, A.~Chakrabarti, V.~Dobrev and S.~Mihov, \emph{{Spectral
  Decomposition and Baxterisation of Exotic Bialgebras and Associated
  Noncommutative Geometries}},
  \href{https://doi.org/10.1142/S0217751X03016100}{\emph{Int. J. Mod. Phys. A}
  {\bfseries 18} (2003) 4201}
  [\href{https://arxiv.org/abs/math/0209321}{{\ttfamily math/0209321}}].

\bibitem{Isaev:1995cs}
A.~Isaev, \emph{{Quantum groups and Yang-Baxter equations}}, {\emph{Sov. J.
  Part. Nucl.} {\bfseries 26} (1995) 501}.

\bibitem{Kulish:2008a}
P.P.~{Kulish}, N.~{Manojlovic} and Z.~{Nagy}, \emph{{Quantum symmetry algebras
  of spin systems related to Temperley-Lieb R-matrices}},
  \href{https://doi.org/10.1063/1.2873025}{\emph{Journal of Mathematical
  Physics} {\bfseries 49} (2008) 023510}
  [\href{https://arxiv.org/abs/0712.3154}{{\ttfamily 0712.3154}}].

\bibitem{Kulish:2009cx}
P.~Kulish, N.~Manojlovic and Z.~Nagy, \emph{{Symmetries of spin systems and
  Birman-Wenzl-Murakami algebra}},
  \href{https://doi.org/10.1063/1.3366259}{\emph{J. Math. Phys.} {\bfseries 51}
  (2010) 043516} [\href{https://arxiv.org/abs/0910.4036}{{\ttfamily
  0910.4036}}].

\bibitem{Crampe:2016nek}
N.~Crampé, L.~Frappat, E.~Ragoucy and M.~Vanicat, \emph{{A New Braid-like
  Algebra for Baxterisation}},
  \href{https://doi.org/10.1007/s00220-016-2780-y}{\emph{Commun. Math. Phys.}
  {\bfseries 349} (2017) 271}
  [\href{https://arxiv.org/abs/1509.05516}{{\ttfamily 1509.05516}}].

\bibitem{Crampe:2020slf}
N.~Crampé and L.~Poulain~d'Andecy, \emph{{Baxterisation of the fused Hecke
  algebra and R-matrices with gl(N)-symmetry}},
  \href{https://arxiv.org/abs/2004.05035}{{\ttfamily 2004.05035}}.

\bibitem{Turaev:1988eb}
V.~Turaev, \emph{{The Yang-Baxter equation and invariants of links}},
  \href{https://doi.org/10.1007/BF01393746}{\emph{Invent. Math.} {\bfseries 92}
  (1988) 527}.

\bibitem{Jones:1989ed}
V.~Jones, \emph{{On knot invariants related to some statistical mechanical
  models}}, {\emph{Pacific J. Math.} {\bfseries 137} (1989) 311}.

\bibitem{jones1990baxterization}
V.~Jones, \emph{Baxterization}, {\emph{International Journal of Modern Physics
  B} {\bfseries 4} (1990) 701}.

\bibitem{Wu_1993}
F.~Wu, \emph{The yang-baxter equation in knot theory},
  \href{https://doi.org/10.1142/S0217979293003486}{\emph{International Journal
  of Modern Physics B} {\bfseries 07} (1993) 3737}
  [\href{https://arxiv.org/abs/https://doi.org/10.1142/S0217979293003486}{{\ttfamily
  https://doi.org/10.1142/S0217979293003486}}].

\bibitem{Vieira:2019vog}
R.S.~Vieira, \emph{{Fifteen-vertex models with non-symmetric $R$ matrices}},
  \href{https://arxiv.org/abs/1908.06932}{{\ttfamily 1908.06932}}.

\bibitem{Beisert:2008tw}
N.~Beisert and P.~Koroteev, \emph{{Quantum Deformations of the One-Dimensional
  Hubbard Model}},
  \href{https://doi.org/10.1088/1751-8113/41/25/255204}{\emph{J. Phys. A}
  {\bfseries 41} (2008) 255204}
  [\href{https://arxiv.org/abs/0802.0777}{{\ttfamily 0802.0777}}].

\bibitem{Beisert_2012}
N.~Beisert, W.~Galleas and T.~Matsumoto, \emph{A quantum affine algebra for the
  deformed hubbard chain},
  \href{https://doi.org/10.1088/1751-8113/45/36/365206}{\emph{Journal of
  Physics A: Mathematical and Theoretical} {\bfseries 45} (2012) 365206}.

\bibitem{Delduc:2014kha}
F.~Delduc, M.~Magro and B.~Vicedo, \emph{{Derivation of the action and
  symmetries of the $q$-deformed $AdS_{5} \times S^{5}$ superstring}},
  \href{https://doi.org/10.1007/JHEP10(2014)132}{\emph{JHEP} {\bfseries 10}
  (2014) 132} [\href{https://arxiv.org/abs/1406.6286}{{\ttfamily 1406.6286}}].

\bibitem{Sutherland:1970}
B.~Sutherland, \emph{Two-dimensional hydrogen bonded crystals without the ice
  rule}, {\emph{Journal of Mathematical Physics} {\bfseries 11} (1970) 3183}.

\bibitem{Sutherland:1975mcqs}
B.~Sutherland, \emph{Model for a multicomponent quantum system},
  \href{https://doi.org/10.1103/PhysRevB.12.3795}{\emph{Phys. Rev. B}
  {\bfseries 12} (1975) 3795}.

\bibitem{Bethe_1931_theorie}
H.~Bethe, \emph{Zur theorie der metalle}, {\emph{Zeitschrift f{\"u}r Physik}
  {\bfseries 71} (1931) 205}.

\bibitem{Idzumi:1994kx}
M.~Idzumi, T.~Tokihiro and M.~Arai, \emph{{Solvable nineteen vertex models and
  quantum spin chains of spin one}},
  \href{https://doi.org/10.1051/jp1:1994245}{\emph{J. Phys. I(France)}
  {\bfseries 4} (1994) 1151}.

\bibitem{Martins:2013ipa}
M.~Martins, \emph{{Integrable three-state vertex models with weights lying on
  genus five curves}},
  \href{https://doi.org/10.1016/j.nuclphysb.2013.05.014}{\emph{Nucl. Phys. B}
  {\bfseries 874} (2013) 243}
  [\href{https://arxiv.org/abs/1303.4010}{{\ttfamily 1303.4010}}].

\bibitem{Martins:2015ufa}
M.~Martins, \emph{{An integrable nineteen vertex model lying on a
  hypersurface}},
  \href{https://doi.org/10.1016/j.nuclphysb.2015.01.018}{\emph{Nucl. Phys. B}
  {\bfseries 892} (2015) 306}
  [\href{https://arxiv.org/abs/1410.6749}{{\ttfamily 1410.6749}}].

\bibitem{Shastry_1986}
B.S.~Shastry, \emph{Exact integrability of the one-dimensional hubbard model},
  \href{https://doi.org/10.1103/PhysRevLett.56.2453}{\emph{Phys. Rev. Lett.}
  {\bfseries 56} (1986) 2453}.

\bibitem{Beisert:2011wq}
N.~Beisert, W.~Galleas and T.~Matsumoto, \emph{{A Quantum Affine Algebra for
  the Deformed Hubbard Chain}},
  \href{https://doi.org/10.1088/1751-8113/45/36/365206}{\emph{J. Phys. A}
  {\bfseries 45} (2012) 365206}
  [\href{https://arxiv.org/abs/1102.5700}{{\ttfamily 1102.5700}}].

\bibitem{Beisert:2015msa}
N.~Beisert, M.~de~Leeuw and P.~Nag, \emph{{Fusion for the one-dimensional
  Hubbard model}},
  \href{https://doi.org/10.1088/1751-8113/48/32/324002}{\emph{J. Phys. A}
  {\bfseries 48} (2015) 324002}
  [\href{https://arxiv.org/abs/1503.04838}{{\ttfamily 1503.04838}}].

\bibitem{deLeeuw:2020ahe}
M.~de~Leeuw, C.~Paletta, A.~Pribytok, A.L.~Retore and P.~Ryan,
  \emph{{Classifying nearest-neighbour interactions and deformations of AdS}},
  \href{https://doi.org/10.1103/PhysRevLett.125.031604}{\emph{Phys. Rev. Lett.}
  {\bfseries 125} (2020) 031604}
  [\href{https://arxiv.org/abs/2003.04332}{{\ttfamily 2003.04332}}].

\bibitem{Stouten:2017azy}
E.~Stouten, P.W.~Claeys, J.-S.~Caux and V.~Gritsev, \emph{{Integrability and
  duality in spin chains}}, \href{https://doi.org/10.1103/PhysRevB.99.075111,
  10.1103/PhysRevB.99.169902}{\emph{Phys. Rev.} {\bfseries B99} (2019) 075111}
  [\href{https://arxiv.org/abs/1712.09375}{{\ttfamily 1712.09375}}].

\bibitem{Stouten:2018lkr}
E.~Stouten, P.W.~Claeys, M.~Zvonarev, J.-S.~Caux and V.~Gritsev,
  \emph{{Something interacting and solvable in 1D}},
  \href{https://doi.org/10.1088/1751-8121/aae8bb}{\emph{J. Phys.} {\bfseries
  A51} (2018) 485204} [\href{https://arxiv.org/abs/1804.10935}{{\ttfamily
  1804.10935}}].

\bibitem{drinfeld1983constant}
V.G.~Drinfeld, \emph{Constant quasiclassical solutions of the yang--baxter
  quantum equation},  in \emph{Doklady Akademii Nauk}, vol.~273, pp.~531--535,
  Russian Academy of Sciences, 1983.

\bibitem{deLeeuw:2021ufg}
M.~de~Leeuw, A.~Pribytok, A.L.~Retore and P.~Ryan, \emph{{Integrable
  deformations of AdS/CFT}},
  \href{https://arxiv.org/abs/2109.00017}{{\ttfamily 2109.00017}}.

\bibitem{Sun:1995}
X.-d.~Sun, S.-k.~Wang and K.~Wu, \emph{Classification of six-vertex-type
  solutions of the colored yang--baxter equation}, {\emph{Journal of
  Mathematical Physics} {\bfseries 36} (1995) 6043}.

\bibitem{Izergin:1981}
A.G.~Izergin and V.E.~Korepin, \emph{{The Inverse Scattering Method Approach to
  the Quantum Shabat-Mikhailov Model}},
  \href{https://doi.org/10.1007/BF01208496}{\emph{Commun. Math. Phys.}
  {\bfseries 79} (1981) 303}.

\bibitem{Khachatryan:2014cfj}
S.~Khachatryan, \emph{{On the solutions to the multi-parametric Yang-Baxter
  equations}},
  \href{https://doi.org/10.1016/j.nuclphysb.2014.04.008}{\emph{Nucl. Phys. B}
  {\bfseries 883} (2014) 629}
  [\href{https://arxiv.org/abs/1311.4994}{{\ttfamily 1311.4994}}].

\bibitem{deLeeuw:2020xrw}
M.~de~Leeuw, C.~Paletta, A.~Pribytok, A.L.~Retore and P.~Ryan,
  \emph{{Yang-Baxter and the Boost: splitting the difference}},
  \href{https://arxiv.org/abs/2010.11231}{{\ttfamily 2010.11231}}.

\bibitem{Klebanov:2004ya}
I.R.~Klebanov and J.M.~Maldacena, \emph{{Superconformal gauge theories and
  non-critical superstrings}},
  \href{https://doi.org/10.1142/S0217751X04020865}{\emph{Int. J. Mod. Phys. A}
  {\bfseries 19} (2004) 5003}
  [\href{https://arxiv.org/abs/hep-th/0409133}{{\ttfamily hep-th/0409133}}].

\bibitem{Sax_2011}
O.O.~Sax and B.~Stefa{\'{n}}ski, \emph{Integrability, spin-chains and the
  {AdS}3/{CFT}2 correspondence},
  \href{https://doi.org/10.1007/jhep08(2011)029}{\emph{Journal of High Energy
  Physics} {\bfseries 2011} (2011) }.

\bibitem{Zamolodchikov:1992zr}
A.B.~Zamolodchikov and A.B.~Zamolodchikov, \emph{{Massless factorized
  scattering and sigma models with topological terms}},
  \href{https://doi.org/10.1016/0550-3213(92)90136-Y}{\emph{Nucl. Phys. B}
  {\bfseries 379} (1992) 602}.

\bibitem{Fendley:1993jh}
P.~Fendley and H.~Saleur, \emph{{Massless integrable quantum field theories and
  massless scattering in (1+1)-dimensions}},  in \emph{{Summer School in
  High-energy Physics and Cosmology (Includes Workshop on Strings, Gravity, and
  Related Topics 29-30 Jul 1993)}}, pp.~301--332, 9, 1993
  [\href{https://arxiv.org/abs/hep-th/9310058}{{\ttfamily hep-th/9310058}}].

\bibitem{Bombardelli:2018jkj}
D.~Bombardelli, B.~Stefanski and A.~Torrielli, \emph{{The low-energy limit of
  AdS$_{3}$/CFT$_{2}$ and its TBA}},
  \href{https://doi.org/10.1007/JHEP10(2018)177}{\emph{JHEP} {\bfseries 10}
  (2018) 177} [\href{https://arxiv.org/abs/1807.07775}{{\ttfamily
  1807.07775}}].

\bibitem{Gomez:2007zr}
C.~Gomez and R.~Hernandez, \emph{{Quantum deformed magnon kinematics}},
  \href{https://doi.org/10.1088/1126-6708/2007/03/108}{\emph{JHEP} {\bfseries
  03} (2007) 108} [\href{https://arxiv.org/abs/hep-th/0701200}{{\ttfamily
  hep-th/0701200}}].

\bibitem{Young:2007wd}
C.A.S.~Young, \emph{{q-deformed supersymmetry and dynamic magnon
  representations}},
  \href{https://doi.org/10.1088/1751-8113/40/30/033}{\emph{J. Phys. A}
  {\bfseries 40} (2007) 9165}
  [\href{https://arxiv.org/abs/0704.2069}{{\ttfamily 0704.2069}}].

\bibitem{Stromwall:2016dyw}
J.~Stromwall and A.~Torrielli, \emph{{AdS$_{3}$/CFT$_{2}$ and q-Poincar\'e
  superalgebras}},
  \href{https://doi.org/10.1088/1751-8113/49/43/435402}{\emph{J. Phys. A}
  {\bfseries 49} (2016) 435402}
  [\href{https://arxiv.org/abs/1606.02217}{{\ttfamily 1606.02217}}].

\bibitem{Borsato:2017icj}
R.~Borsato, J.~Str\"omwall and A.~Torrielli, \emph{{$q$-Poincar\'e invariance
  of the AdS$_3$/CFT$_2$ $R$-matrix}},
  \href{https://doi.org/10.1103/PhysRevD.97.066001}{\emph{Phys. Rev. D}
  {\bfseries 97} (2018) 066001}
  [\href{https://arxiv.org/abs/1711.02446}{{\ttfamily 1711.02446}}].

\bibitem{Fontanella:2016opq}
A.~Fontanella and A.~Torrielli, \emph{{Massless sector of AdS$_3$ superstrings:
  A geometric interpretation}},
  \href{https://doi.org/10.1103/PhysRevD.94.066008}{\emph{Phys. Rev. D}
  {\bfseries 94} (2016) 066008}
  [\href{https://arxiv.org/abs/1608.01631}{{\ttfamily 1608.01631}}].

\bibitem{OhlssonSax:2018hgc}
O.~Ohlsson~Sax and B.~Stefa\'nski, \emph{{Closed strings and moduli in
  AdS$_{3}$/CFT$_{2}$}},
  \href{https://doi.org/10.1007/JHEP05(2018)101}{\emph{JHEP} {\bfseries 05}
  (2018) 101} [\href{https://arxiv.org/abs/1804.02023}{{\ttfamily
  1804.02023}}].

\bibitem{Fontanella:2019ury}
A.~Fontanella, O.~Ohlsson~Sax, B.~Stefa\'nski and A.~Torrielli, \emph{{The
  effectiveness of relativistic invariance in AdS$_{3}$}},
  \href{https://doi.org/10.1007/JHEP07(2019)105}{\emph{JHEP} {\bfseries 07}
  (2019) 105} [\href{https://arxiv.org/abs/1905.00757}{{\ttfamily
  1905.00757}}].

\bibitem{Maldacena:1998uz}
J.M.~Maldacena, J.~Michelson and A.~Strominger, \emph{{Anti-de Sitter
  fragmentation}},
  \href{https://doi.org/10.1088/1126-6708/1999/02/011}{\emph{JHEP} {\bfseries
  02} (1999) 011} [\href{https://arxiv.org/abs/hep-th/9812073}{{\ttfamily
  hep-th/9812073}}].

\bibitem{Strominger:1998yg}
A.~Strominger, \emph{{AdS(2) quantum gravity and string theory}},
  \href{https://doi.org/10.1088/1126-6708/1999/01/007}{\emph{JHEP} {\bfseries
  01} (1999) 007} [\href{https://arxiv.org/abs/hep-th/9809027}{{\ttfamily
  hep-th/9809027}}].

\bibitem{Chamon:2011xk}
C.~Chamon, R.~Jackiw, S.-Y.~Pi and L.~Santos, \emph{{Conformal quantum
  mechanics as the CFT$_1$ dual to AdS$_2$}},
  \href{https://doi.org/10.1016/j.physletb.2011.06.023}{\emph{Phys. Lett. B}
  {\bfseries 701} (2011) 503}
  [\href{https://arxiv.org/abs/1106.0726}{{\ttfamily 1106.0726}}].

\bibitem{Fendley:1990cy}
P.~Fendley, \emph{{A Second supersymmetric S matrix for the perturbed
  tricritical Ising model}},
  \href{https://doi.org/10.1016/0370-2693(90)91160-D}{\emph{Phys. Lett. B}
  {\bfseries 250} (1990) 96}.

\bibitem{Korepanov_1993tetrahedral}
I.~Korepanov, \emph{Tetrahedral zamolodchikov algebras corresponding to
  baxter's l-operators}, {\emph{Communications in mathematical physics}
  {\bfseries 154} (1993) 85}.

\bibitem{Korepanov:1994rc}
I.G.~Korepanov, \emph{{Vacuum curves, classical integrable systems in discrete
  space-time and statistical physics}}, {\emph{Zap. Nauchn. Semin.} {\bfseries
  235} (1996) 272} [\href{https://arxiv.org/abs/hep-th/9312197}{{\ttfamily
  hep-th/9312197}}].

\bibitem{Umeno_1998fermionic}
Y.~Umeno, M.~Shiroishi and M.~Wadati, \emph{Fermionic r-operator and
  integrability of the one-dimensional hubbard model}, {\emph{Journal of the
  Physical Society of Japan} {\bfseries 67} (1998) 2242}.

\bibitem{Mitev:2012vt}
V.~Mitev, M.~Staudacher and Z.~Tsuboi, \emph{{The Tetrahedral Zamolodchikov
  Algebra and the ${AdS_5\times S^5}$ S-matrix}},
  \href{https://doi.org/10.1007/s00220-017-2905-y}{\emph{Commun. Math. Phys.}
  {\bfseries 354} (2017) 1} [\href{https://arxiv.org/abs/1210.2172}{{\ttfamily
  1210.2172}}].

\bibitem{deLeeuw:2020bgo}
M.~de~Leeuw, C.~Paletta, A.~Pribytok, A.L.~Retore and A.~Torrielli, \emph{{Free
  Fermions, vertex Hamiltonians, and lower-dimensional AdS/CFT}}, {\emph{JHEP}
  {\bfseries 02} (2021) 191}
  [\href{https://arxiv.org/abs/2011.08217}{{\ttfamily 2011.08217}}].

\bibitem{sax2013massless}
O.O.~Sax, B.S.~jr and A.~Torrielli, \emph{On the massless modes of the
  ads3/cft2 integrable systems},  2013.

\bibitem{Fendley:1993pi}
P.~Fendley and K.A.~Intriligator, \emph{{Exact N=2 Landau-Ginzburg flows}},
  \href{https://doi.org/10.1016/0550-3213(94)90006-X}{\emph{Nucl. Phys. B}
  {\bfseries 413} (1994) 653}
  [\href{https://arxiv.org/abs/hep-th/9307166}{{\ttfamily hep-th/9307166}}].

\bibitem{Bombardelli:2016scq}
D.~Bombardelli, \emph{{S-matrices and integrability}},
  \href{https://doi.org/10.1088/1751-8113/49/32/323003}{\emph{J. Phys. A}
  {\bfseries 49} (2016) 323003}
  [\href{https://arxiv.org/abs/1606.02949}{{\ttfamily 1606.02949}}].

\bibitem{Jimbo:1989}
M.~Jimbo, \emph{Introduction to the yang-baxter equation},
  \href{https://doi.org/10.1142/S0217751X89001503}{\emph{Int. Journ. of Modern
  Physics A} {\bfseries 04} (1989) 3759}
  [\href{https://arxiv.org/abs/https://doi.org/10.1142/S0217751X89001503}{{\ttfamily
  https://doi.org/10.1142/S0217751X89001503}}].

\bibitem{Fateev:1996ea}
V.A.~Fateev, \emph{{The sigma model (dual) representation for a two-parameter
  family of integrable quantum field theories}},
  \href{https://doi.org/10.1016/0550-3213(96)00256-8}{\emph{Nucl. Phys. B}
  {\bfseries 473} (1996) 509}.

\bibitem{Fateev:2018yos}
V.A.~Fateev and A.V.~Litvinov, \emph{{Integrability, Duality and Sigma
  Models}}, \href{https://doi.org/10.1007/JHEP11(2018)204}{\emph{JHEP}
  {\bfseries 11} (2018) 204}
  [\href{https://arxiv.org/abs/1804.03399}{{\ttfamily 1804.03399}}].

\bibitem{Torrielli:2021hnd}
A.~Torrielli, \emph{{A study of integrable form factors in massless
  relativistic AdS\_3}},  \href{https://arxiv.org/abs/2106.06874}{{\ttfamily
  2106.06874}}.

\bibitem{Hoare:2014oua}
B.~Hoare, \emph{{Towards a two-parameter q-deformation of AdS$_3 \times S^3
  \times M^4$ superstrings}},
  \href{https://doi.org/10.1016/j.nuclphysb.2014.12.012}{\emph{Nucl. Phys. B}
  {\bfseries 891} (2015) 259}
  [\href{https://arxiv.org/abs/1411.1266}{{\ttfamily 1411.1266}}].

\bibitem{Fontanella:2019baq}
A.~Fontanella and A.~Torrielli, \emph{{Geometry of Massless Scattering in
  Integrable Superstring}},
  \href{https://doi.org/10.1007/JHEP06(2019)116}{\emph{JHEP} {\bfseries 06}
  (2019) 116} [\href{https://arxiv.org/abs/1903.10759}{{\ttfamily
  1903.10759}}].

\bibitem{Zarembo:2010sg}
K.~Zarembo, \emph{{Strings on Semisymmetric Superspaces}},
  \href{https://doi.org/10.1007/JHEP05(2010)002}{\emph{JHEP} {\bfseries 05}
  (2010) 002} [\href{https://arxiv.org/abs/1003.0465}{{\ttfamily 1003.0465}}].

\bibitem{Delduc:2018xug}
F.~Delduc, B.~Hoare, T.~Kameyama, S.~Lacroix and M.~Magro,
  \emph{{Three-parameter integrable deformation of $\mathbb{Z}_4$ permutation
  supercosets}}, \href{https://doi.org/10.1007/JHEP01(2019)109}{\emph{JHEP}
  {\bfseries 01} (2019) 109}
  [\href{https://arxiv.org/abs/1811.00453}{{\ttfamily 1811.00453}}].

\bibitem{Lukyanov:2012zt}
S.L.~Lukyanov, \emph{{The integrable harmonic map problem versus Ricci flow}},
  \href{https://doi.org/10.1016/j.nuclphysb.2012.08.002}{\emph{Nucl. Phys. B}
  {\bfseries 865} (2012) 308}
  [\href{https://arxiv.org/abs/1205.3201}{{\ttfamily 1205.3201}}].

\bibitem{Klimcik:2008eq}
C.~Klimcik, \emph{{On integrability of the Yang-Baxter $ \sigma $-model}},
  \href{https://doi.org/10.1063/1.3116242}{\emph{J. Math. Phys.} {\bfseries 50}
  (2009) 043508} [\href{https://arxiv.org/abs/0802.3518}{{\ttfamily
  0802.3518}}].

\bibitem{Klimcik_2014}
C.~Klimčík, \emph{Integrability of the bi-yang–baxter $ \sigma $-model},
  \href{https://doi.org/10.1007/s11005-014-0709-y}{\emph{Letters in
  Mathematical Physics} {\bfseries 104} (2014) 1095–1106}.

\bibitem{Borsato:2017qsx}
R.~Borsato and L.~Wulff, \emph{{On non-abelian T-duality and deformations of
  supercoset string sigma-models}},
  \href{https://doi.org/10.1007/JHEP10(2017)024}{\emph{JHEP} {\bfseries 10}
  (2017) 024} [\href{https://arxiv.org/abs/1706.10169}{{\ttfamily
  1706.10169}}].

\bibitem{Hoare:2016wsk}
B.~Hoare and A.A.~Tseytlin.

\bibitem{Arutyunov:2015mqj}
G.~Arutyunov, S.~Frolov, B.~Hoare, R.~Roiban and A.A.~Tseytlin, \emph{{Scale
  invariance of the $\eta$-deformed $AdS_5\times S^5$ superstring, T-duality
  and modified type II equations}},
  \href{https://doi.org/10.1016/j.nuclphysb.2015.12.012}{\emph{Nucl. Phys. B}
  {\bfseries 903} (2016) 262}
  [\href{https://arxiv.org/abs/1511.05795}{{\ttfamily 1511.05795}}].

\bibitem{Sfetsos:2013wia}
K.~Sfetsos, \emph{{Integrable interpolations: From exact CFTs to non-Abelian
  T-duals}}, \href{https://doi.org/10.1016/j.nuclphysb.2014.01.004}{\emph{Nucl.
  Phys. B} {\bfseries 880} (2014) 225}
  [\href{https://arxiv.org/abs/1312.4560}{{\ttfamily 1312.4560}}].

\bibitem{Sfetsos:2014cea}
K.~Sfetsos and D.C.~Thompson, \emph{{Spacetimes for $\lambda$-deformations}},
  \href{https://doi.org/10.1007/JHEP12(2014)164}{\emph{JHEP} {\bfseries 12}
  (2014) 164} [\href{https://arxiv.org/abs/1410.1886}{{\ttfamily 1410.1886}}].

\bibitem{Delduc:2013qra}
F.~Delduc, M.~Magro and B.~Vicedo, \emph{{An integrable deformation of the
  $AdS_5 \times S^5$ superstring action}},
  \href{https://doi.org/10.1103/PhysRevLett.112.051601}{\emph{Phys. Rev. Lett.}
  {\bfseries 112} (2014) 051601}
  [\href{https://arxiv.org/abs/1309.5850}{{\ttfamily 1309.5850}}].

\bibitem{Hollowood:2014qma}
T.J.~Hollowood, J.L.~Miramontes and D.M.~Schmidtt, \emph{{An Integrable
  Deformation of the $AdS_5 \times S^5$ Superstring}},
  \href{https://doi.org/10.1088/1751-8113/47/49/495402}{\emph{J. Phys. A}
  {\bfseries 47} (2014) 495402}
  [\href{https://arxiv.org/abs/1409.1538}{{\ttfamily 1409.1538}}].

\bibitem{Delduc_2013}
F.~Delduc, M.~Magro and B.~Vicedo, \emph{On classical q-deformations of
  integrable $ \sigma $-models},
  \href{https://doi.org/10.1007/jhep11(2013)192}{\emph{Journal of High Energy
  Physics} {\bfseries 2013} (2013) }.

\bibitem{Seibold:2019dvf}
F.K.~Seibold, \emph{{Two-parameter integrable deformations of the $AdS_3 \times
  S^3 \times T^4$ superstring}},
  \href{https://doi.org/10.1007/JHEP10(2019)049}{\emph{JHEP} {\bfseries 10}
  (2019) 049} [\href{https://arxiv.org/abs/1907.05430}{{\ttfamily
  1907.05430}}].

\bibitem{Bocconcello:2020qkt}
M.~Bocconcello, I.~Masuda, F.K.~Seibold and A.~Sfondrini, \emph{{S matrix for a
  three-parameter integrable deformation of AdS$_{3}\times$ S$^{3}$ strings}},
  \href{https://doi.org/10.1007/JHEP11(2020)022}{\emph{JHEP} {\bfseries 11}
  (2020) 022} [\href{https://arxiv.org/abs/2008.07603}{{\ttfamily
  2008.07603}}].

\bibitem{Garcia:2021iox}
J.M.N.~Garc\'\i{}a and L.~Wyss, \emph{{Three-parameter deformation of
  \ensuremath{\mathbb{R}} $ $ S$^{3}$ in the Landau-Lifshitz limit}},
  \href{https://doi.org/10.1007/JHEP07(2021)028}{\emph{JHEP} {\bfseries 07}
  (2021) 028} [\href{https://arxiv.org/abs/2102.06419}{{\ttfamily
  2102.06419}}].

\bibitem{Seibold:2021lju}
F.K.~Seibold, S.J.~van Tongeren and Y.~Zimmermann, \emph{{On quantum
  deformations of $AdS_3 \times S^3 \times T^4$ and mirror duality}},
  \href{https://arxiv.org/abs/2107.02564}{{\ttfamily 2107.02564}}.

\bibitem{Beisert:2014hya}
N.~Beisert and M.~de~Leeuw, \emph{{The RTT realization for the deformed
  $\mathfrak {gl}(2\vert2)$ Yangian}},
  \href{https://doi.org/10.1088/1751-8113/47/30/305201}{\emph{J. Phys. A}
  {\bfseries 47} (2014) 305201}
  [\href{https://arxiv.org/abs/1401.7691}{{\ttfamily 1401.7691}}].

\bibitem{Stolin1997}
A.~Stolin and P.P.~Kulish, \emph{New rational solutions of yang-baxter equation
  and deformed yangians},
  \href{https://doi.org/10.1023/A:1021460515598}{\emph{Czechoslovak Journal of
  Physics} {\bfseries 47} (1997) 123}.

\bibitem{Ipsen:2018fmu}
A.C.~Ipsen, M.~Staudacher and L.~Zippelius, \emph{{The one-loop spectral
  problem of strongly twisted $ \mathcal{N} $ = 4 Super Yang-Mills theory}},
  \href{https://doi.org/10.1007/JHEP04(2019)044}{\emph{JHEP} {\bfseries 04}
  (2019) 044} [\href{https://arxiv.org/abs/1812.08794}{{\ttfamily
  1812.08794}}].

\bibitem{Drummond:2007gt}
J.M.~Drummond, G.~Feverati, L.~Frappat and E.~Ragoucy, \emph{{Super-Hubbard
  models and applications}},
  \href{https://doi.org/10.1088/1126-6708/2007/05/008}{\emph{JHEP} {\bfseries
  05} (2007) 008} [\href{https://arxiv.org/abs/hep-th/0703078}{{\ttfamily
  hep-th/0703078}}].

\bibitem{Drummond:2007sa}
J.~Drummond, G.~Feverati, L.~Frappat and E.~Ragoucy, \emph{{Generalised
  integrable Hubbard models}},  in \emph{{International Workshop on Recent
  Advances in Quantum Integrable Systems (RAQIS 07) Annecy-le-Vieux, France,
  September 11-14, 2007}}, 2007
  [\href{https://arxiv.org/abs/0712.1940}{{\ttfamily 0712.1940}}].

\bibitem{Maassarani_1998_su_n_HM}
Z.~{Maassarani}, \emph{{The su(n) Hubbard model}},
  \href{https://doi.org/10.1016/S0375-9601(97)00977-8}{\emph{Physics Letters A}
  {\bfseries 239} (1998) 187}
  [\href{https://arxiv.org/abs/cond-mat/9709252}{{\ttfamily
  cond-mat/9709252}}].

\bibitem{Maassarani_1998_su_n_XX}
Z.~{Maassarani} and P.~{Mathieu}, \emph{{The su(N) XX model}},
  \href{https://doi.org/10.1016/S0550-3213(98)80004-7}{\emph{Nuclear Physics B}
  {\bfseries 517} (1998) 395}
  [\href{https://arxiv.org/abs/cond-mat/9709163}{{\ttfamily
  cond-mat/9709163}}].

\bibitem{Sklyanin:1987bi}
E.K.~Sklyanin, \emph{{Boundary conditions for integrable equations}},
  \href{https://doi.org/10.1007/BF01078038}{\emph{Funct. Anal. Appl.}
  {\bfseries 21} (1987) 164}.

\bibitem{Sklyanin:1988yz}
E.~Sklyanin, \emph{{Boundary Conditions for Integrable Quantum Systems}},
  \href{https://doi.org/10.1088/0305-4470/21/10/015}{\emph{J. Phys. A}
  {\bfseries 21} (1988) 2375}.

\bibitem{Alfimov_2020_OSPSigma}
M.~Alfimov, B.~Feigin, B.~Hoare and A.~Litvinov, \emph{{Dual description of
  $\eta$-deformed OSP sigma models}},
  \href{https://doi.org/10.1007/JHEP12(2020)040}{\emph{JHEP} {\bfseries 12}
  (2020) 040} [\href{https://arxiv.org/abs/2010.11927}{{\ttfamily
  2010.11927}}].

\bibitem{Affleck:2021jls}
I.~Affleck, D.~Bykov and K.~Wamer, \emph{{Flag manifold sigma models}: {Spin
  chains and integrable theories}},
  \href{https://doi.org/10.1016/j.physrep.2021.09.004}{\emph{Phys. Rept.}
  {\bfseries 953} (2022) 1} [\href{https://arxiv.org/abs/2101.11638}{{\ttfamily
  2101.11638}}].

\bibitem{Bykov:2021dbk}
D.~Bykov, \emph{{Sigma models as Gross\textendash{}Neveu models}},
  \href{https://doi.org/10.1134/S0040577921080018}{\emph{Teor. Mat. Fiz.}
  {\bfseries 208} (2021) 165}
  [\href{https://arxiv.org/abs/2106.15598}{{\ttfamily 2106.15598}}].

\bibitem{Bykov:2022aka}
D.~Bykov, \emph{{Integrable sigma models on Riemann surfaces}},
  \href{https://arxiv.org/abs/2202.12805}{{\ttfamily 2202.12805}}.

\bibitem{Gohmann_2000}
F.~Göhmann and V.E.~Korepin, \emph{Solution of the quantum inverse problem},
  \href{https://doi.org/10.1088/0305-4470/33/6/308}{\emph{Journal of Physics A:
  Mathematical and General} {\bfseries 33} (2000) 1199}.

\bibitem{Reshetikhin:1990ep}
N.~Reshetikhin, \emph{{Multiparameter quantum groups and twisted
  quasitriangular Hopf algebras}},
  \href{https://doi.org/10.1007/BF00626530}{\emph{Lett. Math. Phys.} {\bfseries
  20} (1990) 331}.

\bibitem{Artamonov_2015_SemisimpleHA}
V.A.~Artamonov, \emph{Semi-simple hopf algebras with restrictions on
  irreducible modules of dimension exceeding 1}, {\emph{St Petersburg
  Mathematical Journal} {\bfseries 26} (2015) 207}.

\bibitem{Bargheer:2008jt}
T.~Bargheer, N.~Beisert and F.~Loebbert, \emph{{Boosting Nearest-Neighbour to
  Long-Range Integrable Spin Chains}},
  \href{https://doi.org/10.1088/1742-5468/2008/11/L11001}{\emph{J. Stat. Mech.}
  {\bfseries 0811} (2008) L11001}
  [\href{https://arxiv.org/abs/0807.5081}{{\ttfamily 0807.5081}}].

\bibitem{Bargheer:2009xy}
T.~Bargheer, N.~Beisert and F.~Loebbert, \emph{{Long-Range Deformations for
  Integrable Spin Chains}},
  \href{https://doi.org/10.1088/1751-8113/42/28/285205}{\emph{J. Phys.}
  {\bfseries A42} (2009) 285205}
  [\href{https://arxiv.org/abs/0902.0956}{{\ttfamily 0902.0956}}].

\bibitem{Abramowitz_1988_Handbook}
M.~Abramowitz, I.A.~Stegun and R.H.~Romer, \emph{Handbook of mathematical
  functions with formulas, graphs, and mathematical tables},  1988.

\end{thebibliography}\endgroup

%
%
%

%
%
%

\end{document}